\documentclass[twoside,12pt]{article}
\usepackage[dvips]{graphicx} % personal option by myo
\usepackage{amsmath}

\newcommand{\nc}{\newcommand}           % new command
\newcommand{\renc}{\renewcommand}       % re-new command
\newcommand{\braket}[1]{ \langle #1 \rangle}
\newcommand{\nonum}{\nonumber \\}
\newcommand{\bv}[1]{\mbox{\boldmath $#1$}}

\nc{\vc}[1]     {\mbox{\boldmath $#1$}} % boldmath(vector)
\nc{\bras}[1]   {\langle#1|}            % <#1|
\nc{\kets}[1]   {|#1\rangle}            % |#1>
\nc{\bra}       {\langle}            % <
\nc{\ket}       {\rangle}            % >
\nc{\HO}        {\hat{O}}   % Operator with wide hat
\begin{document}

\title{Recent development of complex scaling method\\
for many-body resonances and continua in light nuclei}

\author{Takayuki\ Myo,$^{1,2}$ Yuma\ Kikuchi,$^3$ Hiroshi\ Masui,$^4$ Kiyoshi\ Kat\=o$^5$\\
$^1$General Education, Faculty of Engineering, Osaka Institute of Technology,\\ Osaka 535-8585, Japan\\
$^2$Research Center for Nuclear Physics (RCNP), Osaka University,\\ Ibaraki 567-0047, Japan\\
$^3$Nishina Center for Accelerator-based Science, The Institute of \\ Physical and Chemical Research (RIKEN), Wako 351-0198, Japan\\
$^4$Information Processing Center, Kitami Institute of Technology,\\ Kitami 090-8507, Japan\\
$^5$Nuclear Reaction Data Centre, Faculty of Science, Hokkaido University,\\ Sapporo 060-0810, Japan
}

\maketitle

\begin{abstract} 
The complex scaling method (CSM) is a useful similarity transformation of the Schr\"odinger equation, 
in which bound-state spectra are not changed but continuum spectra are separated into resonant and non-resonant 
continuum ones. Because the asymptotic wave functions of the separated resonant states are regularized by the CSM, 
many-body resonances can be obtained by solving an eigenvalue problem with the $L^2$ basis functions. 
Applying this method to a system consisting of a core and valence nucleons, we investigate many-body resonant states in weakly bound 
nuclei very far from the stability lines. Non-resonant continuum states are also obtained with the discretized 
eigenvalues on the rotated branch cuts. Using these complex eigenvalues and eigenstates in CSM, we construct 
the extended completeness relations and Green's functions to calculate strength functions and breakup 
cross sections. Various kinds of theoretical calculations and comparisons with experimental data are presented. 
\end{abstract}

%\eject
\tableofcontents

\section{Introduction}

%% background with unstable nuclei
~~~~~
In the atomic nucleus, the properties of the unbound states are fundamental to the nuclear structures and reactions.
The recent experimental developments in the field of unstable nuclear physics, starting from the discovery of neutron halo structure 
in the neutron-rich nuclei such as $^6$He and $^{11}$Li, 
have shown the various interesting phenomena related to the unbound states of nuclei \cite{tanihata85,tanihata13}.
In unstable nuclei, a few extra nucleons are bound to the system with small binding energies.
This fact indicates that unstable nuclei can easily emit one or two nucleons with small excitation energies around a few MeV.
This, in turn, implies that the position of the lowest threshold is very close to the ground state and that the coupling effect of the continuum states becomes important even in the ground state.
This property of unstable nuclei is quite different from that of stable nuclei, in which the average binding energy is about 8 MeV per nucleon \cite{tanihata96}.
One of the interesting features of the unstable nuclei is their so-called Borromean nature, in which no two-body subsystem has the bound states.
With this feature, the constituents of the system can have a bound state in the three-body case and the lowest threshold is of a three-body emission, not of a two-body one.
This condition requires both experimental and  theoretical studies of the unbound states in the subsystem.
The physics of unstable nuclei is extended to the understanding of the scattering properties of the nuclei.
There are many experiments to investigate the scattering states of unstable nuclei, such as the observation of new resonances in the spectroscopy, 
the various responses to an external Coulomb field, and the breakup reactions of an unstable nucleus as a projectile.
In theory, the unified description of structures and reactions is essential to the unstable nuclear physics.
The resonances embedded in the scattering states provide important information on the structures of the compound system 
in addition to the scattering observables such as cross sections. 

%% CSM oveview
Nuclear resonances are described by applying the $R$-matrix theory \cite{lane1958}, which was developed by Kapur and Peierls \cite{kapur1938}, 
and characterized using the resonance energy and width \cite{wigner47}. 
They are often expressed by a complex energy and theoretically calculated as a pole of the $S$-matrix. 
However, it is difficult for such conventional methods to treat many-body resonances and non-resonant continuum states. 
Here, we refer to the states decaying into more than two-body constituents as ``many-body resonances." 
A significant development in the treatment of resonances from two-body systems to many-body systems has been brought about via the complex scaling method (CSM) \cite{ho83,moiseyev98,moiseyev11}.
It is expected that the CSM plays an important role not only in investigations of the problems of many-body resonances but also 
in the description of the non-resonant many-body continuum states. These states are indispensable to produce the nuclear reaction phenomena.
The aim of this article is to provide a brief review of the CSM in nuclear physics and its applications to many-body resonant and continuum states of light unstable nuclei.

%% CSM history1, ABC
Originally, the CSM was proposed by Aguilar, Combes, and Balslev in 1971 \cite{aguilar71,balslev71}. 
Simon advocated this method as a direct approach of obtaining many-body resonances \cite{simon72}. 
The use of ``direct" implies that the resonance wave functions are directly obtained with complex energy eigenvalues of the quantum many-body system 
by solving an eigenvalue problem of the complex-scaled Schr\"odinger equation, $H^\theta\Psi^\theta=E\Psi^\theta$ with a real scaling angle $\theta$.
In the CSM, the boundary condition of the outgoing wave \cite{siegert39} is implemented automatically for resonances. 
The energy eigenvalues of the bound and resonant states are shown to correspond to the poles of the $S$-matrix: 
For a bound state, the energy $E_B$ is obtained using a real and negative eigenvalue and invariant with respect to the scaling transformation.
For a resonance, the resonance energy $E_r$ and the decay width $\Gamma$ are obtained using the complex energy eigenvalues $E=E_r-i\Gamma/2$. 
In the CSM, the resonances are described by square-integrable ($L^2$) wave functions, the norms of which are definable, 
similar to the bound states.
This property of the resonance wave functions enables the application of the CSM to various theoretical approaches describing the nucleus.

%% CSM history2, atomic, molecular, and nuclear physics
The CSM has thus far been extensively applied to atomic and molecular physics and has been reviewed by Ho \cite{ho83} and Moiseyev \cite{moiseyev11} in the framework of the non-Hermitian quantum mechanics.
Furthermore, the CSM has been developed for obtaining scattering cross sections as well as resonance positions \cite{moiseyev98}. 
There is a discussion of the relation between the energy of the resonance pole and the peak energy of the cross section \cite{klaiman10}.
In atomic and molecular physics, recent development of the theory of resonances has been reported; 
for example, Hatano {\it et al.} discuss the resonance theory for discrete models such as quantum dots \cite{klaiman11} and also for the complete set in the open quantum systems \cite{hatano14}.
Moiseyev has developed the scattering collision theory using complex scaling in his review work \cite{moiseyev11},
where several methods of describing resonances are explained such as in terms of the level density and phase shifts, in addition to the CSM.
Moiseyev and his collaborators have also developed the description of resonances in terms of the wave-packet propagation within the time-dependent approach \cite{goldzak12}. 

In atomic and molecular physics, the dominant interaction in the system is the long-range Coulomb force.
On the other hand, in nuclear physics, the dominant interaction is the short-range one, the nucleon-nucleon ($NN$) interaction,
which is determined to reproduce the observables of the two-nucleon systems.
A typical range of the $NN$ interaction is given by the one-pion exchange as around 1.4 fm \cite{pieper01}.
Owing to the short-range nature of the $NN$ interaction, we can avoid the problem arising from the long-range interaction except for the proton-proton Coulomb part. 
For example, we can expand the nuclear many-body wave functions with the finite-range basis sets which are often analytical like harmonic oscillator and Gaussian functions.
These functions are taken to cover the range of the nuclear interaction inside the nuclei.
For scattering states, we can decompose the wave function into the inner part of the interaction region and the outer non-interacting asymptotic part. 
This decomposition is useful to describe resonances.
In this review, we apply this approach to the Lippmann-Schwinger equation with complex scaling to describe the three-body scattering process of nuclei \cite{kikuchi10,kikuchi11}.

In 1971, Gyarmati and Vertse \cite{gyarmati71,gyarmati72} showed that the complex-scaled wave functions for resonances provide the same matrix elements 
as those obtained using the convergence factor method \cite{zeldovich61,romo68,homma97}, which was introduced to regularize the singular behavior of resonance wave functions in the asymptotic region.
After that, the CSM has often been used in the nuclear cluster model \cite{kruppa88,csoto94}; in this model, the relative motion between clusters are transformed in the CSM.
The CSM applications have been increasing in recent years particularly, in the cluster physics and unstable nuclear physics.

%% Berrgren, extended completeness relation, Gamow shell model
Resonant states are usually described using scattering solutions of the Schr\"odinger equation.
The properties of the states have been discussed in the nuclear reaction theory, as seen in a series of papers by Humblet and Rosenfeld \cite{humblet61}.
Using a different approach, Berggren attempted to describe the resonant state as an extended concept of the bound state \cite{berggren70,berggren96}.
He proposed an extended completeness relation \cite{berggren68,berggren92} which includes explicitly the resonance poles of the $S$-matrix by introducing deformed contours on the complex momentum plane. 
Recently, the method of complex contour deformation has been adopted in preparing the single particle basis states 
in the shell model like approaches to describe weakly bound states
\cite{betan02,michel02,okolowicz03,hagen05,volya06,hagen07,rotureau06,jaganathen12}.

In the CSM, rotation of the momentum axis corresponds to one kind of contour deformation 
for describing the poles of bound and resonant states in the same region of the complex momentum plane. 
Therefore, the CSM can treat bound and resonant states equally including many-body resonances.
It can be understood that the CSM is based on the generalization of the idea of Berggren for resonances, which are regarded as the extensions of the bound states.

\begin{figure}[t]
\centering
\includegraphics[width=10.0cm,clip]{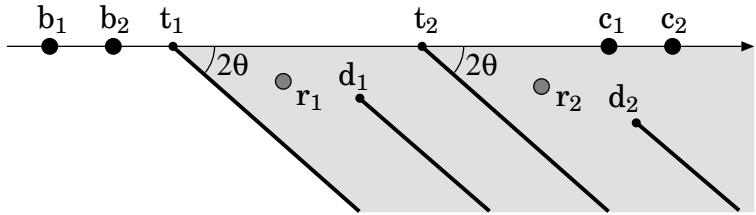}
\caption{Schematic energy spectrum of the complex-scaled Hamiltonian $H^\theta$ in the complex energy plane.
Symbols ($b_1$, $b_2$), ($c_1$, $c_2$), ($r_1$, $r_2$), ($t_1$, $t_2$), and ($d_1$,~$d_2$) represent the eigenvalues of bound states, 
bound states embedded in the continuum, resonances, thresholds with real energies, and thresholds with complex energies of the resonances, respectively. 
The lines with the angle of $2\theta$ from the thresholds indicate the rotated branch cuts.}
\label{fig:CSM_ABC}
\end{figure}

%% basis properties of CSM for bound & resonance
Various kinds of bound states and resonances are generated by the Hamiltonian $H$
and are obtainable by diagonalizing the complex-scaled Hamiltonian $H^\theta$ using the $L^2$ basis functions. 
In Fig.~\ref{fig:CSM_ABC}, we show a schematic distribution of the eigenvalues of the many-body Hamiltonian $H^\theta$. 
Along with the bound states ($b_1$, $b_2$) located below the lowest threshold, the bound states embedded in continuum states are obtained ($c_1$, $c_2$), if they exist. 
Above the thresholds, resonances are obtained with the complex eigenvalues ($r_1$, $r_2$).
In the three-body system, two-body and three-body resonances are obtained above the two-body and three-body thresholds, respectively. 
In the CSM, it is not required to employ the asymptotic boundary conditions for the resonance wave functions, 
since such an asymptotic behavior of the resonant state is properly taken into account in the complex-scaled $L^2$ wave functions. 
Beyond the three-body systems, the many-body eigenvalue distributions are generally explained by the extension of the three-body case.

%%%%%%%%%%%%%%%%%%%%%%%%%%%%%%
%%%%%%%%%%%%%%%%%%%%%%%%%%%%%%

%% basis properties of CSM for continuum states
Another important advantage of the CSM is with regard to the continuum states. 
For the original Hamiltonian $H$ obtained without applying the CSM, all continuum (scattering) states start from the thresholds with real energies and these states are degenerate on the real energy axis. 
On the other hand, for the complex-scaled Hamiltonian $H^\theta$, these continuum states are distributed on the rotated cuts starting from different thresholds with a common angle of $2\theta$, as seen in Fig.~\ref{fig:CSM_ABC}.
The continuum spectra of $H^\theta$ are classified into two types of thresholds: 
(1) the thresholds starting with the real energies ($t_1$, $t_2$) for decaying into two clusters (two-body threshold), three clusters (three-body threshold), and so on; and  
(2) the thresholds with complex energies ($d_1$, $d_2$) for decaying into resonating clusters. 
All branch cuts are rotated with the same angle by using the common scaling angle $\theta$ 
for every coordinate and conjugate momentum of the system.
This feature of the CSM enables the decomposition of the various non-resonant continuum states from the degenerated states with a non-scaled Hamiltonian $H$.
The decomposition of the states brings unique identification of all types of continuum states on the $2\theta$-rotated lines, as shown in Fig.~\ref{fig:CSM_ABC}. 

%% cluster physics with CSM
The CSM is becoming increasingly important in nuclear physics as the focus of studies moves from stable to unstable nuclei and
from low-excitation to high-excitation energies.
In stable nuclei, the resonances appear in the excited energy region above the thresholds for a nucleon or an $\alpha$ particle to be emitted.
The CSM application has numerous in nuclear cluster physics. The $\alpha$ cluster states often exist just near the $\alpha$ particle threshold and 
some of them can be resonances. The famous Hoyle state in $^{12}$C can decay into $3\alpha$ particles with very small decay widths \cite{kurokawa05}.
In the 4$N$ nuclei of $^{8}$Be, $^{12}$C, $^{16}$O, and $^{20}$Ne, the $\alpha$ cluster states have been investigated in detail theoretically and experimentally \cite{horiuchi12}. In the few-nucleon system, the CSM has been applied recently to solve the scattering problem \cite{carbonell14}.

%% many-body resonances
Among the unstable nuclei, neutron-rich nuclei have been extensively studied \cite{tanihata13}.
In particular, physics of neutron halo nuclei is one of the most important topics of investigation in nowadays.
In neutron halo nuclei, one or two neutrons are decoupled from the core nuclei such as in the case of $^6$He forming with $\alpha$+$n$+$n$.
Most of the halo nuclei have only one bound state with a halo structure, and so, the excitation of the nuclei requires the study of unbound states beyond the two-body cases.
Multi-neutron resonances with a core should be treated to clarify their specific properties such as configuration, responses, and coupling with continuum states.
In proton-rich nuclei, the Coulomb repulsion produces a repulsive effect in the energies, and then, the number of resonances is generally larger than that in the neutron-rich cases \cite{charity10,charity11}.
The CSM is applied to examine the many-body resonance phenomena in neutron/proton-rich nuclei. 
In this article, we show the example of He isotopes of $^{5\mbox{-}8}$He and their mirror nuclei of $^5$Li, $^6$Be, $^7$B, and $^8$C.
We predict the many-body resonances of these isotopes \cite{myo11b,myo12b}.

% COSM and GSM
We employ the cluster-orbital shell model (COSM) approach \cite{suzuki88} to construct the many-body basis function.
The many-body resonant and non-resonant continuum states of the total system are obtained as the eigenvectors of 
the complex-scaled (CS) Hamiltonian through diagonalization by using the COSM basis functions.
We call this procedure as the complex-scaled COSM (CS-COSM) approach.
On the other hand, on basis of the shell model, the many-body basis function is constructed from a product of the single-particle states to which the Berggren representation is applied.
An example of such the approaches is the Gamow shell model (GSM) \cite{michel02}.
In the CS-COSM and GSM, the many-body basis functions are prepared in a different way. 
It is meaningful to examine whether the results obtained in two different approaches agree with each other.
To this end, we investigate the contributions of resonant poles and non-resonant continuum states 
to the wave functions obtained in the CS-COSM \cite{aoyama06,masui06,masui14}
and perform a comparison to the GSM approaches for Oxygen and Helium isotopes \cite{masui06,masui14,masui12}.

%% CS-ME
In addition to the energy and decay width of resonances, 
the matrix elements of the resonances are important topics of discussion, because these quantities have an effect on any scattering observables.
We have investigated the reliability of complex-scaled matrix elements for various operators in addition to the energy eigenvalues \cite{homma97}.
Based on the results, we have applied the matrix elements obtained by using the CSM to calculate the transition strengths such as $E1$, $E2$ and $T$-matrix \cite{myo98,matsumoto10,ogata13}.

%% strength function
The CSM provides with a natural extension of the completeness relation consisting of bound and scattering (unbound) states 
to the bound and resonance and non-resonant continuum states for many-body systems.
In the CSM, we can extract the individual effects of not only the resonances but also the non-resonant continuum states in the scattering observables.
In the many-body system, such as a core+$n$+$n$ system, the binary subsystem can also form bound states or resonances. 
In addition to the real three-body states of core+$n$+$n$ components, ``binary states consisting of core+$n$'' +$n$ components can coexist in the three-body scattering states.
By the extended completeness relation for many-body scattering solutions, it is possible to construct the complex-scaled Green's function of the system, 
which is essential to evaluate the continuum level density (CLD) \cite{kruppa98,kruppa99,suzuki05}, the strength function from the response function \cite{myo98,myo01}, 
the Lippmann-Schwinger (LS) equation, and the $T$-matrix for the nuclear reaction problems \cite{kikuchi10,kikuchi11,kruppa07,kikuchi09}. 
As one of the approach for obtaining the transition strength for the three-body scattering states, such as $^6$He into $\alpha$~+~$n$~+~$n$, 
we apply the complex scaling to extend the completeness relation and introduce the complex-scaled Green's function. 

%% CSLS
The complex-scaled Green's function enables us to describe the consistent many-body scattering wave function within the space of the $L^2$ basis functions.
We have developed the method for finding complex-scaled solutions of the Lippmann-Schwinger equation (CSLS) \cite{kikuchi10,kikuchi11}.
In CSLS, the complex-scaled Green's function is applied to the Lippmann-Schwinger formalism for the short-range interaction.
In this article, we report several applications of CSLS to the reactions in which the three-body scattering states are concerned.
The Coulomb and nuclear breakup reactions of two-neutron halo nuclei are discussed to investigate the excitation and breakup mechanisms of these nuclei \cite{kikuchi13a}.
As another application of CSLS, we demonstrate the elastic scattering and radiative capture reaction of the $\alpha$~+~$d$ system \cite{kikuchi11}.
We discuss the effects of the three-body structure of $\alpha$~+~$p$~+~$n$, such as the deuteron breakups and the rearrangement, using the 
the complex-scaled three-body Green's function, on the $\alpha$~+~$d$ scattering.

A summary and a perspective are presented in the last section.
The present status of the developments and applications of the CSM in nuclear physics is summarized.
The remaining subjects of the CSM are also given as future perspectives in relation to the unified description of the quantum many-body unbound states. 
\section{Unified treatments of bound, resonant and continuum states in CSM} \label{sec:method}

In this section, the basic framework of CSM is explained from the two-body system to many-body case.
The complex-scaled eigenstates are categorized as the bound, resonant and continuum states,
which consist of the extended completeness relation.
Using the complex-scaled eigenstates, we introduce the complex-scaled Green's function,
which is essential to evaluate the scattering observables and also to extract the explicit roles of the resonant and continuum states.
We give the example of the usage of the complex-scaled Green's function in the calculation of the strength functions.

%%%%%%%%%%%%%%%%%%%%%%%%%%%%%%
\subsection{Complex scaling method and the ABC theorem} \label{sec:ABC}

%%%%%%%%%%%%%%%%%%%%%%%%%%%%%%
\begin{figure}[b]
\centering
\includegraphics[width=14.0cm,clip]{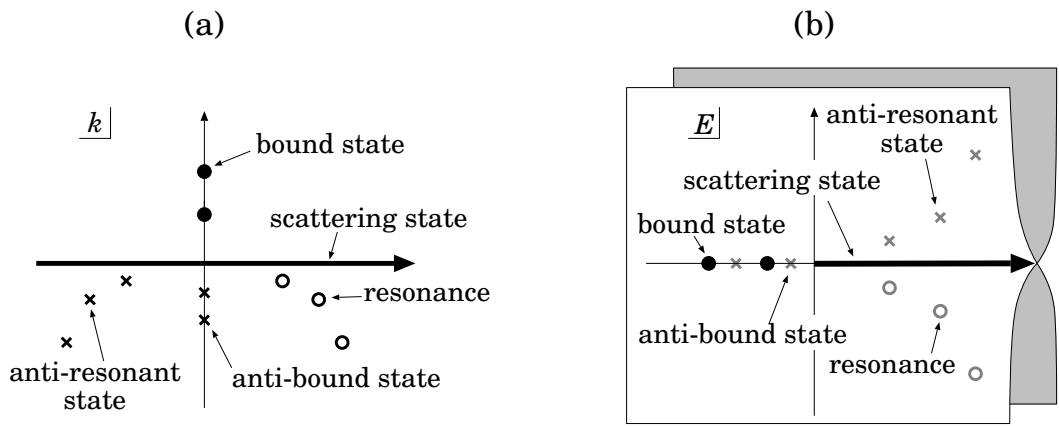}
\caption{$S$-matrix poles in the (a) momentum and (b) energy planes. 
The solid arrows indicate the scattering states with a real momentum and a positive energy.}
\label{fig:sec2_pole-k}
\end{figure}
%%%%%%%%%%%%%%%%%%%%%%%%%%%%%%

%% general pole distribution
Figure~\ref{fig:sec2_pole-k} shows a schematic pole distribution of the $S$-matrix for a single channel two-body system on the complex momentum and energy planes.
The energy plane consists of two Riemann sheets. 
The discrete solutions are classified into bound states, anti-bound states, resonances, and anti-resonances in the complex momentum plane as follows: 
\begin{center}
\renewcommand{\arraystretch}{1.25}
\begin{tabular}{c|c|c|c}
     bound states & anti-bound states          & resonances                   & anti-resonances \\  \hline 
     $k_B=i\gamma_b$ & $k_{AB}=-i\gamma_{ab}$  & $k_{R}=\kappa_r-i\gamma_r$   & $k_{AR}=-\kappa_r-i\gamma_r$ 
\end{tabular}\\ 
\end{center}
where  $\kappa_r$ and $\gamma_r$, $\gamma_b$, $\gamma_{ab}$ are all positive.
For bound, anti-bound and resonant states, the asymptotic wave functions are proportional to the outgoing wave $e^{ik_pr}$, 
where the momentum $k_p$ at every pole is given by $k_p=k_B$,~$k_{AB}$, and $k_R$, respectively 
Therefore, only the bound states have a damping form of the radial wave function $\psi_{k_B} \sim e^{-\gamma_br}$. 
For anti-resonances, the incoming wave is adapted with momentum $k_{AR}$. 
In addition, there exists the virtual states on the negative imaginary axis for the $s$-wave case.

%% CSM
We explain the complex scaling method proposed by Aguilar, Balslev, and Combes \cite{aguilar71,balslev71}.
They introduced the transformation $U(\theta)$ with a scaling angle $\theta$ for the radial coordinate $\vc{r}$ and its conjugate momentum $\vc{k}$ as
\begin{eqnarray}
      U(\theta)\, \vc{r}\, U^{-1}(\theta)
&=&  \vc{r}\ e^{i\theta}, \hspace*{2cm} 
      U(\theta)\, \vc{k}\, U^{-1}(\theta)
~=~  \vc{k}\ e^{-i\theta},
\end{eqnarray}
where $U(\theta) U^{-1}(\theta)=1$.
The Schr\"odinger equation, $H\Psi=E\Psi$, is transformed as
\begin{eqnarray}
       H^\theta \Psi^\theta
&=&    E^\theta \Psi^\theta,\qquad
       \label{eq:sec2_cs-eigen}
      \\
      H^\theta
&=&   U(\theta)HU^{-1}(\theta),
      \label{eq:sec2_cs-ham}
\end{eqnarray}
Here, for the case of one degree of freedom, the complex-scaled wave function is defined as $\Psi^\theta=U(\theta)\Psi=e^{\frac{3}{2}i\theta}\Psi(\vc{r}e^{i\theta})$. 
The factor $e^{\frac{3}{2}i\theta}$ comes from the Jacobian of the coordinate transformation for $\vc{r}$.
The asymptotic wave functions for the poles with momenta $k=k_p$ are described by the outgoing waves for the radial part, 
$e^{ik_p r e^{i\theta}}$. 
The wave functions of the resonances, which originally displayed a divergent behavior as
$e^{ik_R\cdot r}=e^{i(\kappa_r-i\gamma_r)r}=e^{\gamma_r r}\cdot e^{i\kappa_r r}$, behave as 
\begin{eqnarray}
    e^{ik_R\cdot re^{i\theta}}
&=& e^{i(\kappa_r-i\gamma_r) r e^{i\theta}}
~=~ e^{ir(\kappa_r-i\gamma_r) (\cos{\theta}+i\sin{\theta})} 
    \nonumber \\
&=& e^{(-\kappa_r\sin{\theta}+\gamma_r\cos{\theta})r}\cdot e^{i(\kappa_r\cos{\theta}+\gamma_r\sin{\theta})r}. 
\end{eqnarray}
This equation shows that the divergent behavior of the resonance wave functions is regularized when 
the scaling angle $\theta$ is larger than the angle $\theta_r=\tan^{-1}(\frac{\gamma_r}{\kappa_r})$ for the resonance position $\kappa_r-i\gamma_r$. 

The wave functions for these resonances become normalizable by integration along the complex-scaled axis $re^{i\theta}$ 
in the same way as in which the bound states become normalizable.
Therefore, we can obtain these resonances with $\theta_r<\theta$, together with the bound states, by the diagonalization of the Hamiltonian with a finite number of $L^2$ basis functions. 

% ABC
Properties of the solutions of the complex-scaled Schr\"odinger equation are explained in the so-called ABC theorem given by Aguilar, Combes, and Balslev\cite{aguilar71,balslev71}, as follows
\begin{enumerate}
\itemsep=0.1cm
\item The resonance solutions are described by the square-integrable functions, like bound states. 
\item The energies of the bound states are invariant with respect to the scaling.
\item The continuum spectra start at the threshold energies of decays of the system into its subsystems and are rotated clockwise by an angle of $2\theta$ from the positive real energy axis.
\end{enumerate}

When we solve Eq.~(\ref{eq:sec2_cs-eigen}) to obtain the resonances, 
the resonance wave function $\Psi^\theta$ is often to be expanded in terms of the $L^2$ basis functions with a finite number, such as Gaussian functions or
the harmonic oscillator basis functions. The number of the basis states is determined to converge the solutions.
In this case, the complex-scaled Hamiltonian matrix elements are diagonalized, and we obtain the complex energy eigenvalues that are discretized for bound, resonant, non-resonant continuum states. 

If $\Psi^\theta(k)$ is an eigensolution of Eq.~(\ref{eq:sec2_cs-eigen}) with momentum $k$, 
its conjugate solution is given by $\tilde{\Psi}^\theta(k)=\Psi^\theta(-k^*)$ for the bi-orthogonal state \cite{moiseyev11,berggren68,aoyama06}. 
With the bi-orthogonal solutions, the matrix elements for the arbitrary operator under the complex scaling are expressed as 
\begin{eqnarray}
	\bra \tilde{\Phi}(k)|\hat{O} | \Psi(k')\ket
&=&	\bra U(\theta)\tilde{\Phi}(k)| U(\theta)\hat{O}U^{-1}(\theta) | U(\theta)\Psi(k')\ket
        \nonumber\\
&=&     \bra \tilde{\Phi}^\theta(k)|\hat{O}^\theta | \Psi^\theta(k')\ket ,
        \label{eq:sec2_cs-me}
        \nonumber \\
        \hat{O}^\theta 
&=&     U(\theta)\hat{O} U^{-1}(\theta). 
\end{eqnarray}

To solve Eq.~(\ref{eq:sec2_cs-eigen}), we expand the wave functions $\Psi^\theta(k,\vc{r})$ to a finite number of
$L^2$ basis functions, {$u_i(\vc{r})$ for $i=1,2,\ldots, N$ }, such as harmonic oscillator basis states and the Gaussian basis functions
\begin{eqnarray}
      \Psi^\theta(k,\vc{r})
&=&   \sum_i^{N} c_i(k,\theta)\, u_i(\vc{r}).
\end{eqnarray}
The coefficients $c_i(k,\theta)$ and the discrete spectra are obtained by solving the eigenvalue problem
\begin{eqnarray}
      \sum_i^{N}H_{ij}^\theta\, c_j(k,\theta)
&=&   E^\theta\, c_i(k,\theta),
      \\
    H_{ij}^\theta
&=& \bra \tilde{u}_i | H^\theta | u_j \ket ,
\end{eqnarray}
where $H_{ij}^\theta$ are the matrix elements of the complex-scaled Hamiltonian given in Eq.~(\ref{eq:sec2_cs-ham}).
The matrix elements for the arbitrary operator in Eq.~(\ref{eq:sec2_cs-me}) can be calculated using the coefficients $c_i(k,\theta)$ as
\begin{eqnarray}
	\bra \tilde{\Phi}^\theta(k)|\hat{O}^\theta | \Psi^\theta(k')\ket 
&=&     \sum_{i,j} c_i(k,\theta)\, c_j(k',\theta)\,  \bra \tilde{u}_i | \hat{O}^\theta | u_j \ket 
        \label{eq:csme}
\end{eqnarray}

%%%%%%%%%%%%%%%%%%%%%%%%%%%%%%
\begin{figure}[t]
\centering
\includegraphics[width=7.5cm,clip]{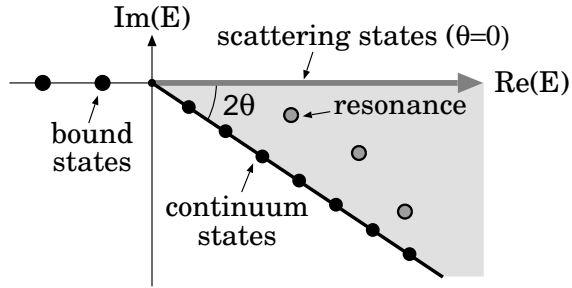}
\caption{Schematic eigenvalue distribution of $H^\theta$ in the single-channel two-body system. 
Continuum states are discretized on the $2\theta$-line as solid circles.}
\label{fig:sec2_pole-e}
\end{figure}
%%%%%%%%%%%%%%%%%%%%%%%%%%%%%%

A schematic distribution of the eigenvalues of $H^\theta$ as asserted by the ABC theorem is shown in Fig.~\ref{fig:sec2_pole-e}. 
In addition to the eigenvalues of the bound states and resonances, the discretized spectra of the non-resonant continuum states are obtained with the eigenvalues on the $2\theta$-line.

In practical applications of the CSM, the interaction forms are limited because the transformed interaction $V(re^{i\theta})$ should maintain analyticity. 
For a Gaussian potential, the CSM can be applicable with $\displaystyle \theta < \pi/4$ . 
When the interaction is not analytic, the exterior scaling transformation becomes useful, 
in which the contour of the coordinate integration is changed to avoid the region where the interaction is not analytic.
The detailed of this method is explained in the review given by Moiseyev \cite{moiseyev11}.

For the description of resonances in the CSM, we adopt the basis expansion method using the Gaussian functions; 
this method has been widely used in the cluster model analysis \cite{aoyama06,hiyama03} and also 
in the cluster orbital shell model approach \cite{suzuki88} for many-body resonances, as is explained in \S \ref{sec:many}.
In addition to the Gaussian functions, the other basis states have been extensively employed, such as
 the complex-range Gaussian basis function \cite{ohtsubo13}, the exponential type basis, and so-called the Hyllerass-type \cite{ho12}.
There are several approaches to describe the resonances, particularly for the three-body case using the CSM
such as the hyperspherical harmonics method \cite{garrido03}.

%%%%%%%%%%%%%%%%%%%%%%%%%%%%%%%%%%%%%%%%%%%%%%%%%%
\subsection{Three-body resonances} \label{sec:three-Body}
\begin{figure}[t] 
\centering
\includegraphics[width=15cm,clip]{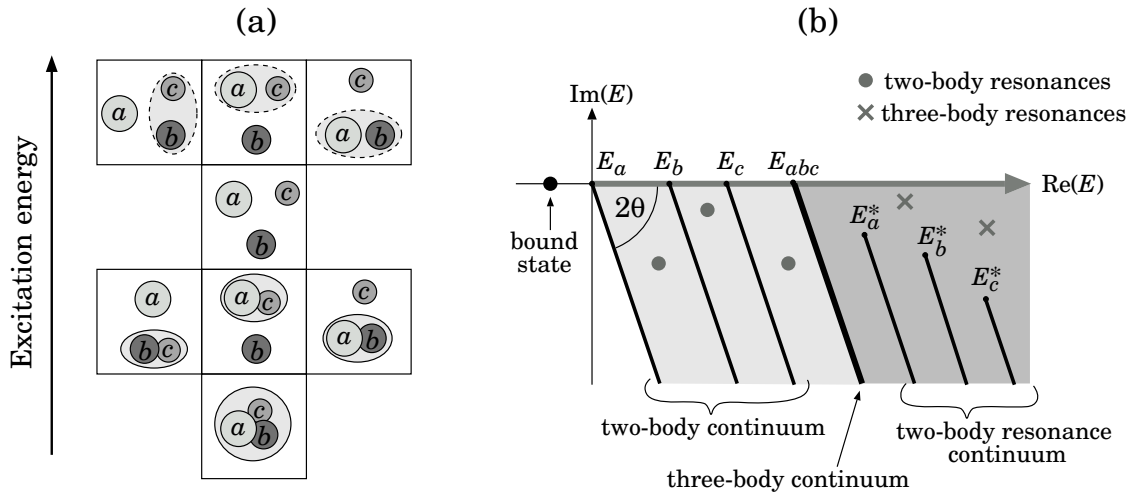}
\caption{Schematic energy eigenvalue distribution for a three-body system in CSM.}
\label{fig:sec2_three-body}
\end{figure}

~~~~
The CSM can be used to describe the many-body resonances and non-resonant continuum states.
In this subsection, we consider a three-body case. 
We assume a system consisting of three clusters, $a$, $b$, and $c$.
In Fig.~\ref{fig:sec2_three-body}, we show a schematic energy eigenvalue distribution of CSM applied to the $a+b+c$ system,
which is governed by the ABC theorem, similar to the two-body case, as was previously explained.
The left side of Fig.~\ref{fig:sec2_three-body} shows a schematic image of the three-body system with increasing excitation energy.

In the figure, a bound ground state of the three-body system, $(abc)$, is assumed, and the three-body threshold energy is shown as $E_{abc}$. 
Every two-body subsystem can have bound states and resonances. 
The threshold energies of two-body decays into $a+(bc)$, $b+(ca)$, and $c+(ab)$ are described by the binding energies of $(bc)$, $(ca)$, and $(ab)$ subsystems, 
which are indicated in Fig.~\ref{fig:sec2_three-body} (b) as $E_{a}$, $E_{b}$, and $E_{c}$, respectively. 
These energies of the two-body bound states, together with those of the three-body bound states, are obtained on the real energy axis 
as the eigenvalues of the complex-scaled three-body Hamiltonian $H^\theta$. 
Besides the three-body bound states, continuum spectra appear on the $2\theta$-lines starting from each real threshold. 
Furthermore, when the two-body subsystems have resonances, their energies are obtained as complex eigenvalues in the CSM; 
these energies are shown as $E_{a}^*$, $E_{b}^*$, and $E_{c}^*$ in Fig.~\ref{fig:sec2_three-body} (b). 
From these resonant thresholds, two-body continuum spectra are obtained. 
The so-called resonant thresholds corresponding to these resonance energies are the origins of the straight lines describing the continuum states.
Although the two-body resonances can occur above each kind of two-body decay threshold, three-body resonances occur only above the three-body threshold. 
The two-body resonances exhibit a variety of structures owing to the complicated structure of the multi-fold Riemann sheets. 

%%%%%%%%%%%%%%%%%%%%%%%%%%%%%%
%%%%%%%%%%%%%%%%%%%%%%%%%%%%%%

\begin{figure}[t]
\centering
\includegraphics[width=8.0cm,clip]{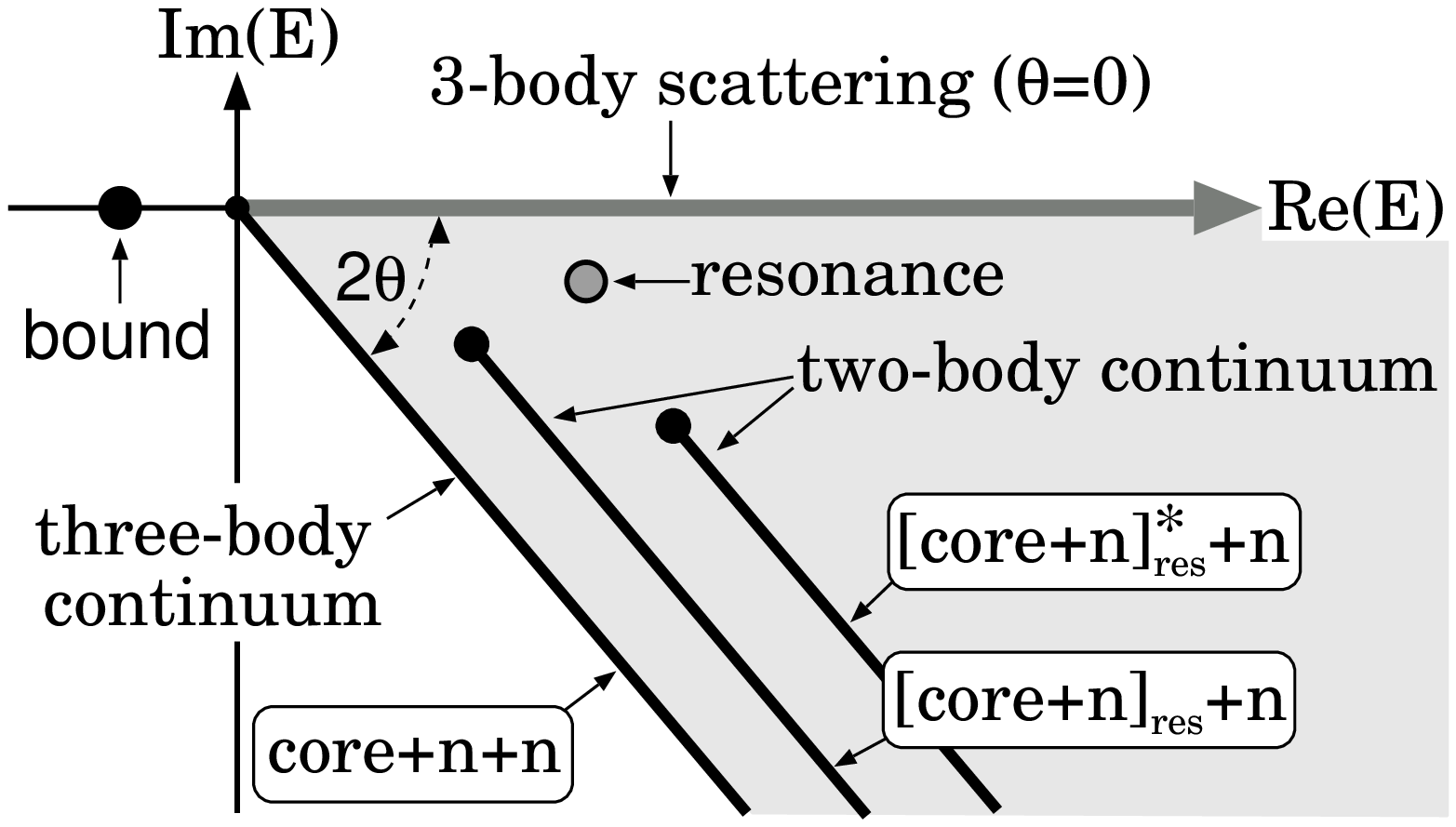}\hspace*{0.8cm}
\includegraphics[width=7.5cm,clip]{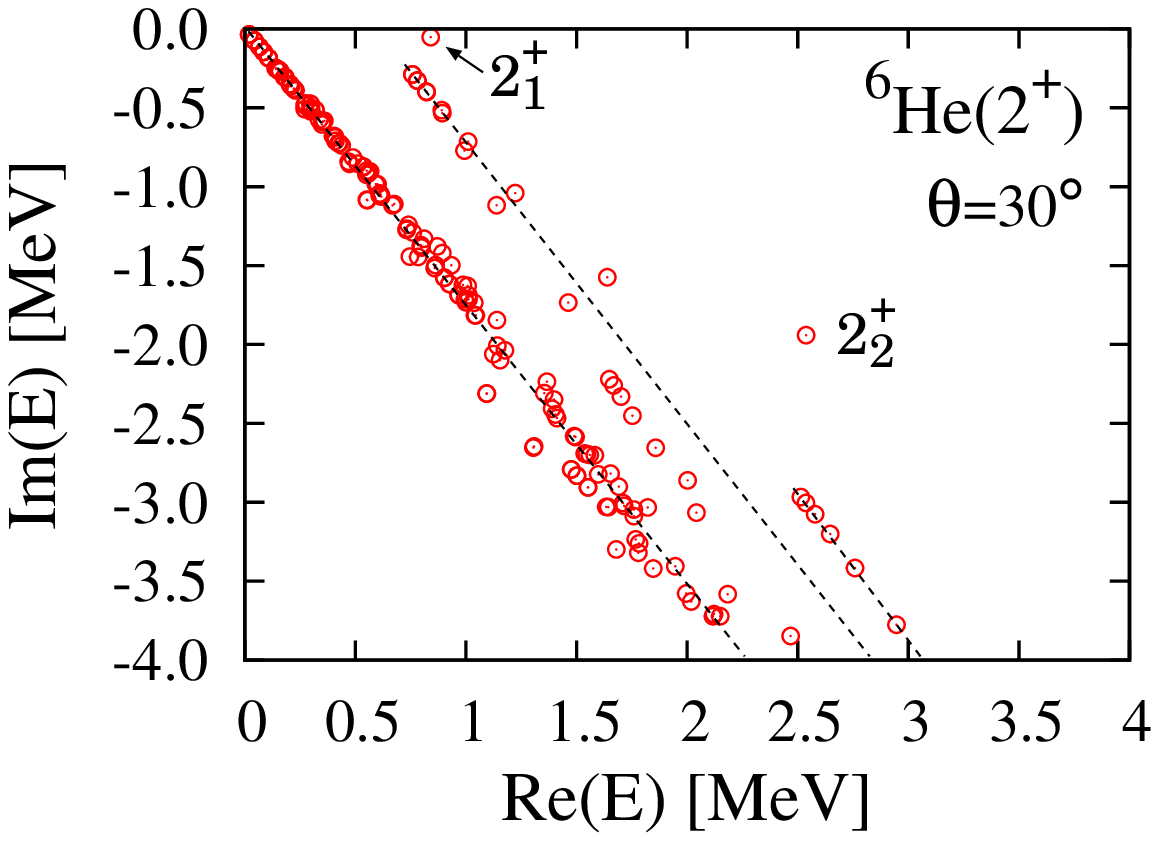}
\caption{
Left: Schematic energy eigenvalue distribution of the Borromean system consisting of core+$n$+$n$.
Right: Energy eigenvalues of the $^6$He($2^+$) states in the complex energy plane measured from the $^4$He+$n$+$n$ threshold \cite{myo09b}.
The three schematic dotted lines correspond to the $^4$He+$n$+$n$ and $^5$He($3/2^-$,$1/2^-$)+$n$ continuum states, in order from left to right, respectively.}
\label{fig:sec2_ene6_2}
\end{figure}

We consider the case of the Borromean three-body system often observed in neutron-rich nuclei such as core+$n$+$n$.
The Borromean system consists of three clusters and has no bound states in any two-body subsystems. 
This fact means that the three-body threshold is the lowest threshold in the Borromean system.
All the Borromean systems observed so far have very few bound states, and many excited states are observed as resonances at low excitation energies. 
Owing to the small separation energy of the ground state, a Borromean system is believed to break up easily by weak perturbation. 
The three-body continuum states appear on the $2\theta$-line from the three-body threshold as shown in the left side of Fig.~\ref{fig:sec2_ene6_2}.
All states above the three-body threshold are resonances including the ground states of the subsystems. 
Two-body continuum spectra also start from two-body resonance thresholds. 

An example of the eigenvalue distribution of $^6$He($2^+$) obtained in the CSM is shown in Fig. \ref{fig:sec2_ene6_2}.
The three dotted lines indicate the $2\theta$ lines for the rotated two- and three-body continuum states.
In the numerical calculation, the continuum energies are discretized and then are not completely aligned with the $2\theta$ line \cite{aoyama06}.
In this figure, we can easily identify the locations of the three-body resonances of $^6$He ($2^+_1$ and $2^+_2$) and,
further, of the two kinds of continuum states of $^5$He($3/2^-$,$1/2^-$)+$n$ and $^4$He+$n$+$n$.

%%%%%%%%%%%%%%%%%%%%%%%%%%%%%
%%%%%%%%%%%%%%%%%%%%%%%%%%%%%%

%%%%%%%%%%%%%%%%%%%%%%%%%%%%%%%%%%%%%%%%%%%%%%%%%%%%%%%%%%%%%%%%%%%%%%%%%%%
\subsection{Extended completeness relation in CSM}
\label{sec:ECR}

\begin{figure}[b]
\centering
\includegraphics[width=16.5cm,clip]{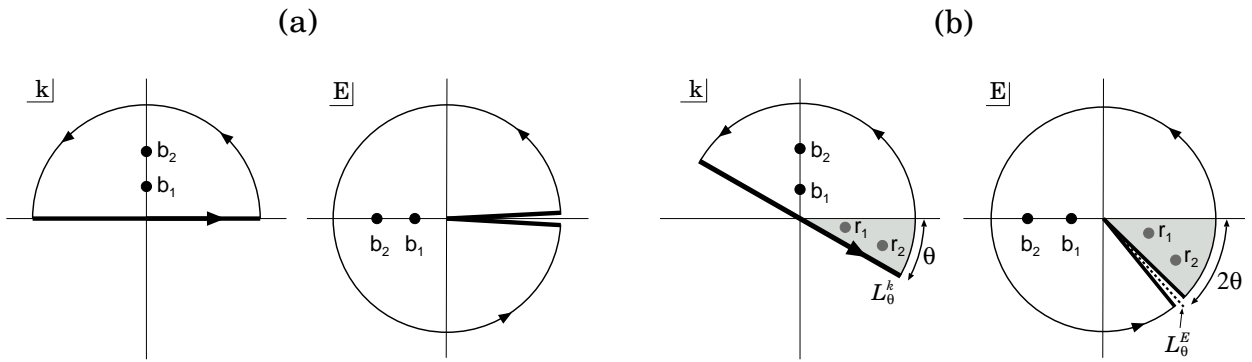}
\caption{The Cauchy integral contours in the momentum and energy planes for the completeness relation (a) without and (b) with the CSM.
The circles $b_1$, $b_2$ and $r_1$, $r_2$ are the poles of bound and resonant states, respectively.}
\label{fig:sec2_pole-ke}
\end{figure}

Bound and scattering (continuum) states form a complete set that is represented by the completeness relation with real eigenenergies/momenta \cite{newton60}
\begin{eqnarray}
{\bf 1}&=&\sum_b^{n_b} |\Psi_b\ket\bra\Psi_b|+\int_0^\infty dE|\Psi_E\ket\bra\Psi_E|, \nonumber \\
       &=&\sum_b^{n_b} |\Psi_b\ket\bra\Psi_b|+\int_{-\infty}^{+\infty} dk|\Psi_k\ket\bra\Psi_k|, \label{eq2-3-1}
     \label{eq:sec2_ECR}
\end{eqnarray}
where $\Psi_b$ and $\Psi_E$ are the bound states (the discrete negative part of the energy spectrum) and continuum states (the continuous positive part), respectively, 
on the first Riemann sheet of the energy plane. The number of bound states is given as $n_b$.
The continuum states ($\Psi_k$, $\Psi_{-k}$) in the momentum representation belong to the states on the real $k$ axis. 
Therefore, integration over the $k$ axis corresponds to that along the rims of the cut of the first Riemann sheet of the energy plane, as shown in Fig.~\ref{fig:sec2_pole-ke} (a). In the case of a potential problem, 
the mathematical proof of the completeness relation given by Eq.\ (\ref{eq:sec2_ECR}) was proposed by Newton \cite{newton60} using the Cauchy theorem. 

%%%%%%%%%%%%%%%%%%%%%%%%%%%%%%
%%%%%%%%%%%%%%%%%%%%%%%%%%%%%%

For the solutions of the complex-scaled Hamiltonian $H^\theta$, it is important to consider the completeness relation. 
A mathematical proof for the completeness relation in the CSM was given by Giraud {\it et al.} \cite{giraud03,giraud04} for the single-channel and coupled-channel cases.
In the CSM, the momentum axis is rotated down by $\theta$, and the poles of resonances can enter the semicircle for the Cauchy's integration shown in Fig.~\ref{fig:sec2_pole-ke} (b). 
Therefore, the resonances are explicitly included in the completeness relation of the complex-scaled Hamiltonian $H^\theta$ as follows.
\begin{eqnarray}
       {\bf 1}
&=&   \sum_b^{n_b}|\Psi_b^\theta \ket \bra \tilde{\Psi}_b^\theta|+\sum_r^{n_r^\theta}|\Psi_r^\theta \ket \bra \tilde{\Psi}_r^\theta|
    + \int_{L_\theta^E} dE|\Psi_E^\theta \ket \bra \tilde{\Psi}_E^\theta|
      \nonumber \\
&=&   \sum_b^{n_b}|\Psi_b^\theta \ket \bra \tilde{\Psi}_b^\theta|+\sum_r^{n_r^\theta}|\Psi_r^\theta \ket \bra \tilde{\Psi}_r^\theta|
    + \int_{L_\theta^k} dk|\Psi_k^\theta \ket \bra \tilde{\Psi}_k^\theta|, 
    \label{eq:sec2_ECR-CSM}
\end{eqnarray}
where $\Psi_b^\theta$ and $\Psi_r^\theta$ are the complex-scaled bound and resonant states, respectively.
The resonant states that enter the semicircle rotated down by $\theta$ in the momentum plane are taken into consideration. Their number is given by $n_r^\theta$. 
Furthermore, the continuum states $\Psi_E^\theta$ and $\Psi_k^\theta$ are located on the rotated cut $L^E_\theta$ of the Riemann plane and on the rotated momentum axis $L_\theta^k$, respectively.
Hereafter, we refer to the relation in Eq.~(\ref{eq:sec2_ECR-CSM}) as the extended completeness relation (ECR) \cite{myo98}.

%%%%%%%%%%%%%%%%%%%%%%%%%%%%%%%%%%%%%%%%%%%%%%%%%%%%%
%%%%%	Separation of The Strength Function	%%%%%
\subsection{Green's function with CSM}
\label{sec:sec2_green}

We define the complex-scaled Green's function here, expanded in terms of ECR using the CSM. 
This function shows a broad applicability, particularly for obtaining the scattering observables 
in the many-body scattering states, such as the three-body breakup reaction of Borromean nuclei.
Applying ECR to calculations of physical quantities makes it possible to investigate the individual contributions of bound, resonant, and non-resonant continuum states. 
Berggren {\em et al.} \cite{berggren68,berggren92} have studied the division of the strength function into the above three contributing states 
by considering various kinds of ECR with different proportions of these contributing states. 
In their investigation, they discussed the validity of each type of the pole expansion and did not examine the non-resonant continuum term. 
It is necessary to perform a more comprehensive study including the continuum contribution.

Green's function is essential to calculate physical quantities such as transition strengths in terms of the scattering states having the continuous energy $E$. 
We transform Green's function ${\cal G}(E)$ in the CSM to obtain the scattering observables beyond the calculation of the energy eigenvalues of resonances.
We define the complex-scaled Green's function ${\cal G}^\theta(E)$ as
\begin{eqnarray}
        {\cal G}^\theta(E)	
&=&	U(\theta) {\cal G}(E) U^{-1}(\theta)
~=~	\frac{ {\bf 1} }{ E-H^\theta }
~=~	\sum_{~\nu} \hspace{-0.57cm}\int
	\frac{|\Psi_\theta^\nu\rangle \langle \tilde{\Psi}_\theta^\nu|}{E-E_\nu^\theta} , 
	\label{eq:sec2_green}
\end{eqnarray}
where the eigenvalue $E_\nu^\theta$ is associated with the wave function $\Psi_\theta^\nu$.
In the coordinate representation,
%%%%%%%%%%%%%%%%%%%%%%%%%%%%%%%
%%%%%	Complex Scaled Green Function 
\begin{eqnarray}
	{\cal G}^\theta(E,\vc{ r},\vc{ r}')
&=&	\left\langle	\vc{ r}		\left|
         \frac{{\bf 1}}{E-H^\theta} \right|
	\vc{ r}'		 \right\rangle
	.
	\label{eq:sec2_green1}
\end{eqnarray}
%%%%%%%%%%%%%%%%%%%%%%%
Substituting Eq.~(\ref{eq:sec2_ECR-CSM}) into Eq.~(\ref{eq:sec2_green1}), we can divide Green's function into three terms, as
%%%%%%%%%%%%%%%%%%%%%%%
\begin{eqnarray}
	{\cal G}^\theta(E,\vc{ r},\vc{ r}')
&=&	\sum_b^{n_b}
	\frac{	\Psi^\theta		(\vc{ r},k_B) \,
		\{\tilde{\Psi}^*(\vc{ r}',k_B) \}^\theta 
	}
	{ E -	E_B}
+	\sum_r^{n_r^\theta}
	\frac{ 	\Psi^\theta		(\vc{ r},k_R) 	
		\,
		\{\tilde{\Psi}^* (\vc{ r}',k_R) \}^\theta
	}
	{ E -	E_R}
\nonumber\\
& & 	+ \int_{L_\theta^k} dk_\theta
	\frac{ 	\Psi^\theta		(\vc{ r},k_\theta)\,
		\{\tilde{\Psi}^*(\vc{ r}',k_\theta) \}^\theta
	}
	{ E -	E^\theta} ,
	\label{eq:sec2_cs-green}
\end{eqnarray}
%%%%%%%%%%%%%%%%%%%%%%%
where $E_B$ and $E_R=E_r- i \Gamma/2 $ are the energy eigenvalues of the bound and resonant states, respectively.
The complex-scaled Green's function is widely used in the CSM.
It is noted that the last continuum term in Eq.~(\ref{eq:sec2_cs-green}) can be decomposed into several parts in the many-body case as shown in Fig. \ref{fig:sec2_ene6_2}.
It is possible to investigate the each contribution definitely not only of the discrete states but also of the various continuum states \cite{myo01}.

In this article, we show our recent progress in using Green's function to obtain the physical quantities related to the scattering observables,
such as the continuum level density (CLD) \cite{suzuki05}, strength function \cite{myo98}, Lippmann-Schwinger equation \cite{kikuchi10}, 
and the $T$-matrix calculation in the reaction theory \cite{kikuchi13a}.

%%%%%%%%%%%%%%%%%%%%%%%%%%%%%%

In this section, we explain the case of the strength function $S(E)$.
The strength function $S(E)$ is expressed in terms of the response function $R(E)$ as
\begin{eqnarray}
	S_\lambda(E)
&=&	\sum_\nu
	\bra \tilde{\Psi}_0   | \HO^{\dagger}_\lambda |\Psi_\nu	\ket
	\bra \tilde{\Psi}_\nu | \HO_\lambda	|\Psi_0		\ket
	\delta(E - E_\nu)
~=~	-\frac1{\pi}{\rm Im} R_\lambda(E),
        \label{eq:sec2_strength}
	\\
	R_\lambda(E)
&=&	\int d{\vc r} d\vc{r}' 	\;
	\tilde{\Psi}_0^*(\vc{ r}) 		\;
	\HO^{\dagger}_\lambda 		\;
	{\cal G}(E,\vc{ r},\vc{ r}')	\;
	\HO_\lambda 			\;
	\Psi_0(\vc{ r}'),
	\label{eq:sec2_response}
\end{eqnarray}
%%%%%%%%%%%%%%%%%%%%%%%
where $E$ is the energy on the real axis and $|\Psi_0\ket$, $|\Psi_\nu\ket$, and $\HO_\lambda$ are the initial states, 
final states, and an arbitrary transition operator of rank $\lambda$, respectively. The quantities $E_\nu$ are the energies of the final state. 
In this expression, we assume that the bound (initial) and final states form a complete set of the Hamiltonian $H$:
\begin{eqnarray}
      {\bf 1}
&=& \sum_{\nu} \hspace{-0.57cm}\int |\Psi_\nu\ket\bra\tilde{\Psi}_\nu|,
\end{eqnarray}
where the summation includes the initial state with $\nu=0$ as an element of the complete set.

%%%%%%%%%%%%%%%%%%%%%%%%%%%%%%%%%%%%%%%%%%%%%%%%%%%%%%%%%%%%%
%%%%%	Complex Scaled Response and Green Function	%%%%%
The complex-scaled initial wave functions $\Psi^\theta_0$, the Hamiltonian $H^\theta$, and the transition operator $\HO_\lambda^\theta$, are used to express the response function as
\begin{eqnarray}
	R_\lambda(E)
&=&	\int d\vc{ r} d\vc{ r}' 		\;
	\{\tilde{\Psi}_0^*(\vc{ r})\}^\theta 	\;
	(\HO^{\dagger}_\lambda)^\theta		\;
	{\cal G}^\theta(E,\vc{ r},\vc{ r}')	\;
	\HO^\theta_\lambda				\;
	\Psi_0^\theta(\vc{ r}')
	.
	\label{eq:sec2_response2}
\end{eqnarray}
 From Eq.~(\ref{eq:sec2_response2}), we obtain the following relations for the response function:
%%%%%%%%%%%%%%%%%%%%%%%
\begin{eqnarray}
%%%%%%%%%%%%%%%%%%%%%%%
%%%%%	Division
%\hspace*{-1.5cm}
	R_\lambda(E)&=&
%	\hspace{0.1cm}
%=	\hspace{0.1cm}
	R_{\lambda,B}(E) 
+	R_{\lambda,R}^\theta(E)
+	R_{\lambda,k}^\theta(E),
	\label{eq:sec2_response-3}
	\\
%%%%%%%%%%%%%%%%%%%%%%%
%%%%%	Bound part
	 R_{\lambda,B}(E)
&=&	\sum_b^{n_b}
	\frac{
		\bra \tilde{\Psi}_0^\theta	|
		(\HO^{\dagger}_\lambda)^\theta 	|
		\Psi_b^\theta   \ket
		\bra \tilde{\Psi}_b^\theta	|
		\HO_\lambda^\theta		|
		\Psi_0^\theta \ket 
	}
	{ E - E_B}
	,
	\label{eq:sec2_response-b}
	\\
%%%%%%%%%%%%%%%%%%%%%%%
%%%%%	Resonance part
	R_{\lambda,R}^\theta(E)
&=&	\sum_{r}^{n_r^\theta}
	\frac{
		\bra \tilde{\Psi}_0^\theta	|
		(\HO^{\dagger}_\lambda)^\theta 	|
		\Psi_r^\theta  \ket
		\bra \tilde{\Psi}_r^\theta	|
		\HO_\lambda^\theta		|
		\Psi_0^\theta \ket
	}
	{ E - E_R}
	,
	\label{eq:sec2_response-r}
	\\
%%%%%%%%%%%%%%%%%%%%%%%
%%%%%	Continuum part
	R_{\lambda,k}^\theta(E)
&=&	\int_{L_\theta^k} dk_\theta
	\frac{
		\bra \tilde{\Psi}_0^\theta	|
		(\HO^{\dagger}_\lambda)^\theta 	|
		\Psi_{k_\theta} \ket
		\bra \tilde{\Psi}_{k_\theta}|
		\HO_\lambda^\theta 		|
		\Psi_0^\theta \ket 
	}
	{ E - E_\theta}
	.
	\label{eq:sec2_response-k}
\end{eqnarray}
The strength function $S_\lambda(E)$ with the initial state $\Psi_0$ using the operator $\HO_\lambda$
is defined in terms of Green's function as 
\begin{eqnarray}
	S_\lambda(E)
&=&     -\frac1{\pi}\ \sum_{~\nu} \hspace{-0.57cm}\int\ {\rm Im}\left[  \frac{
	\bras{\tilde{\Psi}_0^\theta}  (\HO^\dagger_\lambda)^\theta \kets{\Psi_\nu^\theta}
	\bras{\tilde{\Psi}_\nu^\theta} \HO_\lambda^\theta          \kets{\Psi_0^\theta}
        }{E-E^\nu_\theta}
        \right] .
	\label{eq:sec2_strength3}
\end{eqnarray}
The function $S_\lambda(E)$ includes all of the resonant and continuum components of the final states,
and then is similarly decomposed as
%%%%%%%%%%%%%%%%%%%%%%%%%
\begin{eqnarray}
	S_{\lambda}  (E)
&=&	S_{\lambda,B}(E)
+	S_{\lambda,R}^\theta(E)
+	S_{\lambda,k}^\theta(E).
\label{eq:sec2_strength-3}
\end{eqnarray}
%%%%%%%%%%%%%%%%%%%%%%%%%%%%%%%%%%%%%%
%%%%%	Explanation of each term %%%%%
The matrix elements of the complex-scaled operator are independent of $\theta$ \cite{homma97}. 
The $\theta$-dependence of $R_{\lambda,R}^\theta(E)$ and $R_{\lambda,k}^\theta(E)$ originates from $n_r^\theta$, $L^k_\theta$, and $E_\theta$.
The strength function $S_{\lambda}(E)$ is an observable with positive definite and also independent of $\theta$.

%%%%%%%%%%%%%%%%%%%%%%%%%%%%%%%%%%%%%
%%%%%   Merits in Calculation   %%%%%
We can calculate each term of Eq.~(\ref{eq:sec2_strength-3}). 
Owing to the decomposition of the final state contributions, we can unambiguously investigate which state determines the structure observed in the strength function. 
This is a prominent feature of the CSM and is applicable to the many-body cases such as the Borromean three-body systems, as shown in \S\ref{sec:many}.
                                                                               
%%%%%%%%%%%%%%%%%%%%%%%%%%%%%%
\begin{figure}[t] 
\centering
\includegraphics[width=9.0cm,clip]{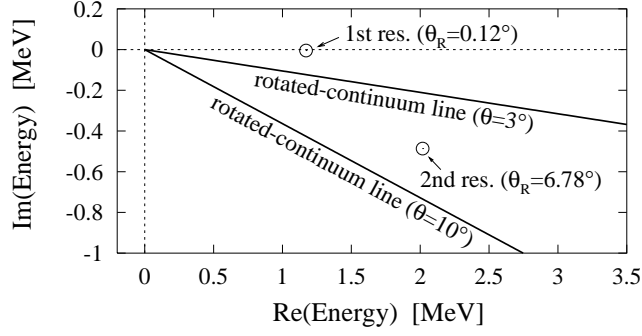}
\caption{Positions of the first and second resonances of $1^-$ states in the complex energy plane with rotated-continuum lines using $\theta$=3$^\circ$ and 10$^\circ$.}
\label{fig:sec2_csoto-pole}
\end{figure}
%%%%%%%%%%%%%%%%%%%%%%%%%%%%%%

We demonstrate the calculation of the strength function using the CSM.
We show the strength function of the $E1$ transition in the simple potential model \cite{homma97,myo98,csoto90} as
\begin{eqnarray}
H&=&-\frac{1}{2}\nabla^2+V(r),
\hspace*{0.8cm}
V(r)~=~-8e^{-0.16r^2}+4e^{-0.04r^2}.
\label{eq:sec2_Gyar}
\end{eqnarray}
In this model, we calculate the $J^\pi=0^+$ and 1$^-$ states.
For $J^\pi=1^-$, with the scaling angle $\theta$ selected as 3$^\circ$, we obtain a bound state ($E_{b.s.}$=$-$0.68 MeV), 
a resonance ($E_1=1.17-i0.49\times10^{-2}$ MeV), and the non-resonant continuum states.
As seen in Fig.~\ref{fig:sec2_csoto-pole}, for such a small $\theta$, we cannot obtain the second resonance.
On the other hand, when $\theta$ is selected as $10^\circ$, one more resonance ($E_2=2.02-i0.49$ MeV) is divided from the continuum states.
                                                                                
%%%%%%%%%%%%%%%%%%%%%%%%%%%%%%
\begin{figure}[t]
\centering
\includegraphics[width=13.0cm,clip]{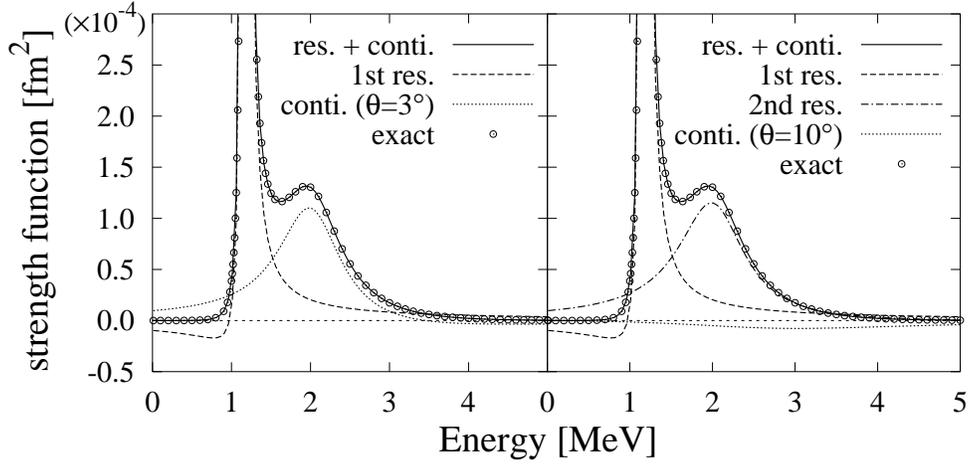}
\caption{$E1$ transition strength ($0^+ \to 1^-$) using the schematic potential with $\theta$=3$^\circ$ (left) and 10$^\circ$ (right). 
The dashed and dash-dotted curves represent the components of the first and second resonances, respectively. 
The dotted curve represent the continuum term, and the solid curves represent the sum of the all contributions. The open circles represent the exact calculations.}
\label{fig:sec2_csoto-E1}
\end{figure}
%%%%%%%%%%%%%%%%%%%%%%%%%%%%%%

In Fig.~\ref{fig:sec2_csoto-E1}, we show the dipole strengths from the ground $0^+$ state into 1$^-$ state. 
We compare the strengths with the exact results (open circles). A sharp peak is observed just above 1 MeV. 
The dashed curve denotes the strength to the first resonance. We conclude that the sharp peak arises from the first resonance. 
The left part of the figure is the case of small $\theta$ of $3^\circ$, and shows 
the continuum contribution (dotted curve) forming a second peak at 2 MeV. 
The right part of the figure shows the case where the larger $\theta$ of 10$^\circ$ is selected to obtain the second resonance; 
here, the contribution of the second resonance (dash-dotted curve) becomes a major component, 
whereas the residual continuum contribution (dotted curve) becomes very small. 
Thus, we conclude that the two major peaks of the $E1$ strengths come from two 1$^-$ resonances.
                                                                                
\section{Many-body resonances and non-resonant continuum states} \label{sec:many}

In this section, we show the results of application of the CSM to the spectroscopy of unstable nuclei.
We investigate the many-body resonances beyond the two-body case observed in the unstable nuclei.
We select the application to the neutron-rich He isotopes and their mirror proton-rich nuclei, 
most of the states of which are unbound owing to the weak binding nature of valence nucleons to the $\alpha$ particle.
In the theoretical model, we treat the $\alpha$ particle as a core nucleus and solve the motions of valence neutrons for the neutron-rich He isotopes and of valence protons for the proton-rich case.
We also give the results of the coupled channel model of $^{11}$Li with multi-configuration of the $^9$Li core.
This model dynamically explains the halo formation of $^{11}$Li and the importance of the tensor and pairing correlations are discussed.

%%%%%%%%%%%%%%%%%%%%%%%%%%%%%%%%%%%%%%%%%%%%%%%%%%%%%%%%%%%%%%%%%%%%%%%%%%%%%%%%%
\subsection{CS-COSM for resonant and non-resonant continuum states} \label{sec:COSM}
We briefly explain the theoretical method of describing the resonance and non-resonant continuum states of He isotopes and their mirror nuclei on the basis of the cluster model with an $\alpha$ core.
To describe the unbound states, it is important to treat the boundary condition of the multi-particle emission and to solve the relative motion of each particle or cluster.
For this purpose, the cluster model is suitable; we use the cluster orbital shell model (COSM) \cite{suzuki88} with complex scaling, called as CS-COSM, hereafter.
In CS-COSM, it is easy to extend the model to include the excitations of the core nucleus in terms of the multi-configuration representation \cite{myo07b}.

We use the CS-COSM using the following Hamiltonian with $\theta=0$ for the system consisting of $\alpha$ and valence nucleons \cite{myo09b,myo07a,myo10}:
\begin{eqnarray}
	H_{\rm CS-COSM}
&=&	\sum_{i=1}^{N_{\rm v}+1} t_i - T_G + \sum_{i=1}^N V^{\alpha N}_i + \sum_{i<j}^{N_{\rm v}}   V^{NN}_{ij} ,
	\\
&=&	\sum_{i=1}^{N_{\rm v}} \left[ \frac{\vc{p}^{ 2}_i}{2\mu} + V^{\alpha N}_i \right] + \sum_{i<j}^{N_{\rm v}} \left[ \frac{\vc{p}_i\cdot \vc{p}_j}{(A_c+1)\mu} + V^{NN}_{ij} \right] ,
        \label{eq:sec3_COSM_ham}
\end{eqnarray}
where $t_i$ and $T_G$ are the kinetic energies of each particle ($N$ and $\alpha$) and of the center of mass of the total system, respectively.
The operator $\vec{p}_i$ is the relative momentum between a valence nucleon and $\alpha$. 
The reduced mass $\mu$ is $A_c m/(A_c+1)$, where $m$ is the nucleon mass and $A_c$, the mass number being 4 of the $\alpha$ core. 
The number $N_{\rm v}$ is a valence nucleon number around $\alpha$.

In the CS-COSM, the total wave function $\Psi^{JT}_{\rm CS\mbox{-}COSM}$ with mass number $A$, a spin $J$, and an isospin $T$ is represented by the superposition of the various configurations $\Phi^{JT}_c (A)$ as
\begin{eqnarray}
    \Psi_{\rm CS\mbox{-}COSM}^{JT}
&=& \sum_c C^J_c\, \Phi^{JT}_c(A),
    \label{eq:sec3_COSM-WF0}
    \qquad
    \Phi^{JT}_c
~=~ \prod_{i=1}^{N_{\rm v}} a^\dagger_{\kappa_i}|0\rangle , 
    \label{eq:sec3_COSM-WF1}
\end{eqnarray}
where the vacuum $|0\rangle$ represents the $\alpha$ particle.
The creation operator $a^\dagger_{\kappa}$ denotes the single particle state of a valence nucleon above $^4$He,
with the quantum number $\kappa=\{n,\ell,j,t_z\}$ in a $jj$ coupling scheme.
Here, the index $n$ is used to distinguish the different radial component of the single particle state.
The $z$ component of the isospin of each nucleon is given as $t_z$. 
We describe the radial component of the single nucleon wave function using the Gaussian expansion method \cite{aoyama06,hiyama03}.
The index $c$ represents the set of $\kappa$ as $c=\{\kappa_1,\ldots,\kappa_{N_{\rm v}}\}$.
The expansion coefficients $\{C_c^J\}$ in Eq.~(\ref{eq:sec3_COSM-WF0}) are determined 
by the diagonalization of the Hamiltonian matrix elements.

%%%%%%%%%%%%%%%%%%%%%%%%%%%%%%
\begin{figure}[t]
\centering
\includegraphics[width=14.5cm,clip]{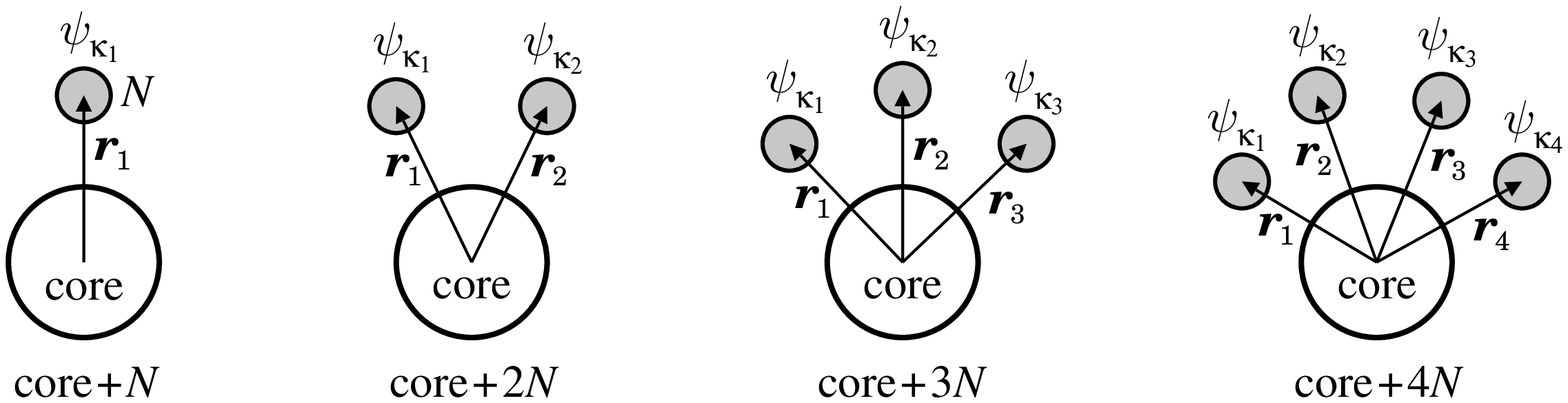}
\caption{Sets of the spatial coordinates in the CS-COSM for the core nucleus plus valence nucleons system.}
\label{fig:sec3_COSM}
\end{figure}
%%%%%%%%%%%%%%%%%%%%%%%%%%%%%%

The coordinate representation of a single particle state corresponding to $a^\dagger_{\kappa}$ is given by 
$\psi_{\kappa}(\vc{r})$ and is a function of the relative coordinate $\vc{r}$ between the center of mass of the $\alpha$ core
and a valence nucleon \cite{suzuki88}.
The schematic illustration of the coordinate set is shown in Fig.~\ref{fig:sec3_COSM} for a general case.
In this model, the radial part of $\psi_{\kappa}(\vc{r})$ is expanded with the Gaussian basis functions for each orbit as
\begin{eqnarray}
    \psi_{\kappa}(\vc{r})
&=& \sum_{k=1}^{N_{\ell j}} d^k_{\kappa}\ \phi_{\ell j t_z}^k(\vc{r},b_{\ell j}^k),
    \label{eq:sec3_COSM-base1}
    \\
    \phi_{\ell j t_z}^k(\vc{r},b_{\ell j}^k)
&=& {\cal N}\, r^{\ell} e^{-(r/b_{\ell j}^k)^2/2} [Y_{\ell}(\hat{\vc{r}}),\chi^\sigma_{1/2}]_{j}\ \chi^\tau_{t_z},
    \label{eq:sec3_Gauss}
	\\
    \langle \psi_{\kappa}|\psi_{\kappa'} \rangle 
&=& \delta_{\kappa,\kappa'},
    \label{eq:sec3_COSM-base2}
\end{eqnarray}
where the index $k$ is used to distinguish the range parameter $b_{\ell j}^k$ of the Gaussian basis functions with the number $N_{\ell j}$. 
The normalization factor of the basis is given by ${\cal N}$.
The coefficients $\{d^k_{\kappa}\}$ in Eq.~(\ref{eq:sec3_COSM-base1}) are determined using the Gram-Schmidt orthonormalization,
and hence, the basis states $\psi_{\kappa}$ are orthogonal to each other, as shown in Eq.~(\ref{eq:sec3_COSM-base2}).
The numbers of the independent radial bases of $\psi_{\kappa}$ are at most $N_{\ell j}$ and are determined to converge the physical solutions.
The same method, using Gaussian bases as a single-particle basis, is employed in the tensor-optimized shell model \cite{myo09a,myo11a}.
The antisymmetrization between a valence nucleon and $\alpha$ is treated on the orthogonality condition model \cite{aoyama06}, 
in which the single particle state $\psi_{\kappa}$ is imposed to be orthogonal to the $0s$ state occupied by an $\alpha$ core in the present case.
The length parameters $b_{\ell j}^k$ are chosen in geometric progression \cite{aoyama06,myo09b}.

We can easily apply the CSM to the Hamiltonian in Eq.~(\ref{eq:sec3_COSM_ham}) and the COSM wave function in Eq.~(\ref{eq:sec3_COSM-WF0}).
In the CS-COSM, all of the relative coordinates $\vc{r}_i$ between $\alpha$ and a valence nucleon are transformed into $\vc{r}_i\, e^{i\theta}$
for $i=1,\ldots,N_{\rm v}$; here, $\theta$ is a scaling angle.
The Hamiltonian in Eq.~(\ref{eq:sec3_COSM_ham}) is transformed into the complex-scaled Hamiltonian $H^\theta$, and the complex-scaled Schr\"odinger equation is given as
\begin{eqnarray}
    H_{\rm CS\mbox{-}COSM}^\theta \Psi^{JT,\, \theta}_{\rm CS\mbox{-}COSM}
&=& E_{JT}^\theta \Psi^{JT,\, \theta}_{\rm CS\mbox{-}COSM}
\end{eqnarray}
with the eigenstates $\Psi^{JT,\, \theta}_{\rm CS\mbox{-}COSM}$.
The expansion coefficients are determined by solving the eigenvalue problem of $H^\theta$ with the CS-COSM basis functions,
\begin{eqnarray}
    \Psi_{\rm CS\mbox{-}COSM}^{JT,\,\theta}
&=& \sum_c C^{JT,\, \theta}_c \Phi^{JT}_c(A) .
\end{eqnarray}
The energy eigenvalues $E_{JT}^\theta$ are obtained on a complex energy plane for each spin $J$ and isospin $T$.
In the CS-COSM, we adopt a finite number of the basis states, which provides a discretized representation of the continuum states in the CSM.

We explain the case of $^7$He with the isospin $T=3/2$, which has no bound states, consisting of $\alpha$ and three valence neutrons.
The various states of $^7$He consist of the four-body extended completeness relation \cite{myo09b} as
\begin{eqnarray}
	{\bf 1}
&=&	\sum_{~\nu} \kets{\Psi^\theta_\nu}\bras{\tilde{\Psi}^\theta_\nu} , 
        \\
&=&	\{\mbox{Four-body resonances of $^7$He}\}
        \nonumber\\
&+&	\{\mbox{Two-body continuum states consisting of $^6$He$^{(*)}$+$n$}\}
        \nonumber\\
&+&	\{\mbox{Three-body continuum states consisting of $^5$He$^{(*)}$+$n$+$n$}\}
        \nonumber\\
&+&	\{\mbox{Four-body continuum states consisting of $\alpha$+$n$+$n$+$n$}\} , 
\end{eqnarray}
where 
$\{ \Psi^\nu_\theta,\tilde{\Psi}^\nu_\theta \}$ form a set of biorthogonal bases with a state $\nu$.
This relation is necessary to calculate the strength distribution into the four-body unbound states of $^7$He.

%%%%%%%%%%%%%%%%%%%%%%%%%%%%%%%%%%%%%%%%%%%%%%%%%%%%
\subsection{He isotopes and their mirror nuclei} \label{sec:He-COSM}

In this part, we show the application of the CS-COSM to the analysis of the structures of He isotopes and their mirror nuclei.
We solve the motions of valence nucleons around the $\alpha$ particle not only for bound states but also for resonant and non-resonant continuum states using the CSM.
The mirror symmetry is also discussed between the proton-rich and neutron-rich nuclei in relation to the role of the Coulomb interaction in proton-rich nuclei.

\subsubsection{Energy levels of He isotopes and their mirror nuclei} \label{sec:ene_COSM}

Before discussions on the individual nuclei, we show a systematics of energy levels observed experimentally and calculated by CS-COSM for He isotopes and their mirror nuclei.  
In the Hamiltonian of the CS-COSM, two kinds of interactions between core-$N$ and $N$-$N$ are necessary.
In the present analysis with an $\alpha$ core, the $\alpha$-$n$ interaction $V^{\alpha n}$ is given by the microscopic KKNN potential \cite{aoyama06,kanada79}, 
in which the tensor correlation of $\alpha$ is renormalized using the resonating group method for the $\alpha$+$N$ scattering.
We use the Minnesota potential \cite{tang78} as the nuclear part of $V^{NN}$ in addition to the Coulomb interaction.
In the wave function, the $\alpha$ core is treated as the $(0s)^4$ configuration of a harmonic oscillator wave function.
For the single-particle states in the CS-COSM, we take the angular momenta $\ell\le 2$ and adjust the two-neutron separation energy of $^6$He($0^+$) to the experimental value of 0.975 MeV \cite{myo09b,myo07b,myo10}.

\begin{figure}[t]
\centering
\includegraphics[width=8.2cm,clip]{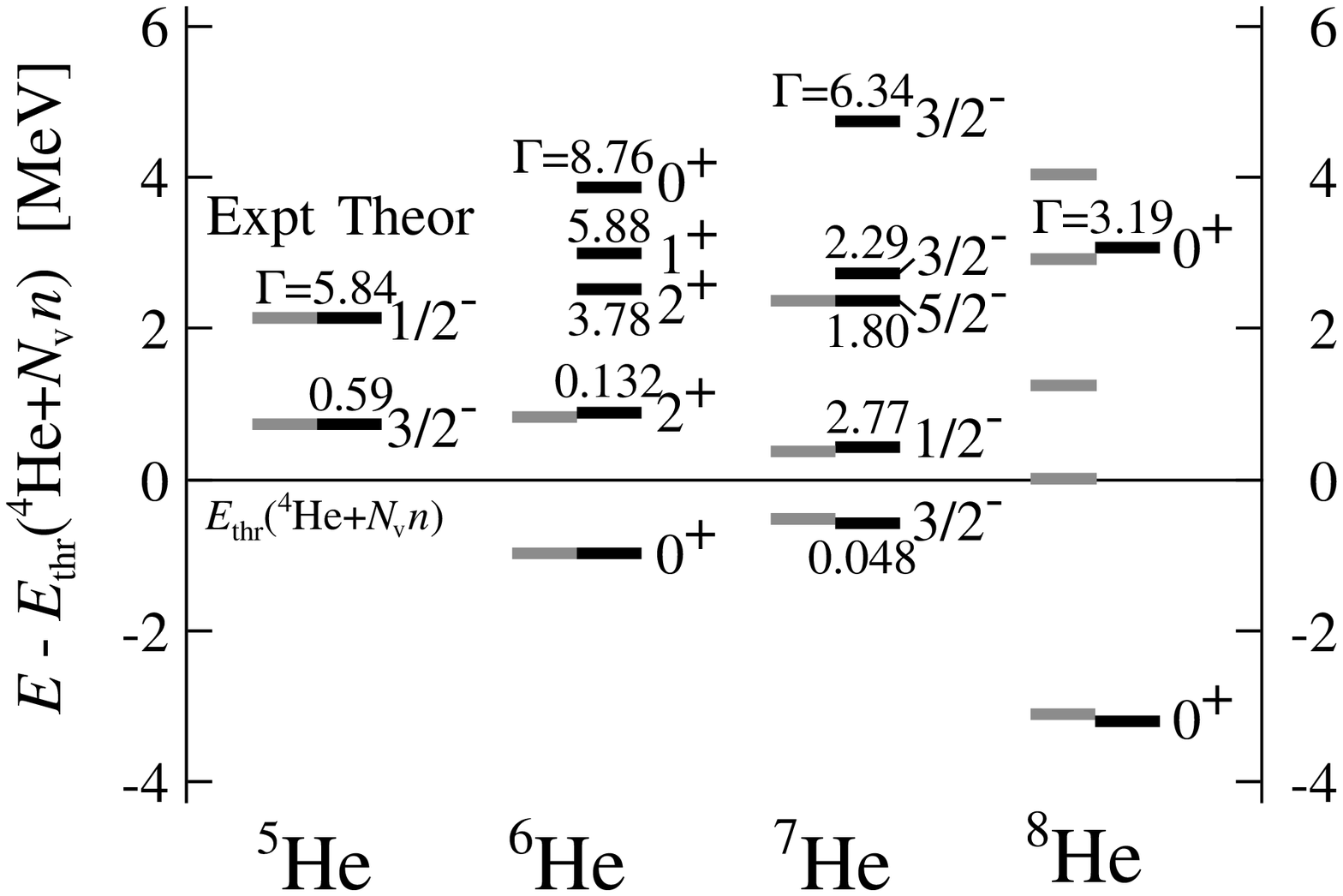}\hspace*{0.8cm}
\includegraphics[width=8.2cm,clip]{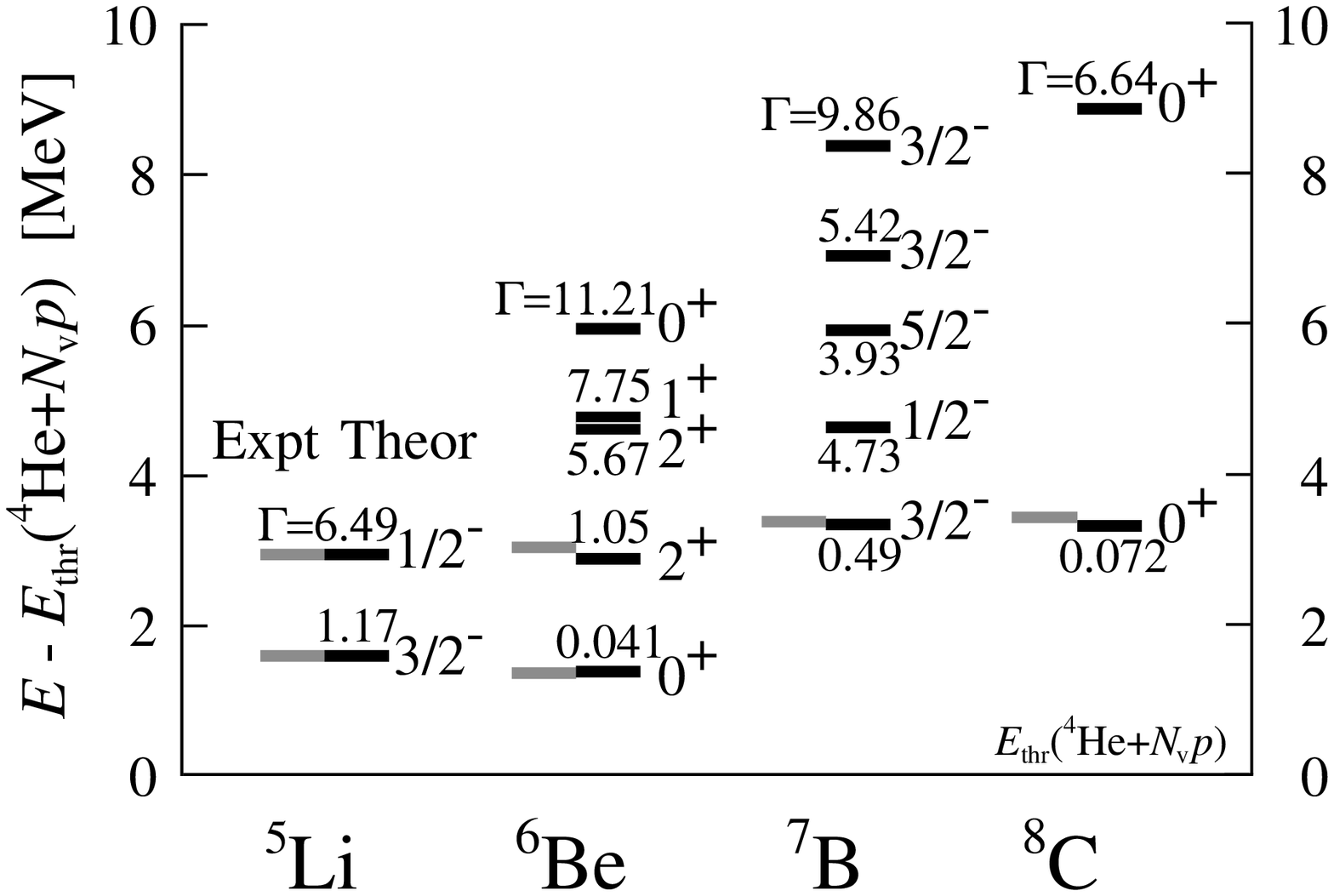}
\caption{
Energy levels of He isotopes (left) and their mirror nuclei (right) measured from the $\alpha$ particle emission. Units are in MeV.
The black and gray lines correspond to the theoretical and experimental results, respectively. Small numbers denote theoretical decay widths. 
For $^8$He and $^8$C, only the $0^+$ states are shown in theory. }
\label{fig:sec3_ene_COSM}
\end{figure}

\begin{figure}[t]
\begin{minipage}[l]{8.5cm}
\centering
\includegraphics[width=9.0cm,clip]{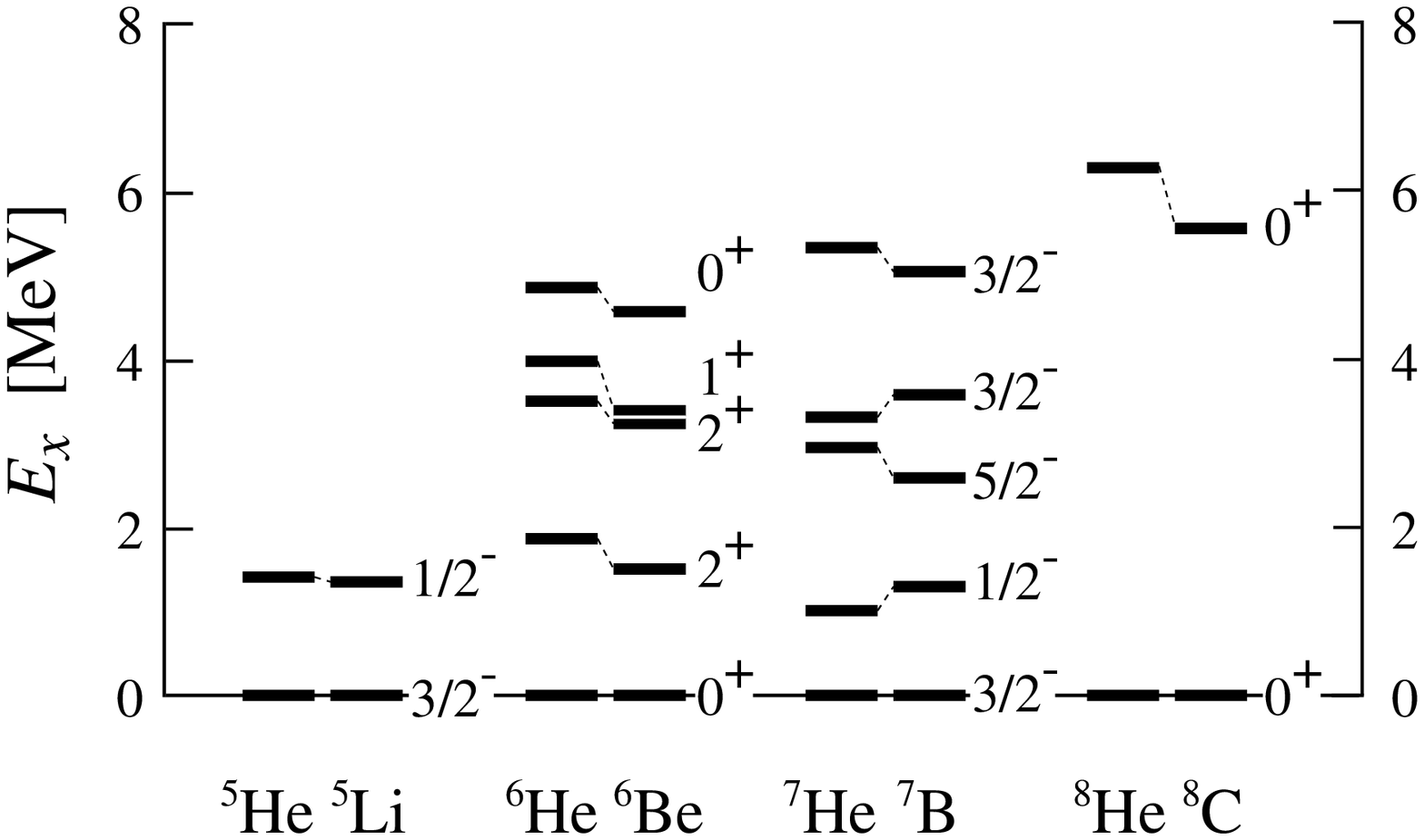}
\caption{Excitation energy spectra (MeV) of mirror nuclei of $A=5,6,7$, and 8.}
\label{fig:sec3_excite_COSM}
\end{minipage}
\hspace*{0.7cm}
\begin{minipage}[r]{9.0cm}
\centering
\includegraphics[width=8.8cm,clip]{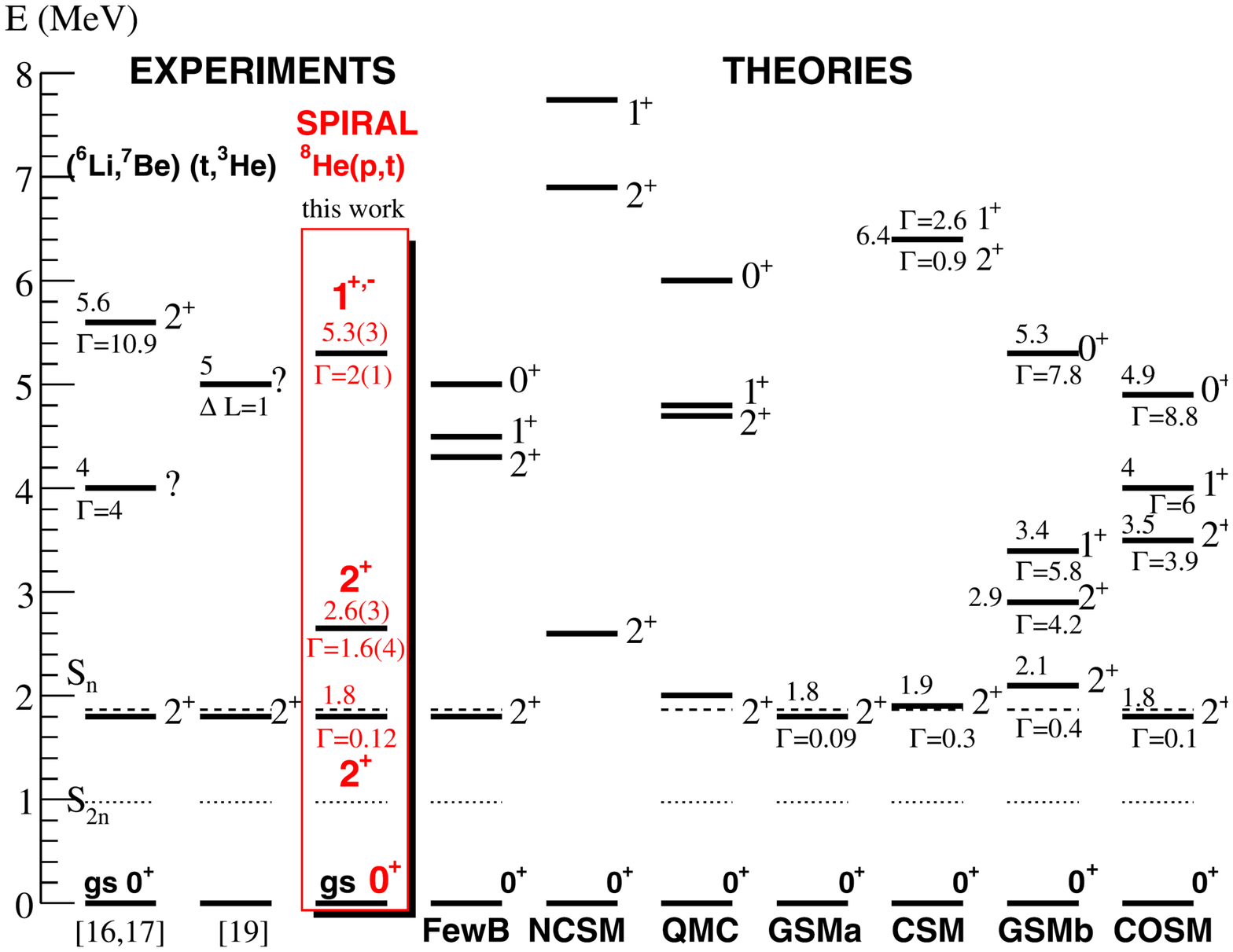}
\caption{Experimental results of $^6$He resonances taken from Fig. 6 of Mougeot's paper \cite{mougeot12}.}
\label{fig:sec3_6He_exp}
\end{minipage}
\end{figure}

Figure \ref{fig:sec3_ene_COSM} shows the energy levels of He isotopes and their mirror nuclei with the CS-COSM, 
measured from the energy of the $\alpha$ particle.
The small numbers near the energy levels represent the decay widths of the states.
A good agreement can be observed between the theoretically and experimentally obtained energy positions up to a five-body case of $^8$He and $^8$C, 
in which only the $0^+$ states are shown.
It is found that the order of energy levels is the same for the neutron-rich and proton-rich sides.
Further, there are several theoretical predictions of the excited states.
In Fig. \ref{fig:sec3_excite_COSM}, we compare the energy spectra of neutron-rich and proton-rich nuclei.
A good symmetry is confirmed between the corresponding nuclei.
The differences in the excitation energies for individual energy levels are less than 1 MeV \cite{myo12b}.
Experimentally, the search for new resonances in drip-line nuclei are the current topic of research in unstable nuclei.
We show the example of the $^6$He($1^+$) results of SPIRAL \cite{mougeot12} in comparison with the results obtained using various theories including our CS-COSM (simply COSM) in Fig.~\ref{fig:sec3_6He_exp}.

The RMS radii of the matter and charge for the ground states of $^6$He and $^8$He are calculated in the CS-COSM, 
where the explicit form of the radius operator is provided in Ref. \cite{suzuki88}.
The values are shown in Table~\ref{tab:sec3_radius_COSM} and reproduce the recent experiments.
Hence, the CS-COSM wave functions describe the specially extended distributions of neutrons in halo and skin structures observed in the He isotopes. 

%%%%%%%%%%%%%%%%%%%%%%%%%%%%%%%%%%%%%%%%%%%%%%%%%%%%%%%%%%
%%%%%%%%%%%%%%%%%%%%%%%%%%%%%%%%%%%%%%%%%%%%%%%%%%%%%%%%%%
%%%%%%%%%%%%%%%%%%%%%%%%%%%%%%%

%%%%%%%%%%%%%%%%%%%%%%%%%%%%%%
\nc{\lw}[1]{\smash{\lower1.5ex\hbox{#1}}}
\begin{table}[thb]
\caption{Matter ($R_{\rm m}$) and charge ($R_{\rm ch}$) RMS radii of $^6$He and $^8$He
in comparison with the experimental values; a\cite{tanihata92}, b\cite{alkazov97}, c\cite{kiselev05}, and d\cite{mueller07}.
Units are in fm.}
\label{tab:sec3_radius_COSM}
\centering
%\small
\begin{tabular}{r p{2.0cm} p{5.0cm}}
\hline
                     & Present  & Experiments        \\ 
\hline
\lw{$^6$He}~~~$R_{\rm m}$  &  2.37  & 2.33(4)$^{\rm a}$~~~~2.30(7)$^{\rm b}$~~~~2.37(5)$^{\rm c}$ \\
              $R_{\rm ch}$ &  2.01  & 2.068(11)$^{\rm d}$ \\
\hline
\lw{$^8$He}~~~$R_{\rm m}$  &  2.52  & 2.49(4)$^{\rm a}$~~~~2.53(8)$^{\rm b}$~~~~2.49(4)$^{\rm c}$ \\
              $R_{\rm ch}$ &  1.92  & 1.929(26)$^{\rm d}$ \\
\hline
\end{tabular}
\end{table}
%%%%%%%%%%%%%%%%%%%%%%%%%%%%%%

%%%%%%%%%%%%%%%%%%%%%%%%%%%%%%%%%%%%%%%%
\subsubsection{$^6$He and $^6$Be}\label{sec:A=6}

%%%%%%%%%%%%%%%%%%%%%%%%%%%%%%
\begin{table}[thb]
\centering
\begin{tabular}{cc}
\begin{minipage}{0.4\hsize}
\centering
\caption{Components of the ground states of $^{6}$He and $^6$Be.}
\vspace*{0.3cm}
\label{tab:sec3_comp6_0}
\begin{tabular}{cccc}
\hline
Config.         &  $^6$He($0^+_1$) & $^6$Be($0^+_1$) \\ 
\hline
 $(p_{3/2})^2$  &  0.917           & $0.918-i0.006$  \\
 $(p_{1/2})^2$  &  0.043           & $0.041+i0.000$  \\
 $(1s_{1/2})^2$ &  0.009           & $0.010+i0.006$  \\
 $(d_{5/2})^2$  &  0.024           & $0.024+i0.000$  \\
 $(d_{3/2})^2$  &  0.007           & $0.007+i0.000$  \\
\hline
\end{tabular}
\end{minipage}
\hspace*{0.8cm}
\begin{minipage}{0.45\hsize}
\centering
\caption{Dominant components of the $2^+_1$ states of $^{6}$He and $^6$Be.}
\vspace*{0.3cm}
\label{tab:sec3_comp6_2}
\begin{tabular}{cccc}
\hline
Config.               & $^6$He ($2^+_1$) & $^6$Be($2^+_1$) \\ 
\hline
 $(p_{3/2})^2$        & $0.898+i0.013$   & $0.891+i0.030$  \\
 $(p_{3/2})(p_{1/2})$ & $0.089-i0.013$   & $0.097-i0.024$  \\
\hline
\end{tabular}
\end{minipage}
\end{tabular}
\end{table}
%%%%%%%%%%%%%%%%%%%%%%%%%%%%%%
%%%%%%%%%%%%%%%%%%%%%%%%%%%%%%
\begin{table}[thb]  % 6He/Src05.4/02_0+ & 6Be/Src05/02_0+
\caption{Radial properties of the ground states of $^6$He and $^6$Be in units of fm,
in comparison with the experimental values for $^6$He; a\cite{tanihata92}, b\cite{alkazov97}, c\cite{kiselev05}, and d\cite{mueller07}.}
\label{tab:sec3_radius6}
\centering
\begin{tabular}{cccc}
\hline
                        & $^6$He     &  $^6$Be         \\ \hline
$R_{\rm m}$             &~~~2.37~~~  &  2.80 + $i$0.17 \\ 
$R_p$                   & 1.82       &  3.13 + $i$0.20 \\
$R_n$                   & 2.60       &  1.96 + $i$0.08 \\
$R_{\rm ch}$            & 2.01       &  3.25 + $i$0.21 \\
$r_{NN}$                & 4.82       &  6.06 + $i$0.35 \\
$r_{{\rm c}\mbox{-}2N}$ & 3.15       &  3.85 + $i$0.37 \\
$\theta_{NN}$           & 74.6       &  75.3           \\
\hline
\end{tabular}
\end{table}
%%%%%%%%%%%%%%%%%%%%%%%%%%%%%%

We compare the structures of the mirror nuclei, $^6$He and $^6$Be. 
The $^6$He nucleus is a two-neutron halo nucleus and the $^6$Be nucleus is the unbound system. 
At their ground states, the detailed components of each configuration are listed in Table~\ref{tab:sec3_comp6_0},
which are the squared values of the amplitudes $\{C^J_c\}$ defined in Eq.~(\ref{eq:sec3_COSM-WF0}).
It is to be noted that the squared amplitude of the various configurations in the resonance wave functions becomes a complex number and 
its real part can give a physical meaning when the imaginary parts has a relatively small value.
The summation of the imaginary part becomes zero owing to the normalization of the states.
It is found that the ground states of two nuclei show a similar mixing of configurations, dominated by the $p$-shell configurations.
In Table~\ref{tab:sec3_comp6_2}, the configurations of the $2^+_1$ states of $^6$He and $^6$Be are shown.
A good correspondence is seen for the two dominant configurations of the $2^+_1$ states in two nuclei.
These results indicate that the mirror symmetry is maintained well in the configurations between $^6$Be and $^6$He.

Comparison of the spatial properties between $^6$He and $^6$Be is important for discussing the effect of the Coulomb repulsion in $^6$Be,
although the radius of $^6$Be becomes complex number because of the property of resonance.
Using Eq.~(\ref{eq:csme}), we obtain the radius of resonance.
We show the results of $^6$He and $^6$Be in Table \ref{tab:sec3_radius6}, for the matter ($R_{\rm m}$), proton ($R_p$), neutron($R_n$), and charge ($R_{\rm ch}$) parts; 
the relative distances between valence nucleons ($r_{NN}$) and between the $\alpha$ core and the center of mass of two valence nucleons ($r_{{\rm c}\mbox{-}2N}$); 
the opening angle between two nucleons ($\theta_{NN}$) at the center of mass of $\alpha$.
It is found that the values of $^6$Be are almost real, and so, the real parts can be regarded to represent the properties of the radius of $^6$Be.
The distances between valence protons and between the core and $2p$ in $^6$Be are larger than those for $^6$He by 26\% and 22\%, respectively. 
This result is considered to be caused by the Coulomb repulsion between the three constituents of $\alpha$+$p$+$p$ for $^6$Be.

%%%%%%%%%%%%%%%%%%%%%%%%%%%%%%%%%%%%%%%%
\subsubsection{$^7$He and $^7$B}\label{sec:A=7}

We discuss the structures of $^7$He and its mirror nucleus $^7$B within the $\alpha$+3$N$ four-body picture \cite{myo11b}. 
They are both unbound nuclei. Our CS-COSM predicts five resonances for each nucleus.
For $^7$He, the $3/2^-$ ground state is obtained by 0.40 MeV above the $^6$He ground state,
which agrees with the recent experiments of 0.44 MeV and 0.36 MeV \cite{skaza06}.
This state is the two-body resonance located in the energy range between the thresholds of $^6$He+$n$ and $\alpha$+$3n$. 
The other four states are the four-body resonances above the $\alpha$+$3n$ threshold energy. 
For $^7$B, the resonances are all located above the $\alpha$+$3p$ threshold, as shown in Fig.~\ref{fig:sec3_ene_COSM}, and are interpreted as four-body resonances.
The energy of the $^7$B ground state is obtained as $E_r=3.35$ MeV and agrees with the recent experimental value of $E_r=3.38(3)$ MeV \cite{charity11}.
The decay width is 0.49 MeV, which is good and slightly smaller than the experimental value of 0.80(2) MeV.
There is no experimental evidence of the excited states of $^7$B and the further experimental data are desirable.

It is meaningful to discuss the mirror symmetry between $^7$He and $^7$B consisting of the $\alpha$ core and three valence neutrons or protons.
We show the spectroscopic factors ($S$ factors) of one-nucleon removal from each nucleus, namely, the $^6$He-$n$ components of $^7$He and the $^6$Be-$p$ components of $^7$B.
These quantities are important to examine the coupling behaviors between the $A=6$ daughter nuclei and the last nucleon.
Before showing the results, we explain the calculation of the $S$ factors in the case of the $^6$He-$n$ components for $^7$He.
\begin{eqnarray}
    S^{J,\nu}_{J',\nu'}
&=& \sum_{\kappa} S^{J,\nu}_{J',\nu',\kappa}\, ,
\qquad
    S^{J,\nu}_{J',\nu',\kappa}
~=~ \frac{1}{2J+1} \langle \widetilde{\Phi}^{J'}_{\nu'}|| a_\kappa ||\Psi^J_\nu \rangle^2\, ,
    \label{eq:sec3_sfac}
\end{eqnarray}
where $a_{\kappa}$ is the annihilation operator for the valence neutron with the state $\kappa$. 
$J$ and $J'$ are the spins for $^7$He and $^6$He, respectively.
The index $\nu$ ($\nu'$) indicates the eigenstate of $^7$He ($^6$He).
In this expression, the $S$-factors $S^{J,\nu}_{J',\nu'}$ are allowed to be complex values.

The sum rule value of $S$ factors can be considered.
When we count all the obtained complex $S$-factors, not only of the resonant states but also of the non-resonant continuum states in the final states,
the summed value of $S$-factors satisfies the associated particle number being a real value.
For the $^6$He-$n$ decomposition of $^7$He, the summed value of the $S$-factors $S^{J,\nu}_{J',\nu'}$ in Eq.~(\ref{eq:sec3_sfac}), by considering all the $^6$He states, is given as
\begin{eqnarray}
    \sum_{J',\nu'}\ S^{J,\nu}_{J',\nu'}
&=& \sum_{\kappa,m}\ 
    \langle \widetilde{\Psi}^{JM}_\nu|a^\dagger_{\kappa,m} a_{\kappa,m}| \Psi^{JM}_\nu \rangle
~=~ 3\ ,
    \label{eq:sec3_sf-sum}
\end{eqnarray} 
where we use the completeness relation of $^6$He as
\begin{eqnarray}
    1
&=& \sum_{J',M'}\sum_{\nu'}\hspace*{-0.57cm}\int |\Phi^{J'M'}_{\nu'}\rangle \langle \widetilde{\Phi}^{J'M'}_{\nu'}|.
    \label{eq:sec3_3-ECR}
\end{eqnarray}
Here, the labels $M$ ($M'$) and $m$ are the $z$ components of the angular momenta of the wave functions of $^7$He ($^6$He) and 
of the creation and annihilation operators of valence neutrons, respectively.
The summed value of $S$-factors of $^6$He+$n$ over the various $^6$He states becomes the number of valence neutrons in $^7$He.

%%%%%%%%%%%%%%%%%%%%%%%%%%%%%%
\begin{table}[b]
\centering
\begin{tabular}{cc}
\begin{minipage}{0.42\hsize}
\centering
\caption{$S$-factors of the $^6$He-$n$ components in $^7$He. Details are described in the text.}
\label{tab:sec3_sf_He7}
\vspace*{0.3cm}
\begin{tabular}{cccc}
\hline
          & $^6$He($0^+_1$)-$n$ &  $^6$He($2^+_1$)-$n$ \\ \hline
$3/2^-_1$ & $0.63+i0.08$        &  $1.60-i0.49$   \\
$3/2^-_2$ & $0.00-i0.01$        &  $0.97+i0.01$   \\
$3/2^-_3$ & $0.01+i0.00$        &  $0.04-i0.01$   \\ 
$1/2^- $  & $0.95+i0.03$        &  $0.07-i0.02$   \\
$5/2^- $  & $0.00+i0.00$        &  $1.00+i0.01$   \\
\hline
\end{tabular}
\end{minipage}
\hspace*{1.0cm}
\begin{minipage}{0.42\hsize}
\centering
\caption{$S$-factors of the $^6$Be-$p$ components in $^7$B. Details are described in the text.}
\label{tab:sec3_sf_B7}
\vspace*{0.3cm}
\begin{tabular}{cccc}
\hline
          & $^6$Be($0^+_1$)-$p$ &  $^6$Be($2^+_1$)-$p$ \\ \hline
$3/2^-_1$ & $0.51+i0.02$        &~~$2.35-i0.15$   \\ 
$3/2^-_2$ & $0.02-i0.01$        &~~$0.96-i0.01$   \\
$3/2^-_3$ & $0.00+i0.01$        & $-0.01-i0.06$   \\ 
$1/2^- $  & $0.93-i0.02$        &~~$0.10-i0.01$   \\
$5/2^- $  & $0.00+i0.00$        &~~$1.04-i0.01$   \\
\hline
\end{tabular}
\end{minipage}
\end{tabular}
\end{table}
%%%%%%%%%%%%%%%%%%%%%%%%%%%%%%

\begin{figure}[t]
\centering
\includegraphics[width=7.8cm,clip]{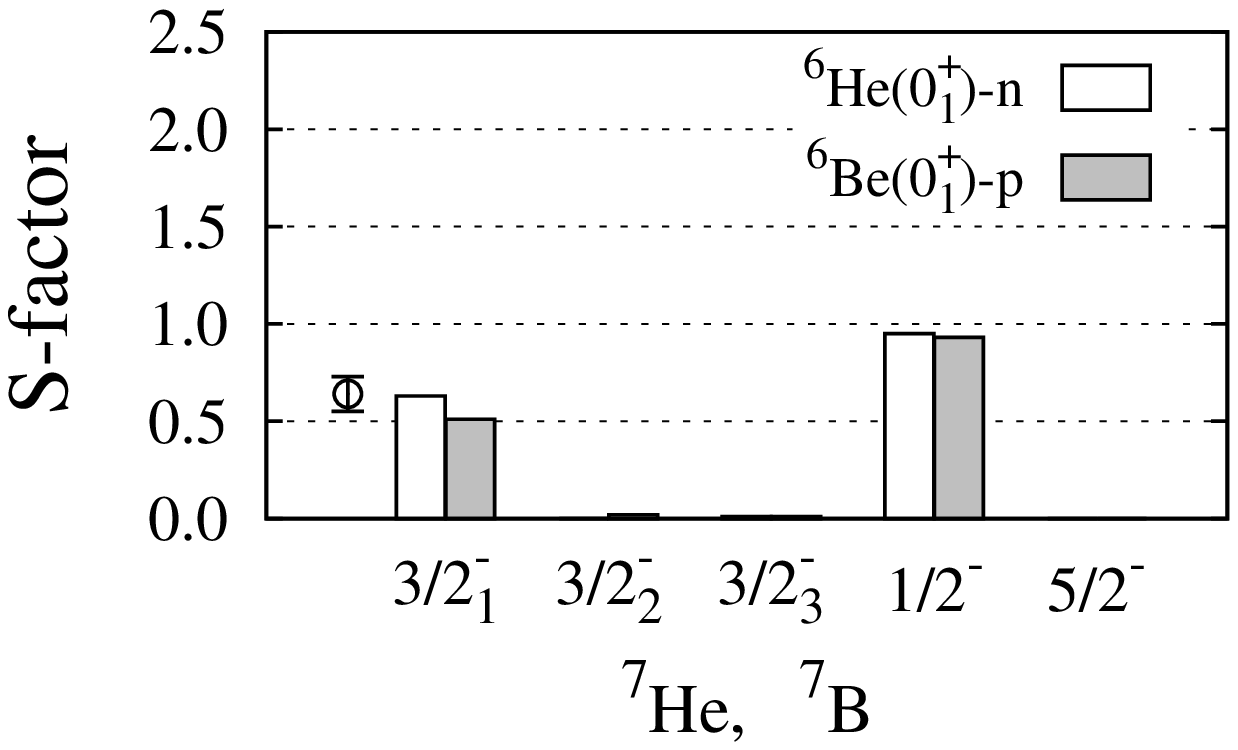}\hspace*{1.0cm}
\includegraphics[width=7.8cm,clip]{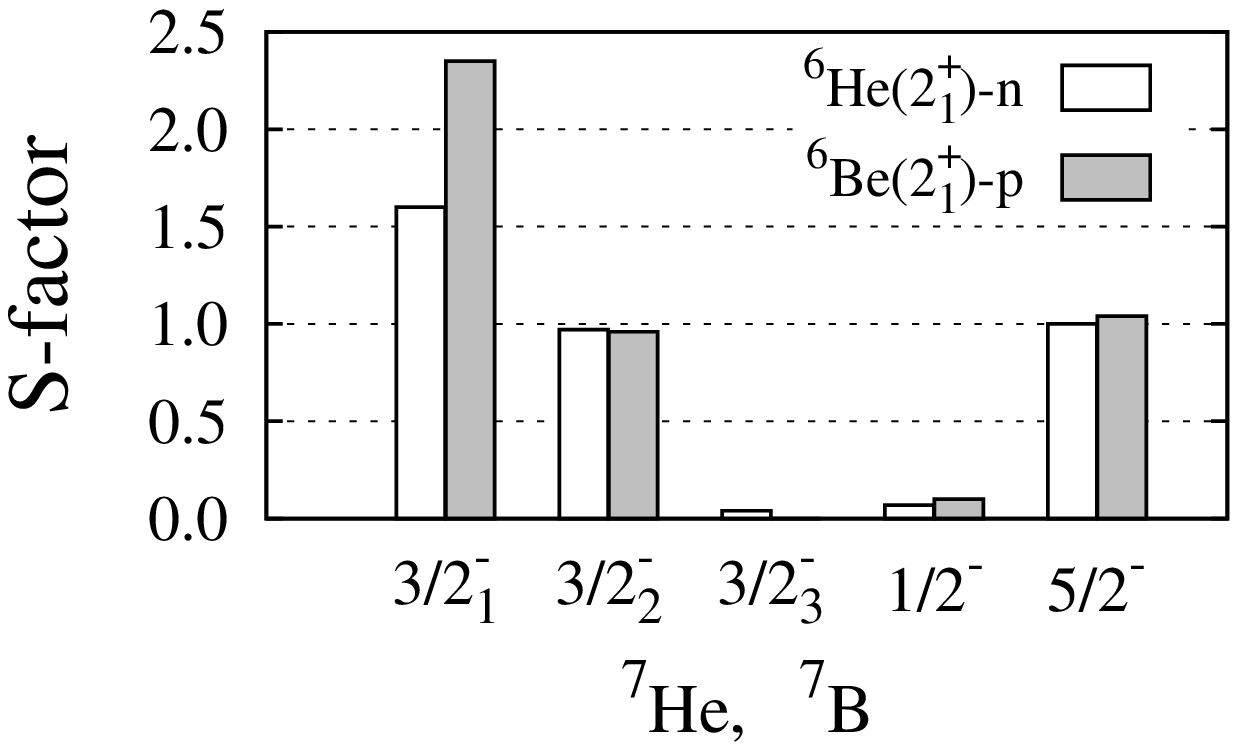}
\caption{$S$-factors of $^7$B and $^7$He, in which the $A=6$ daughter nuclei are the $0^+_1$ (left) and $2^+_1$ (right)  states.
The open circle denotes the experimental value for the $^7$He($3/2^-$) state \cite{beck07}.}
\label{fig:sec3_sfac}
\end{figure}

\begin{figure}[t]
\centering
\includegraphics[width=10.0cm,clip]{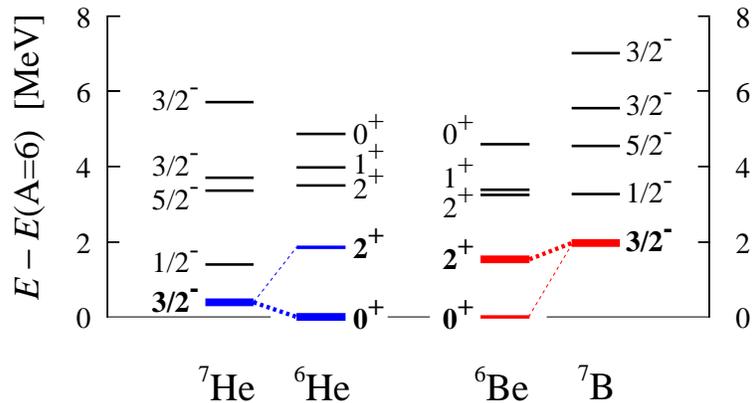}
\caption{(Color online) Relation between the excitation energies in $A$=7 and $A$=6 cases.}
\label{fig:sec3_ene_A7A6}
\end{figure}

We list the results of the $S$-factors of $^7$He and $^7$B in Tables~\ref{tab:sec3_sf_He7} and \ref{tab:sec3_sf_B7}, respectively.
Most of the components are found to show almost the real values in $^7$He and $^7$B.
Accordingly, the comparison of the real parts of the $S$-factors for $^7$He and $^7$B is shown in Fig.~\ref{fig:sec3_sfac}. 
From this figure, a sizable difference can be observed between the ground states of $^7$He and $^7$B in terms of the components including the $A=6$($2^+_1$) states.
The $^6$Be($2^+_1$)-$p$ component of $^7$B, obtained as $2.35$, is larger than the $^6$He($2^+_1$)-$n$ component of $^7$He, as $1.60$, by 47\% for the real part.
The other four excited states show similar behavior between the two nuclei  in Fig.~\ref{fig:sec3_sfac}.
These results indicate that the mirror symmetry breaking occurs only in their ground states.
The reason for the difference in the $2^+$ coupling is that the ground state of $^7$B is located close to the $^6$Be($2^+_1$) state by 0.45 MeV in the energy, as shown in Fig.~\ref{fig:sec3_ene_COSM} and also in Fig.~\ref{fig:sec3_ene_A7A6},
where the decay widths of the two states are rather small in comparison with the values of other resonances.
This situation does not occur in $^7$He; the energy difference between $^7$He($3/2^-_1$) and $^6$He($2^+_1$) is obtained as 1.46 MeV, 
as shown in Fig.~\ref{fig:sec3_ene_A7A6}. 
The small energy difference between $^7$B and $^6$Be($2^+_1$) enhances the $^6$Be($2^+_1$)-$p$ component of $^7$B 
because of the increase of coupling to the open channel of the $^6$Be($2^+_1$)+$p$ threshold.
On the other hand, the $^6$Be($0^+_1$)-$p$ component of $^7$B becomes smaller than the value of $^7$He by 24 \%, as shown in Fig.~\ref{fig:sec3_sfac},
because the energy difference between the ground states of $^7$B and $^6$Be is obtained as 1.97 MeV, larger than that in the case of $^7$He, 0.40 MeV.
The difference of the $S$-factors in $^7$He and $^7$B originates from the Coulomb repulsion, which acts to shift the entire energies of the $^7$B states up.

In conclusion, the mirror symmetry is broken in the ground states of $^7$He and $^7$B, 
while the excited states of the two nuclei retain the symmetry.
It is desired to experimentally observe the $2^+$ components of $A=6$ nuclei for $^7$He and $^7$B and examine the mirror symmetry in two nuclei.

%%%%%%%%%%%%%%%%%%%%%%%%%%%%%%
%%%%%%%%%%%%%%%%%%%%%%%%%%%%%%%

%%%%%%%%%%%%%%%%%%%%%%%%%%%%%%
%%%%%%%%%%%%%%%%%%%%%%%%%%%%%%%

In relation to the $S$-factors, we calculate the strength function $S(E)$ for the one-neutron removal of $^7$He$(3/2^-_1)$ to form $^6$He
using Eq.~(\ref{eq:sec3_strength2}), as a function of the energy $E$ of $^6$He \cite{myo09b}. 
We explain the one-neutron removal strength of $^7$He to form $^6$He in detail.
This is a function of the real energy of $^6$He, $E$.
We first introduce the three-body Green's function ${\cal G}(E,\vc{\eta},\vc{\eta}')$ of $^6$He, which is used to derive the strength function.
The procedure is the same as that for the electric transitions \cite{myo01,myo03}.
The coordinates $\vc{\eta}$ and $\vc{\eta}'$ represent the set of $\vc{r}_i$ ($i=1,\ldots,X$) in Fig.~\ref{fig:sec3_COSM}.
Here, we introduce the complex-scaled Green's function ${\cal G}^\theta(E,\vc{\eta},\vc{\eta}')$ of $^6$He as
\begin{eqnarray}
        {\cal G}^\theta(E,\vc{\eta},\vc{\eta}')
&=&     \left\bra \vc{\eta} \left|
        \frac{ {\bf 1} }{ E-H_\theta }\right|\vc{\eta}' \right\ket
~=~     \sum_\nu\hspace*{-0.57cm}\int\
        \frac{\Phi^\theta_\nu(\vc{\eta})\ [\tilde{\Phi}^*_\nu(\vc{\eta}')]^\theta}{E-E_\nu^\theta}
~=~     \sum_\nu\hspace*{-0.57cm}\int\
        {\cal G}^\theta_\nu(E,\vc{\eta},\vc{\eta}')\, .
        \label{eq:sec3_green_many}
\end{eqnarray}
In the derivation in Eq.~(\ref{eq:sec3_green_many}), we insert the ECR of $^6$He given in Eq.~(\ref{eq:sec3_3-ECR}) with the CSM,
where the $^6$He eigenenergy, $E_\nu^\theta$, corresponds to the eigenstate $\Phi^\theta_\nu$.

Next, the strength function $S(E)$ for the annihilation operator $a_{\kappa}$ with the nucleon state $\kappa$ 
is defined using Green's function in a usual case without the CSM as
\begin{eqnarray}
        S(E) 
&=&     \sum_\kappa S_\kappa(E) , \qquad
        \label{eq:sec3_spec}
        S_\kappa(E) 
~=~     \sum_\nu \hspace*{-0.57cm}\int\
        \bras{\tilde{\Psi}_0}a^\dagger_\kappa \kets{\Phi_\nu}\bras{\tilde{\Phi}_\nu} a_\kappa \kets{\Psi_0}\
        \delta(E-E_\nu) .
        \label{eq:sec3_strength0}
\end{eqnarray}
The wave function $\Psi_0$ is the ground state of $^7$He.
To calculate the strength function $S_\kappa(E)$ in Eq.~(\ref{eq:sec3_strength0}), 
we operate the complex scaling on $S_\kappa(E)$ and use the complex-scaled Green's function of Eq.~(\ref{eq:sec3_green_many}) as
\begin{eqnarray}
        S_\kappa(E)
&=&     -\frac1{\pi}\ {\rm Im}
        \left[
        \int d\vc{\eta} d\vc{\eta}'                       \:
        [\tilde{\Psi}_0^*(\vc{\eta})]^\theta (a^\dagger_\kappa)^\theta\:
        \right.
        {\cal G}^\theta(E,\vc{\eta},\vc{\eta}')            \:
        a_\kappa^\theta \Psi^\theta_0(\vc{\eta}')
        \biggr]
~=~     \sum_\nu\hspace*{-0.57cm}\int\ S_{\kappa,\nu}(E)\, ,
        \label{eq:sec3_strength1}
        \\
        S_{\kappa,\nu}(E)
&=&     -\frac1{\pi}\ {\rm Im}
        \left[   \frac{
                \bras{\tilde{\Psi}_0^\theta}  (a^\dagger_\kappa)^\theta \kets{\Phi_\nu^\theta}
                \bras{\tilde{\Phi}_\nu^\theta} a_\kappa^\theta          \kets{\Psi_0^\theta}
          }{E-E_\nu^\theta}\right] .
        \label{eq:sec3_strength2}
\end{eqnarray}
In Eq.~(\ref{eq:sec3_strength2}), the strength function is calculated using the one-neutron removal matrix elements
$\bras{\tilde{\Phi}_\nu^\theta} a_\kappa^\theta\kets{\Psi^\theta_0}$.
Thus, the one-neutron removal strength $S_\kappa(E)$ is obtained as a function of the real energy $E$ of $^6$He.
When we discuss the structures of $S_\kappa(E)$, it is useful to decompose $S_\kappa(E)$
into each component $S_{\kappa,\nu}(E)$ by using the complete set of the final state $\nu$ of $^6$He.
We can categorize $\nu$ of $^6$He using the ECR in Eq.~(\ref{eq:sec3_3-ECR}).

%%%%%%%%%%%%%%%%%%%%%%%%%%%%%%
\begin{figure}[t]
\centering
\includegraphics[width=7.0cm,clip]{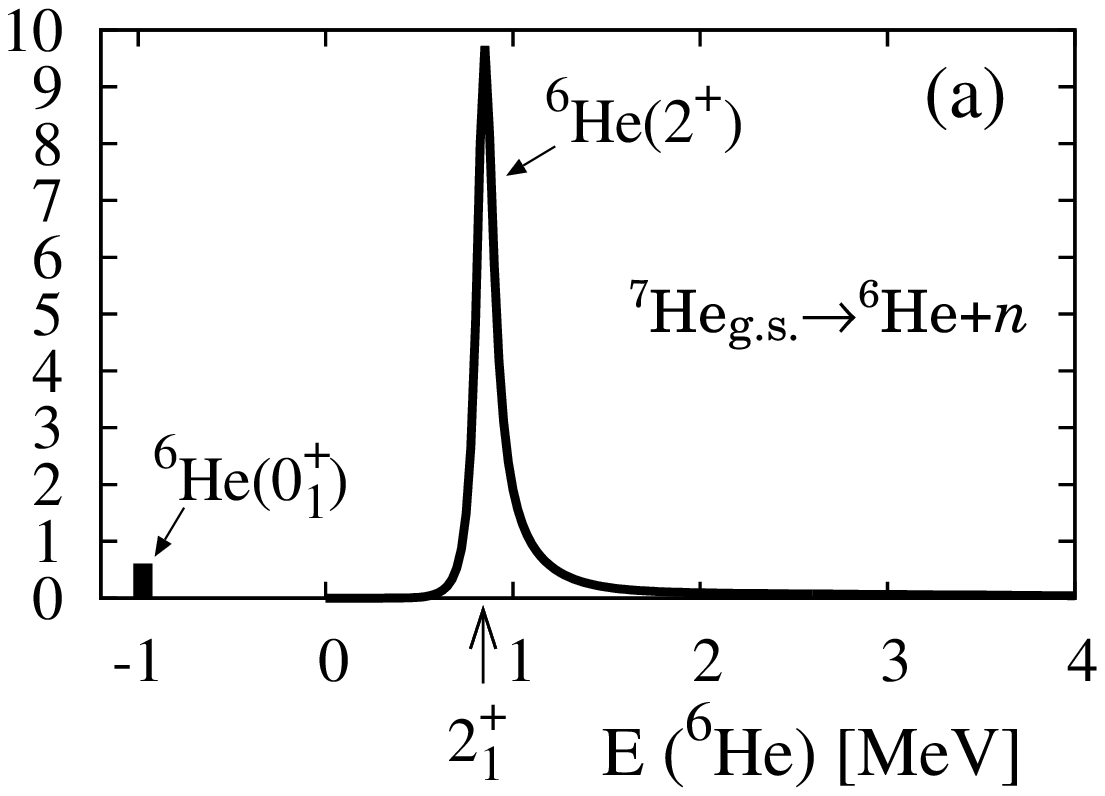}\hspace*{1.0cm}
\includegraphics[width=8.2cm,clip]{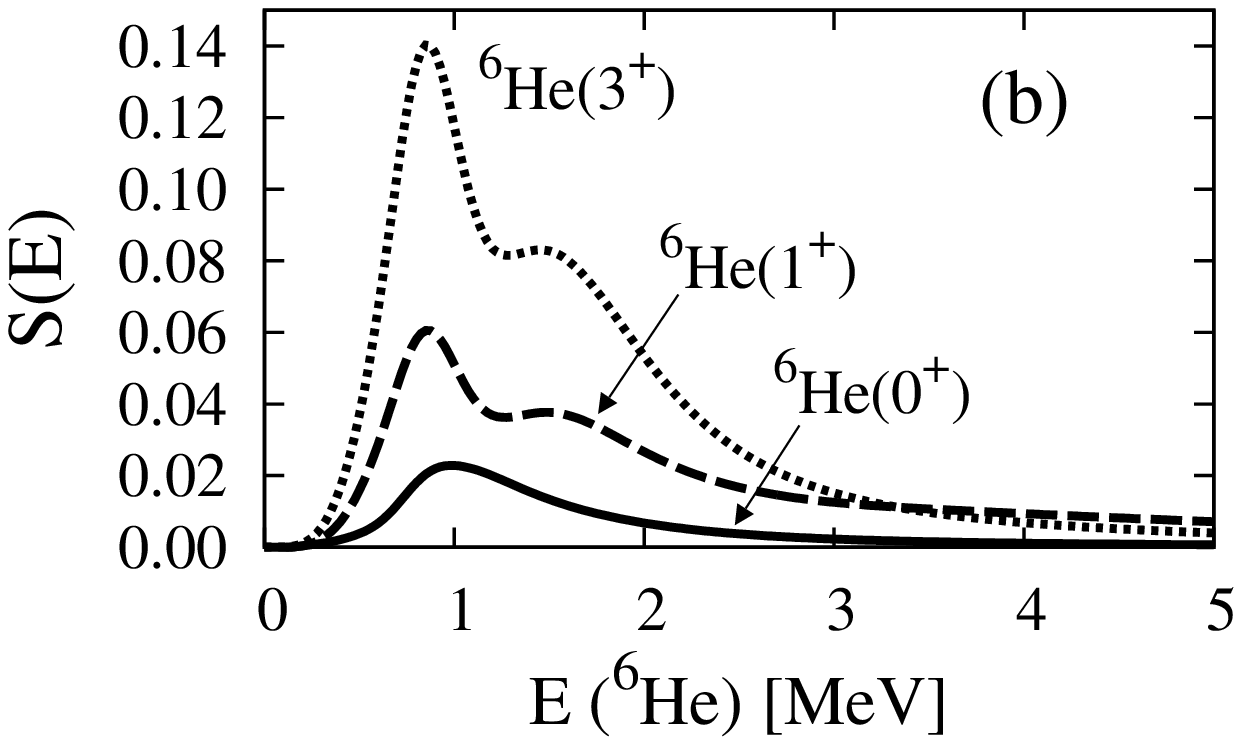}
\caption{
One-neutron removal strength of $^7$He$_{\rm g.s.}$ to form $^6$He with the energy $E$, measured from the $\alpha$+$n$+$n$ threshold energy.
The vertical arrow indicates the $2^+_1$ energy of $^6$He.
The strength to the ground $0^+_1$ state of $^6$He is shown by a histogram at the corresponding energy of $-0.98$ MeV.}
\label{fig:sec3_removal}
\end{figure}

%%%%%%%%%%%%%%%%%%%%%%%%%%%%%
\begin{figure}[ht]
\centering
\includegraphics[width=7.0cm,clip]{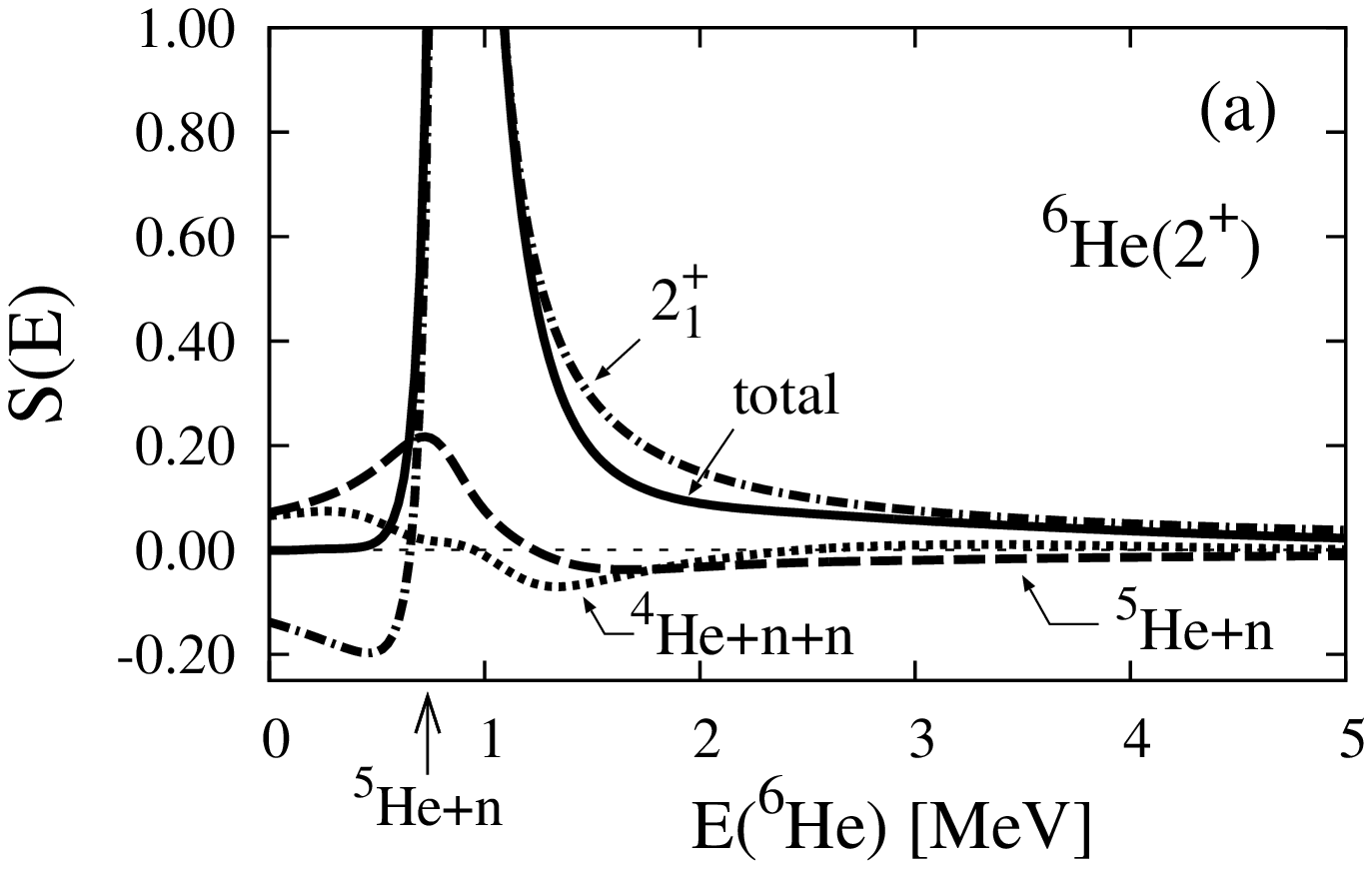}\hspace*{1.0cm}
\includegraphics[width=7.0cm,clip]{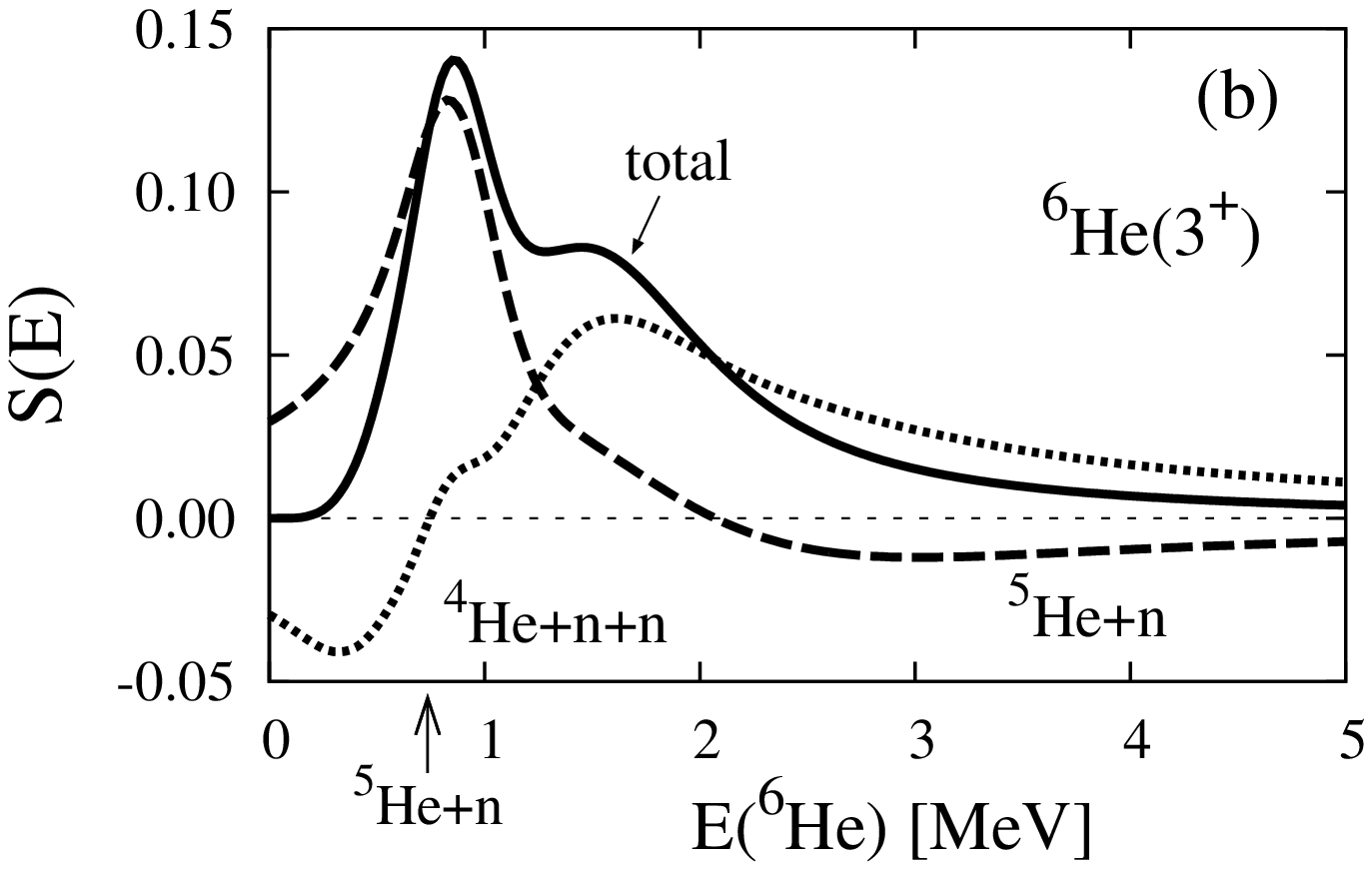}
\caption{Decomposition of the $^6$He($2^+$, $3^+$) components of the strength functions.
The vertical arrow indicates the threshold energy (0.74 MeV) of the $^5$He($3/2^-$)+$n$ channel in $^6$He.}
\label{fig:sec3_sep1}
\end{figure}
%%%%%%%%%%%%%%%%%%%%%%%%%%%%%%%

In Fig.~\ref{fig:sec3_removal}, we show the one-neutron removal strength $S(E)$ of $^7$He$(3/2^-)$ to form $^6$He with a spin $J$, 
where the $^6$He energy is measured from the $\alpha$+$n$+$n$ threshold.
In Fig.~\ref{fig:sec3_removal} (a), the dominant component comes from the $2^+$ state, the strength of which peaks at the resonance energy 0.84 MeV
of $^6$He($2^+_1$).
For the continuum energy region above the $\alpha$+$n$+$n$ threshold, 
the $0^+$, $1^+$, and $3^+$ states have small contributions to the strengths, as shown in Fig.~\ref{fig:sec3_removal} (b). 
From these results, it is concluded that the one-neutron removal strength of $^7$He is dominantly exhausted 
by the $^6$He($2^+_1$) resonance above the $\alpha$+$n$+$n$ threshold energy.

We discuss the detailed structures seen in the strengths in Fig.~\ref{fig:sec3_removal}. 
The origin of the $2^+$ peak is the $^6$He($2^+_1$) resonance, 
whose matrix element $(1.60-i0.49)$ shown in Table \ref{tab:sec3_sf_He7} is so large that 
the strength function $S(E)$ produces a sharp peak owing to the small decay width of 0.132 MeV of $^6$He($2^+_1$).
In Fig.~\ref{fig:sec3_sep1} (a), the $2^+$ strength distribution is decomposed into the three kinds of the components:
$^6$He($2^+_1$) resonance, $^5$He($3/2^-$)+$n$ continuum state, and $\alpha$+$n$+$n$ continuum state.
It is clearly and explicitly shown that the sharp peak originates from the $^6$He($2^+_1$) resonance.
The residual $2^+$ continuum strengths are found to be small.
Among these components, the $^5$He($3/2^-$)+$n$ two-body continuum component shows a peak at around 0.75 MeV. 
This energy coincides with the position of the $^5$He+$n$ threshold energy (0.74 MeV), and 
the peak reflects the threshold effect of the $^5$He+$n$ open channel.
The $\alpha$+$n$+$n$ contribution is small and does not produce a definite structure in the distribution.
In Fig.~\ref{fig:sec3_sep1} (b), the strengths of $^6$He($3^+$) are decomposed into two kinds of continuum components.
It is found that the strengths of the $^6$He($3^+$) unbound states show some structures, 
which come from the two different continuum components of $^5$He+$n$ and $\alpha$+$n$+$n$.
In summary, the one-neutron removal strength of $^7$He to form $^6$He is successfully obtained.
In the results, the important contribution of $^6$He($2^+_1$) is clearly shown.

%%%%%%%%%%%%%%%%
\subsubsection{$^8$He and $^8$C}\label{sec:A=8}

We compare the structures of the mirror states of $^8$He and $^8$C as the five-body states with an $\alpha$ core.
In this analysis, we discuss the structures of the $0^+$ states of two nuclei.
We obtain two $0^+$ states for each nucleus, both of which are five-body resonances in $^8$C, as shown in Fig.~\ref{fig:sec3_ene_COSM}.
For $^8$He, the binding energy of the ground state is obtained as 3.22 MeV from the $\alpha$+$4n$ threshold, which agrees with the experimental value of 3.11 MeV.
FOr $^8$C, the ground state energy is obtained as $E_r=3.32$ MeV, which agrees with the recent experimental $E_r=3.449(30)$ MeV \cite{charity11}.
The decay width is obtained as 0.072 MeV, which is slightly smaller than the experimental value of 0.130(50) MeV.
There is no experimental evidence of the excited states of $^8$C and the further experimental data are desirable.

%%%%%%%%%%%%%%%%%%%%%%%%%%%%%%
\begin{table}[t]
\centering
\begin{minipage}[l]{7.8cm}
\centering
\caption{Dominant parts of the complex squared amplitudes $(C^J_c)^2$ of the ground states of $^8$He and $^8$C.}
\label{tab:sec3_comp8_1}
\begin{tabular}{cccc}
\hline
Configuration              &  $^8$He($0^+_1$) & $^8$C($0^+_1$) \\ \hline
 $(p_{3/2})^4$             &  0.860           & $0.878-i0.005$ \\
 $(p_{3/2})^2(p_{1/2})^2$  &  0.069           & $0.057+i0.001$  \\
 $(p_{3/2})^2(1s_{1/2})^2$ &  0.006           & $0.010+i0.003$  \\
 $(p_{3/2})^2(d_{3/2})^2$  &  0.008           & $0.007+i0.000$ \\
 $(p_{3/2})^2(d_{5/2})^2$  &  0.042           & $0.037+i0.000$\\
 other 2p2h                &  0.011           & $0.008+i0.000$\\
\hline
\end{tabular}
\end{minipage}
\hspace*{0.5cm}
\begin{minipage}{9.5cm}
\centering
%%%%%%%%%%%%%%%%%%%%%%%%%%%%%%
%%%%%%%%%%%%%%%%%%%%%%%%%%%%%%
\caption{Dominant parts of the complex squared amplitudes $(C^J_c)^2$ of the $0^+_2$ states of $^8$He and $^8$C.}
\label{tab:sec3_comp8_2}
\centering
\begin{tabular}{cccc}
\hline
Configuration              &  $^8$He($0^+_2$)  & $^8$C($0^+_2$) \\ \hline
 $(p_{3/2})^4$             &~~$0.020-i0.009$   &~~$0.044+i0.007$ \\
 $(p_{3/2})^2(p_{1/2})^2$  &~~$0.969-i0.011$   &~~$0.934-i0.012$ \\
 $(p_{3/2})^2(1s_{1/2})^2$ &$-0.010-i0.001$   &$-0.001+i0.000$\\
 $(p_{3/2})^2(d_{3/2})^2$  &~~$0.018+i0.022$   &~~$0.020+i0.003$ \\
 $(p_{3/2})^2(d_{5/2})^2$  &~~$0.002+i0.000$   &~~$0.002+i0.001$ \\
\hline
\end{tabular}
\end{minipage}
\end{table}
%%%%%%%%%%%%%%%%%%%%%%%%%%%%%%

We discuss the properties of the configuration mixing of $^8$He and $^8$C \cite{myo12b,myo10}.
For their ground states, we list the main configurations with their complex squared amplitudes $(C^J_c)^2$ obtained from Eq. (\ref{eq:sec3_COSM-WF0}), 
 in Table \ref{tab:sec3_comp8_1}. 
It is found that the $(p_{3/2})^4$ configuration dominates the total wave function with a squared amplitude of 0.86 for $^8$He and 0.88 in the real part for $^8$C.
The 2p2h excitations from the $p_{3/2}$ orbit are a summation of about 0.14 for $^8$He and about 0.12 for $^8$C.
Among the 2p2h components, the $p_{1/2}$ and $d_{5/2}$ orbits well contribute to the ground states.
These results indicate that the $jj$-coupling picture and the $p_{3/2}$ sub-closed property are well established in the ground states of $^8$He and $^8$C.
From the values of the squared amplitudes, the trend of the configuration mixing in the ground states of $^8$He and $^8$C is found to be quite similar,
which means a good mirror symmetry between the two states.
The RMS radius of the $^8$C ground state is obtained as $2.81-i$0.08 fm, the real part of which is larger than 2.52 fm of $^8$He, as shown in Table. \ref{tab:sec3_radius_COSM}.
This is due to the Coulomb repulsion of four valence protons of $^8$C, similar to the comparison between $^6$He and $^6$Be.

We discuss the excited $0^+_2$ resonant states of $^8$He and $^8$C.
The dominant configurations of four valence neutrons/protons are listed in Table \ref{tab:sec3_comp8_2}.
In these states, the $(p_{3/2})^2(p_{1/2})^2$ configuration commonly dominates the total wave function with a large squared amplitude of around 0.95 for the real part,
whereas $(p_{3/2})^4$ is very small in the two nuclei.
Hence, the $0^+_2$ states of $^8$He and $^8$C corresponds to the 2p2h excitation from the ground states
and can be described prominently within a single configuration.

The coupling properties of four valence nucleons around $\alpha$ are discussed from the viewpoint of the nucleon pair numbers \cite{myo12b}.
We discuss the structures of the two $0^+$ states of $^8$He and $^8$C.
We calculate the complex pair numbers $P({J^\pi},S)$ of four valence neutrons/protons in $^8$He and $^8$C, 
which are defined by using the matrix element of the operator as
\begin{eqnarray}
    P({J^\pi},S) 
&=& \langle\sum_{\kappa \le \kappa'} A^\dagger_{J^\pi,S}(\kappa \kappa')A_{J^\pi,S}(\kappa \kappa')\rangle.
\end{eqnarray}
Here, $\kappa$ and $\kappa'$ are quantum numbers for the single particle states; and 
$A^\dagger_{J^\pi,S}$ ($A_{J^\pi,S}$) is the creation (annihilation) operator of a nucleon pair with the coupled angular momentum, parity $J^\pi$, and the coupled intrinsic spin $S$.
The summation of $P({J^\pi},S)$ over all $J^\pi$ and $S$ becomes six, the pair number of four protons.
The complex pair numbers are useful to understand the properties of the four nucleons coupled with $\alpha$ from the viewpoint of pair configuration.
It is noted that $P({J^\pi},S)$ represents the component of the one pair of nucleons having ($J^\pi$, $S$) quantum numbers, which is taken from four valence nucleons.
In fact, when $^8$He($0^+$) is decomposed into $2n(0^+)$ and $^6$He($0^+$), this component consists of two $0^+$ pairs of $2n$ regarding $^6$He as $\alpha$+$2n(0^+)$. 
For the $2n(2^+)$ and $^6$He($2^+$) decomposition of $^8$He($0^+$), there are two $2^+$ pairs.

%%%%%%%%%%%%%%%%%%%%%%%%%%%%%%
\begin{figure}[t]
\centering
\includegraphics[width=7.2cm,clip]{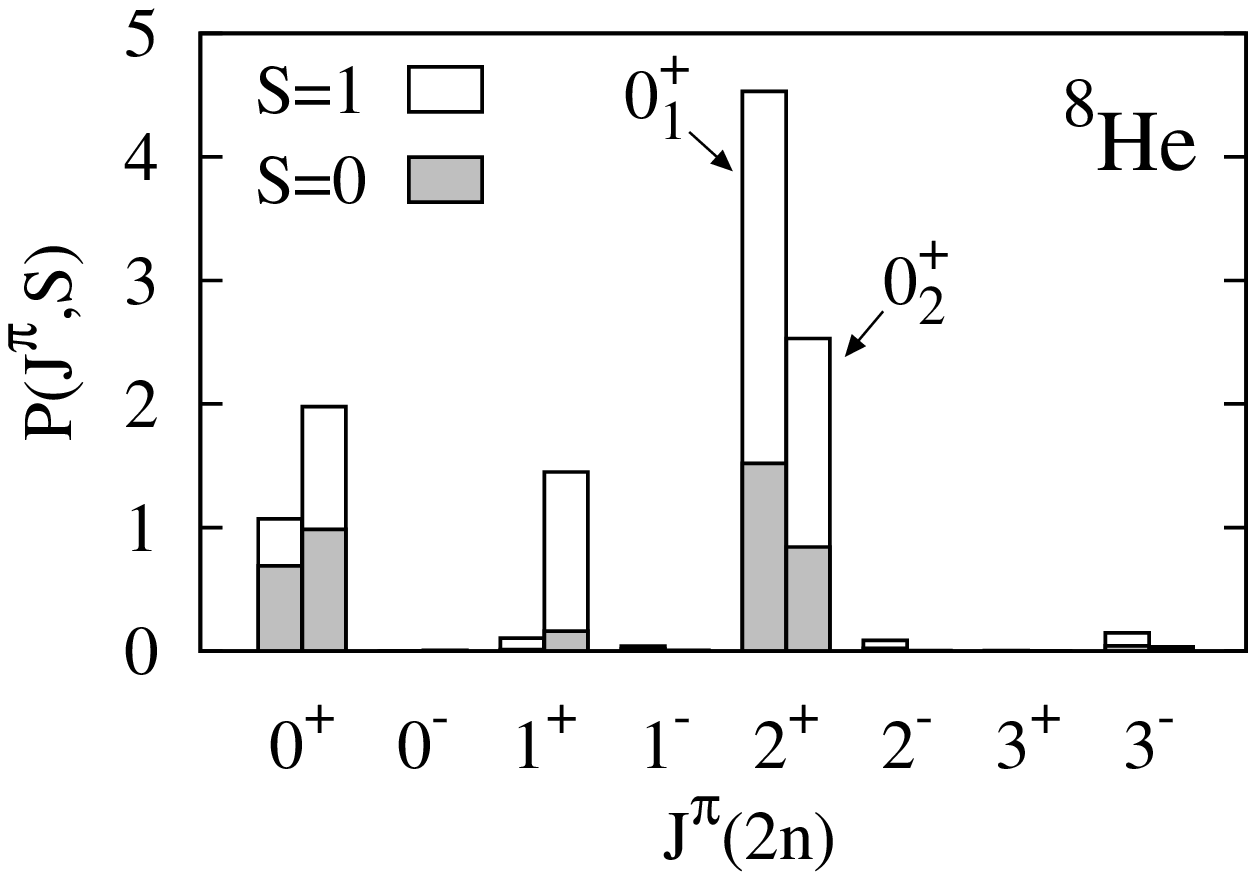}\hspace*{1.5cm}
\includegraphics[width=7.2cm,clip]{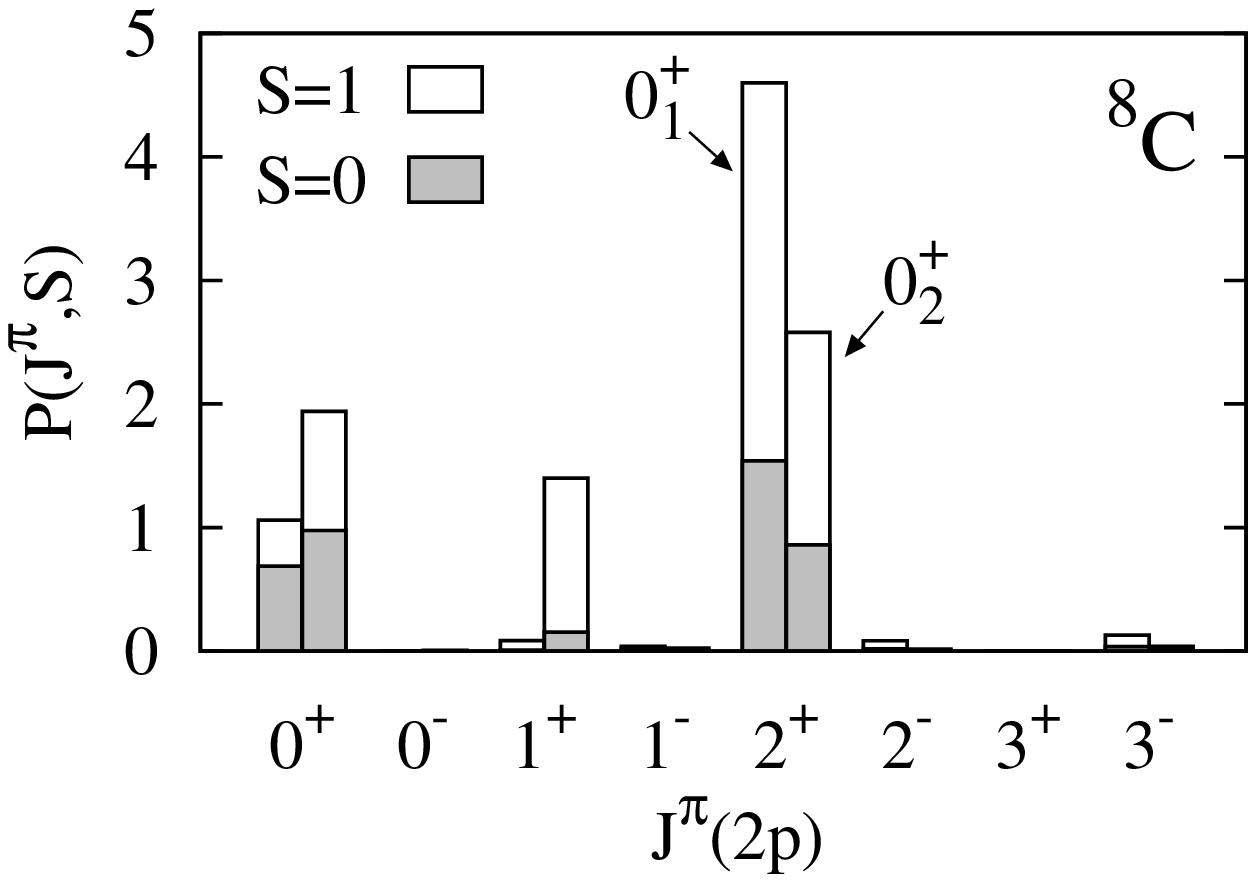}
\caption{Real parts of the complex pair numbers of valence neutrons $P({J^\pi},S)$ in the $^8$He($0^+_1$, $0^+_2$) (left) 
and $^8$C($0^+_1$, $0^+_2$) states (right), decomposed into the $S=0$ (shaded) and $S=1$ components (blank).}
\label{fig:sec3_pair_8C}
\end{figure}
%%%%%%%%%%%%%%%%%%%%%%%%%%%%%%

We show the real parts of the pair number distributions in Fig.~\ref{fig:sec3_pair_8C}.
It is found that the $2^+$ nucleon pairs are dominant in the ground states of $^8$He and $^8$C.
In the case where the CFP decomposition is used, this $J^\pi$ distribution is consistent with the $(p_{3/2})^4$ configuration with a large mixing in $^8$He and $^8$C, as shown in Table \ref{tab:sec3_comp8_1}.
The $0^+_2$ state has several $0^+$ and $1^+$ components of nucleon pairs as well as the $2^+$ one.
This is also consistent with the $(p_{3/2})^2(p_{1/2})^2$ configuration with a dominant mixing in $^8$He and $^8$C.

From the pair number analysis, the importance of the $2^+$ pair in the ground states of $^8$He and $^8$C is shown.
This result was suggested in the experiment of $^8$He \cite{korsheninnikov03} and is obtained in the $^6$He+$n$+$n$ three-body analysis \cite{adahchour06}. 
On the other hand, the other experiments \cite{chulkov05,keeley07} suggest the different results which indicate the contribution of the $p_{1/2}$ orbit in the $^8$He ground state.
For $^8$C, recent experiments \cite{charity10,charity11} show that the decay of the $^8$C ground state can go through the $^6$Be($0^+_1$)+$2p$ channel with the high probability as 0.92(5),
while in Fig. \ref{fig:sec3_pair_8C}, the large amount of the $^6$Be($2^+_1$)+2$p$ decomposition is obtained theoretically for the $^8$C ground state.
This difference indicates that the theoretical description of the decay process of $^8$C into the $^6$Be + $2p$ channel is further to be considered.

To summarize the mirror symmetry, the structures of $^8$He and $^8$C are similar in terms of the properties of the pair numbers of valence nucleons.
This result indicates that the mirror symmetry is well retained in two nuclei for the $0^+$ states.

%%%%%%%%%%%%%%%%%%%%%%%%%%%%%%%%%%%%%%%%%%%%%%%%%%%%%%%%%%%%%
The property of the $0^+$ pairs of nucleons in the $0^+_2$ states of $^8$He and $^8$C is interesting 
in relation to the dineutron-like cluster correlation in $^8$He suggested in AMD \cite{enyo07}.
For $^8$He, we calculate the monopole transition into unbound states and investigate the effect of the $0^+_2$ resonance in the strength.
Recently, Yamada {\it et al.} discussed the relation between the excited clustering state and its monopole strength from the ground state \cite{yamada08}.
The enhancement of the monopole strength can occur in the clustering states, because of the concentration of the strength to the relative motion of the intercluster.
It is meaningful to investigate the monopole strength into $^8$He($0^+_2$) in relation to dineutron structure.
With regard to the strength, it is also important to investigate the effects of the continuum states in addition to that of resonance.
We consider not only the $0^+_2$ resonance of $^8$He but also all the continuum states of $^7$He+$n$, $^6$He+$2n$, $^5$He+$3n$, and $\alpha$+$4n$ with five-body ECR in Eq. (\ref{eq:sec2_ECR-CSM}).

%%%%%%%%%%%%%%%%%%%%%%%%%%%%%
\begin{figure}[t]
\centering
\includegraphics[width=8.0cm,clip]{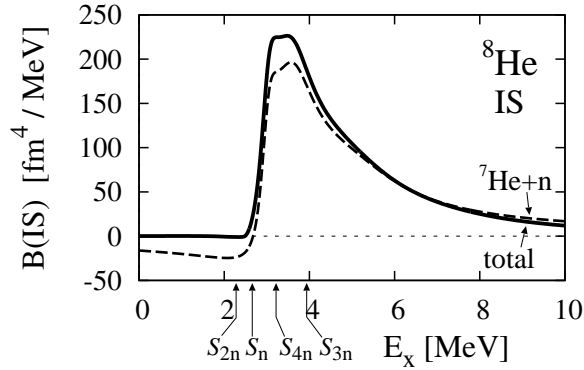}\hspace*{1.0cm}
\caption{Monopole transition strength of $^8$He for isoscalar (IS) response as functions of the excitation energy of $^8$He.
The threshold energies of $^7$He+$n$, $^6$He+$2n$, $^5$He+$3n$, and $\alpha$+$4n$ are shown by arrows with $S_n$, $S_{2n}$, $S_{3n}$ and $S_{4n}$, respectively.
The dotted line is the $^7$He+$n$ component.}
\label{fig:sec3_monopole}
\end{figure}
%%%%%%%%%%%%%%%%%%%%%%%%%%%%%%

In Fig.~\ref{fig:sec3_monopole}, the isoscalar (IS) monopole strength of $^8$He is shown \cite{myo10}.
It is found that the strength results in a low enhancement just above 3 MeV in the excitation energy.
There is no clear signature of the $0^+_2$ state around its excitation energy of 6.3 MeV in the strength.
This result is understood from the single particle structures of the $0^+_2$ state of $^8$He.
In the $0^+_2$ state, the $p_{1/2}$ orbit of valence neutrons is largely mixed as shown in Table \ref{tab:sec3_comp8_2}.
This orbit cannot be excited from the $p_{3/2}$ orbit mixed in the ground state by the monopole operator.
As a result, the monopole strength to the $0^+_2$ state of $^8$He becomes negligible. 
Instead, the continuum strength provides the main contribution.

We further decompose the monopole strength shown in Fig.~\ref{fig:sec3_monopole} into the components of the several continuum states.
It is found that the strength is dominantly exhausted by the $^7$He($3/2^-_1$)+$n$ components.
This result implies a sequential breakup process of $^8$He$_{\rm g.s.}$$\to$$^7$He($3/2^-_1$)+$n$$\to$$^6$He$_{\rm g.s.}$+$n$+$n$ in the monopole excitation.
Similarly, a large contribution of the sequential process via $^7$He+$n$ is reported in the Coulomb breakup experiment of $^8$He \cite{iwata00}, 
which is dominated by the $E1$ transition.

%%%%%%%%%%%%%%%%%%%%%%%%%%%%%%%%%%%%%%%%%%%%%%%%%%%%%%%%%%%%%
\subsection{$^{11}$Li with tensor and pairing correlations}\label{sec:11Li_cc}

We discuss the structure of $^{11}$Li, a famous two-neutron halo nucleus \cite{tanihata85}, on the basis of the coupled three-body model of $^9$Li+$n$+$n$, 
which is extended from the three-body CS-COSM.
In the model, the configuration mixing of the $^9$Li core is performed to take into account the tensor and pairing correlation,
which are important to explain the breaking of the neutron magic number of $N=8$ in $^{11}$Li.
In this section, we focus on the ground state properties of $^{11}$Li, mainly, the mechanism of the halo formation in $^{11}$Li with a large $s^2$ mixing.
A detailed analysis of the Coulomb breakup of $^{11}$Li into three-body scattering states is shown in terms of the complex-scaled solutions of the Lippmann-Schwinger equation \cite{kikuchi13a} in \S \ref{sec:CSLS}.

%%%%%%%%%%%%%%%%%%%%%%%%%%%%%%%%%%%%%%%%%%%%%%%%%%%%%%%%%%%%%
\subsubsection{CS-COSM with configuration mixing in core}
We introduce the extended core+$n$+$n$ of the CS-COSM including the core excitations \cite{myo07b,myo08}. 
We derive the three-body Hamiltonian from the $A$-nucleon system. 
The $A$-body Hamiltonian $H_A$ is given as 
\begin{eqnarray}
    H_A
&=& \sum_{i=1}^{A}t_i - T_{G} + \sum_{i<j}^A v_{ij}=T+V,
\label{eq:ham_A}
\end{eqnarray}
where $T =\sum t_i - T_{G}$ is the internal kinetic energy operator of the system in which the center-of-mass motion, $T_{G}$ is subtracted; and $V=\sum v_{ij}$ is the two-body interaction. 
We decompose the Hamiltonian of an $A$-nucleon system into a core part with mass number $A_c$ and $N_{\rm v}$ valence neutrons, as $A=A_c+N_{\rm v}$ .
The relative coordinates between the core and the valence nucleons are shown in Fig. \ref{fig:sec3_COSM} for the three-body case. 
The kinetic energy term $T$ is rewritten using the reduced mass $\mu$ between the core and a neutron as
\begin{eqnarray}
T&=& T_c
+ \sum_{i=1}^{N_{\rm v}} \frac{\vc{p}_i^2}{2\mu}
+ \sum_{i<j}^{N_{\rm v}} \frac{\vc{p}_i\cdot\vc{p}_j}{(A_c+1)\mu}~.
\end{eqnarray}
Here, $T_c=\sum_{i=1}^{A_c}t_i - T^{c}_{G}$ is the kinetic energy of the core nucleus. 
The potential term $V$ is similarly decomposed as
\begin{eqnarray}
    V 
&=& \sum_{i<j}^{A_c} v_{ij} + \sum_{i=1}^{N_{\rm v}}\sum_{j=1}^{A_c} v_{ij} 
  + \sum_{i<j}^{N_{\rm v}} v_{ij}
~=~ V_c + \sum_{i=1}^{N_{\rm v}} V_i + \sum_{i<j}^{N_{\rm v}} v_{ij},
  \label{eq:sec3_interaction}
\end{eqnarray}
where $V_i=\sum_{j=1}^{A_c} v_{ij} $ is the mean field potential for each valence neutron ($i=1,\ldots, N_{\rm v}$) 
and $V_c$ is the potential term in the core nucleus. The Hamiltonian $H_A$ in Eq~.(\ref{eq:ham_A}) is rewritten as
\begin{eqnarray}
     H_A
&=&  \Biggl[T_c + \sum_{i=1}^{N_{\rm v}} \frac{\vc{p}_i^2}{2\mu}+
     \sum_{i<j}^{N_{\rm v}} \frac{\vc{p}_i\cdot\vc{p}_j}{(A_c+1)\mu}\Biggr]+
     \Bigl[V_c + \sum_{i=1}^{N_{\rm v}} V_i + \sum_{i<j}^{N_{\rm v}} v_{ij}\Bigr]
     \\
&=&  H_c + \sum_{i=1}^{N_{\rm v}} \Bigl[\frac{\vc{p}_i^2}{2\mu}+V_i\Bigr]
      + \sum_{i<j}^{N_{\rm v}} \Bigl[ v_{ij}
      + \frac{\vc{p}_i\cdot\vc{p}_j}{(A_c+1)\mu}\Bigr],
\label{eq:sec3_core-COSM}
\end{eqnarray}
where the first term $H_c=T_c+V_c$ is the Hamiltonian of the core nucleus, and 
the second term is the single-particle Hamiltonian for the relative motion between the single nucleon and the core.  
This Hamiltonian defines the orbitals for the valence nucleons around the core nucleus.
The third term is the two-body operator, which produces the coupling between valence nucleons, such as the dineutron correlation in $^{11}$Li.

For the wave function, we start with the core part $\Phi(A_c)$ and write the Schr\"odinger equation as
\begin{eqnarray}
      \Phi(A_c)
&=&   \sum_{\alpha} C_\alpha\phi_\alpha(A_c),
\label{eq:sec3_corewf}
\\
H_c\Phi(A_c)&=&E_c\Phi(A_c)~,
\label{eq:sec3_core}
\end{eqnarray}
where the index $\alpha$ is used as a label to distinguish between various configurations of the core nucleus with amplitudes $C_\alpha$,
which are determined by the variational equations with the energy $E_c$ using Eq.~(\ref{eq:sec3_core}). 
We employ the shell-model-like basis wave function for $\phi_\alpha(A_c)$.

The total wave function of the $A$-nucleon system of $^{11}$Li and the corresponding Schr\"odinger equation are given as
\begin{eqnarray}
     \Psi(A)
&=& {\cal A} \left \{\sum_\alpha \phi_\alpha(A_c)\chi_\alpha(nn) \right \},
\label{eq:sec3_11Li}
\\
     H_A\Psi(A)
&=&  E\Psi(A)~,
\label{eq:sec3_11Li2}
\end{eqnarray}
where the total Hamiltonian $H_A$ is given in Eq.~(\ref{eq:sec3_core-COSM}).
We omit the angular momentum coupling between the core nucleus and the valence neutrons for simplicity.  
The operator ${\cal A}$ is the antisymmetrizer between the core nucleons and the valence neutrons.
The mixing amplitudes of the core configurations $\alpha$ are included in the two-neutron wave functions $\chi_\alpha(nn)$.  
In the description of $\chi_\alpha(nn)$ for the loosely bound two neutrons in $^{11}$Li, we apply the so-called hybrid-VT model, which is the extension of COSM.
The details of the model are given in Refs. \cite{aoyama06,myo07b}.
In order to solve Eq.~(\ref{eq:sec3_11Li2}), we employ the orthogonality condition model (OCM) \cite{aoyama06,saito77}.

For the calculation of the matrix element of the Hamiltonian in Eq.~(\ref{eq:sec3_core-COSM}), we fold the Hamiltonian by using the wave function of the core nucleus.  
In the coupled-channel three-body OCM, we obtain the following equation for the valence neutrons $\chi_\alpha(nn)$,
\begin{eqnarray}
\sum_{\beta} \Biggl[
   H^c_{\alpha\beta}+ \sum_{i=1}^{N_{\rm v}} \Bigl\{\frac{\vc{p}_i^2}{2\mu}\delta_{\alpha\beta} + V^F_{i,\alpha\beta}
+  v^{\rm PF}_i \delta_{\alpha\beta} \Bigr\} 
+  \sum_{i<j}^{N_{\rm v}} \Bigl\{ v_{ij} +\frac{\vc{p}_i\cdot\vc{p}_j}{(A_c+1)\mu}
  \Bigr\}\delta_{\alpha\beta} \Biggr]\chi_\beta(nn) 
&=& E\chi_\alpha(nn),
\label{eq:sec3_Li-COSM2}
\end{eqnarray}
\begin{eqnarray}
    H^c_{\alpha\beta}
&=& \langle \phi_\alpha(A_c)|H_c|\phi_\beta(A_c) \rangle,
\qquad
    V^F_{i,\alpha\beta}
~=~ \langle \phi_\alpha(A_c)|V_i|\phi_\beta(A_c)\rangle , 
\end{eqnarray}
where the term $ v^{\rm PF}$ is the Pauli potential to remove the Pauli-forbidden states from the relative motion between $^9$Li and $n$
\cite{myo07a}.

%%%%%%%%%%%%%%%%%%%%%%%%%%%%%%%%%%%%%%%%%%%%%%%%%%%%%%%%%%%%%
\subsubsection{$^9$Li core with tensor-optimized shell model for $^{11}$Li}
In the calculation, we express $^9$Li($3/2^-$) by a multi-configuration in terms of the tensor-optimized shell model (TOSM)~\cite{myo07b,myo11a,myo05,myo12a}. 
\begin{eqnarray}
  \Phi(^{9}\mbox{Li})
&=&\sum_i^{N} C_i\, \phi^{3/2^-}_i,
  \label{eq:sec3_WF9}
\end{eqnarray}
where $N$ is the configuration number and we consider up to the 2p2h excitations within the $0p$ shell for $\Phi^{3/2^-}_i$.
Based on the TOSM studies on light nuclei \cite{myo07b,myo05}, we adopt the spatially modified harmonic oscillator wave function 
as a single particle orbit and 
treat the length parameters $b_\kappa$ of every orbit $\kappa$ of $0s$, $0p_{1/2}$, and $0p_{3/2}$ as the independent variational parameters. 
This variation is be important to optimize the tensor correlation~\cite{myo05,sugimoto04,ogawa06}. 
We solve the variational equation for the Hamiltonian of $^9$Li and determine $\{C_i\}$ in Eq.~(\ref{eq:sec3_WF9}) as well as 
the length parameters $\{b_{\kappa}\}$ of the $0s$ and $0p$-orbits. 
The variation in the energy expectation value with respect to the total wave function $\Phi(^9{\rm Li})$ is given by
\begin{eqnarray}
\delta\frac{\bra\Phi|H(\mbox{$^9$Li})|\Phi\ket}{\bra\Phi|\Phi\ket}&=&0\ .
\end{eqnarray}

We explain the interactions for $^{11}$Li employed in the Hamiltonians in Eq.~(\ref{eq:sec3_Li-COSM2}),
the details of which are given in the previous works \cite{myo07b,myo08}.
In the study on $^{11}$Li, we focus on the tensor correlation, which is newly considered to solve the $s$-$p$ shell gap problem. 
To do this, we extend the three-body model of $^{11}$Li to incorporate the tensor correlation, particularly for the $^9$Li part. 
In the present study, we aimed to investigate $^{11}$Li using the experimental information on $^9$Li and $^{10}$Li as much as possible. 
We explain the interactions in three terms; $V_c$ of $^9$Li in Eq.~(\ref{eq:sec3_interaction}), core-$n$ $V^F_i$, and $v_{ij}$ of last two neutrons in the Hamiltonian given in Eq.~(\ref{eq:sec3_Li-COSM2}).

For the potential between the last two neutrons in $^{11}$Li, we use a realistic interaction AV8$^\prime$ because the model space of the two neutrons has no restriction. 
The $^9$Li-$n$ potential, $V^F_i$, is given by folding the $G$-matrix interaction \cite{furutani80,hasegawa71} using the $^9$Li density obtained with the TOSM. 
No phenomenological state-dependence is used in the $^9$Li-$n$ potential, such as the use of a deeper potential only for the $s$-wave.
We introduce one parameter $\delta$ as in the expression $(1+\delta)V^F_{(2)}$, which is used to adjust the second-range strength of the $G$-matrix interaction in the calculation of the $^9$Li-$n$ potential.
The parameter $\delta$ can reflects the dependence on the starting energy in the $G$-matrix calculation, originating from the tensor interaction \cite{aoyama06,furutani80}. 
In the calculation, $\delta$ is determined and is used to reproduce the two-neutron separation energy of $^{11}$Li as 0.31 MeV.
For the interaction used in the $^9$Li core; $V_c$ in $H(^9{\rm Li})$ in Eq.~(\ref{eq:sec3_interaction}),
we use the $G$-matrix proposed by Akaishi \cite{myo07b,myo05,akaishi04,ikeda04} for $V_c$, which is constructed from the AV8$^\prime$ potential with a large momentum space.

%%%%%%%%%%%%%%%%%%%%%%%%%%%%%%
\begin{figure}
\begin{minipage}{0.52\textwidth}
\begin{center}
\includegraphics[width=9.0cm,clip]{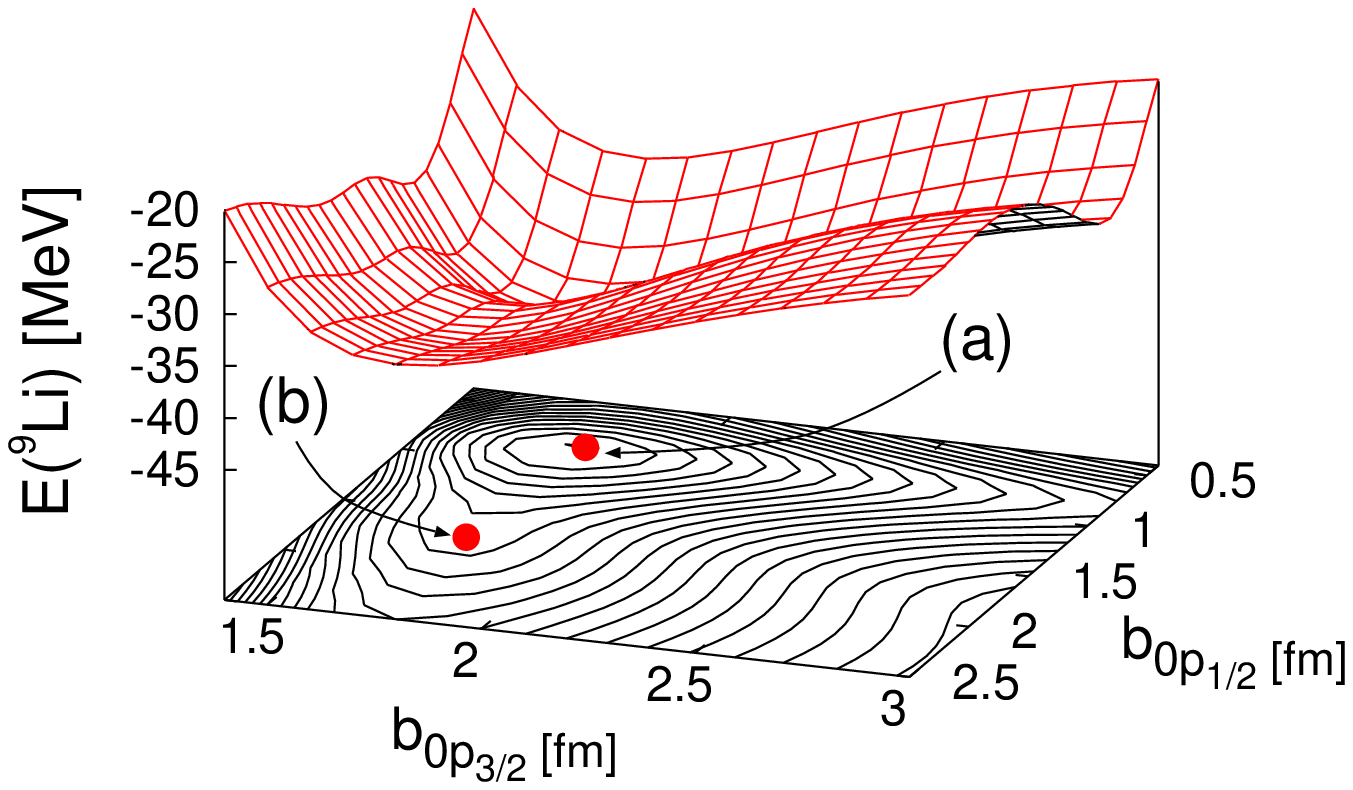}
\caption{(Color online) Energy surface of $^9$Li with respect to the two length parameters of the $0p$ orbits \cite{myo07b}.
Two energy minima are indicated by (a) and (b).}
\label{9Li_ene}
\end{center}
\end{minipage}
\hspace*{0.3cm}
\begin{minipage}{0.45\textwidth}
\begin{center}
\makeatletter
\def\@captype{table}
\makeatother
\renc{\baselinestretch}{1.15}
\begin{tabular}{c|ccccc}
\hline
                                                   & \multicolumn{3}{c}{Present}  \\
                                                   &   (a)   &  (b)     & (a+b)   \\
\noalign{\hrule height 0.5pt}
 Energy [MeV]                                      & $-43.8$ & $-37.3$  & $-45.3$ \\
 $\langle V_T\rangle $ [MeV]                       & $-22.6$ &~~$-1.8$  & $-20.7$ \\
\noalign{\hrule height 0.5pt}
 $R_m$                 [fm]                        &  2.30   & 2.32     & $2.31$  \\
\noalign{\hrule height 0.5pt}
 0p0h                                              & 91.2    &  60.1    & $82.9$  \\
 $(0p_{3/2})^{-2}_{01}(0p_{1/2})^2_{01}$           & 0.03    &  37.1    & ~$9.0$  \\
 $(0s_{1/2})^{-2}_{10}(0p_{1/2})^2_{10}$           & 8.2     &   1.8    & ~$7.2$  \\
\hline
\end{tabular}
\caption{Properties of $^9$Li with configuration mixing. The states (a) and (b) correspond to the energy minima. The state (a+b) is obtained by superposing (a) and (b).}
\label{tab:sec3_9Li}
\end{center}
\end{minipage}
\end{figure}
%%%%%%%%%%%%%%%%%%%%%%%%%%%%%%

%%%%%%%%%%%%%%%%%%%%%%%%%%%%%%%%%%%%%%%%%%%%%%%%%
%\subsubsection{$^{9}$Li with TOSM}

We briefly summarize the results of the $^9$Li properties in TOSM \cite{myo07b} which give a dynamical influence on the motion of last neutrons in $^{11,10}$Li.
In Fig.~\ref{9Li_ene}, we display the energy surface of $^9$Li as functions of the length parameters of two $0p$ orbits, where $b_{0s}$ is already optimized as 1.45 fm. 
There are two energy minima, (a) and (b), which have a nearly similar $b_{0p_{3/2}}$ value of 1.8 fm and a small (0.85 fm) and a large (1.8 fm) of $b_{0p_{1/2}}$ value, respectively. 
The properties of the two minima are listed in Table \ref{tab:sec3_9Li} with the dominant 2p2h configurations and their probabilities. 
The minimum (a) shows a large tensor contribution. The largest probabilities of the 2p2h configurations are 
given by $(0s)^{-2}_{10}(0p_{1/2})^2_{10}$ for the minimum (a) to produce the tensor correlation, similar to the results in Refs.~\cite{myo11a,myo05},  
and by $(0p_{3/2})^{-2}_{01}(0p_{1/2})^{2}_{01}$ for the minimum (b) to explain the $0p$ shell pairing correlation.
Here, the subscripts 01 or 10 represent the spin and isospin for the two-nucleon pair, respectively.
These results indicate that the minima (a) and (b) represent the different correlations of the tensor and pairing characters, respectively. 
Table~\ref{tab:sec3_9Li} shows the results of the superposition of minima (a) and (b); this superposition is done to obtain a $^9$Li wave function, while including the tensor and pairing correlations.
The matter RMS radius is 2.31 fm which agree with the experimental value of 2.32 $\pm$ 0.02 fm \cite{tanihata88}.

%%%%%%%%%%%%%%%%%%%%%%%%%%%%%%%%%%%%%
\subsubsection{Pauli-blocking effect in $^{10}$Li and $^{11}$Li}

%%%%%%%%%%%%%%%%%%%%%%%%%%%%%%%
\begin{figure}[t]
\centering
\includegraphics[width=17.5cm,clip]{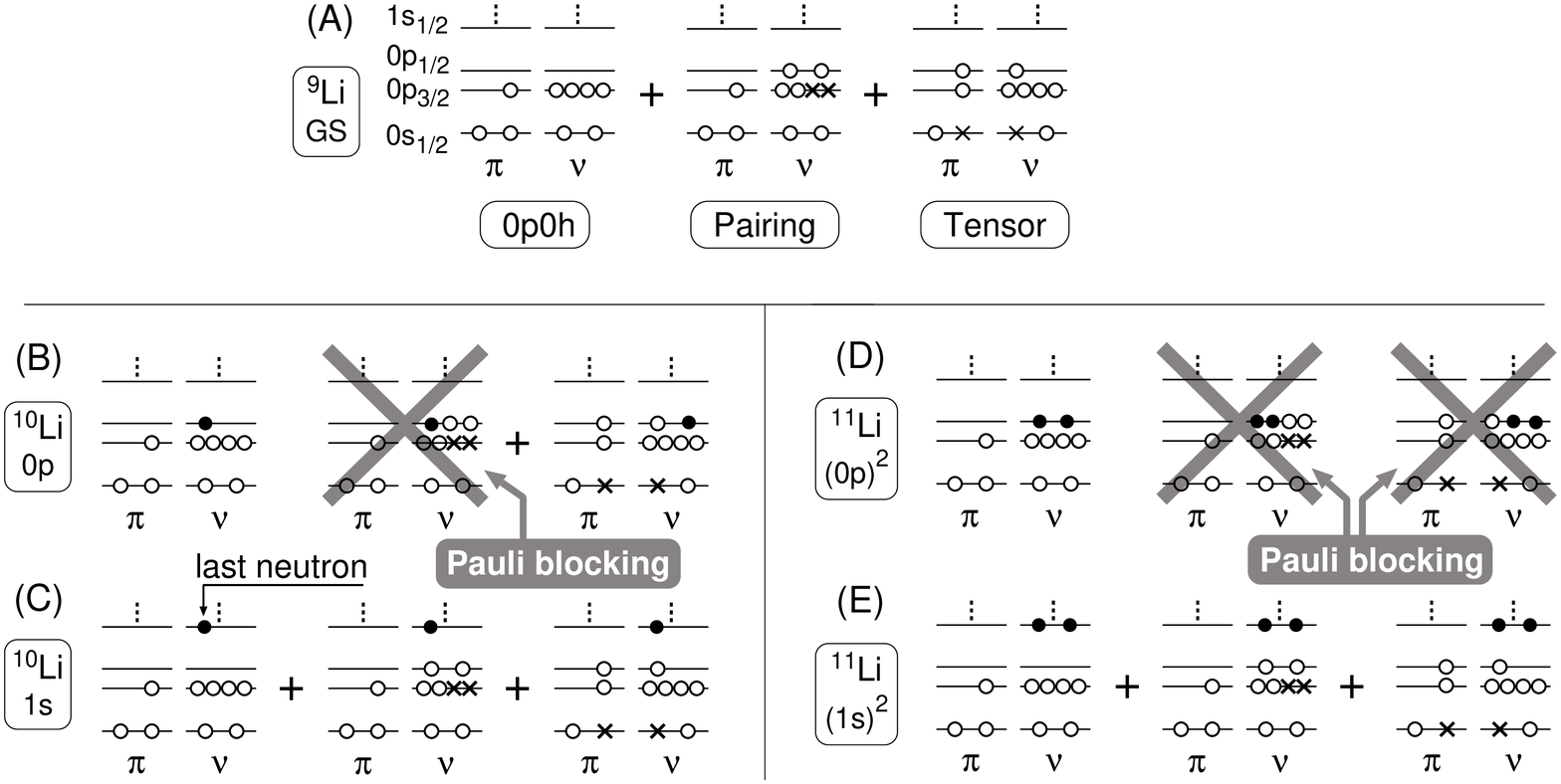}
\caption{Schematic illustration for Pauli-blocking in $^{10,11}$Li.}
\label{fig:sec3_Li-Pauli}
\end{figure}
%%%%%%%%%%%%%%%%%%%%%%%%%%%%%%%

Considering the results of the configuration mixing of $^9$Li, we discuss the Pauli-blocking effects in $^{10}$Li and $^{11}$Li in Fig.~\ref{fig:sec3_Li-Pauli}.  
For Fig.~\ref{fig:sec3_Li-Pauli}(A), the $^9$Li ground state with the label, GS, consists of the 0p0h and the 2p2h states, which are mixed by the tensor and pairing correlations.

We consider $^{10}$Li by adding one neutron to $^9$Li.
For Fig.~\ref{fig:sec3_Li-Pauli}(B), when the last neutron occupies the $0p_{1/2}$ orbit, the 2p2h excitation of the pairing correlation in $^9$Li is blocked by Pauli principle. 
The tensor correlation is partially blocked, but not fully.
As a result, the correlation energy of $^9$Li is partially lost in $^{10}$Li. 
On the other hand, for Fig.~\ref{fig:sec3_Li-Pauli}(C), showing the $1s$ state of $^{10}$Li, the Pauli-blocking does not occur strongly and 
$^9$Li can gain the correlation energy by the mixing of the 2p2h excitations. 
Hence, the energy difference between the $p$ and $s$ states of $^{10}$Li becomes small, which can explain the inversion phenomenon \cite{myo07b,kato99}.

For $^{11}$Li, a similar consideration is made by adding two neutrons to $^9$Li.
The blocking effect is expected for $^{11}$Li, the details of which are given in a previous paper \cite{myo07b}.
For Fig.~\ref{fig:sec3_Li-Pauli}(D), when two neutrons occupy the $0p_{1/2}$-orbit, the 2p2h excitations in $^9$Li are blocked by Pauli principle.
Particularly, the blocking of the tensor correlation is expected to work more strongly in the $^{11}$Li case than the $^{10}$Li case, 
because of the presence of the last two neutrons in the $p_{1/2}$ orbit in the $^{11}$Li case. 
For Fig.~\ref{fig:sec3_Li-Pauli}(E), showing $(1s)^2$ of two neutrons, the Pauli-blocking does not occur, similar to the $1s$ state of $^{10}$Li.  
Then the relative energy between the $(0p)^2$ and $(1s)^2$ configurations of $^{11}$Li becomes small sufficiently to explain the breaking of the magicity in $^{11}$Li.  
It is interesting to investigate how these blocking effects relate to the problem of the $s$-$p$ shell gap in $^{10}$Li and $^{11}$Li.

%%%%%%%%%%%%%%%%%%%%%%%%%%%%%%%%%%%%%
\subsubsection{Properties of $^{10}$Li}

We describe $^{10}$Li with a coupled $^9$Li+$n$ model, in which the $^9$Li-$n$ folding potential is determined using the two-neutron separation energy of $^{11}$Li as 0.31 MeV.
The two-body resonances are described in the CSM. As shown in Table \ref{tab:sec3_10Li}, when the TOSM is used on $^9$Li, 
the dual $p$-wave resonances are obtained near the $^9$Li+$n$ threshold energy with the coupling to the $^9$Li spin (3/2$^-$). 

For the $s$-wave states, the scattering lengths $a_s$ for the $^9$Li+$n$ system show negative values. 
In particular, the $2^-$ state shows a large negative $a_s$ value of $-17.4$ fm, which is comparable to the value of the $n$-$n$ system ($-18.5$ fm) \cite{teramond87}.
This result suggests the existence of a virtual state near the $^9$Li+$n$ threshold energy, and the inversion phenomenon in $^{10}$Li is nicely explained in the present model with the TOSM.   
The $d$-wave resonance states of $^{10}$Li are also predicted in Table~\ref{tab:sec3_10Li_d}.

For comparison, we describe $^{10}$Li without the excitations of $^9$Li (``inert core''), i.e., we adopt only the single 0p0h configuration in $^9$Li. 
There is no Pauli blocking effect of 2p2h configurations on the $p$-wave states shown in Fig.~\ref{fig:sec3_Li-Pauli} (B).  
In this case, we adjust the $\delta$ parameter of the potential strength as $(1+\delta)V^F$ \cite{myo08}.
From Table \ref{tab:sec3_10Li}, the $p$-wave resonances are obtained just above the $^9$Li+$n$ threshold energy, and $a_s$ shows small positive values, 
which do not suggest the virtual states of $^{10}$Li.
The $s$-$p$ shell gap is large in $^{10}$Li.
These results indicate that Pauli-blocking reasonably works to describe the properties of $^{10}$Li.

%%%%%%%%%%%%%%%%%%%%%%%%%%%%%%%%%
\begin{figure}[thb]
\begin{minipage}[thb]{11.0cm}
\centering
{\makeatletter\def\@captype{table} 
\caption{Resonance energies $E_r$ and the decay widths $\Gamma$ of the $p$-wave resonances of $^{10}$Li, in units of MeV, measured from the $^9$Li+$n$ threshold,
using TOSM and the inert core for $^9$Li. The $s$-wave scattering lengths $a_s$ are shown in units of fm.}
\label{tab:sec3_10Li} 
\begin{tabular}{ccc}
\hline\noalign{\smallskip}
                              &    TOSM      &  inert core    \\
\hline\noalign{\smallskip}
$(E_r,\Gamma)(1^+)$ [MeV]~    &~ (0.22,~ 0.09)~ &  (0.03,~ 0.005) \\
$(E_r,\Gamma)(2^+)$ [MeV]~    &~ (0.64,~ 0.45)~ &  (0.33,~ 0.20)  \\
$a_s(1^-)$ [fm]               &  $ -5.6$     &    1.4         \\
$a_s(2^-)$ [fm]               &  $-17.4$     &    0.8         \\
\hline\noalign{\smallskip}
\end{tabular}
\makeatother}
\end{minipage}
\hspace*{0.5cm}
\begin{minipage}[thb]{6.0cm}
\centering
{\makeatletter\def\@captype{table} 
\caption{$E_r$ and $\Gamma$ of the $d$-wave resonance of $^{10}$Li, in units of MeV.}
\label{tab:sec3_10Li_d} 
\begin{tabular}{cc}
\hline\noalign{\smallskip}
                        &    TOSM        \\
\hline\noalign{\smallskip}
$1^-~(E_r,\Gamma)$~    &~ (5.84,~ 5.16) \\
$2^-~(E_r,\Gamma)$~    &~ (5.81,~ 5.20) \\
$3^-~(E_r,\Gamma)$~    &~ (6.57,~ 6.31) \\
$4^-~(E_r,\Gamma)$~    &~ (5.30,~ 3.84) \\
\hline\noalign{\smallskip}
\end{tabular}
\makeatother}
\end{minipage}
\end{figure}
%%%%%%%%%%%%%%%%%%%%%%%%%%%%%%%%%%%%%%%%%

%%%%%%%%%%%%%%%%%%%%%%%%%%%%%%%
\begin{figure}[thb]
\centering
\includegraphics[width=10.5cm,clip]{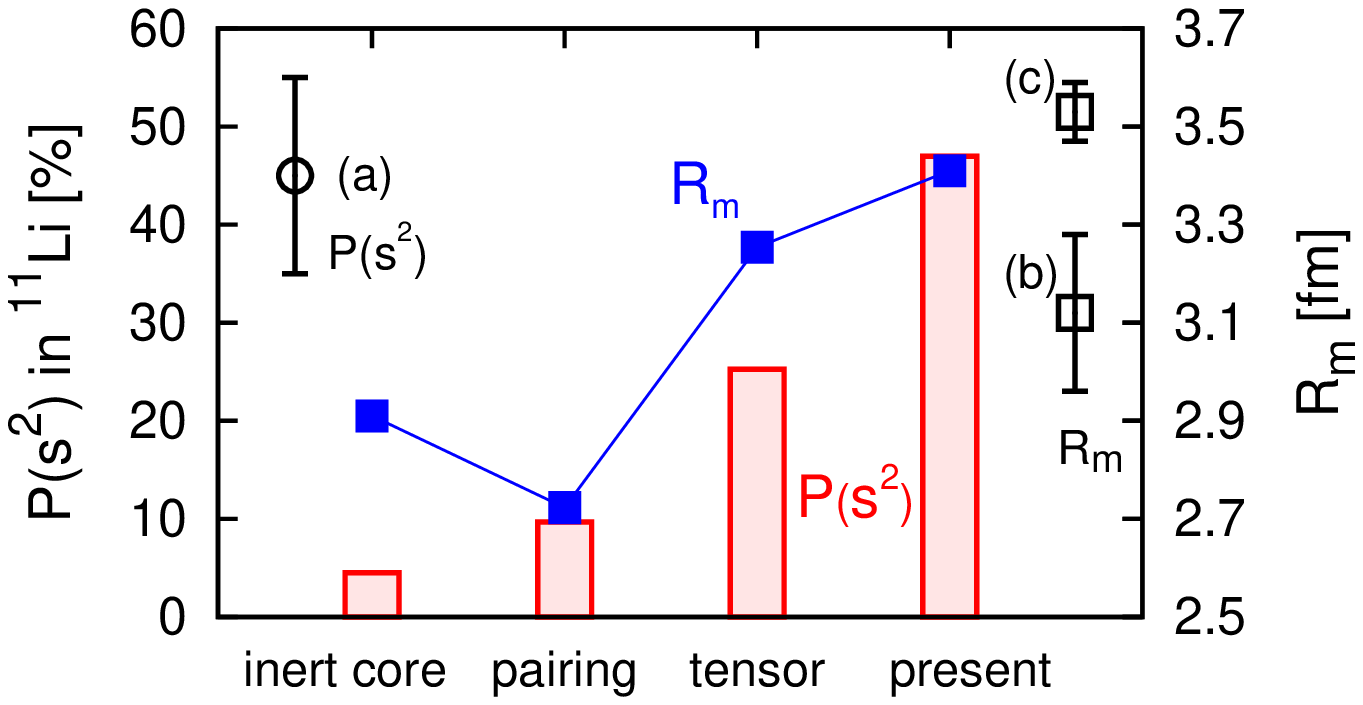}
\caption{
$(1s)^2$ probability $P(s^2)$ and matter RMS radius $R_m$ of $^{11}$Li expressed using four models, in comparison with the experimental values
((a)~\cite{simon99}, (b)~\cite{tanihata88}, and (c)~\cite{tostevin97}).}
\label{fig:sec3_11Li}
\end{figure}
%%%%%%%%%%%%%%%%%%%%%%%%%%%%%%%

%------------------------------------
\subsubsection{Properties of $^{11}$Li}\label{sec:11Li}

We describe $^{11}$Li using a coupled $^9$Li+$n$+$n$ model.
The ground state properties of $^{11}$Li are shown in Table~\ref{tab:sec3_11Li}, 
along with the respective configuration $P((nlj)^2)$ of halo neutrons and the various RMS radii.
When the TOSM is employed for $^9$Li, a large $P((1s)^2)$ value of 46.9 \% and a large matter radius $R_m$ of 3.41 fm are obtained, which are sufficient to explain the observations. 
The case of the ``inert core'' has a small $P((1s)^2)$ of 4.3\% and small $R_m$ values of 2.99 fm, which disagree with the observations. 
From the comparison between TOSM and inert core, the tensor and pairing correlations in $^9$Li are found to play important roles in breaking the magicity and creating the halo structure of $^{11}$Li, in addition to their role in the $s$-$p$ inversion phenomenon of $^{10}$Li.
In Fig.~\ref{fig:sec3_11Li}, we consider the individual correlation of tensor and pairing in $^9$Li in the calculation of $^{11}$Li. 
It is found that $P(s^2)$ becomes larger in the tensor case than in the pairing case.  
Finally, both blocking effects enhance $P(s^2)$ more and result in almost an equal amount of $(1s)^2$ and $(0p)^2$ configurations for $^{11}$Li. 

%%%%%%%%%%%%%%%%%%%%%%%%%%%%%%
\begin{table}[t]
\caption{Ground state properties of $^{11}$Li with $S_{2n}=0.31$ MeV. Two kinds of the $^9$Li descriptions of TOSM and inert core are shown.
Details are described in the text.}
\label{tab:sec3_11Li} 
\begin{center}
\renc{\baselinestretch}{1.15}
\begin{tabular}{lccllc}
\hline
                           &~~TOSM~~~& inert core & Expt. \\
\hline\noalign{\smallskip}
$P((p_{1/2})^2))$ [\%]   &  $42.7$  &  90.6      & ---  \\
$P((1s_{1/2})^2)$        &  $46.9$  & ~~4.3      & 45$\pm$10\cite{simon99} \\
\hline\noalign{\smallskip}
$R_m $ [fm]                &  3.41    &   2.99     & 3.12$\pm$0.16\cite{tanihata88},~~3.53$\pm$0.06\cite{tostevin97},~~3.71$\pm$0.20\cite{dobrovolsy06} \\
$R_p $                     &  2.34    &   2.24     & 2.88$\pm$0.11\cite{tanihata88} \\
$R_n $                     &  3.73    &   3.23     & 3.21$\pm$0.17\cite{tanihata88} \\
$R_{ch}   $                &  2.44    &   2.34     & 2.467$\pm$0.037\cite{sanchez06},~~~2.423$\pm$0.034\cite{puchalski06} \\
\hline\noalign{\smallskip}
\end{tabular}
\end{center}
\end{table}
%%%%%%%%%%%%%%%%%%%%%%%%%%%%%%

%%%%%%%%%%%%%%%%%%%%%%%%%%
\begin{figure}[t]
\centering
\includegraphics[width=13.0cm,clip]{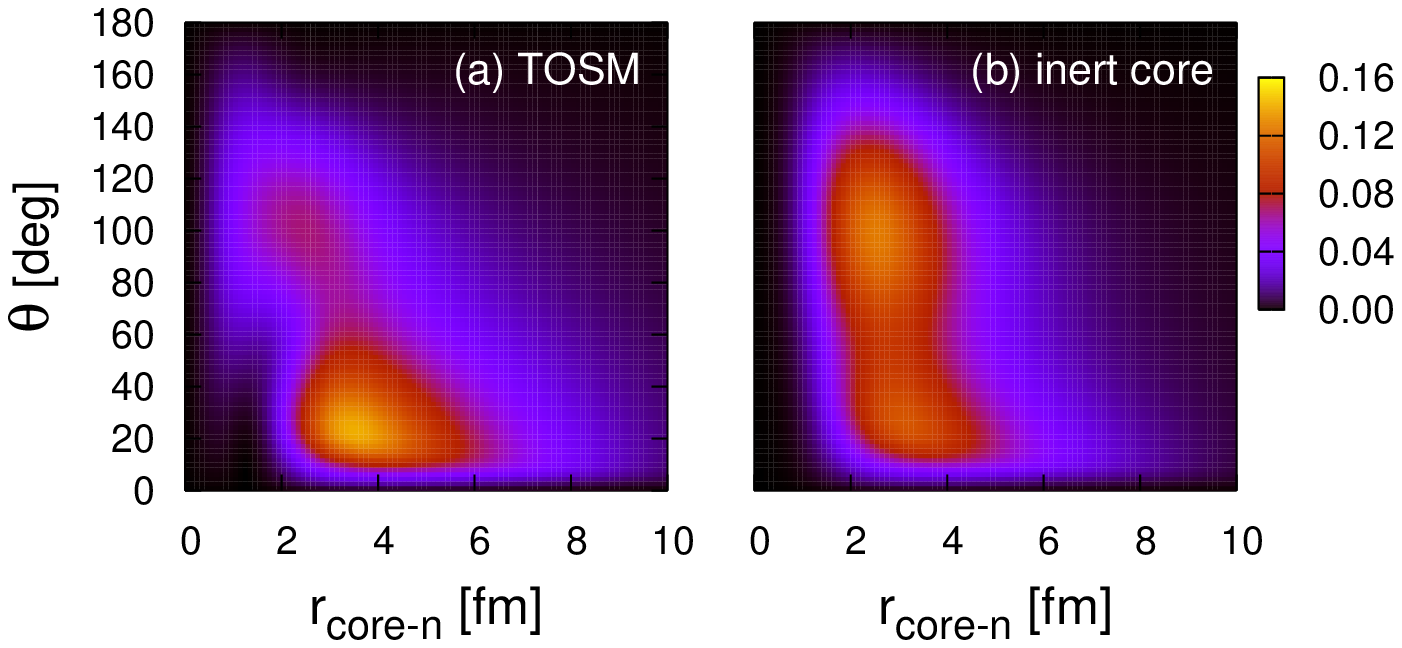}
\caption{(Color online) Two-neutron correlation density $\rho_{nn}(r_{{\rm core-}n}, \theta)$ for $^{11}$Li \cite{myo08}.  
Panel (a) shows the case of the calculation with the TOSM of $^9$Li and (b) shows the inert core case of $^9$Li.}
\label{fig:sec3_2n-density}
\end{figure}
%%%%%%%%%%%%%%%%%%%%%%%%%%

To see the spatial correlations of the halo neutrons in $^{11}$Li \cite{myo08,esbensen92,hagino05,matsuo05,zhukov93,nielsen01}, 
we calculate the density distribution of halo neutrons $\rho_{nn}(r,\theta)$ in $^{11}$Li as functions of the $^9$Li-$n$ distance $r$ and the opening angle $\theta$ between two neutrons.
In Fig.~\ref{fig:sec3_2n-density}, in the TOSM case (a), the dineutron configuration provides a maximum value of the density, although the density of neutrons is widely distributed.  
In contrast, the inert core case (b) with a small $s^2$ component does not show the enhancement of the dineutron configuration and the cigar-type configuration coexists with the dineutron type.
These results mean the importance of the $s^2$ component to the formation of the dineutron clustering configuration.
\section{ Comparison to the GSM approach}
In the previous section, 
the COSM approach \cite{suzuki88} has been discussed very promising
for the treatment of multi-valence particle systems around a
core-nucleus and for the description of the many-body unbound states.
The Hamiltonian of the COSM is given by Eq.~(\ref{eq:sec3_COSM_ham}),
which is the same as the translational invariant shell model.
The single-particle wave functions are expressed by the solutions
of the first term of Eq.~(\ref{eq:sec3_COSM_ham}),
which includes not only the bound state
but also resonant and continuum states.

In order to treat the resonant and continuum states
of the single-particle wave functions,
the so called Gamow shell model (GSM) has been
developed recently~\cite{betan02,
  michel02,hagen07,rotureau06,jaganathen12,volya09,betan14};
in this model,
the completeness relation discussed by Newton~\cite{newton60}
is generalized so as to include resonance poles
in a deformed contour of the Cauchy integration.
This ECR was proposed
by Berggren~\cite{berggren70,berggren96,berggren68,berggren92}.
The CSM also provides us with one of the deformed (rotated) contours,
as discussed in~\ref{sec:ECR}.
In our COSM calculations of the CSM (CS-COSM),
we employ the Gaussian basis functions and express
a single-particle wave function by the expansion of these Gaussian
basis functions.
The advantage of Gaussian basis functions is
the easy and analytical calculations of 
the Talmi-Moshinsky transformation coefficients
to obtain the matrix elements of the inter-nucleon
interactions in the second term of  Eq.~(\ref{eq:sec3_COSM_ham}).

In this section, we apply the CS-COSM to the
$sd$-shell nuclei in order to confirm the reliability
and extensibility of our core+$N_{V} \! \cdot \! (n$ or $p)$ model in descriptions
of spatially extended nuclear systems and reaction phenomena.
For this purpose, first, we show the radii
and energies of the oxygen isotopes and the mirror nuclei for $N_{V} \leq 4$
($^{17}$F, $^{18}$Ne, $^{20}$Mg)~\cite{masui06}.
Based on the results for the isotopes and the mirrors,
we discuss the way in which the CS-COSM can describe the radii
of many-body weakly bound systems and, simultaneously,
the effect of the change of the core size
on the total system through the semi-microscopic interaction
between the core and a valence nucleon~\cite{masui06}.
Next, we compare the CS-COSM to the GSM.
For reproducing the long tail of the weakly bound system
in the asymptotic region, it is necessary to take into account
the unbound components correctly.
Hence, we show the results $^{18}$O and $^{6}$He
as examples~\cite{masui14,masui07}
for stable and unstable nuclear systems, respectively.
Finally, we show the application of the CS-COSM to
the nucleon capture reactions of $^{17,18}$O~\cite{yamamoto09}.
In the model space of $^{16}$O$+n(+n)$,
the scattering states are described with
the CS-COSM wave functions. 

It is presented that the CS-COSM approach
is very suitable to describe not only spatially
extended bound states that are strongly coupled with continuum states
but also the nucleon capture reactions
in the case of a system with a relatively large core
plus a few valence nucleons.

\subsection{Energy and radius of oxygen isotopes}
Before showing the comparison with the GSM,
we discuss the dynamical effect of the core size on the radius of the total
system in the weakly bound systems.
Applying the CS-COSM to oxygen isotopes in the model space
of the $^{16}$O core plus neutrons,
we examine whether the change of the radii
can be explained by spatial extension of valence nucleons on the core.

In drip-line nuclei, where very small  binding energies are observed,
the nuclear radius extends with a long asymptotic tail of the wave function.
The structure, in which nucleus with a small binding energy,
has an extremely large tail of the wave function
is known as the ``halo'' structure~\cite{tanihata85}.
In a halo nucleus,
one or two nucleons are loosely bound with a relatively
stable core nucleus.
The typical examples of the halo structure are
$^{6}$He, $^{11}$Be and $^{11}$Li.
Recently, other heavier nuclei with the halo structure have been
observed~\cite{gaudefroy12,tanaka10,kobayashi12,nakamura09,takechi12}.

Unlike the typical halo nuclei,
there are some exceptional cases where a nucleus has
a large binding energy and a large nuclear size, simultaneously.
The oxygen isotopes have been considered to belong to the above
case based on the observation of their nuclear size~\cite{ozawa01b}.
In Ref.~\cite{ozawa01b},
the observed nuclear size allows the empirical $A^{1/3}$-law,
and radii of $^{23}$O and $^{24}$O show enhanced values as compared to
the empirical one.
On the other hand,
a smaller radius of $^{23}$O has also been recently
observed~\cite{kanungo11}.

From a theoretical point of view,
the enhancement of the radius from the empirical $A^{1/3}$-value at $^{23}$O 
and the systematics of the change of the radius for $^{16-24}$O
have not been described thus far~\cite{nakada06,abuibrahim09,hagen09}.
For determining the binding energy and radius of the oxygen isotopes,
we consider the change of the core size and its effect on the valence
nucleon model space to be key mechanisms.

Using the Gaussian expansion method (GEM)~\cite{hiyama03}
for finding the wave function,
we calculate the energy and radius of oxygen isotopes
by introducing a size-parameter dependence of the core nucleus
in the CS-COSM framework~\cite{masui06}.
In the calculation,
we use a semi-microscopic approach~\cite{kaneko91}
for determining the potential  between the core and valence nucleons.
Based on the semi-microscopic point of view,
the potential between the core and the $i$th valence nucleon
is constructed by a folding procedure
of the core wave function as follows~\cite{kaneko91}:
\begin{equation}
  \label{eq:sec4_core_n_hamiltonian_11}
   \sum_{m \in \mbox{\tiny Core}}
  \left\bra
    \Phi_{C} [b] \,
    \left|
      \frac{}{}
  v_{i m} \,
  \right|
      {\cal A'}
  \left\{
    \frac{}{}
  |  \Phi_{C} [b] \ket \,
  |  \Phi_{V} \ket 
\right\}
\right\ket
 \simeq 
  \Bigl[
  V^{d}_{i} [b] + V^{ex}_{i} [b]
  + v_{i}^{PF} [b]
\Bigr]
| \Phi_{V} \ket 
  \mbox{ .}
\end{equation}
Here, the wave function of the core $| \Phi_{C}[b] \ket$
is assumed to be the harmonic oscillator (h.o.) wave function
with the size parameter $b=1.723$ fm,
which is chosen to reproduce the observed radius of $^{16}$O,
i.e., $R_{\rm rms} = 2.53$ fm~\cite{ozawa01b}.
We take the size parameter $b$ as a variational parameter
in the calculation of the isotopes in order to investigate
the effect of the change in the core size on the radius~\cite{masui06}.
$| \Phi_{V} \ket $ is the wave function for the valence nucleons.
In Eq.~(\ref{eq:sec4_core_n_hamiltonian_11}),
$V_{i}^{d}$ and $V_{i}^{ex}$
are the direct and exchange parts, respectively, 
of the folding potential~\cite{kaneko91},
and $v_{i}^{PF}$ is the Pauli-potential to eliminate
to the Pauli forbidden states
in the relative motion between the core and valence nucleons.

\begin{figure}[t]
    \centering
  \includegraphics[width=10.0cm,clip]{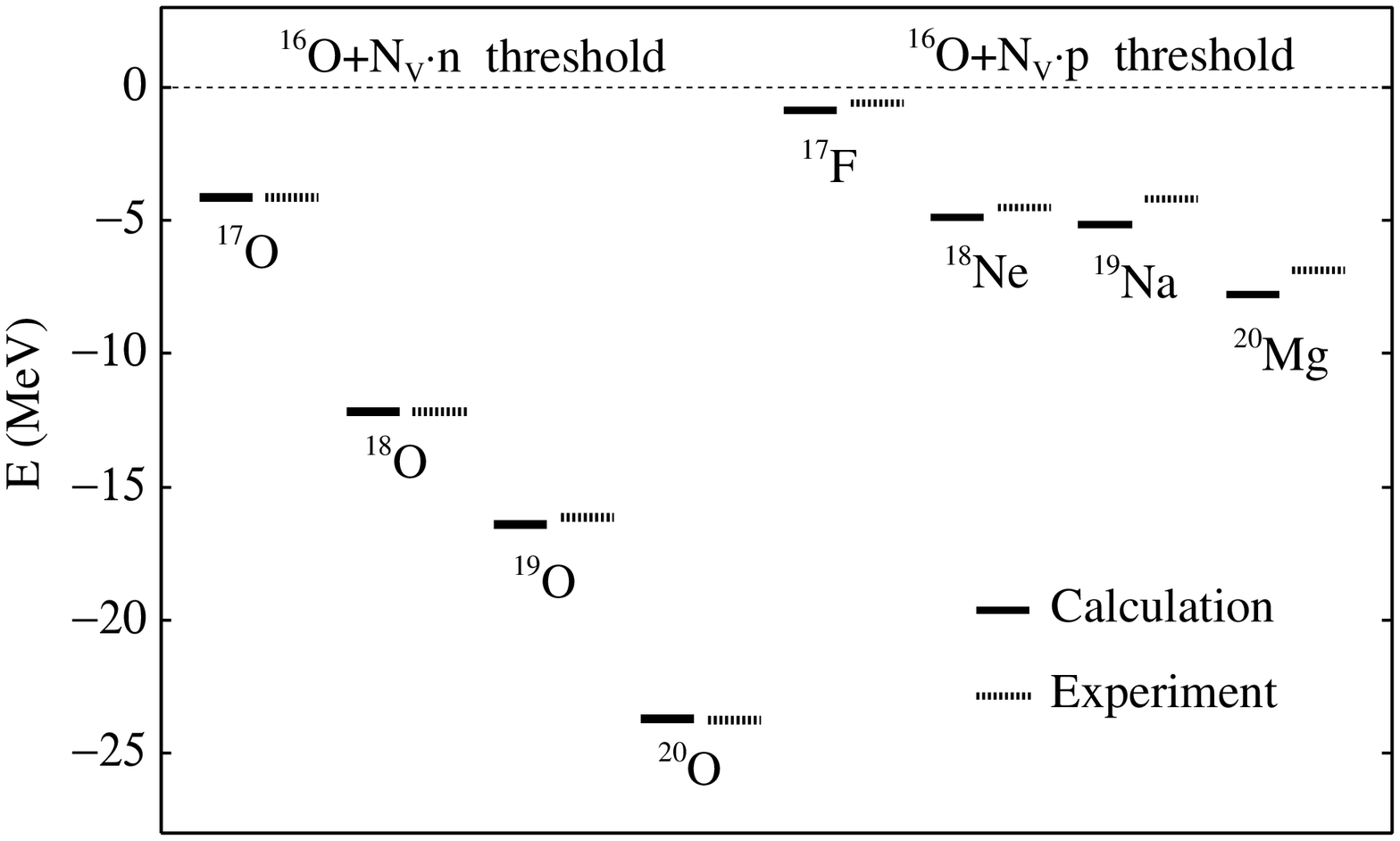}
  \caption{Calculated energies with respect to the
    $^{16}$O+$N_{V} \cdot  n$ and +$N_{V} \cdot  p $ thresholds~\cite{masui06}
    and the experimental values~\cite{ajzenberg86,ajzenberg87}
    for the oxygen isotopes and the mirrors.}
  \label{fig:sec4_B_E_16O_XN}
\end{figure}

We calculate the energy and radius of the oxygen isotopes
with the fixed core-size parameter ($b = 1.723$ fm)
and the mirror nuclei with the $^{16}$O-core plus protons~\cite{masui06}.
The calculated energies of the oxygen isotopes and the mirrors
are shown in the left- and right-half regions
of Fig.~\ref{fig:sec4_B_E_16O_XN}.
The calculated energies agree with 
the experimental ones~\cite{ajzenberg86,ajzenberg87}
in both isotopes and the mirrors.
Hence, the potential and basis function
can be considered as appropriate
to describe the oxygen isotopes and the mirrors
for $N_{V} \leq 4$.

\begin{figure}[t]
    \centering
 \includegraphics[width=10.5cm,clip]{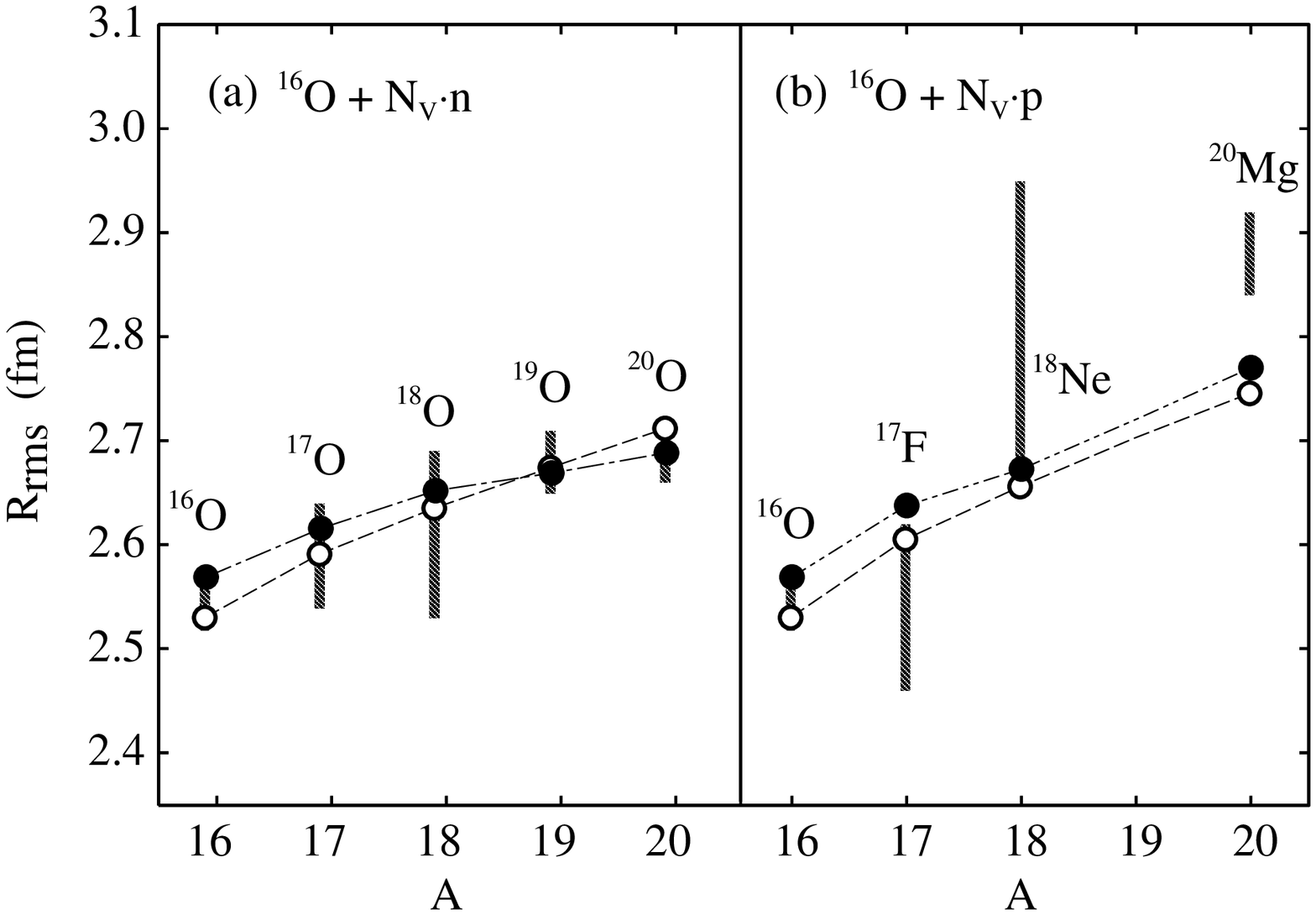}
  \caption{Calculated r.m.s. radii for (a)$^{16}$O+$N_{V} \cdot  n$
    and (b)$^{16}$O+$N_{V} \cdot p$~\cite{masui06}.
    Open and solid circles are
    results for a fixed- and varied-$b$ size parameters, respectively.
    Thick lines are the experimental values~\cite{ozawa01b}.
    The dotted lines are drawn as guides to the eye.}
  \label{fig:sec4_R_rms_16O_XN_fixed_b}
\end{figure}

The calculated radii of the isotopes and the mirror nuclei
with the fixed core-size parameter 
are shown in Fig.~\ref{fig:sec4_R_rms_16O_XN_fixed_b}.
In the fixed-$b$ calculation,
the calculated radii agree with the observed ones~\cite{ozawa01b}
within the error bars except for $^{20}$Mg.
This result indicates that 
an enhancement of the radius at a certain nucleus
cannot be reproduced by using only the fixed core-size parameter.

The nuclear radius is determined by the size of the core as well as 
the spatial extension of the wave function of the valence nucleons.
Hence, it can be considered that the change in the core size
affects the total radius of a nucleus.
Also, the optimum value of the core-size parameter $b$
depends on other quantities, such as
the number of valence nucleons and the Coulomb interaction.
Hence, the optimum core size for the isotopes and the mirrors
changes with the increase in the number of valence nucleons.

\begin{table}[t]
  \centering
  \caption{The energy calculated using the $^{16}$O-core
    and the radii for the fixed- and varied-$b$ parameters.
    The coefficients $c_{1}^{2}$ and $c_{2}^{2}$ correspond to those in Eq.~(\ref{eq:sec4_0+_expand}).}
  \label{tab:sec4_radius_O-iso}
  \begin{tabular}{rlllcc}
    \hline
    & & E (MeV) & $R_{\rm rms}$ (fm)
    & $c_{1}^{2}$ & $c_{2}^{2}$ \\ 
    \hline
    $^{18}$O: &  Fixed-$b$
    & $-12.18$ & 2.64 & 0.836 & 0.103 \\
    & Varied-$b$
    & $-12.00$ & 2.65 & 0.832 & 0.106 \\
    & Expt.\cite{ozawa01b,ajzenberg87}
    & $-12.19$
    & $2.61\pm0.08$ & --- & --- \\
    \hline
    $^{18}$Ne: &  Fixed-$b$
    & $-4.94$
    & 2.66 & 0.824 & 0.115 \\
    & Varied-$b$
    & $-4.68$
    & 2.68 & 0.812 & 0.125 \\
    & Expt.\cite{ozawa01b,ajzenberg87}
    & $-4.52$
    & $2.81\pm0.14$ & --- & --- \\
    \hline
  \end{tabular}
\end{table}

In order to observe the precise effect
of the change of the core size,
we perform calculations for $N_{V}=2$ systems,
$^{18}$O ($^{16}$O+$2n$) and $^{18}$Ne ($^{16}$O+$2p$)~\cite{masui06}.
The calculated energies and radii
of the $0^{+}$ ground states of $^{18}$O and $^{18}$Ne
are shown in Table~\ref{tab:sec4_radius_O-iso}.
The optimum values are $ b=1.74$ fm
for both $^{18}$O and $^{18}$Ne, 
and the calculated radii are
$R_{\rm rms} = 2.65$ fm and $2.68$ fm, respectively.

Even though the adopted values of $b$ are the same
for $^{18}$O and $^{18}$Ne,
the difference in the radii between the fixed- and  varied-$b$ calculations
of the $^{18}$Ne case is larger than that for the $^{18}$O case.
To investigate the mechanism underlying the radii 
of $^{18}$O and $^{18}$Ne,
we expand the $0^{+}_{1}$ wave function
with the single-particle states of the $^{16}$O+$N$ system as
\begin{equation}
  \label{eq:sec4_0+_expand}
| 0^{+} \ket  =  c_{1}| (d_{5/2})^{2} \ket
+ c_{2}| (s_{1/2})^{2} \ket + \cdots
\mbox{ .}
\end{equation}
The calculated $c_{k}^{2}$ are shown in Table~\ref{tab:sec4_radius_O-iso}.
In $^{18}$O, the difference of $c_{i}^{2}$ in fixed-$b$ and varied-$b$
calculations is not significant in comparison with the results of $^{18}$Ne. 
In  $^{18}$Ne, the value of $c_{2}^{2}$,
which is the squared amplitude of $(s_{1/2})^{2}$ configurations,
increases from $0.115$ in the fixed-$b$ calculation to $0.125$ in the varied-$b$ one.
Oppositely,  $c_{1}^{2}$ decreases from $0.824$ to $0.812$.

In the previous study~\cite{masui06},
we showed that the semi-microscopic $^{16}$O$+N$ potential
is sensitive to the change in the core-size parameter.
The potential becomes weaker with larger $b$-values.
Only the $s$-wave potential of the $^{16}$O+$p$ is insensitive
to the change of the $b$-value due to the Coulomb force.
Hence, even though the adopted-$b$ parameter is the same
for both the $^{18}$O and $^{18}$Ne cases,
the $s$-wave component in $^{18}$Ne increases larger than in $^{18}$O.
In the view point of the spatial extension of the wave function, 
the $s$-wave component is broader than the other partial waves.
Therefore, the radius of $^{18}$Ne with the varied-$b$ calculation
becomes larger than that of the fixed-$b$ calculation.

To determine the dependence on the nucleon-number,
we perform the varied-$b$ calculation
for $N_{V} \leq 4$.
The calculated radii are shown in Fig.~\ref{fig:sec4_R_rms_16O_XN_fixed_b}(a)
for the oxygen isotopes ($^{16}$O+$N_{V} \cdot n$ systems)
and in Fig.~\ref{fig:sec4_R_rms_16O_XN_fixed_b}(b)
for the mirror nuclei ($^{16}$O+$N_{V} \cdot p$ systems).
On comparing the results for oxygen isotopes in
Fig.~\ref{fig:sec4_R_rms_16O_XN_fixed_b}(a),
we find that the changes in the radii obtained
by the varied-$b$ calculation
with the increase in the number of valence neutrons
are smaller than those obtained by the fixed-$b$ calculation.
On the other hand,
on comparing results shown in Fig.~\ref{fig:sec4_R_rms_16O_XN_fixed_b}(b),
the increase in the radii obtained by the varied-$b$ calculation
is sizable for $^{20}$Mg,
while there is no substantial increase in the fixed-$b$ calculation values.
This is due to the difference in the optimum values in
the varied-$b$ calculation for the isotopes and the mirrors,
as shown in Table~\ref{tab:sec4_Rrms_Neo_COSM}.

For the oxygen isotopes,
the optimum value of the size parameter, $b_{\rm op}$ becomes smaller
as the number of valence neutrons increases.
In contrast, for the mirror nuclei,
the optimum value of $b$ remains the same  
in $^{18}$Ne and $^{20}$Mg, as shown in Table~\ref{tab:sec4_Rrms_Neo_COSM}.
Therefore, the radius of $^{20}$Mg becomes considerably
larger than that of $^{20}$O.

\begin{table}[t]
  \centering
  \caption{$R_{\rm rms}$ and $b_{\rm op}$ 
    obtained by the varied-$b$ calculations.
    Labels (n) and (p) indicate the oxygen isotopes and the mirrors, respectively.
     All values are in fm.}
  \label{tab:sec4_Rrms_Neo_COSM}
  \begin{tabular}{rcccc}
    \hline
     $N_{V}$  & 1 & 2 & 3 & 4 \\
    \hline
    $b_{\rm op}$: (n) & $1.76$ & $1.74$ & $1.72$ & $1.71$ \\
         (p)   & $1.76$ & $1.74$ & $-$   & $1.74$ \\
    \hline
     $R_{\rm rms}$: (n) & $2.62$ & $2.65$ & $2.67$ & $2.69$ \\
               (p) & $2.64$ & $2.68$ & $-$  & $2.77$ \\
    \hline
  \end{tabular}
\end{table}

To summarize so far,
the core-size parameter in the varied-$b$ calculation,
which is energetically determined
by the core and valence-nucleon systems,
can be different values for oxygen isotopes and the mirrors.
In the model space of $N_{V} \leq 4$,
we showed the optimized values of $b$ for the mirror nuclei are
obtained as systematically larger than those for the oxygen isotopes.
Using the adopted-$b$ values, the difference of the radius between
the oxygen isotope and the mirror becomes larger
as the number of valence nucleon increases.
We consider the change of the core size is
an important mechanism to determine the nuclear radius
through the semi-microscopic core+$N$ potential,
and the effect is different for the oxygen isotopes and the mirror nuclei.
Hence, it has a possibility to describe 
the sudden change of the radius at $^{23}$O~\cite{ozawa01b}
in terms of the change of the core size
and its effect on the valence nucleon model space.

\subsection{Comparison to GSM for  $^{18}$O and  $^{6}$He}

\subsubsection{Basis set and formalism of calculation in GSM}
The importance of the continuum states in the shell model picture
has been discussed for
many years~\cite{okolowicz03,fano61,mahaux69,bennaceur00}.
The first step for inclusion of the continuum states
involves dividing the model space
into a physical space (bound states)
and an unphysical one (continuum states).
The coupling of the physical and unphysical spaces
is considered the calculation through the effective interaction.
However, the problem of treating 
the continuum-continuum coupling into the model space
still remains.

The explicit treatments of the continuum-continuum coupling
have been developed by re-defining the single-particle states
with bound and unbound states.
The Gamow shell model (GSM)
is one of such the approaches,
and has been extensively applied to
unstable nuclei~\cite{betan02,michel02,hagen05,hagen07,
rotureau06,jaganathen12,volya09,betan14}.
In unstable nuclear systems,
as shown in the previous sections,
it is indispensable to treat resonant states.
We have discussed that the CS-COSM is also a promising approach
to treat multi-valence particle systems
with a core nucleus and to describe the many-body resonant states.
Hence, it is very interesting to investigate the similarity
and difference between the CS-COSM and the GSM.

The basic idea of the GSM is connected with the Berggren
ensemble~\cite{berggren68},
which finds a mathematical setting in the Rigged Hilbert Space~\cite{gelfand61}. 
The Berggren completeness relation~\cite{berggren92}
replaces the real-energy scattering states in the completeness relation
discussed by Newton~\cite{newton60}
into the resonance part of the spectrum 
and a background of complex-energy continuum states
on the same footing as the bound and scattering spectra.
As the benefit of the explicit inclusion of the non-resonant
continuum and resonant poles,
the contribution of the unbound states to the one- and two-body
matrix elements can be discussed.

\begin{figure}[t]
  \centering 
  \includegraphics[width=11.0cm,clip]{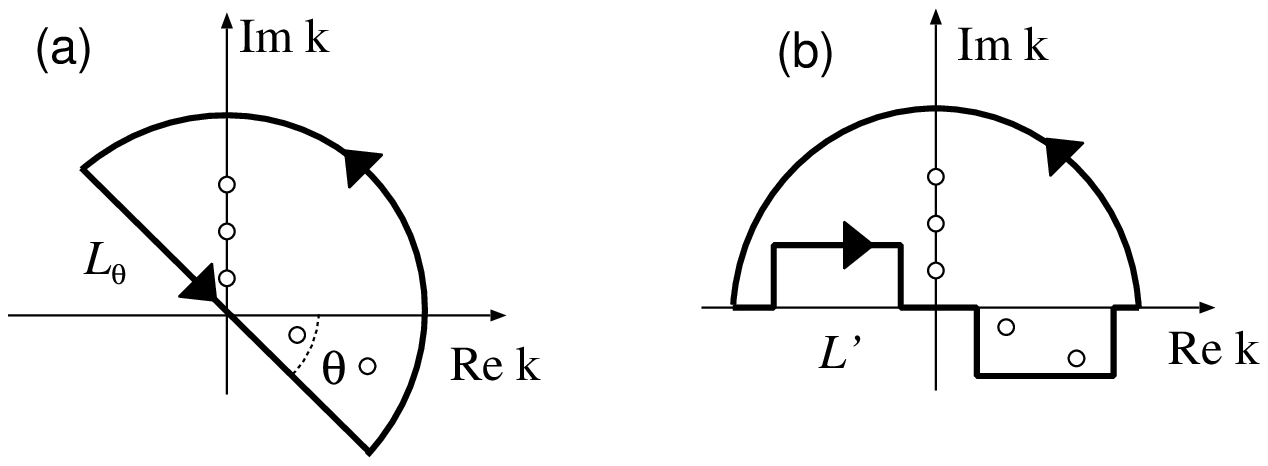}
  \caption{Definition of contour-path on the complex momentum plane
    for (a) CSM and (b) GSM.}
  \label{fig:sec4_contour}
\end{figure}

In the CSM,
we use the extended completeness relation~\cite{myo98} as follows:
\begin{equation}
  \label{eq:sec4_complete_rel_CSM_1}  \bv{1}_{i}
  =   \sum_{n = b, r} |  \phi^{\theta \, (n)}_{i} \ket
  \bra \tilde{\phi}^{\theta \, (n)}_{i}|
  + \oint_{L_{\theta}} dk \, | \phi^{\theta}_{i}(k) \ket
  \bra \tilde{\phi}^{\theta}_{i}(k) |
  \mbox{ .}
\end{equation}
The integration of the closed path of the momentum
is carried out along a rotated semi-circle on the complex momentum plane as 
shown in Fig.~\ref{fig:sec4_contour}(a).

The single-particle wave functions $ \phi^{\theta \, (n)}_{i} $ of the CSM
in Eq.~(\ref{eq:sec4_complete_rel_CSM_1}) are eigenfunctions of the
single particle Hamiltonian given by the first term
in Eq.~(\ref{eq:sec3_COSM_ham}).
They are expressed by the Gaussian expansion with the basis function
like Eq.~(\ref{eq:sec3_COSM-base1}).

In the GSM calculations,
the contour path of the momentum is deformed
so that the resonant poles are included 
in the closed path, as shown in Fig.~\ref{fig:sec4_contour}(b).
In this case, the completeness relation is defined by the bound
 and resonant states and the continuum along the deformed contour path $L'$
 as follows:
\begin{equation}
  \label{eq:sec4_complete_rel_GSM}
  \bv{1}_{i}
  =   \sum_{n = b, r} |  \phi^{(n)}_{i} \ket
  \bra \tilde{\phi}^{(n)}_{i} |
  + \oint_{L'} dk \, | \phi_{i}(k) \ket
  \bra \tilde{\phi}_{i}(k) |
  \mbox{ ,}
\end{equation}
where ``$r$'' in the sum stands for the resonant states enclosed by $L'$.
The path presented in Fig.~\ref{fig:sec4_contour}(b) is one of
examples, in which the contour path on the real momentum axis is deformed
so that two resonant poles are included in the closed path. 
In practice, the contour path is discretized and the integration
for $L'$ is done by taking the sum of the discretized complex momentum.

One of the differences between the CS-COSM and the GSM is
the treatment of the unbound components of the single-particle states.
The CS-COSM approach describes the single-particle wave function
as the linear combination of the Gaussian basis sets.
Thus, the wave function of $N_{V}$ valence nucleons
is expressed by a sum of the  Slater determinant of the Gaussian basis functions as 
\begin{eqnarray}
  \label{eq:sec4_COSM_Basis_001}
 \Phi_{\rm COSM}^{(m)}(\bv{r}_{1} , \bv{r}_{2} \, , \cdots , \bv{r}_{N_{V}} ; JM)
 &  \equiv &
  {\cal A}
  \left\{
    u_{1 }^{(m)}(r_{1}) \,
    u_{2 }^{(m)}(r_{2}) 
    \cdots    
    u_{N_{V} }^{(m)}(r_{N_{V}})
    \,
    | JM^{(m)} \ket
    \right\}
  \mbox{ .}
  \nonum
\end{eqnarray}
Here, ${\cal A}$ is the antisymmetrizer for the valence nucleons.
By taking all possible combinations of $m$,
we can obtain the desired wave functions of
valence nucleons including continuum effects.

In the GSM,
the components of the unbound states are included
to the wave function
in terms of the explicit treatment of the continuum states.
The basis function of the GSM is constructed by a products of
the single-particle states,
$\hat{h}_{i} \psi_{i} = \epsilon_{i} \psi_{i} $,
as follows:
\begin{eqnarray}
  \label{eq:sec4_GSM_Basis_001}
 \Phi_{\rm GSM}^{(k)}(\bv{r}_{1} , \bv{r}_{2} \, , \cdots , \bv{r}_{N_{V}} ; JM )
 &  \equiv &
  {\cal A}
  \left\{
    \psi_{1 }^{(k)}(r_{1}) \,
    \psi_{2 }^{(k)}(r_{2}) 
    \cdots    
    \psi_{N_{V}}^{(k) }(r_{N_{V}})
    \,
    | JM^{(k)} \ket
    \right\}
  \mbox{ .}
  \nonum
\end{eqnarray}

Mathematically, the wave functions of the CS-COSM and the GSM
are identical,
if the model spaces of the basis function are ``complete''.
Two approaches can be connected under the unitary transformation,
$ U \Psi_{\rm COSM}  =  \Psi_{\rm GSM} $,
and we have a relation for the basis functions of
the CS-COSM and the CSM as follows:
\begin{eqnarray}
  \label{eq:sec4_COSM_Basis_003}
   \sum_{m} \, ( U c^{(m)}) \,
  \Phi_{\rm  COSM}^{(m)}
  & = & 
  \sum_{k} \, d^{(k)} \,
  \Phi_{\rm  GSM}^{(k)}
  \mbox{ ,}
\end{eqnarray}
where $c^{(m)}$ and $d^{(k)}$ are expansion coefficients of the wave
function.
Eq.~(\ref{eq:sec4_COSM_Basis_003}) shows the Gaussian basis set
can include the components of the unbound states
through the unitary transformation of the Berggren ensembles.
However, practical calculations of both method are performed
within appropriate basis states and different numerical calculations.

Another difference between the CS-COSM and the GSM is the treatment of two-body matrix
elements (TBME).
To calculate TBME, one needs to perform the Talmi-Moshinsky (TM) transformation
and calculate the coefficients using the basis function.
In the CS-COSM, with the help of the mathematical property of the Gaussian,
the TM-coefficients can be calculated analytically.
In the GSM, since the direct calculation of the TBME using
the unbound states is numerically demanded,
the harmonic oscillator (h.o.)
expansion procedure is applied~\cite{michel08}.
The matrix element between 
$ \Phi^{(i)}_{\rm GSM} $ and $ \Phi^{(j)}_{\rm GSM}$
is obtained through the following steps:
\begin{eqnarray}
  \label{eq:sec4_TBME_GSM_01}
  \bra
    \Phi^{(i)}_{\rm GSM} | O_{12} | \Phi^{(j)}_{\rm GSM} 
    \ket  
   & = &
   \sum_{\alpha, \beta}
   \bra
   \Phi^{(i)}_{\rm GSM} |
   \Phi_{\rm ho}^{(\alpha)}
   \ket
   \bra
   \Phi_{\rm ho}^{(\alpha)} \, |
   O_{12} |
   \Phi_{\rm ho}^{(\beta)}
   \ket
   \bra
   \Phi_{\rm ho}^{(\beta)} |
   \Phi^{(j)}_{\rm GSM} 
   \ket  \nonum
   & = & 
   \sum_{\alpha, \beta}
   d^{*}_{i,\alpha} \,
   d_{j,\beta}
   \bra
   \Phi_{\rm ho}^{(\alpha)} \, |
   O_{12} |
   \Phi_{\rm ho}^{(\beta)}
   \ket
    \mbox{ ,}
\end{eqnarray}
where $ \Phi_{\rm ho}^{(\alpha)} $ are h.o. basis functions
and $d_{i,\alpha}$ is the overlap
between the GSM basis function $\Phi^{(i)}_{\rm GSM}$ and the h.o. basis function:
\begin{equation}
  \label{eq:GSM_comp_02}
d_{i,\alpha} \equiv \bra
   \Phi_{\rm ho}^{(\beta)} \, |
    \Phi^{(i)}_{\rm GSM} 
   \ket
   \mbox{ .}
\end{equation}

The model space of every calculation can be examined
by comparing the components of the calculated wave functions.
We expand the wave function obtained by the CS-COSM
with the single-particle states
and compare the components to those obtained by the GSM.
For this purpose,
we prepare eigenfunctions of the multi-nucleon systems
and complete set of the single-particle states obtained by using the CSM.
In the case of the $0^{+}$ ground state of a core+$N$+$N$ system,
the wave function can be expanded with 
the single-particle states as follows:
\begin{eqnarray}
  \label{eq:sec4_eigen_COSM_18O_2}
  | \Psi_{\rm COSM} (0^{+}) \ket & = &
  \left\{
    \bv{1}_{1} \otimes \bv{1}_{2}
  \right\}
    | \Psi_{\rm COSM} (0^{+}) \ket
  \nonum
  & = & 
  C_{1} \, | (\psi_{0d_{5/2}})^{2} \ket    
  +C_{2} \, | (\psi_{1s_{1/2}})^{2} \ket
  + \cdots
  \mbox{ .}
\end{eqnarray}
Here, $\psi_{n \ell j}$ are single-particle states
of the core+$N$ sub-system.
In the case of two valence-nucleons system,
the combination of the components is of three types:
(a) both states are pole states,
for which we refer to $(C_{k})^{2}$ as ${\cal P}_{(n \ell j)^{2}}$,
(b) one is a pole and the other one is a continuum state,
and (c) both states are continuum states.
In cases (b) and (c), we indicate the number of continuum
states in the basis set and refer to this number as ``${\cal S}\#$''.
Hence, the cases (b) and (c) correspond to
${\cal S}1$ and ${\cal S}2$, respectively.

\subsubsection{Oxygen isotopes}
We show a comparison for $^{18}$O
as an example of the stable nucleus.
In the CS-COSM calculation, the $^{16}$O+$n$ interaction in
Eq.~(\ref{eq:sec4_core_n_hamiltonian_11})
reproduces two bound states ($5/2^+$, $1/2^+$)
and one narrow resonant state ($3/2^+$)
in the $^{17}$O system.
For the state with angular momenta $\ell \geq 3$ and $\ell = 1$;
$p_{3/2}$,  $p_{1/2}$, $f_{7/2}$, $\ldots$, $h_{9/2}$,
neither bound nor resonant states are obtained
in the region of $\theta \le 10^{\circ}$ in the complex momentum plane.
These partial-wave components become the continuum contributions.
We use the Volkov No.2 interaction~\cite{volkov65} for the valence nucleons.
In the GSM calculation~\cite{michel02},
the single-particle states are obtained by solving the $^{16}$O+$n$
system so as to reproduce the three states,
$5/2^+$, $1/2^+$ and $3/2^+$.
For the valence nucleons,
the surface-delta type interaction such as
  $V(r) = -V_{0} \, \delta(\bv{r}_{1} - \bv{r}_{2})
  \delta(r_{1} - R_{0})$
is employed.

The results for $^{18}$O are shown in
Table~\ref{tab:sec4_Contri_continuum_2}.
The main component in $^{18}$O is $(0d_{5/2})^{2}$.
Since the $5/2^{+}$-state is the ground state of $^{17}$O,
the ${\cal P}_{(0d_{5/2})^{2}}$ indicates two valence neutrons
occupy the single-particle orbit of $^{17}$O.
${\cal P}_{(1s_{1/2})^{2}}$ is less than 10 percent
of the total contribution in  the ground state of $^{18}$O.
The imaginary parts of the contributions of these two bound states
are not zero but very small because of a  numerical error
in the expansion of the complex rotated wave function;
these values are less than $10^{-3}$.
The resonant pole contributions have an imaginary part,
which compensates for that of the continuum contributions,
${\cal S}1$ and ${\cal S}2$.

\begin{table}[t]
  \centering
  \caption{Contribution of the poles and continua in $^{18}$O.
    The core-$n$ eigenstates are calculated by CSM~\cite{masui07}
    with a rotational angle $\theta =10 ^{\circ}$.
   The result obtained by the GSM is taken from
   Ref.\protect\cite{michel02}.
   In the table, $\delta$ are  small values of the order of
   $\delta \sim 10^{-3}$.}
  \label{tab:sec4_Contri_continuum_2}
\begin{tabular}{lll}
    \hline
    $(C_{k})^{2}$   & CS-COSM~\cite{masui07}& 
    GSM~\cite{michel02} \\
    \hline
    ${\cal P}_{(0d_{5/2})^{2}}$
    & $0.830 -i \delta$  
    & $0.872 +i \delta$ 
    \\ 
    ${\cal P}_{(1s_{1/2})^{2}}$
    & $0.096  -i \delta$
    & $0.044  -i \delta$
    \\ 
    ${\cal P}_{(0d_{3/2})^{2}}$
    & $0.028 - i 0.005$
    & $0.028  -i 0.007$
    \\ 
     ${\cal S}1$
    & $0.020 + i 0.004$
    & $0.042 +i 0.005$
    \\
    ${\cal S}2$
    & $ 0.026 + i 0.001$
    & $ 0.015 + i 0.002$\\
    \hline
  \end{tabular}
\end{table}

From Table~\ref{tab:sec4_Contri_continuum_2},
we find the CS-COSM and GSM calculations show
similar results for the contributions
of ${\cal P}$ and ${\cal S}$,
even though the $NN$-interactions are different.
Therefore, it can be considered that
the single-particle state picture is dominant
in the ground state,
and the range of the $NN$-interaction,
whether a finite or zero-range one is applied,
is insensitive to determine
the nature of the ground state wave function.

\subsubsection{Helium isotopes}
In the $^{18}$O system,
the dependence of the $NN$-interaction
to the ground state wave function is weak,
and hence the CS-COSM and the GSM give almost the same results~\cite{masui06}.
For $^{6}$He, 
first, we perform the comparison between
the CS-COSM and the GSM using different interactions
for the valence nucleons in the same way as done in $^{18}$O.
We investigate the $NN$-interaction dependence
for the case of the weakly bound systems.
Second, in order to confirm the model spaces for the CS-COSM and the GSM,
we show the comparison by using exactly the same interaction
for the valence nucleons~\cite{masui14}.

We prepare the $0^{+}$ ground state wave function $\Psi_{\rm COSM}
(0^{+})$ of $^{6}$He solved by using the CS-COSM
and expand the wave function with the $^{5}$He ($\alpha$+$n$) system
using the ECR.
Since there is no bound state in the $\alpha$+$n$ system,
we construct the ECR by applying the CSM with a large rotation angle $\theta = 38^{\circ}$.
The $3/2^{-} \, (0p_{3/2})$ and $1/2^{-} \, (0p_{1/2})$ states are obtained
with complex eigenvalues,
$0.74-i0.29$ MeV and $2.11-i2.94$ MeV, respectively.
We calculate two pole-contributions, ${\cal P}_{(0p_{3/2})^{2}}$ and ${\cal P}_{(0p_{1/2})^{2}}$.
Other resonant states for higher angular momenta 
are obtained as very broad resonant states  with a large imaginary part as
$0d_{5/2} = 28.5 -i21.6 $ MeV, $0d_{3/2} = 29.1 -i37.3$ MeV
and $0f_{7/2} = 25.4 -i29.0$ MeV.
Therefore, we do not treat these states as the pole ones
and include them to the continuum-contributions, ${\cal S}1$ and ${\cal S}2$.

For the comparison,
we show the results obtained by the GSM~\cite{michel02,hagen05}.
In the GSM calculations of Refs.~\cite{michel02,hagen05},
the authors employ a surface-delta-type~\cite{michel02}
and a separable-type~\cite{hagen05} interactions.
This is because the calculation of TBME using the single-particle
states, in which resonant and continuum states are included,
becomes numerically demanding task, 
and the h.o. expansion~\cite{michel08} in
Eq.~(\ref{eq:sec4_TBME_GSM_01})
is not introduced in the calculation.
Contrary to these GSM calculations,
we use the Minnesota force~\cite{tang78},
which is a finite-range effective interaction
fitted to the $NN$-scattering phase-shifts.

We show the comparison between the CS-COSM~\cite{masui07}
and the GSM~\cite{michel02,hagen05} in Table~\ref{tab:sec4_Contri_continuum_3}.
In our CS-COSM calculation, the partial waves in the core+$n$ system
are taken up to $\ell_{\rm max} = 5$.
On the other hand, the GSM calculations have been done in
the model space as $\ell =1$.
Due to the difference of the model space,
the calculated contributions of the CS-COSM and the GSM are different each
other, except  for ${\cal P}_{(0p_{3/2})^{2}}$.

\begin{table}[t]
  \centering
  \caption{Contribution of the poles and continua for the $^{6}$He
    case~\cite{masui07}.
    The GSM calculations are taken from
    Refs.~\protect{\cite{michel02,hagen05}}.}
  \label{tab:sec4_Contri_continuum_3}
  \begin{tabular}{crrr}
    \hline
    $(C_{k})^{2}$ &  CS-COSM~\cite{masui07}
    & GSM~\cite{michel02} & GSM~\cite{hagen05}  \\
    \hline
    ${\cal P}_{(0p_{3/2})^{2}}$  
    & $ 1.211  -i0.666 $
    & $0.891  -i0.811$
    & $1.105-i0.832$
    \\ 
     ${\cal P}_{(0p_{1/2})^{2}}$  
    & $ 1.447  +i0.007$
    & $0.004  -i0.079$
    & $0.226-i0.161$  \\
    $S1$  
    & $-2.909 +i0.650$
    & $0.255  + i0.861$
    & $-0.259 + i1.106$   \\
    $S2$  
    & $ 1.251 +i0.009$
    & $-0.150+i0.029$
    &   $-0.072 -i0.113$   \\
    \hline
  \end{tabular}
\end{table}

As shown in Table~\ref{tab:sec4_Contri_continuum_3},
${\cal P}_{(0p_{1/2})^{2}} $ of the GSM is much smaller than that of the CS-COSM.
Since the $0p_{1/2}$-state of $^{5}$He is a broad resonant state,
the coupling to continuum states,
in which higher partial waves are included,
becomes important.
In the GSM calculations, the model space is limited to $\ell =1$. 
Hence, in order to investigate the contribution of inclusion of the higher
partial waves,
we perform the calculation with the $\ell =1$ model space,
where the strength of the $NN$-interaction is adjusted to the $0^{+}$ state.
We expand $(p_{3/2})^{2}$ and $(p_{1/2})^{2}$
into the pole and continuum contributions.
Thus, we add a suffix of $lj$ to ${\cal S}1$ and ${\cal S}2$
and show the results in Table~\ref{tab:sec4_Contri_conti_p32_p12_6He}.

\begin{table}[t]
  \centering
  \caption{Poles and continua contributions
    of the $p_{3/2}$- and $p_{1/2}$-components in $^{6}$He~\cite{hagen05,masui07}.}
  \label{tab:sec4_Contri_conti_p32_p12_6He}
  \begin{tabular}{cclll}
    \hline
    \hline
    $\ell j$ & Contribution  & CS-COSM~\cite{masui07} &  & GSM \cite{hagen05}\\
    & & ($\ell_{\rm max} = 5$) & ($\ell=1$)& ($\ell=1$) \\
    \hline
   $p_{3/2}$ &  ${\cal P}_{(0p_{3/2})^{2}}$ & 
    $ ~~1.211 -i0.666$ &
    $ ~~1.139 -i0.742$ &
    $ ~~1.105 -i0.832$ \\
  &   ${\cal S}1_{p_{3/2}}$ &
    $ -0.252 +i0.692 $ & 
    $ -0.119 +i0.773 $ & 
    $ -0.060 +i0.881  $ \\
  &    ${\cal S}2_{p_{3/2}}$ &
    $  -0.042   -i0.026$ & 
    $  -0.060   -i0.031 $ &
    $ -0.097 -i0.050$ \\
    & Sum  &
    $~~0.917 $ &
    $~~0.960 $ &
    $~~0.948 $ \\
    \hline
   $p_{1/2}$ & ${\cal P}_{(0p_{1/2})^{2}}$ &
    $~~  1.447  +i0.007$  &
    $~~  0.353  -i0.077$  &
    $~~  0.226 -i0.161 $ \\
    &   ${\cal S}1_{p_{1/2}}$ &
    $  -2.658    -i0.042 $ &
    $  -0.534    +i0.065 $ &
    $ -0.198 +i0.224 $ \\
    & ${\cal S}2_{p_{1/2}}$ &
    $~~  1.249 +i0.034 $ &
    $~~  0.221 +i0.012 $ &
    $~~  0.025 -i0.063 $ \\
    & Sum &
    $~~ 0.038 $ &
    $~~0.040 $ &
    $~~ 0.052$ \\
    \hline
    \hline
  \end{tabular}
\end{table}

The $(p_{3/2})^{2}$-component increases from $0.917$ ($\ell_{\rm max} = 5 $)
to $0.960$ ($\ell = 1 $) in the CS-COSM calculations.
In both cases, ${\cal P}_{(0p_{3/2})^{2}}$, ${\cal S}1_{p_{3/2}}$
and ${\cal S}2_{p_{3/2}}$ contributions of
the $\ell_{\rm max} = 5 $ calculation are almost the same 
with the $\ell =1$ ones,
and also similar to those obtained by the GSM.
This shows that the coupling to the higher partial waves
does not affect to the $p_{3/2}$-components in the $0^{+}$ state of $^{6}$He.

On the other hand, though
the $(p_{1/2})^{2}$-components of the CS-COSM calculations with
$\ell_{\rm max} = 5$ and $\ell =1$ are similar each other,
which are $0.038$ and $0.040$,
the poles and continuum contributions,
${\cal P}_{(0p_{1/2})^{2}}$, ${\cal S}1_{p_{1/2}}$ and 
${\cal  S}2_{p_{1/2}}$, change drastically
from $\ell_{\rm max} = 5$ to $\ell =1$.
${\cal P}_{(0p_{1/2})^{2}}$ becomes, for example,
$0.353-i0.077$ ($\ell = 1 )$
from $1.446+i0.007$ ($\ell_{\rm max} = 5 $).
The result of the $\ell = 1 $ calculation of the CS-COSM
is similar to the GSM result,
although the $NN$-interactions are different.

In the case of $\ell =1$, the coupling to the higher partial waves
does not exist in the model space.
Hence, it can be considered that
the coupling affects to the contributions of the continuum and poles
in the broad resonant single-particle states
of the $p_{1/2}$-wave in $^{6}$He,
and it is important to take in to account the continuum states correctly.

\begin{figure}[t]
  \begin{minipage}{0.45\hsize}
  \centering
  \includegraphics[width=9.0cm,clip]{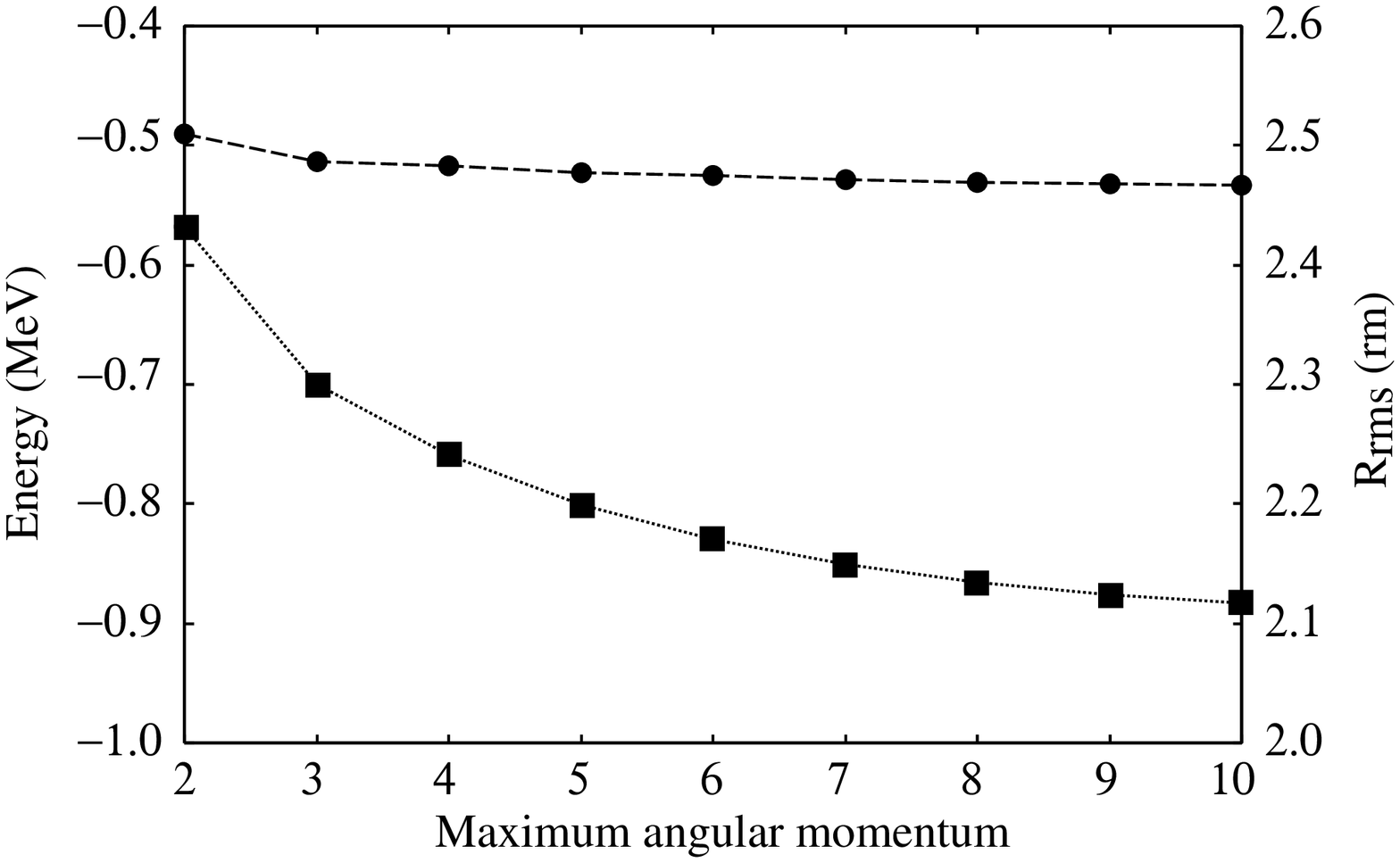}
  \caption{Calculated energy and r.m.s. radius of $^{6}$He.
    Solid squares and circles show the energy and r.m.s. radius,
  respectively.}
  \label{fig:sec4_Rrms_E_6He}
\end{minipage}
\hspace*{6mm}
\begin{minipage}{0.50\hsize}
~~\includegraphics[width=9.0cm,clip]{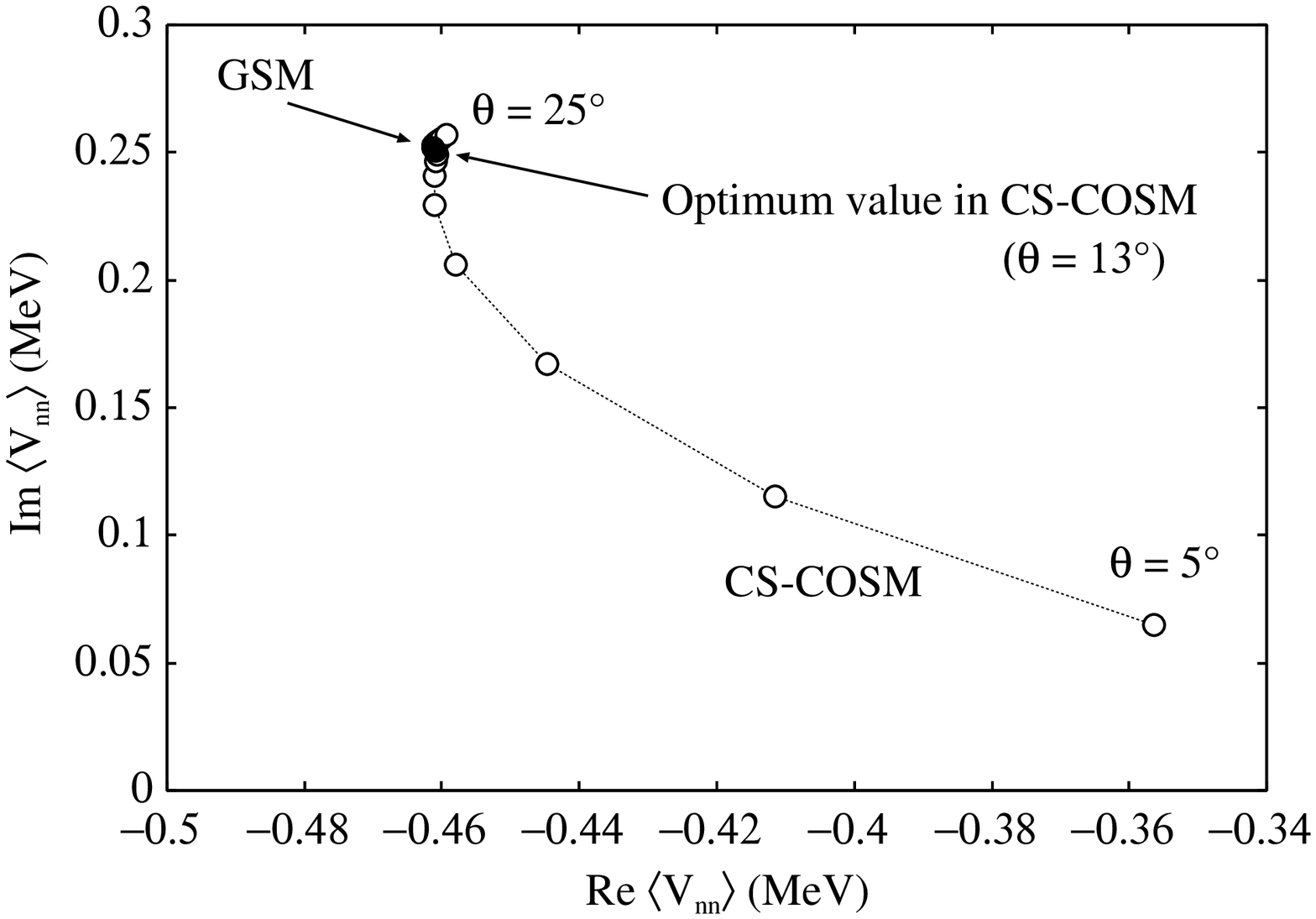}
  \caption{The $\theta$-trajectory of the CSM calculation
    in the CS-COSM and the results of the GSM.
    For the GSM calculation, the optimized result is shown.}   
  \label{fig:sec4_ho_exp_L0_L5}
  \end{minipage}
\end{figure}

The importance of the higher partial waves has been
pointed out in Ref.~\cite{aoyama95}.
The rearrangement channel of the coordinate system,
which is called the ``T-base'' component, 
is necessary to describe the nucleon-nucleon correlation
and is important to reproduce the $^{6}$He ground state energy.
In the ``V-base'' coordinate system (the COSM coordinate),
the component of the T-base coordinate system corresponds
to the inclusion of a large number of single-particle states
with high angular momenta.
The convergence is shown in Fig.~\ref{fig:sec4_Rrms_E_6He}.
To simulate the inclusion of the T-base coordinate system,
$\ell_{\rm max} =  5$ is not sufficient to fully describe
the correlation of valence nucleons in the $^{6}$He ground state.

To describe the correlation, a finite-range interaction
for the valence nucleon is necessary to be introduced.
Recently, the h.o. expansion of TBME has been introduced~\cite{michel08}
to calculate the finite-range interaction efficiently in the GSM framework.
We have performed a precise comparison~\cite{masui14}
between the CS-COSM and the GSM using the same $NN$-interaction, the Minnesota
force~\cite{tang78} and an effective three-body force
so that the binding energy of $^{6}$He corresponds to the experiment
in the $\ell_{\rm max} =5$ model space.
The same kind of comparison has been done in Ref.~\cite{kruppa14}.
We show the calculated energies of $^{6}$He only for
$\ell_{\rm max} =1$, $2$ and $5$
in Table~\ref{tab:sec4_6He_energies_0+_2+}.
The correspondence between the CS-COSM and the GSM are excellent in this case.

We also investigate how the presence of the Coulomb force
affects to the calculations of the CS-COSM and the GSM.
In Table~\ref{tab:sec4_6Be_energies_0+_2+},
we show the $^{6}$Be ($\alpha$+$2p$ system) case
for $\ell_{\rm max} =1$, $2$ and $5$.
Different from the $^{6}$He case,
the energy of $2^{+}_{1}$-state shows a small difference
in the convergence of the complex energy.
Hence, we discuss the origin of the difference in the following.

Due to the introduction of the complex scaling for the CS-COSM 
and the Gamow states in the GSM,
we have to solve a non-Hermitian problem for both cases.
Therefore, the standard variational problem is no longer valid,
and it is necessary to find the optimum values of the parameters
by searching for a stationary point of the eigenvalue.
In the CS-COSM approach, the variational parameters are
the complex rotation angle $\theta$ and the parameters
$b_{0}$ and $\gamma$
in a definition of the Gaussian width;
$b_{n_{i}} = b_{0} \gamma^{n_{i}-1}$~\cite{aoyama06,hiyama03}
for the Gaussian basis functions.
In the GSM, on the other hand,
the deformed contour for each $(\ell,j)$ is varied to obtain
the best numerical precision of calculated eigenenergies.
The contour-path and number of discretized continuum states
are the variational parameters.
Further, since the GSM introduces 
the h.o.expansion procedure~\cite{michel08},
the width parameters of h.o. functions,
size parameter $b_{ho}$ and the maximum principal quantum number
$N = 2n+ \ell$ are also the variational parameters
for searching the stationary point of the energy.

\begin{table}
  \begin{minipage}{0.48\hsize}
  \caption{Energies of the ground $0_1^{+}$ and
    the first excited $2_1^{+}$ states of $^{6}$He
    calculated using the CS-COSM and GSM
    approaches~\protect{\cite{masui14}}.
   All units except for the angular momentum are in MeV.}
  \center
   \label{tab:sec4_6He_energies_0+_2+}
  \begin{tabular}{ccccc}
    \hline
    &  $\ell_{\rm max} $
    & CS-COSM &  GSM  \\
    \hline
    & 1 & $-0.117 $  & $-0.116$ \\
  $E(0^{+}_{1})$ 
  & 2 & $-0.737  $   & $-0.737  $ \\
    & 5 & $-0.978  $   & $-0.977  $ \\
    \hline
    & $\ell_{\rm max} $  & CS-COSM & GSM \\
        \hline
        & 1 & $0.805  - i 0.086 $  & $0.804 - i 0.086$ \\
        $E(2^{+}_{1})$
        & 2 & $0.675  - i 0.038 $  & $0.669 - i 0.041$ \\
        & 5 & $0.589 -  i 0.021 $  & $0.577 - i 0.024$ \\
    \hline
  \end{tabular}
\end{minipage}
\hspace*{6mm}
\begin{minipage}{0.45\hsize}
  \caption{Same as Table~\ref{tab:sec4_6He_energies_0+_2+}
    except for the $^{6}$Be case.}
 \center
   \label{tab:sec4_6Be_energies_0+_2+}
  \begin{tabular}{cccccc}
   \hline
    &
   $\ell_{\rm max} $  & CS-COSM & GSM \\
    \hline
    & 1 & $1.932 -i0.152$  &  $1.926 - i 0.146$ \\ 
  $E(0^{+}_{1})$  
   & 2 & $1.490 -i0.046$  &  $1.482 - i 0.041$ \\
    & 5 & $1.285 -i0.031$  &  $1.279 - i 0.024$ \\
    \hline
     &
    $ \ell_{\rm max} $  & CS-COSM & GSM \\
    \hline
    & 1 & $2.741 -i0.703  $ & $2.776 - i 0.711 $ \\
   $E(2^{+}_{1})$ 
    & 2 & $2.614 -i0.559  $ & $2.610 - i 0.596 $ \\
    & 5 & $2.517 -i0.491  $ & $2.495 - i 0.505 $ \\  
    \hline
  \end{tabular}
  \end{minipage}
\end{table}

For the case of a broad resonant state,
which has the same order of the real and imaginary part of the
complex eigenvalue,
the variational parameters might be examined carefully,
since the coupling to the continuum becomes much more important
as discussed in the $^{6}$He ($\ell =1$) case. 
Here, we show the convergence on the stationary point
of the expectation values $\braket{V_{nn}}$ in Fig.~\ref{fig:sec4_ho_exp_L0_L5},
obtained by changing the rotation angle $\theta$
of the CSM calculation in the CS-COSM.
The GSM result is optimized for the
parameters of the h.o. function, $b_{ho}$ and $N_{\rm max}$.
The optimized $\braket{V_{nn}}$ with respect to the rotation angle
$\theta$ in the CS-COSM well corresponds to that of the GSM calculation.

In comparison with the $^{6}$Be case,
the difference is only less than one percent
even for the $2^{+}$-state,
as shown in Table~\ref{tab:sec4_6Be_energies_0+_2+}.
Based on the results of  the comparison between the CS-COSM and the GSM,
in the range of $N_{V} =2$,
we can conclude that both the CS-COSM and GSM
calculations describe the weakly bound states
taking into account the coupling with unbound resonant and continuum states.

%%%%%%%%%%%%%%%%%%%%%%%%%%%%%%%%%%%%%%%%%%%%%%%%%%%%%%%%%%%%%%%%%%%%%%%%%%%%%%%

\subsection{Nucleon capture reactions in CS-COSM}
In the COSM approach, radial wave functions of valence nucleons
do not require assumption and are solved with a many-body Schr\"odinger equation.
In the previous subsection,
we have discussed the CS-COSM was able to
describe the halo and resonant states of $^6$He and $^6$Be
in the same way as that in the GSM calculations.
Next, we apply this CS-COSM to nucleon capture reactions,
which are of interest in astrophysics and are closely connected
with nuclear structures around the threshold energy.
After explaining the framework of calculations,
we discuss the radiative capture reaction cross sections
of $^{16}$O$+n$ and $^{17}$O$+n$.

\subsubsection{Radiative capture cross sections in CS-COSM}
The cross section $\sigma^{\rm cap}_{E \lambda}$
of a radiative capture reaction $A+a\to B+\gamma$
is expressed in the following form using the photo-disintegration
cross section  $\sigma^{\rm dis}_{E \lambda}$ of its inverse reaction:
\begin{eqnarray}
  \sigma_{E\lambda}^{\rm cap}(E)&=&\frac{(2I_{A}+1)(2I_{a}+1)}{2(2I_{B}+1)}
  \frac{k_{cm}^{2}}{k_{\gamma}^{2}}
  \sigma_{E\lambda}^{\rm dis}(E),
\label{eq:sec4_E_lambda_1}
\end{eqnarray}
where
\begin{eqnarray}
  \sigma_{E\lambda}^{\rm dis}(E)&=&
  \frac{(2\pi)^{3}(\lambda+1)}{\lambda [(2\lambda+1)!!]^{2}}
  \biggl(\frac{E_{\gamma}}{\hbar c}\biggr)^{2\lambda -1}
  \frac{dB(EM\lambda;E)}{dE}. \label{eq:sec4_E_lambda_2}
\end{eqnarray}
Here, $k_{\gamma}=E_{\gamma}/\hbar c$ and $k_{cm}$
are the wave numbers of the photon with the multi-polarity
$\lambda$ and of the center-of-mass motion 
between the core $A$ and a valence nucleon $a$, respectively.
The photon energy is given as $E_\gamma=E-E_B$,
where $E$ and $E_B$ are the center-of-mass energy
and the binding energy of $B$, respectively.
Because we measure the energies $E$ and $E_B$
from the threshold of the ``$A+a$'' system,
$E_B$ is a negative value.

Thus, we can calculate the cross section of the radiative
capture reaction through the transition strength, 
\begin{eqnarray}
  \frac{dB(EM\lambda;E)}{dE}
  &=&
  \frac{1}{2J_B+1}
  |\left\langle\Psi_{A+a}(E)||O_\lambda||\Psi_B\right\rangle|^2,
\end{eqnarray}
where $O_\lambda$ is the $\lambda$-pole
electromagnetic transition operator.
While $\Psi_B$ is a bound state of the binding energy
$E_B$, $\Psi_{A+a}(E)$ is an unbound state of the energy $E$.
The transition strength
$|\left\langle\Psi_{A+a}(E)||O_\lambda||\Psi_B\right\rangle|^2$
is calculated using the complex scaled solutions
of the Lippmann-Schwinger (CSLS) equation~\cite{kikuchi10,kikuchi11,kikuchi09},
as discussed in section \ref{sec:CSLS}.
Since the scattering state $\Psi_{A+a}(E)$ can be expressed
as the solution of the CSLS equation,
which is defined as Eq.~(\ref{eq:CSLS_bra}) in section \ref{sec:CSLS},
the transition strength becomes as follows:
\begin{eqnarray}
  |\left\langle\Psi_{A+a}(E)||O_\lambda||\Psi_B\right\rangle|^2
  & &  \nonumber \\
  & &\hspace*{-4cm}
  =
  \left| \left\langle
      \Phi_0(E)||O_\lambda||\Psi_0(B)\right\rangle+\sum_\nu
    \frac{\left\langle\Phi_0|V|\Psi_\nu^\theta(A+a)
      \right\rangle
      \left\langle
        \tilde{\Psi}_\nu^\theta(A+a)||O_\lambda||
        \Psi_0(B)\right\rangle}{E-E_\nu^\theta}\right|^2,
\end{eqnarray}
where $\Phi_0$ is a solution of the asymptotic Hamiltonian $H_0$
and $V$ is the interaction given by $(H-H_0)$.
  
\subsubsection{Nucleon capture cross sections on $^{16,17}$O}
We calculate the radiative capture cross sections
of the $^{16}$O($n,\gamma)^{17}$O and  $^{16}$O($p,\gamma)^{17}$F
reactions in the astrophysical energy region. 
The nuclear reactions of astrophysical interest are closely
connected with nuclear structures around the threshold energy.
The rate of a specific reaction can decisively affect
the production of heavier elements in a stellar nucleosynthesis;
such a reaction is called a ``key reaction'' in the synthesis process.
Although the reaction cross section of $^{17}$O($n,\gamma)^{18}$O
is one of the key reactions in the nucleosynthesis
for the most iron-poor star cases~\cite{yamamoto09}, 
we have no experimental data for this reaction.
A reliable theoretical estimation to the cross section
of the $^{17}$O($n,\gamma)^{18}$O reaction is necessary
for the network calculation to achieve reliable results
of the element abundance.
In order to study the $^{17}$O($n, \gamma)^{18}$O reaction,
the use of an $^{16}$O+$n$+$n$ model is promising.
For this purpose, it is indispensable to examine the reliability
of the $^{16}$O+$N$ model.
We investigate $^{16}$O($n,\gamma)^{17}$O 
and $^{16}$O($p,\gamma)^{17}$F reactions using
a simple $^{16}$O+$N$ ($N$=$n,\ p$) model and develop
the theoretical framework mentioned above. 
 
%*********************************************************************************************************************
\begin{figure}[htbp]
\begin{minipage}{0.48\hsize}
\centering
\includegraphics[width=7.5cm,clip]{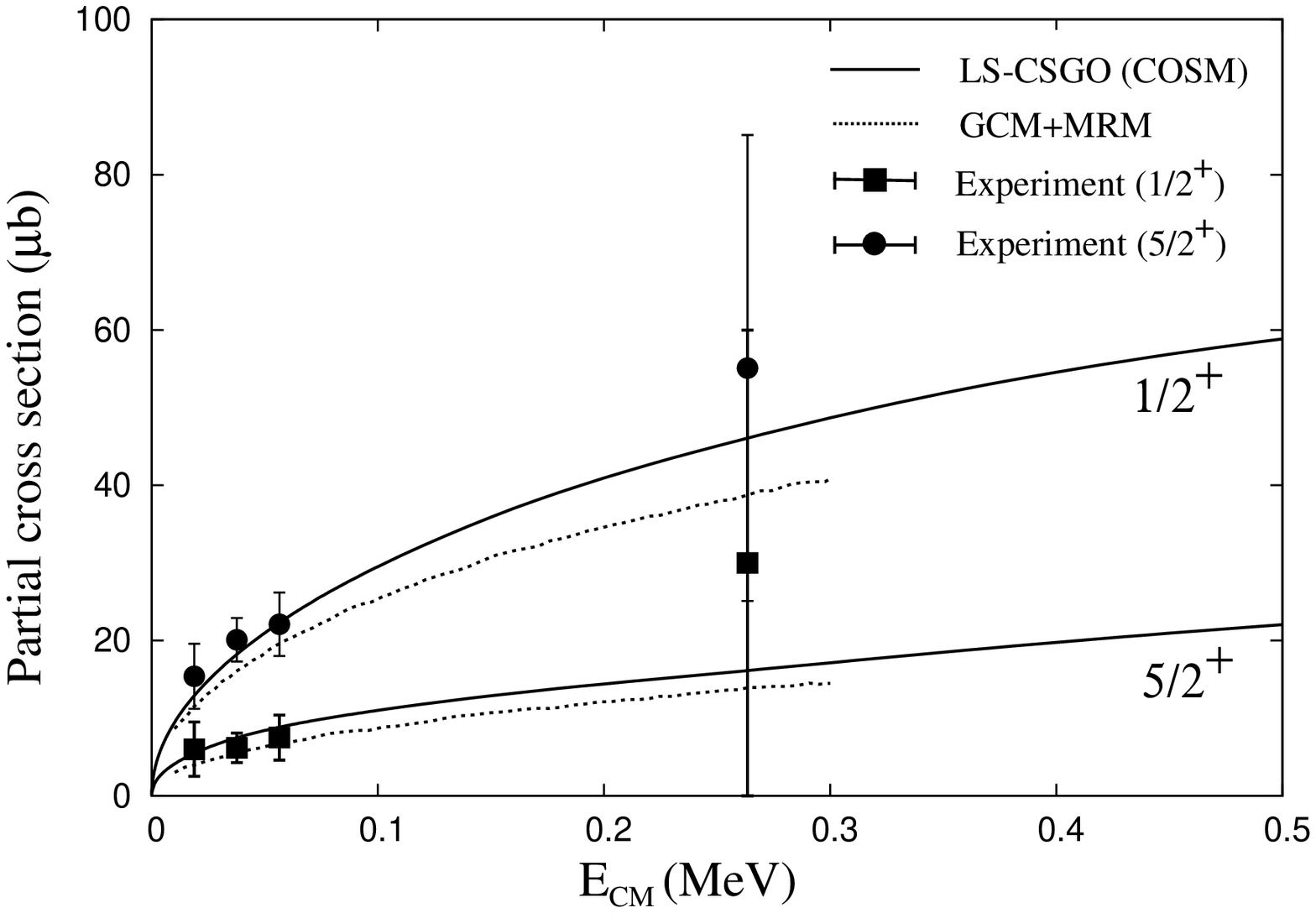}  
  \caption{Partial cross sections of the capture reaction 
$^{16}$O($n,\gamma)^{17}$O for the $E$1 transition. 
The solid lines indicate our calculation results~\cite{yamamoto09} and 
the dotted lines show the results  calculated by GCM+MRM \cite{dufour05} for 
the transitions from the scattering state to the 
5/2$^{+}$ and  1/2$^{+}$ states. 
The squares and circles denote experimental data \cite{igashira95} 
for the ground state, 5/2$^{+}$, and the first excited state, 1/2$^{+}$, 
respectively. }
  \label{fig:sec4_cs_ng}
\end{minipage}
\hspace*{8mm}
\begin{minipage}{0.45\hsize}
  \centering
  \includegraphics[width=7.5cm,clip]{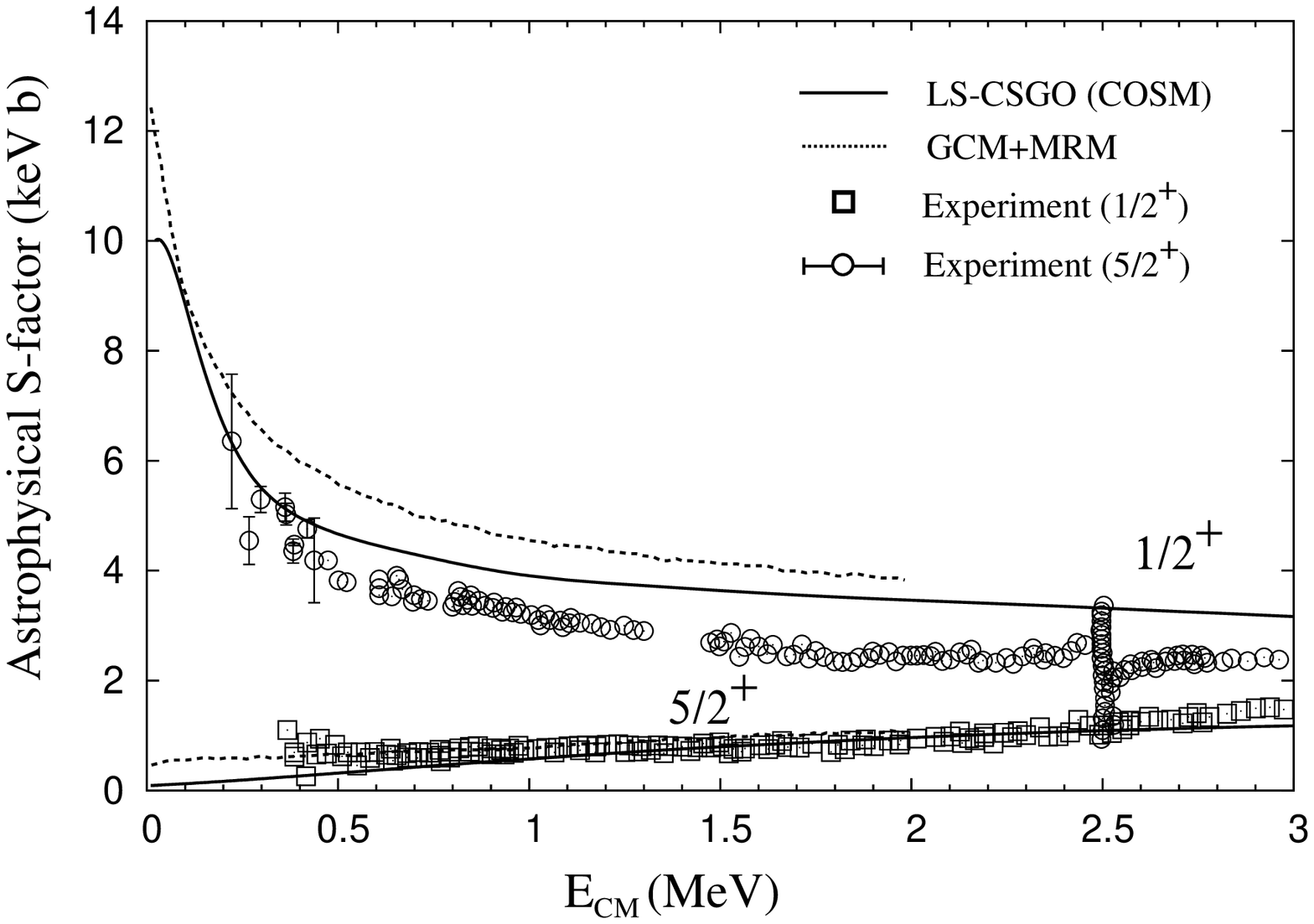}
   \caption{Astrophysical $S$-factor in the reaction
 $^{16}$O($p,\gamma)^{17}$F. 
 The solid lines show our calculation
 results\cite{yamamoto09} and the dashed lines
 show the results calculated by GCM+MRM~\cite{baye98}
 for the capture reaction to 
 the $5/2^{+}$ and $1/2^{+}$ states. 
 The open squares and circles denote experimental data taken from
 Ref.~\protect\cite{morlock97}. }
   \label{fig:sec4_16opg_asf}
\end{minipage}
\end{figure}

A detailed explanation of our model is given in
Ref.~\cite{yamamoto09},
where energies \cite{tilley93} of the ground $5/2^+$ and
excited $1/2^+$ states and the observed scattering phase shifts
of the $^{16}$O$+n$~\cite{johnson67} and $+p$ \cite{blue65} systems
are well reproduced.
We also compare the present results with those obtained
by the  GCM with the microscopic
$R$-matrix method (GCM+MRM)~\cite{dufour05},
as shown in Fig.~\ref{fig:sec4_cs_ng},
in which we display the partial cross sections of the $E$1 transition
for  the $^{16}$O($n,\gamma)^{17}$O reaction from the scattering
states to the ground and first excited states,
$J^{\pi}=5/2^{+}$ and $1/2^{+}$, respectively.
The solid and dotted lines show the transitions obtained by the
present method and GCM+MRM, respectively.
It is confirmed that the observed partial cross sections
can be well explained by 
the $E$1 transitions from the $p$-wave scattering
states~\cite{igashira95} in continuum energies
to the ground ($5/2^{+}$) and first excited states ($1/2^{+}$).
Our model predicts the cross sections,
which are slightly larger than those obtained by the GCM+MRM,
and reproduces the experimental data observed
in the astrophysical energies.
Since the error bars of the cross sections 
observed at $E_{cm}\sim0.26$ MeV are large,
a reliable theoretical approach is needed.

Therefore, we calculate the astrophysical $S$-factor for
the capture reaction $^{16}$O($p,\gamma)^{17}$F
and compare it with the results obtained by 
the GCM+MRM \cite{baye98}.
The $S$-factor is estimated by the $E$1 transition
from continuum states to the ground ($5/2^{+}$)
and first excited ($1/2^{+}$) states. 
The results are shown in Fig.~\ref{fig:sec4_16opg_asf}.
The experimental data are presented with open squares
and open circles for the $5/2^{+}$ and $1/2^{+}$ states,
respectively.
For the $E$1 transition to the $5/2^{+}$ state,
our calculation accurately reproduces the experimental results
and shows a good correspondence to the results obtained by the GCM+MRM.
Our calculation well reproduces the $E$1 transition reaction
to the $5/2^{+}$ state.
For the $1/2^{+}$ state,
though the both calculations seem to slightly overestimate as
compared with the experimental data,
our COSM calculation shows a better agreement than the GCM+MRM one.
The $E$1 transition to the $1/2^{+}$ state is sensitive
to the structure of the wave function.
From the experiments~\cite{morlock97},
the $1/2^{+}$ state of $^{17}$F is understood
to have a single proton halo structure.

To explain the observed data more quantitatively
and to increase the reliability of the predictions, 
it is important to study whether the excitation of the $^{16}$O-core plays
an important role in forming the halo structure
of the $1/2^{+}$ state and in reducing the calculated $E$1
transition strength using the present model.

\section{Continuum level density}\label{sec:continuum}
The continuum level density (CLD) plays an important role in the description of scattering phenomena and structures of nuclei, since CLD is connected with the scattering $S$-matrix as shown in Refs.\cite{levine69,tsang75,osborn76}. 
A variation of the level densities due to an interaction defines the CLD, and the scattering phase shift is derived with the CLD. Through the CLD, we can see a close relation between level structures and scattering phase shifts. 
In this section, we give a brief explanation of the CLD in relation to the extended completeness relation with complex scaling, 
which provides the decomposition of the scattering phase shifts into resonance and continuum contributions \cite{odsuren14}.
We also show the applications of the CLD to nuclear two- and three-body scattering problems associated with the $\alpha$ cluster. 
The CLD also becomes the basis for the description of the three-body scattering states using the Lippmann-Schwinger equation, shown in section \ref{sec:CSLS}.

Kruppa and Arai \cite{kruppa98,kruppa99,arai99} calculated the CLD using discretized real eigenvalues of Hamiltonians and smoothing techniques based on the Strutinsky procedure \cite{strutinsky67}. They applied the CLD to determine resonance parameters. However, as discussed in Ref.~\cite{arai99}, their results for the CLD exhibit a strong dependence on the smoothing parameters. 

In Ref.~\cite{suzuki05}, we proposed a more direct method to calculate
the CLD using the CSM,
where its own smoothing procedure is included but no artificial smoothing technique such as the Strutinsky procedure is needed.
The concept of the method is based on the extended completeness relation (ECR) \cite{myo98} given in \S \ref{sec:ECR}, originally proposed by Berggren \cite{berggren68}, for bound, resonance, and continuum states in the CSM. The exact proofs of this ECM in the CSM were given for a single-channel system \cite{giraud03} and a coupled-channel system \cite{giraud04} . Green's functions can be expressed using the ECM in terms of discrete eigenvalues in the CSM with a finite number of basis functions. Because the complex-scaled Hamiltonians $H^\theta$ and $H_0^\theta$ have complex eigenvalues, the singularities such as a $\delta$-function are avoided and replaced by Lorentzian functions. It is also shown that the CLD denoted by $\Delta(E)$ can be calculated independent of the scaling angle $\theta$ in the CSM. 

\subsection{CLD and phase shift}
\subsubsection{CLD}
The level density $\rho(E)$ of the Hamiltonian $H$ is defined by
\begin{equation}
\rho(E)=\int\hspace{-0.60cm}\sum_i\delta(E-E_i),\label{eq:sec5-1-1-1}
\end{equation}
where $E_i$ are eigenvalues of $H$, and summation and integration are taken for discrete and continuous eigenvalues, respectively. This definition of the level density is also expressed using Green's function \cite{shlomo92}:
\begin{eqnarray}
\rho(E) &=& -\frac{1}{\pi}{\rm Im}\left\{ {\rm Tr} \left[\frac{1}{E+i0-H}\right]\right\} \nonumber, 
\label{eq:sec5-1-1-2}
\end{eqnarray}
where $+i0$ indicates the limit $+i\epsilon\to +i0$. When the
Hamiltonian is described by a sum of an asymptotic term $H_0$ and the short-range interaction $V$ ($H=H_0+V$), 
the CLD $\Delta(E)$ for an energy $E$ is expressed in terms of balance between the density $\rho(E)$ obtained from the Hamiltonian $H$ and the level density $\rho_{0}(E)$ of continuum states obtained from the asymptotic Hamiltonian $H_0$ as
\begin{eqnarray}
 \Delta(E)  &=&  \rho(E) -  \rho_{0}(E), \nonumber\\
            &=& -\frac{1}{\pi}{\rm Im}\left[{\rm Tr}\left\{\frac{1}{E+i0-H}-\frac{1}{E+i0-H_0}\right\}\right]. 
            \label{eq:sec5-1-1-3}
\end{eqnarray}

This expression indicates that $\Delta(E)$ reflects the influence of the interactions expressed by a difference between $H$ and $H_0$ \cite{kruppa98,kruppa99,shlomo92}. On the other hand, $\Delta(E)$ is known to be connected with the scattering $S$-matrix $S(E)$ as \cite{levine69,tsang75,osborn76}
\begin{eqnarray}
\Delta(E) =\frac{1}{2\pi}{\rm Im}\frac{d}{dE}{\rm ln}\left\{{\rm det}\ S(E)\right\}.
\label{eq:sec5-1-1-4}
\end{eqnarray}
The scattering $S$-matrix for a single channel system is expressed as $S(E)=e^{2i\delta(E)}$, where $\delta(E)$ is the scattering phase shift. Next, in single-channel two-body systems, we have 
\begin{eqnarray}
\Delta(E) =\frac{1}{\pi}\frac{d\delta}{dE}.
\label{eq:sec5-1-1-5}
\end{eqnarray}
Using this expression, we can calculate the phase shift as
\begin{eqnarray}
\delta(E)=\pi\int^{E}_{-\infty}dE'\Delta(E').
\label{eq:sec5-1-1-6}
\end{eqnarray} 

When the interaction $V$ is attractive and produces some bound states, the number of bound states is given by
\begin{equation}
N_B=\int_{-\infty}^0 \Delta(E) dE.
\label{eq:sec5-1-1-7}
\end{equation}
From this result, we can see an example of the Levinson theorem as
\begin{equation}
\delta(0)=N_B\pi.
\label{eq:sec5-1-1-8}
\end{equation}
Furthermore, we see that at resonance energy $E_r$, $\delta(E_r)=\pi/2$ provides a maximum cross section. The level density must increase owing to the attractive interaction, i.e., $\Delta(E_r)= d\delta(E_r)/dE~>~0$. Since $ d\delta(E_r)/dE \sim 1/\Gamma$, where $\Gamma$ is the decay width, the quantity $\Delta(E)$ is also called as a time-delay \cite{taylor72}.

\subsubsection{CS-CLD}
We calculate the CLD using the CSM; here, the CLD is expressed with the complex-scaled Green's function as
\begin{eqnarray}
 \Delta^\theta(E)  &=& -\frac{1}{\pi}{\rm Im}\int d\vc{r}\left\{{\cal G}^\theta(E,\vc{r},\vc{r})-{\cal G}^\theta_0(E,\vc{r},\vc{r})\right\}, 
            \label{eq:sec5-1-2-1}
\end{eqnarray}
where ${\cal G}^\theta(E,\vc{r},\vc{r}')$ is given in Eq. (\ref{eq:sec2_green}) and ${\cal G}^\theta_0(E,\vc{r},\vc{r}')$ is defined as
\begin{eqnarray}
{\cal G}^\theta_0(E,\vc{r},\vc{r}')= \left\langle \vc{r}\left|\frac{1}{E-H_0^\theta }\right|\vc{r}'\right\rangle.
\label{eq:sec5-1-2-2}
\end{eqnarray} 

The complex-scaled Green's functions are expressed using the eigenvalues of the complex-scaled Hamiltonian, $H^\theta$, and the asymptotic Hamiltonian, $H_0^\theta$. The eigenvalue problems of $H^\theta$ and $H_0^\theta$ are solved using a finite number of $L^2$ basis functions. Within the total number $N$ of basis states, the eigenvalues of the complex-scaled Hamiltonian $H^\theta$ are classified as the bound state energies $E_b$ $(b=1,\ 2,\ldots,\ N_B)$, the resonance complex energies $E_r-i\Gamma_r/2$ $(r=1,\ 2,\ldots,\ N_R^\theta)$, and the rotated continuum energies ${\cal E}_c^R-i{\cal E}_c^I$ $(c=1,\ 2,\ldots,\ N-N_B-N_R^\theta)$, as discussed in previous sections. For $H_0^\theta$, the rotated continuum energies are also presented by the discretized eigenvalues, ${\cal E}_c^{0R}-i{\cal E}_c^{0I}$ $(c=1,\ 2, \ldots,\ N)$. Therefore, the CS-CLD is expressed within the $N$-basis functions in the CSM in the following  form:
\begin{eqnarray}
\Delta_N^\theta(E)&= &\sum_b^{N_B}\delta(E-E_b)+\frac{1}{\pi}\sum_r^{N^\theta_R}\frac{\Gamma_r/2}{(E-E_r)^2+\Gamma^2_r/4}\nonumber \\
 & &+\frac{1}{\pi}\sum_c^{N-N_B-N^\theta_R}\frac{{\cal E}_c^I}{(E-{\cal E}_c^R)^2+{{\cal E}_c^I}^2}-
\frac{1}{\pi}\sum_c^N\frac{{\cal E}_c^{0I}}{(E-{\cal E}_c^{0R})^2+{{\cal E}_c^{0I}}^2}.\label{eq:sec5-1-2-3}
\end{eqnarray}  

The resonance term of $E_r-i\Gamma_r/2$ is described by using the Breit-Wigner form. Although the number $N_R^\theta$ of the resonance term depends on the scaling parameter $\theta$, 
every Breit-Wigner form of the resonance is independent of $\theta$. 
On the other hand, the continuum terms (third and fourth ones in Eq.~(\ref{eq:sec5-1-2-3})) are also described by using the Breit-Wigner form as well. 
But each of them depends on $\theta$, because the eigenvalues of the continuum solutions are complex numbers of ${\cal E}_c^R-i{\cal E}_c^I$ and ${\cal E}_c^{0R}-i{\cal E}_c^{0I}$ on the ``$2\theta$-lines." 

\subsubsection{Discretization of continuum states}
\begin{figure}[b]
  \begin{minipage}[b]{0.50\linewidth}
   \begin{center}
    \includegraphics[width=7.0cm]{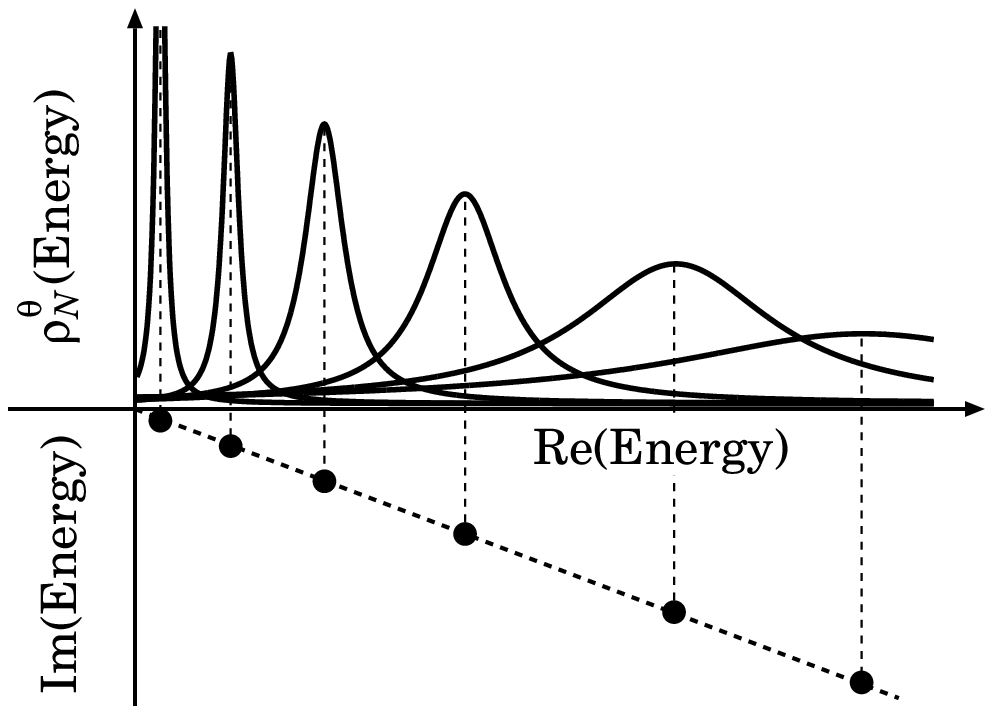}
    \caption{Schematic energy eigenvalue distribution (black circles) for a complex-scaled Hamiltonian and contributions to the level density (solid lines). }  
    \label{fig:sec5-1-3-1}
   \end{center}
  \end{minipage}
  \hspace{0.3cm}
  \begin{minipage}[b]{0.45\linewidth}
  \vspace*{-1cm}
   \begin{center}
        \includegraphics[width=9.0cm]{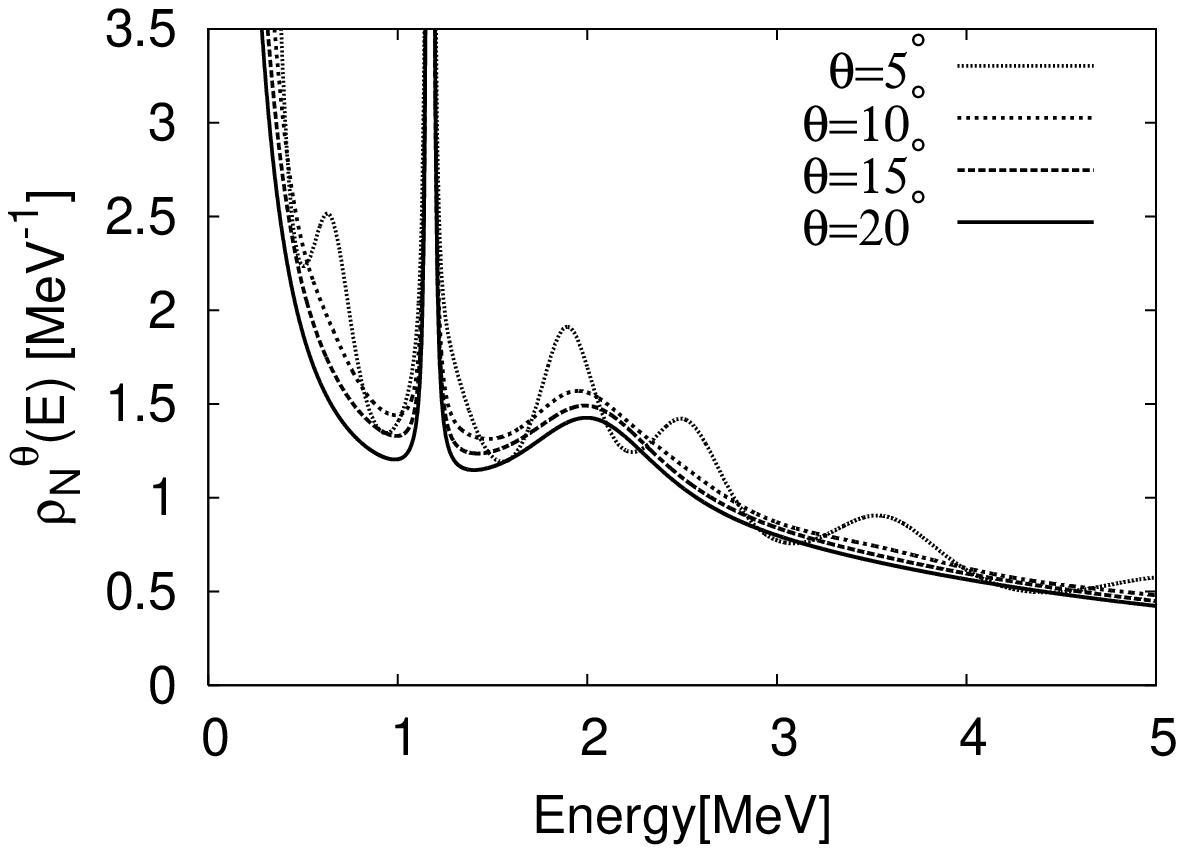}
            \caption{Level density $\rho^\theta_N(E)$ calculated for different values of $\theta$.} 
    \label{fig:sec5-2-2}
   \end{center}
%   \vskip-\lastskip
  \end{minipage}
 \end{figure}

The third and fourth terms in Eq.~(\ref{eq:sec5-1-2-3}) describe the contributions from the rotated continuum states. In the present method, the continuum states are expressed with a finite number of basis functions. The continuum energies of $H^\theta$ and $H_0^\theta$ are discretized as complex numbers on the ``$2\theta$-lines." We can show that the CS-CLD, being a function of the real energy $E$, becomes independent of $\theta$ and well converges to a smooth curve when the appropriate basis number $N$ and $\theta$ are employed \cite{suzuki05,suzuki08}. The energy eigenvalues in the continuum are rotated and discretized in the complex energy plane, as shown by black circles in Fig.~\ref{fig:sec5-1-3-1}. The CS-CLD defined by Eq.~(\ref{eq:sec5-1-1-2}) is expressed using the complex scaled level densities as
\begin{eqnarray}
\Delta^\theta_N(E)= \rho_N^\theta(E) - \rho_{0,N}^\theta(E),
\label{eq:sec5-1-3-1}
\end{eqnarray}
where
\begin{eqnarray}
\rho^\theta_N(E)&=&\sum_b^{N_B}\delta(E-E_b)+\frac{1}{\pi}\sum_r^{N^\theta_R}\frac{\Gamma_r/2}{(E-E_r)^2+\Gamma^2_r/4}+\frac{1}{\pi}\sum_c^{N-N_B-N^\theta_R}\frac{{\cal E}_c^I}{(E-{\cal E}_c^R)^2+{{\cal E}_c^I}^2}\ , \\
\rho^\theta_{0,N}(E)&=&\frac{1}{\pi}\sum_c^N\frac{{\cal E}_c^{0I}}{(E-{\cal E}_c^{0R})^2+{{\cal E}_c^{0I}}^2}.
\label{eq:sec5-1-3-2}
\end{eqnarray}
%%%%%%%%%%%%%%%%%%%%%%%%%%%%%%%%%%%%
%%%%%%%%%%%%%%%%%%%%%%%%%%%%%%%%%%%%%%%%%%%

The continuum contribution is automatically smoothed out when we use a sufficiently large number of basis functions for a given scaling parameter $\theta$. 
In the usual basis function method, a smoothing technique such as the Strutinsky procedure \cite{strutinsky67} is required to calculate the CLD \cite{kruppa98,kruppa99}, 
because the continuum is discretized on the real axis, and each continuum contribution has a $\delta$-function form. 
As mentioned above, the present discretization method in the CSM does not require any auxiliary technique like the Strutinsky procedure, 
because no singularity like the $\delta$-function appears. As discussed in a previous paper \cite{suzuki05}, 
this expression of the level density is dependent of the scaling parameter $\theta$, but this $\theta$-dependence disappears in the form of CLD $\Delta_{N}^\theta(E)$. 
This result indicates a cancellation of the $\theta$-dependences in $\rho^{\theta}_{N}(E)$ and $\rho^{\theta}_{0,N}(E)$. 
It is also shown that $\Delta_{N}^\theta (E) $ gives a good description for $\Delta(E)$.

\subsubsection{Phase shifts in CSM}
\begin{figure}[b]
\begin{center} 
  \includegraphics[width=16.0cm]{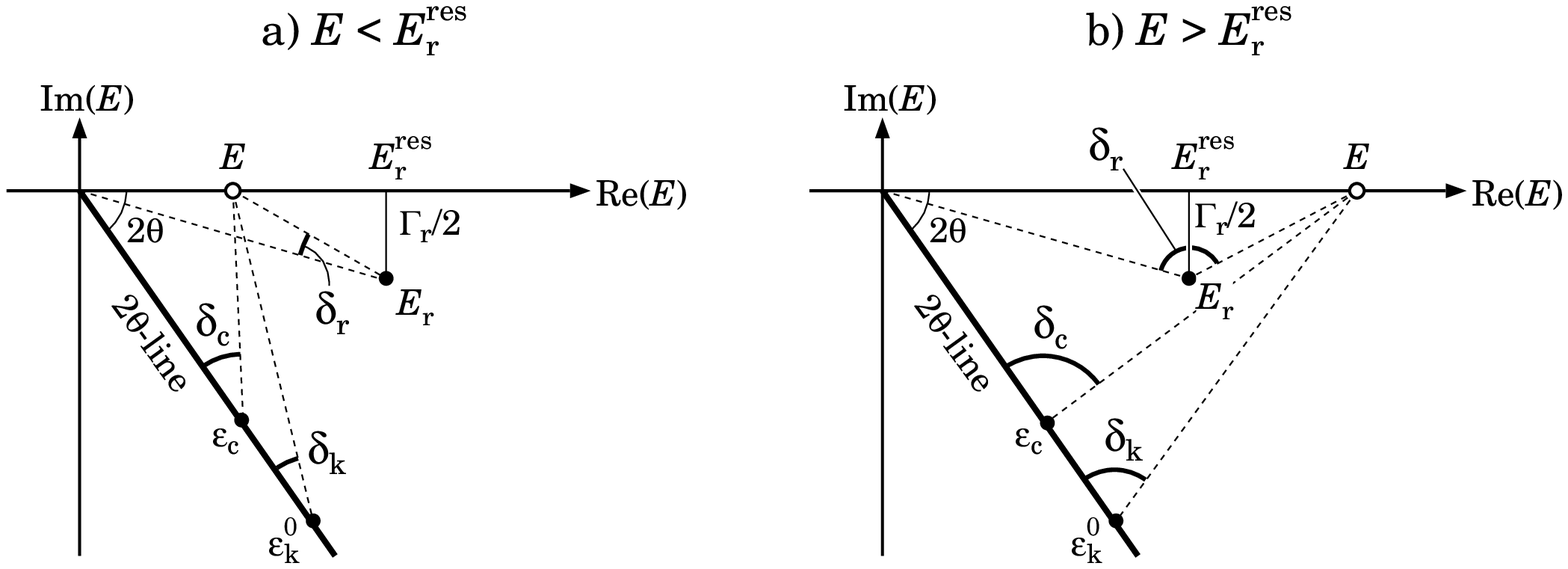}
  \caption{The geometrical interpretation of the phase shifts $\delta_r$, $\delta_c$ and $\delta_k$.}\label{fig:sec5-1-4-1}
\end{center}
\end{figure}

From Eqs.~(\ref{eq:sec5-1-1-6}) and (\ref{eq:sec5-1-1-7}), we can calculate the phase shift \cite{odsuren14} as 
\begin{eqnarray}
\delta_N^\theta(E)&=&\pi\int_{-\infty}^{E} \Delta_N^\theta(E)dE  \nonumber \\
 &=& N_B\pi + \sum_{r=1}^{N_r^\theta}\int_{0}^{E}dE\frac{\Gamma_r/2}{(E-E_r^{res})^2+\Gamma_r^2/4}
+\int_{0}^{E}dE \left[\sum_{c=1}^{N_c^\theta}\frac{\epsilon^i_c}{(E-\epsilon^r_c)^2+\epsilon^{i2}_c}-\sum_{k=1}^{N}\frac{\epsilon^{0}i_k}{(E-\epsilon^{0r}_k)^2+\epsilon^{0i2}_k}\right].\nonumber \\
 &=&N_B\pi+\delta_R(E)+ \delta_C(E).
\label{eq:sec5-1-4-1} 
\end{eqnarray}  
This expression of the phase shift indicates that it consists of the bound state term, resonance phase shifts $\delta_R$, and non-resonance continuum contributions $\delta_C$. The resonance and non-resonance phase shifts are given as 
\begin{eqnarray}
\delta_R(E)=\sum_{r=1}^{N_r^\theta}\delta_r(E), \hspace{1cm} \delta_C(E)=\sum_{c=1}^{N_c^\theta}\delta_c(E)-\sum_{k=1}^{N}\delta_k(E),
\label{eq:sec5-1-4-2}
\end{eqnarray}
where
\begin{eqnarray}
& &\hspace{1cm}\delta_r(E)=\mbox{cot}^{-1}\frac{(E_r^{res}-E)}{\Gamma_r/2}-\mbox{cot}^{-1}\frac{E_r^{res}}{\Gamma_r/2},\nonumber\\
\delta_c(E)&=&\mbox{cot}^{-1}\frac{\epsilon^r_c-E}{\epsilon^i_c}-\mbox{cot}^{-1}\frac{\epsilon^r_c}{\epsilon^i_c},\hspace{1cm} \delta_k(E)=\mbox{cot}^{-1}\frac{\epsilon^{0r}_k-E}{\epsilon^{0i}_k}-\mbox{cot}^{-1}\frac{\epsilon^{0r}_k}{\epsilon^{0i}_k}
\label{eq:sec5-1-4-3}
\end{eqnarray}
The geometrical interpretation of the phase shifts $\delta_r$, $\delta_c$, and $\delta_k$ are shown in Fig.~\ref{fig:sec5-1-4-1}.
%%%%%%%%%%%%%%%%%%%%%%%%%%%%%%%%%%%%%%%%%%%%%%%%%%%%%%%%%%%%%%%%%%%%%%%%%%%%%%%%%%%%%%%%%%%%%%%%%%%%%%%%%%
%%%%%%%%%%%%%%%%%%%%%%%%%%%%%%%%%%%%%%%%%%%%%%%%%%%%%%%%%%%%%%%%%%%%%%%%%%%%%%%%%%%%%%%%%%%%%%%%%%%%%%%%%%%%%

\subsection{Examples of several systems}
\subsubsection{Simple potential model}
Applying the CSM to a simple potential model \cite{csoto90} given by the Hamiltonian (\ref{eq:sec2_Gyar}), we can solve the eigenvalue problem easily. As basis functions, we employ the Gaussian expansion method developed by Kamimura {\it et. al.} \cite{hiyama03}, where we use the parameters $N=30$, $b_0=0.2$, and $\gamma=1.2$ for the Gaussian size parameters $b_i=b_0\gamma^{i-1}$ $(i=1,2,\ldots,N)$.

The resonance eigenvalues for $0^+$ and $1^-$ states are presented in Ref. \cite{homma97}.  Using the eigenvalues including bound and continuum states, we calculate the level density $\rho_{N}^{\theta}(E)$ given by Eq.~(\ref{eq:sec5-1-3-2}) and show the result for $J^\pi=0^+$ in Fig.~\ref{fig:sec5-2-2}. Although an oscillation behavior is seen at $\theta=5^\circ$, this oscillation is smoothed when $\theta$ increases to larger than $10^\circ$. Even at $\theta=5^\circ$, the oscillation may disappear if we employ a large size of basis functions, resulting in the intervals between discretized continuum eigenvalues becoming smaller than their imaginary parts. However, it is easier to consider a larger value of $\theta$ so as to increase the imaginary parts of the discretized continuum eigenvalues. The intervals between the discretized continuum eigenvalues depend on the size $N$ of basis functions. 
A critical value of $\theta$ may be defined by the scaling angle at which the imaginary parts of the discretized continuum eigenvalues become larger than the intervals of the eigenvalues. Such a critical $\theta$ depends on $N$, and hence, we express this critical $\theta$ value as $\theta_N$.  When $\theta$ becomes larger than $10^\circ$ in the present simple potential case, $\rho_{N}^{\theta}(E)$ shows the same behavior as seen in Fig.~\ref{fig:sec5-2-2}; hence, we can put $\theta_N\approx 10^\circ$. For $\theta>\theta_N$, the only absolute values of $\rho_{N}^{\theta}(E)$ depend on $\theta$.

\begin{figure}[tbp]
  \begin{minipage}[b]{0.50\linewidth}
   \begin{center}
    \includegraphics[width=9.2cm]{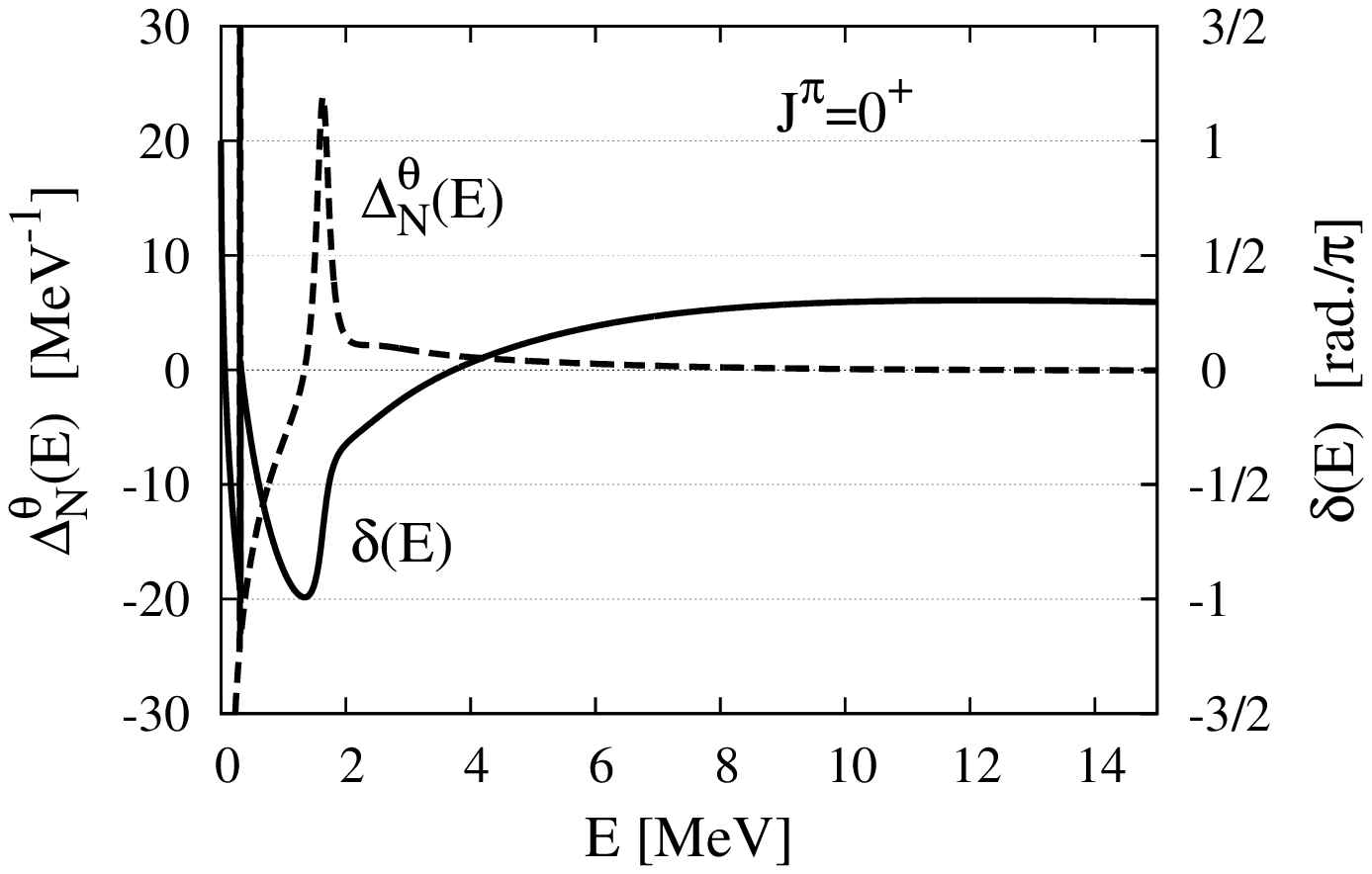}
    \caption{Calculated CS-CLD $\Delta_N^\theta(E)$ and scattering phase shift $\delta(E)$ with $\theta=20^\circ$ for the $J^\pi=0^+$ state of the Hamiltonian given by Eq. (\ref{eq:sec2_Gyar}).}  
    \label{fig:sec5-2-1-1}
   \end{center}
  \end{minipage}
  \hspace{0.4cm}
  \begin{minipage}[b]{0.47\linewidth}
   \begin{center}
    \includegraphics[width=8.5cm]{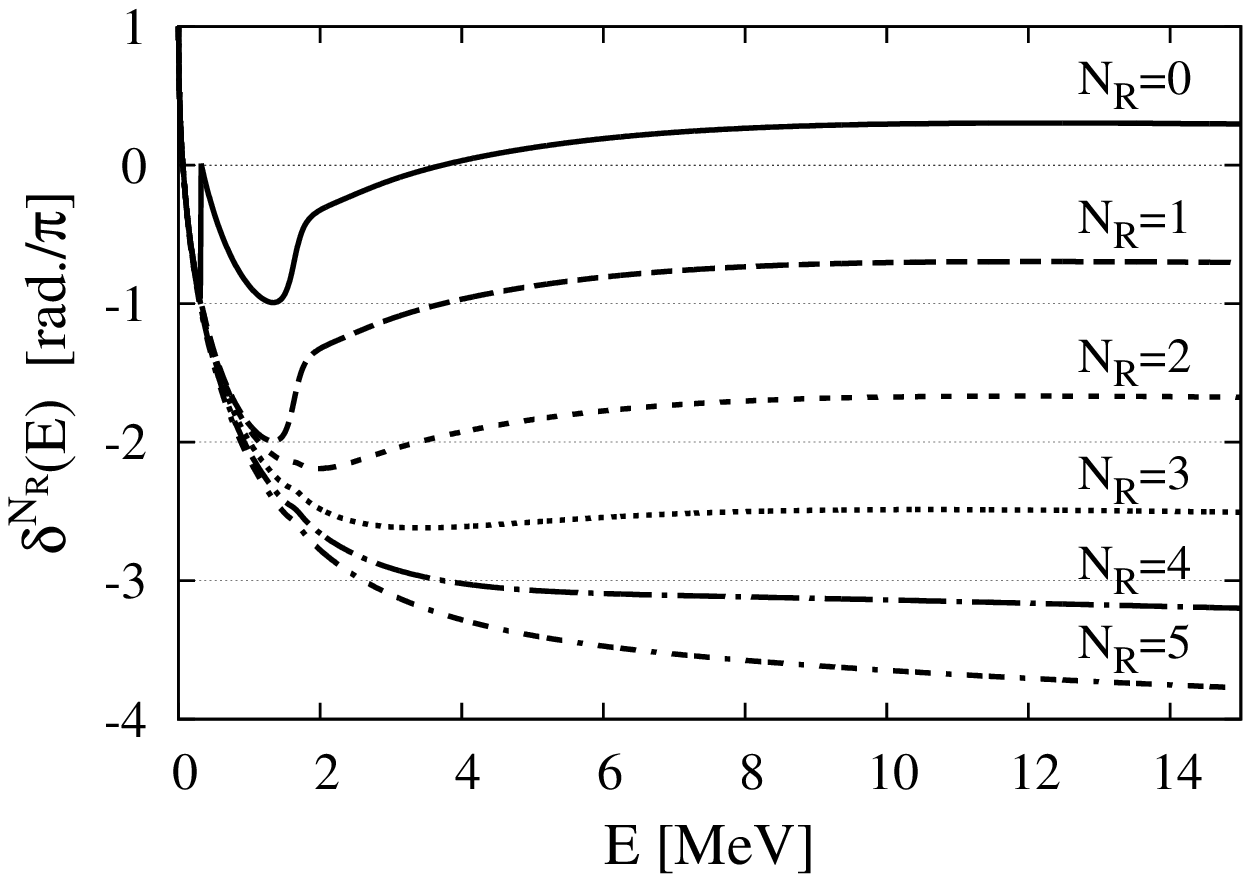}
    \caption{Scattering phase shift $\delta^{N_R} (E)$ in Eq.~(\ref{eq:subtract}) subtracting the resonance contributions for the $J^\pi=0^+$ state.} 
    \label{fig:sec5-2-1-2}
   \end{center}
%   \vskip-\lastskip
  \end{minipage}
 \end{figure}

This $\theta$-dependence of the absolute values $\rho_N^\theta(E)$ can be canceled through subtraction of $\rho^{\theta}_{0,N}(E)$; in other words, we show that the CLD $\Delta_{N}^{\theta}(E)$ defined in Eq.~(\ref{eq:sec5-1-3-1}) has no $\theta$-dependence for $\theta\ge\theta_N$. In Fig.~4 of Ref.\cite{suzuki05}, the CS-CLD $\Delta_{N}^{\theta}(E)$ calculated for $\theta=10^\circ,\ 15^\circ$, and $20^\circ$ are presented to be almost identical.  In Fig.~\ref{fig:sec5-2-1-1}, we show the calculated CS-CLD and phase shift, which indicate the existence of two sharp resonances at a very low energy and around 2 MeV. To observe the resonance contributions to the phase shift in detail, we calculate the phase shifts from which resonance terms are eliminated as
\begin{eqnarray}
\delta^{N_R}(E)=\delta(E)-\sum_{r=0}^{N_R}\delta_r(E), \label{eq:sec5-2-1-1}
\label{eq:subtract}
\end{eqnarray}
where $\delta_r(E)$ is the resonance phase shift of the $r$-th resonance term given in Eq.~(\ref{eq:sec5-1-4-3}). The results are presented in Fig.~\ref{fig:sec5-2-1-2}, where we can see the resonance effects from other resonance eigenvalues with large imaginary parts in addition to the contributions from the sharp resonant states. Although their effects are not concentrated at a narrow energy region, every resonance decreases the phase shift by $\pi$ asymptotically.
%%%%%%%%%%%%%%%%%%%%%%%%%%%%%%%%%%%%

\subsubsection{$\alpha+n$ system}
\begin{figure}[t]
  \begin{center}
    \includegraphics[width=15.5cm]{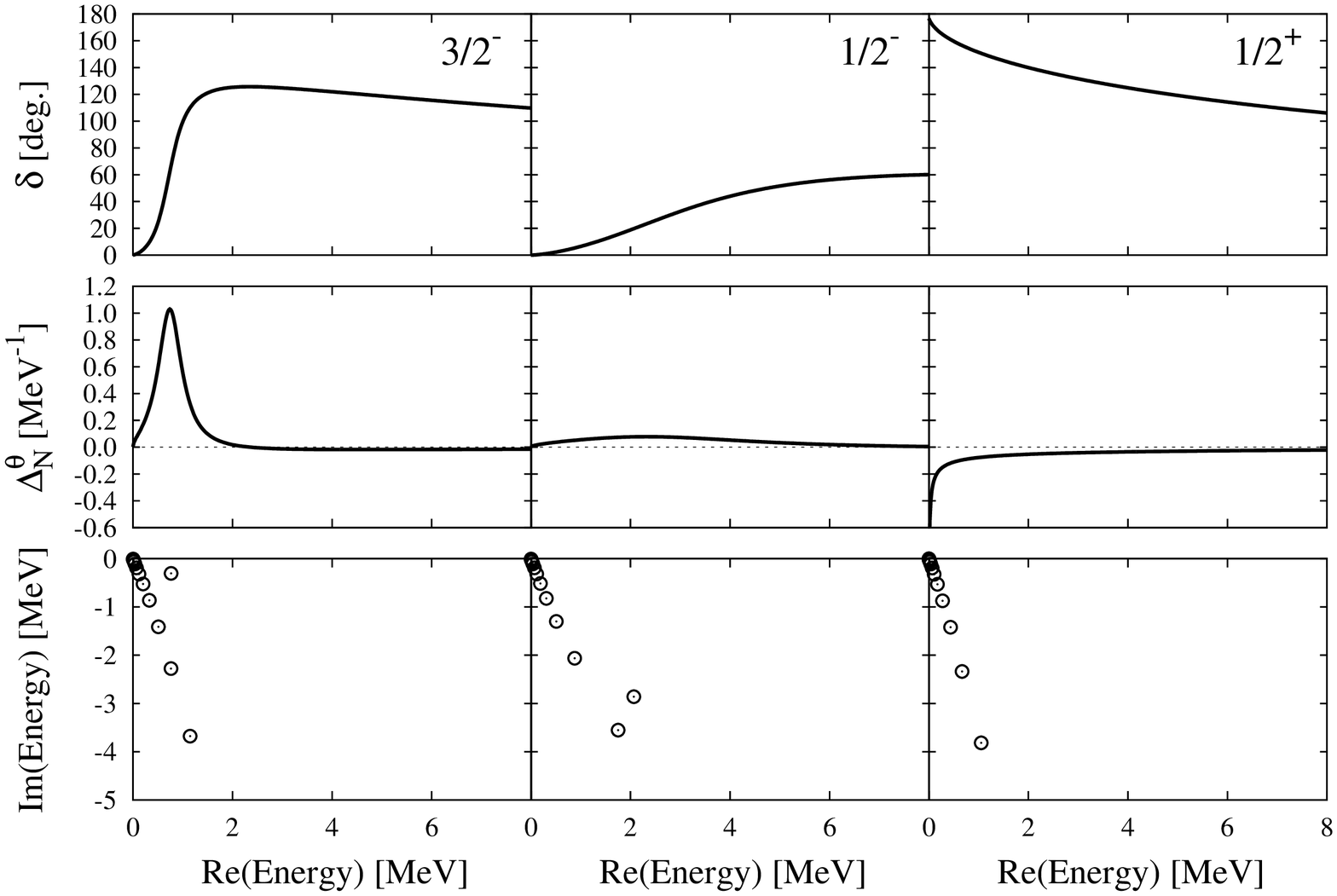}
\caption{The $\alpha+n$ system with $J^\pi=3/2^-,1/2^-, 1/2^+$ states.
Upper: scattering phase shifts, middle: continuum level density, and lower: energy eigenvalue distributions with $\theta$ being 35$^\circ$.}
   \label{fig:sec5-2-2-1}
  \end{center}
\end{figure}

As a realistic example, we show the results of a $^5$He=$\alpha+n$ system. 
For the interaction between $\alpha$ and n, the KKNN interaction \cite{kanada79} is used.
Using the same basis set as the case of the simple model, we calculate the energy eigenvalues of the complex-scaled Hamiltonian with $\theta=35^\circ$, and the results for three states of $3/2^-$, $1/2^-$, and $1/2^+$ are shown in the lowest panels of Fig.~\ref{fig:sec5-2-2-1}. 
We can see that $3/2^-$ and $1/2^-$ states have one respective resonance pole corresponding to the observed resonances of $^5$He. The $1/2^+$ state has one Pauli forbidden state but no resonance. 
The resonant structures of $^5$He have been discussed in detail using the CSM by Aoyama {\it et al.} \cite{aoyama95}. 
In addition to the resonances, the discretized continuum solutions are also obtained along the ``$2\theta$-line". 
Several continuum solutions are off the ``$2\theta$-line". 
As a reason, it is considered that the couplings between the continuum states and the resonance are not correctly described owing to the insufficiency of basis functions. 
However, the resonant solutions are solved with appropriate accuracy and 
the CLD is obtained from these continuum solutions satisfactorily, 
although positions of some continuum solutions are slightly off the ``$2\theta$-line."

Applying Eq.~(\ref{eq:sec5-1-2-3}) to the obtained eigenvalue distribution of the complex-scaled Hamiltonian for $3/2^-$, $1/2^-$, and $1/2^+$ states, we calculate the CLD of the $\alpha+n$ system. The results are shown in the middle panels of Fig.~\ref{fig:sec5-2-2-1}. The $3/2^-$ and $1/2^-$ states have their respective peaks, although the peak of the $1/2^-$ state is not sharp. The $1/2^+$ state has no peak and negative values due to its repulsive nature. The peaks in the CLD of $3/2^-$ and $1/2^-$ states appear at the position with the width corresponding to the resonance energy and width.

Using the obtained CLD, we calculate the phase shift. In the top panels of Fig.~\ref{fig:sec5-2-2-1}, we show the scattering phase shifts of the $3/2^-$, $1/2^-$, and $1/2^+$ states. We compare these results with the exact phase shifts, which well explain the observation \cite{hoop66} and show a very good quantitative agreement within the thickness of lines for every state \cite{suzuki05}.

\subsubsection{$\alpha+\alpha$ system}
In a similar way as that in the $\alpha$+$n$ case, we calculate the CLD and the scattering phase shifts of the $\alpha+\alpha$ system, which includes the Coulomb interaction. Since the Coulomb interaction has a typically long-range character, the asymptotic Hamiltonian $H_0$ involves the Coulomb interaction.
\begin{figure}[t]
  \begin{center}
    \includegraphics[width=14.0cm]{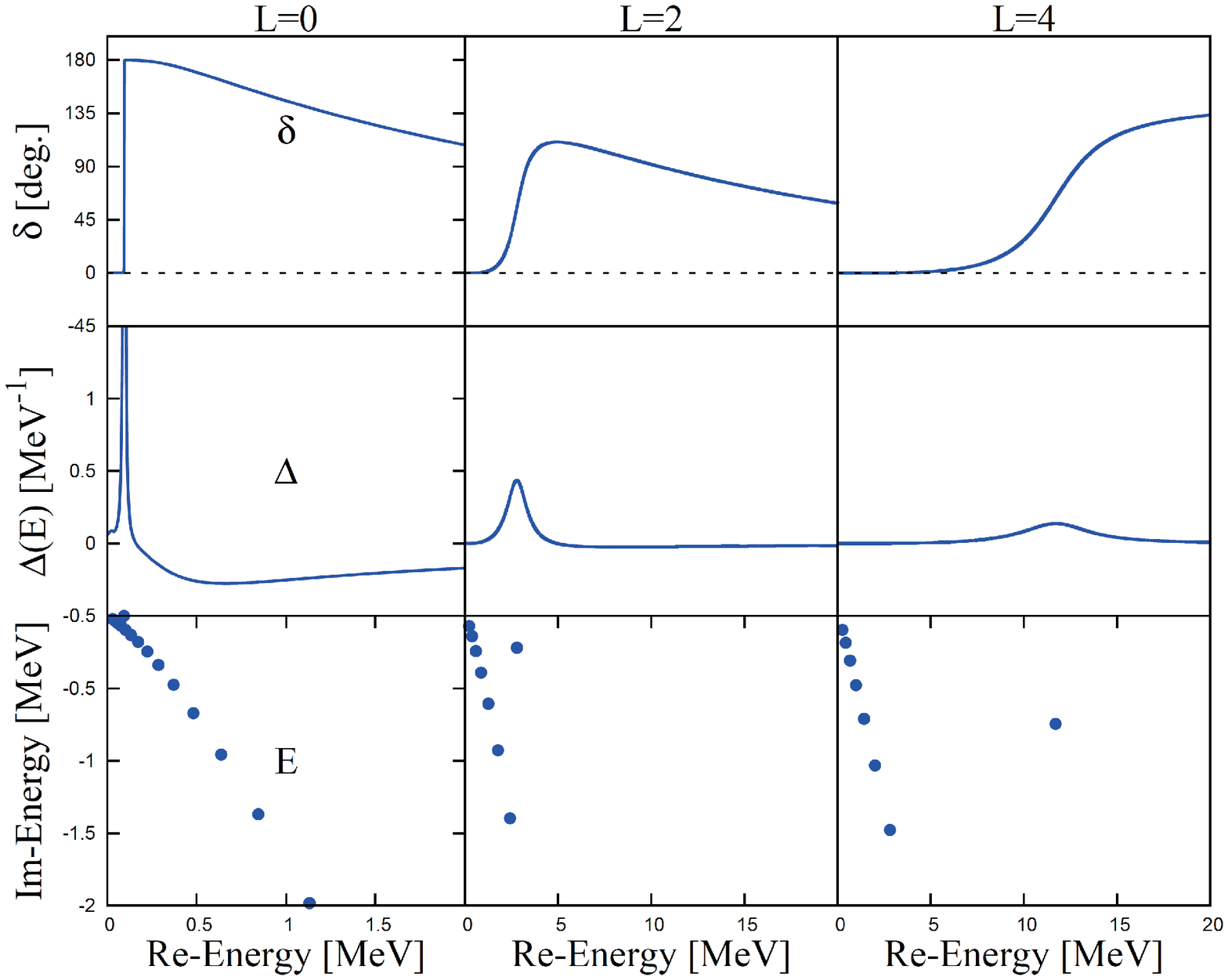}  
   \caption{The phase shifts $\delta$, CS-CLD $\Delta$, and energy spectrum $E$ of the $\alpha$-$\alpha$ system for $J^\pi=0^+,\ 2^+$, and $\ 4^+$ states.}
   \label{fig:sec5-2-3-1}
  \end{center}
\end{figure}
  
The relative motion between two $\alpha$-clusters is described within the OCM framework \cite{saito77}:
\begin{equation}
\left[ T_{rel}+V_{\alpha\alpha}^C(r)+V_{\alpha\alpha}^N(r)+ V_{PF} -E \right]\psi^J_{\rm rel}(\vc{r})=0, \label{eq:sec5-2-3-1}
\end{equation}
where $V_{\alpha\alpha}^C$ and $V_{\alpha\alpha}^N$ are the folding Coulomb and nuclear potentials, respectively, obtained by assuming a $(0s_{1/2})^4$ harmonic oscillator wave function with an oscillator constant $\nu_{\alpha}=0.2675$ fm$^{-2}$ for an $\alpha$ cluster.
When we employ an effective two-nucleon force of a one-range Gaussian form, the folding Coulomb and nuclear potentials are expressed as
\begin{eqnarray}
V_{\alpha\alpha}^C(r)&=&\left(\frac{4e^2}{r}\right){\mbox e}{\mbox r}{\mbox f}
  \left(r\sqrt{\frac{4}{3}\nu_{\alpha}}\right)=\frac{4e^2}{r}\mbox{erf}(0.5972r),
  \label{eq:sec5-2-3-2}
\\
V_{\alpha\alpha}^N(r)&=&2X_D \left[ \frac{2\nu_{\alpha}}{2\nu_{\alpha}+3\mu/2}
  \right]^{3/2} V_{0}\exp\left[-\frac{\nu_{\alpha}\mu}{\nu_{\alpha}+3\mu/4}r^2\right]=-103.0\exp{(-0.2009r^2)},
  \label{eq:sec5-2-3-3}
\end{eqnarray}
where erf$(x)$ is the error function, and $X_D=2.445$, $V_0=-72.98$ MeV, and $\mu=0.46$ fm$^{-2}$ are the folding parameter, the strength, and the range parameters of the Schmid-Wildermuth force \cite{schmid61}, respectively. 
The fourth term $V_{PF}$ in Eq.~(\ref{eq:sec5-2-3-1}) is the so-called Pauli potential, which projects the Pauli forbidden states (PF; $0S,~1S,~0D$  states in this case) from the $\alpha\alpha$ relative motion \cite{kukulin86}. 
We solve the complex-scaled Schr\"odinger equation, Eq.~(\ref{eq:sec5-2-3-1}) in the same way as the calculations mentioned above. 
Using the obtained eigenvalues of $J^\pi=0^+,~2^+$, and $4^+$, we calculate the CLD. In the $\alpha$-$\alpha$ system, the eigenvalues of the asymptotic Hamiltonian $H_0$ are obtained by including the point-Coulomb potential \cite{suzuki05,odsuren14}. The results of the eigenvalues and the CLD are shown in Fig.~\ref{fig:sec5-2-3-1}.  

Integrating the obtained CLD, we acquire the scattering phase shifts and the results are shown in Fig.~\ref{fig:sec5-2-3-1}.  The scattering phase shifts well coincide with the ones obtained from the scattering solutions. The resonance width of the $0^+$ state is very small in comparison with the resonance energy. These results indicate that the present method to calculate CLD is also very powerful even for a long-range interaction such as the Coulomb potential.

\subsubsection{CLD of the 3$\alpha$ system}
The CLD is also calculated for coupled channel systems and three-body systems. In Ref.~\cite{suzuki08}, 
applications of the three-body CLD to the coupled-channel system were discussed. Here, we show the CLD of the 3$\alpha$ system. The three-body CLD has been discussed by Osborn and his co-workers \cite{tsang75,osborn76}. 
The purpose of the present calculation is to see how the 3$\alpha$ resonant states are discretely observed in the level density as a function of the real energy.
The CLD of the $3\alpha$ system is defined as
\begin{eqnarray}
\Delta(E)=-\frac{1}{\pi}\left\{\mbox{Tr}\left[\frac{1}{E-H_{3B}}-\frac{1}{E-H_{3B}^0}-
\left(\frac{1}{E-H_{2B}}-\frac{1}{E-H_{3B}^0}\right)\right]\right\} , 
\label{eq:sec5-2-4-1}
\end{eqnarray}
where $H_{3B}$ is the total Hamiltonian for the $3\alpha$ system with the $\alpha$-$\alpha$ interaction $V_{2\alpha}(r_{ij})=V^N_{\alpha\alpha}+V^C_{\alpha\alpha}$ given in Eqs. (\ref{eq:sec5-2-3-2}) and (\ref{eq:sec5-2-3-3}).
The asymptotic Hamiltonian $H^0_{3B}$ consists of the kinetic energy and the point-Coulomb potential between two $\alpha$-clusters:
\begin{equation}
H^0_{3B}=T_{3\alpha} + \sum_{i<j}^3\frac{2\cdot 2e^2}{r_{ij}} ,
\end{equation} 
where $T_{3\alpha}$ is the relative kinetic energy of $3\alpha$.
The two-body CLD of the system consisting of two interacting $\alpha$'s and one non-interacting $\alpha$ is given by the following Hamiltonian:
\begin{eqnarray}
H_{2B}=T_{3\alpha}+V_{2\alpha}(r_{12})+ \sum_{i=1}^2 \frac{2\cdot 2e^2}{r_{i3}}, \hspace*{0.2cm}
\label{eq:sec5-2-4-2}
\end{eqnarray}
Here, it should be noticed that the resonant states of $^8$Be$+\alpha$ are included in the three-body CLD.

%%%%%%%%%%%%%%%%%%%%%%%%%%%%%%%%%%%%%%%%
\begin{figure}[htb]
\begin{center} 
  \includegraphics[height=0.451\textheight]{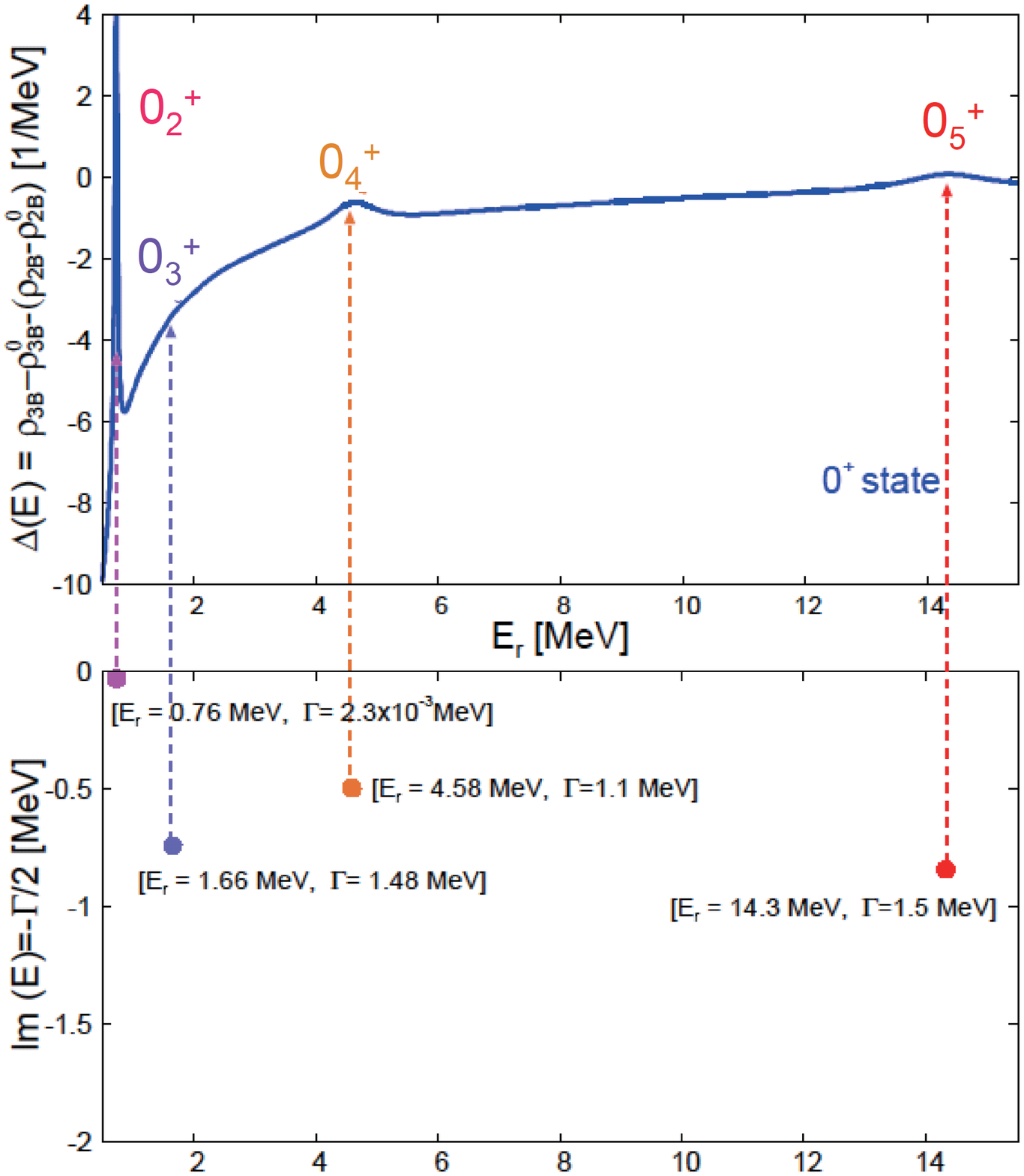}
  \caption{The $3\alpha$ CLD of $0^+$ states of $^{12}$C in the upper panel together with their pole positions in the 
  complex energy plane of the lower panel.}\label{fig:sec5-2-4-1}
\end{center}
\end{figure}
%%%%%%%%%%%%%%%%%%%%%%%%%%%%%%%%%%%%%%%%%

In Fig.~\ref{fig:sec5-2-4-1}, we show the $3\alpha$ CS-CLD of $0^+$ states together with their pole positions in the 
complex energies. The Hoyle state is seen as a sharp peak at the energy just above the $3\alpha$ threshold. 
On the other hand, the broad $0_4^+$ and $0_5^+$ states only have small bumps around resonance energies. 
Although the $0_3^+$ state has the same order as the imaginary value of its complex energy, we cannot see 
a signature clearly in the CLD. This result does not necessary mean that the $0_3^+$ cannot be observed in any degree. 
When its matrix element of, for instance, the electric transition matrix, is large enough to distinguish from 
neighboring continuum states, it is possible to observe experimentally \cite{kato10}. In fact, an observation of a broad $0^+$ state
corresponding to the predicted $0^+_3$ resonance has recently been reported by Itoh {\it et al.}\cite{itoh11}.

\section{Scattering with complex-scaled solutions of the Lippmann-Schwinger equation}\label{sec:CSLS}

In unstable nuclei, valence nucleons are bound to the system with small binding energies.
Owing to this weakly-bound nature, unstable nuclei are broken up with a low-excitation energy together with particle emissions.
The breakup reactions are essential tools to understand the exotic properties of the unstable nuclei~\cite{tanihata13}.
To extract the properties of unstable nuclei from the breakup reactions, a description of the many-body scattering states of weakly-bound systems is required.
Considering this situation, the demand for theoretical approaches is now increasing.

Recently, several kinds of theoretical approaches have been developed to solve the many-body scattering problems on the footing of the bound-state techniques.
Some of the methods are often applied to the few-nucleon scattering phenomena using bare nucleon-nucleon interaction~\cite{carbonell14,efros07}.
We have also proposed the method of describing the many-body scattering states using CSM, referred to as the complex-scaled solutions of the Lippmann-Schwinger equation (CSLS)~\cite{kikuchi10,kikuchi11,kikuchi09,kikuchi13a}.
In CSLS, the scattering states are described by combining the formal solutions of the Lippmann-Schwinger equation with the complex-scaled Green's function.
The complex-scaled Green's function defined in Eq.(\ref{eq:sec2_green1}) automatically satisfies the correct boundary conditions using the complex-scaled eigenstates.
As shown in the previous sections, the complex-scaled Green's function gives the correct CLD of the scattering states of two-body system \cite{suzuki05,suzuki08} and 
further consistently works in the description of the three-body CLD.
On the basis of the method, we can describe the many-body scattering states of weakly-bound nuclei.
In the scatterings, CSLS is an effective method of evaluating the physical quantities as functions of the subsystem energies in a many-body system, which provide useful information to clarify the exotic properties of unstable nuclei through breakup reactions. 
In particular, it is important to understand the internal correlations between the weakly coupled constituents, such as the halo neutrons.

We investigate the reactions related to the three-body scattering states by using CSLS.
In this review, we select the following three topics:
(i) Coulomb breakup reactions of two-neutron halo nuclei, $^6$He and $^{11}$Li,
(ii) nuclear breakup reaction of $^6$He, and
(iii) elastic scattering and radiative capture reaction of the $\alpha$~+~$d$ system.

%%%%%%%%%%%%%%%%%%%%%%%%%%%%%%%%
\subsection{CSLS Formalism}

We first explain the CSLS formalism with the aim of describing three-body scattering states.
The formal solution of the Lippmann-Schwinger equation can be described as
\begin{equation}
\Psi^{(\pm)} = \Phi_0 + \lim_{\varepsilon \to 0}
\frac{1}{E-H\pm i\varepsilon} V \Phi_0,
\label{eq:LS}
\end{equation}
where $\Phi_0$ is a solution of an asymptotic Hamiltonian $H_0$.
The total Hamiltonian is represented as $H$, and the interaction $V$ is given by subtracting $H_0$ from $H$.
The boundary condition of the scattering state is represented by $\pm i\varepsilon$.
Equation~(\ref{eq:LS}) is equivalent to the Schr{\"o}dinger equation and useful to describe the many-body scattering states.

In CSLS, we utilize the complex-scaled Green's function $\mathcal{G}^\theta(E,\vc{r},\vc{r}')$ defined in Eq.~(\ref{eq:sec2_green}), which provides the consistent level densities for many-body systems.
The complex-scaled Green's function is related to the non-scaled Green's function $\mathcal{G}(E,\vc{r},\vc{r}')$ with outgoing boundary conditions as
\begin{equation}
\lim_{\varepsilon \to 0} \frac{1}{E-H+i\varepsilon} = \mathcal{G}(E,\vc{r},\vc{r}') = U(\theta)^{-1} \mathcal{G}^\theta (E,\vc{r},\vc{r}') U(\theta).
\label{eq:GFinCSLS}
\end{equation}
Using the eigenstates of $H^\theta$ and their bi-orthogonal states, $\{\chi^\theta_\nu,\tilde{\chi}^\theta_\nu\}$ \cite{moiseyev11,berggren68,aoyama06}, 
we rewrite Green's function in Eq.~(\ref{eq:GFinCSLS}) as
\begin{equation}
\mathcal{G}(E,\vc{r},\vc{r}') = \sum_\nu\hspace{-0.46cm}\int U(\theta)^{-1} \big| \chi^\theta_\nu \big\ket \frac{1}{E-E^\theta_\nu} \big\bra \tilde{\chi}^\theta_\nu \big| U(\theta).
\label{eq:GreenF}
\end{equation}
Combining the Green's function in Eq.~(\ref{eq:GreenF}), we obtain the outgoing and incoming scattering states, $\Psi^{(+)}$ and $\Psi^{(-)}$, in Eq.~(\ref{eq:LS}) as
\begin{equation}
\big| \Psi^{(+)} \big\ket = \big| \Phi_0 \big\ket + \sum_\nu\hspace{-0.46cm}\int U(\theta)^{-1} \big| \chi^\theta_\nu \big\ket \frac{1}{E-E^\theta_\nu} \big\bra \tilde{\chi}^\theta_\nu \big| U(\theta) V \big| \Phi_0 \big\ket
\end{equation}
and
\begin{equation}
\big\bra \Psi^{(-)} \big| = \big\bra \Phi_0 \big| + \sum_\nu\hspace{-0.46cm}\int \big\bra \Phi_0 \big| V U(\theta)^{-1} \big| \chi^\theta_\nu \big\ket \frac{1}{E-E^\theta_\nu} \big\bra \tilde{\chi}^\theta_\nu \big| U(\theta),
\label{eq:CSLS_bra}
\end{equation}
respectively.
In the derivation of $\Psi^{(-)}$, we assume the Hermiticity of $H$ and $V$.
The operators $U(\theta)$ and $U^{-1}(\theta)$ are processed in the calculation of the matrix elements and do not operate on $\chi^\theta_\nu$ and $\tilde{\chi}^\theta_\nu$.

%%%%%%%%%%%%%%%%%%%%%%%%%%%%%%%%%%%%%%%%%%%%%%%%%%%%%%%%%%%%%%%%%
\subsection{Three-body Coulomb breakup reactions of halo nuclei}

The Coulomb breakup reactions have been performed to investigate the exotic properties of two-neutron halo nuclei such as $^6$He and $^{11}$Li~\cite{aumann99,wang02,ieki93,shimoura95,zinser97,nakamura06}.
The breakup reactions of halo nuclei provide rich information on the structures and responses of the various states excited from the ground state.

To theoretically investigate the Coulomb breakup reactions of two-neutron halo nuclei, it is necessary to describe the core~+~$n$~+~$n$ three-body scattering states given that the two-neutron halo nuclei are the Borromean systems in which no binary subsystems have bound states.
We here describe the three-body scattering states of two-neutron halo nuclei by using CSLS.
In this review, we show the results of the mechanism of Coulomb breakup reactions of $^6$He and $^{11}$Li.
We describe the Coulomb breakup cross sections as functions of the relative energies of binary subsystems.
From the results, we pin down the decay modes of two-neutron halo nuclei in the Coulomb breakup reactions.

In addition to the decay modes, we also investigate the effect of excitations in the $^9$Li core on the Coulomb breakup reaction of $^{11}$Li.
As was explained in \S \ref{sec:11Li_cc}, the configuration mixing of $^9$Li with 2p2h excitations is essential to reproduce the halo properties of $^{11}$Li \cite{myo07b}.
We investigate the $E1$ strength distribution of $^{11}$Li and discuss the role of the 2p2h excitations in $^9$Li core in the Coulomb breakup reactions.

In the Coulomb breakup reactions of two-neutron halo nuclei, we describe the final scattering states by using the core~+~$n$~+~$n$ three-body model and CSLS in Eq.~(\ref{eq:CSLS_bra}).
For the Coulomb breakup, the asymptotic Hamiltonian $H_0$ is defined as
\begin{equation}
H_0 = h_{\rm core} + \sum_{i=1}^3 t_i - T_{\rm c.m.},
\end{equation}
where $h_{\rm core}$ is the internal Hamiltonian for the core nucleus.
The kinetic operators for each particle and for the center-of-mass of the total system are represented as $t_i$ and $T_{\rm c.m.}$, respectively.
The solution of $H_0$ is expressed as
\begin{equation}
\Phi_0 (\vc{k},\vc{K}) = \Phi_{\rm gs}^{\rm core} \otimes \phi_0 (\vc{k},\vc{K}),
\label{eq:asy_halo}
\end{equation}
where $\vc{k}$ and $\vc{K}$ are the asymptotic momenta in a three-body Jacobi coordinate.
Asymptotically, the core nucleus is in its ground state, whose wave function is given by $\Phi_{\rm gs}^{\rm core}$.
For $^6$He, we assume the ground state of the $\alpha$ core as the frozen configuration of $(0s)^4$~\cite{kikuchi09,kikuchi10}.
For $^{11}$Li, we use the ground-state wave function~\cite{kikuchi13a}, which is given in \S\ref{sec:11Li}, as the TOSM wave function.
The asymptotic wave function for the relative motion of the core~+~$n$~+~$n$ three-body system, $\phi_0$, is defined as
\begin{equation}
\phi_0 (\vc{k},\vc{K}) = \frac{1}{(2\pi)^3} e^{i\boldsymbol{k}\cdot\boldsymbol{r}+i\boldsymbol{K}\cdot\boldsymbol{R}},
\end{equation}
where $\vc{r}$ and $\vc{R}$ are conjugates to the relative momentum $\vc{k}$ and $\vc{K}$, respectively.

To describe the scattering states of two-neutron halo nuclei, it is also required to find Green's function given in Eq.~(\ref{eq:GreenF}).
Here, we obtain the eigenstates $\{\chi^\theta_\nu\}$ and their eigenvalues $\{E^\theta_\nu\}$ from Eq.~(\ref{eq:GreenF}) by solving the following complex-scaled Schr{\"o}dinger equation.
\begin{equation}
H^\theta \chi^\theta_\nu = E^\theta_\nu \chi^\theta_\nu.
\label{eq:cs_Scheq}
\end{equation}
To solve Eq.~(\ref{eq:cs_Scheq}), we employ the core~+~$n$~+~$n$ three-body OCM~\cite{aoyama06}.
For the core~+~$n$~+~$n$ three-body system, the total Hamiltonian $H$ is given as
\begin{equation}
H = h_{\rm core} + \sum_{i=1}^3 t_i - T_{\rm c.m.} + \sum_{i=1}^2 V_{{\rm core\mbox{-}}n} (r_i) + V_{n{\rm \mbox{-}}n} + v_{\rm PF}, %\hat{V}_{\rm PF},
\end{equation}
where $V_{{\rm core\mbox{-}}n}$ and $V_{n{\rm \mbox{-}}n}$ are the interactions for core-$n$ and $n$-$n$, respectively.
The coordinate $r_i$ represents the distance between the core nucleus and the $i$-th neutron.
The Pauli potential $v_{\rm PF}$ is the projection operator, which removes the Pauli forbidden states from the relative motion between the core nucleus and neutrons~\cite{kukulin86}.
For the $^6$He case, we use the KKNN potential~\cite{kanada79} and the Minnesota force~\cite{tang78} as $V_{{\rm core\mbox{-}}n}$ and $V_{n{\rm \mbox{-}}n}$, respectively.
For $^{11}$Li, we use the same Hamiltonian as that used in \S\ref{sec:11Li}.
By applying CSM to the total Hamiltonian $H$, we obtain the complex-scaled Hamiltonian $H^\theta$ in Eq.~(\ref{eq:cs_Scheq}).

\subsubsection{Coulomb breakup cross sections of $^\mathbf{6}$He and $^\mathbf{11}$Li}
We show the Coulomb breakup cross sections with respect to the excitation energies of $^6$He and $^{11}$Li.
The target is Pb, and the incident energies of $^6$He and $^{11}$Li projectiles are 240 and 70 MeV/nucleon, respectively.
The Coulomb breakup cross sections are calculated from the $E1$ strength distributions and the equivalent photon method~\cite{bertulani88}
using the following equation.
\begin{equation}
\frac{d^6\sigma}{d\vc{k} d\vc{K}} = \frac{16\pi^3}{9\hbar c}N (E_\gamma) \frac{d^6 B(E1)}{d\vc{k} d\vc{K}},
\label{eq:cs_six}
\end{equation}
where the virtual photon number with the photon energy $E_\gamma$ is given as $N(E_\gamma)$~\cite{bertulani88}.
The $E1$ strength distribution is calculated using the CSLS solutions in Eq. (\ref{eq:CSLS_bra}) as
\begin{equation}
\frac{d^6 B(E1)}{d\vc{k} d\vc{K}} = \frac{1}{2J_0+1} \left|\big\bra \Psi^{(-)} (\vc{k},\vc{K}) || O(E1) || \Psi_0 \big\ket\right|^2,
\end{equation}
where $O(E1)$ represents the operator of the $E1$ transition.
The wave function and the total spin of the ground state of the core~+~$n$~+~$n$ are given as $\Psi_0$ and $J_0$, respectively.

From Eq.~(\ref{eq:cs_six}), the differential cross sections with respect to the excitation energies $E$ are described as
\begin{equation}
\frac{d\sigma}{dE} = \int \int d\vc{k} d\vc{K} \frac{d^6\sigma}{d\vc{k} d\vc{K}}
\delta\left(E-\frac{\hbar^2k^2}{2\mu}-\frac{\hbar^2K^2}{2M}\right),
\end{equation}
where $\mu$ and $M$ are reduced masses corresponding to the momenta $\vc{k}$ and $\vc{K}$, respectively.

\begin{figure}[tb]
\includegraphics[width=8.5cm,clip]{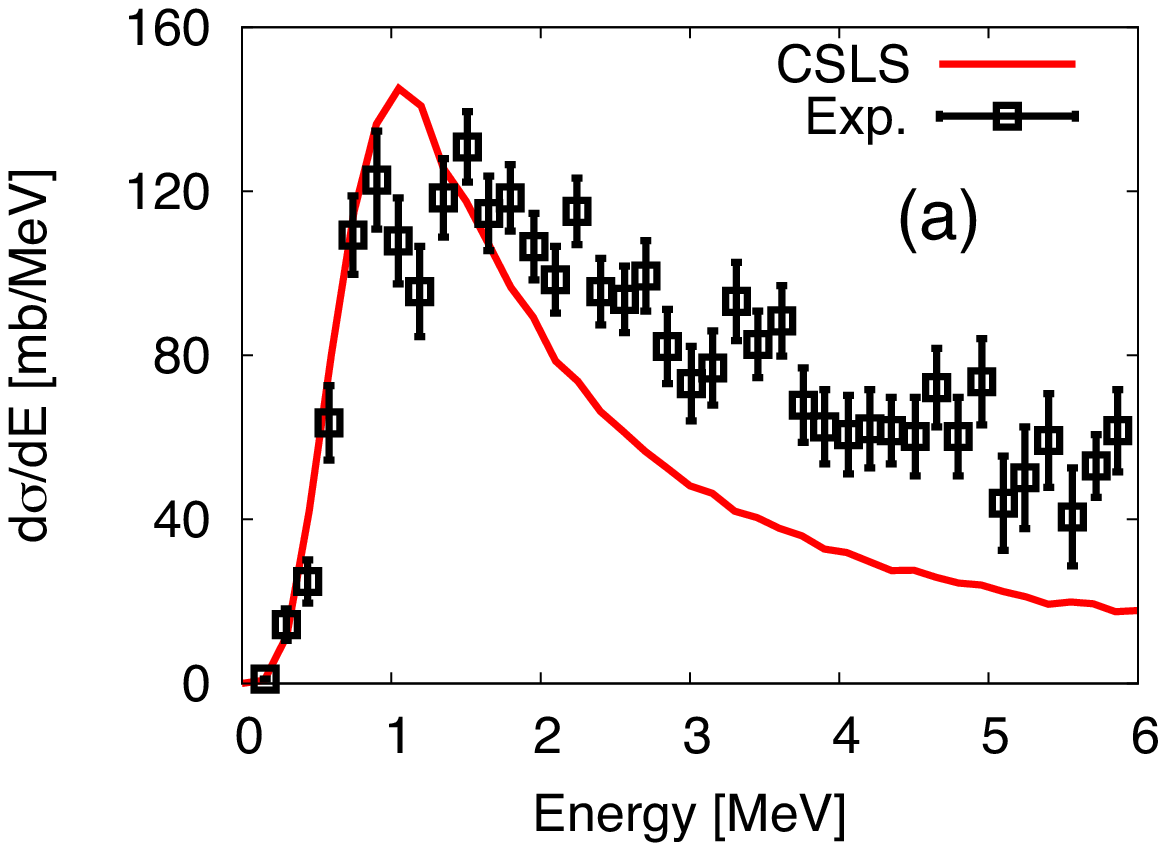}
\hfill
\includegraphics[width=8.5cm,clip]{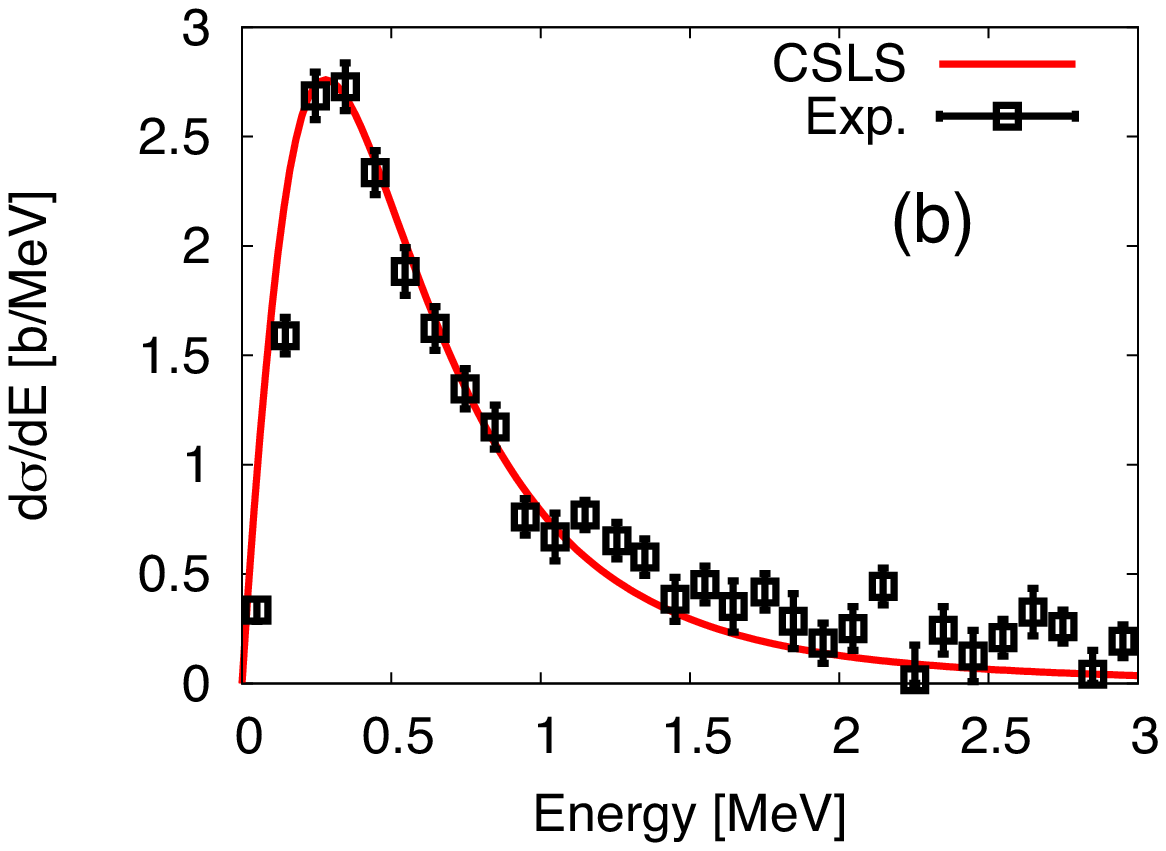}
\caption{(Color online)
Coulomb breakup cross sections measured from the three-body breakup thresholds.
The panels (a) and (b) represent the results for $^6$He and $^{11}$Li, respectively.
The experimental data for $^6$He and $^{11}$Li are taken from Refs.~\cite{aumann99} and \cite{nakamura06}, respectively, and shown as open squares with error bars.}
\label{fig:CS}
\end{figure}

In Fig.~\ref{fig:CS}, we show the cross sections for $^6$He and $^{11}$Li measured from the core~+~$n$~+~$n$ threshold energies in comparison with the experiments.
For $^6$He, there exists a low-energy enhancement in the distribution at around 1 MeV and the cross section gradually decreases with the excitation energy.
The result fairly reproduces the observed cross section \cite{aumann99}, especially in the low excitation energy region below $E\sim 2$ MeV.
The height and position of the low-energy enhancement in the strength agree well with the experimental results~\cite{aumann99}.
For $^{11}$Li, the calculated results show good agreement with the experimental results~\cite{nakamura06} in terms of shape and magnitude over the whole energy region.
The cross section shows a low-lying enhancement at around 0.25 MeV, which rapidly decreases as the energy increases.

Furthermore, it is confirmed in both cases of $^6$He and $^{11}$Li that the low-lying enhancements in the cross sections are dominated by strong final-state interactions (FSIs) in the dipole excited states~\cite{kikuchi10,kikuchi13a}.
This fact indicates that the Coulomb breakup cross sections reflect the characteristics of the final three-body scattering states and that the information on the ground-state structure of halo nuclei is masked by the strong FSI.
To clarify the mechanisms of the Coulomb breakup reactions, it is necessary to investigate the decay modes of $^6$He and $^{11}$Li in detail, as these decay modes provide the information on the correlations in the final states.
This analysis is shown in the following sections, based on the invariant mass spectra of the binary subsystems.

%%%%%%%%%%%%%%%%%%%%%%%%%%%%%%%%%%%%%%%%%%%%%%%%%%%%%%%%%%%%%%%%%%%%%%%%%%%%%%%%%%%%%%%%%%%%%%%%%%%%%%%%%
\subsubsection{Effects of 2p2h excitations due to the tensor and pairing correlations in $^\mathbf{9}$Li}

\begin{figure}[t]
\centering\includegraphics[width=8.5cm,clip]{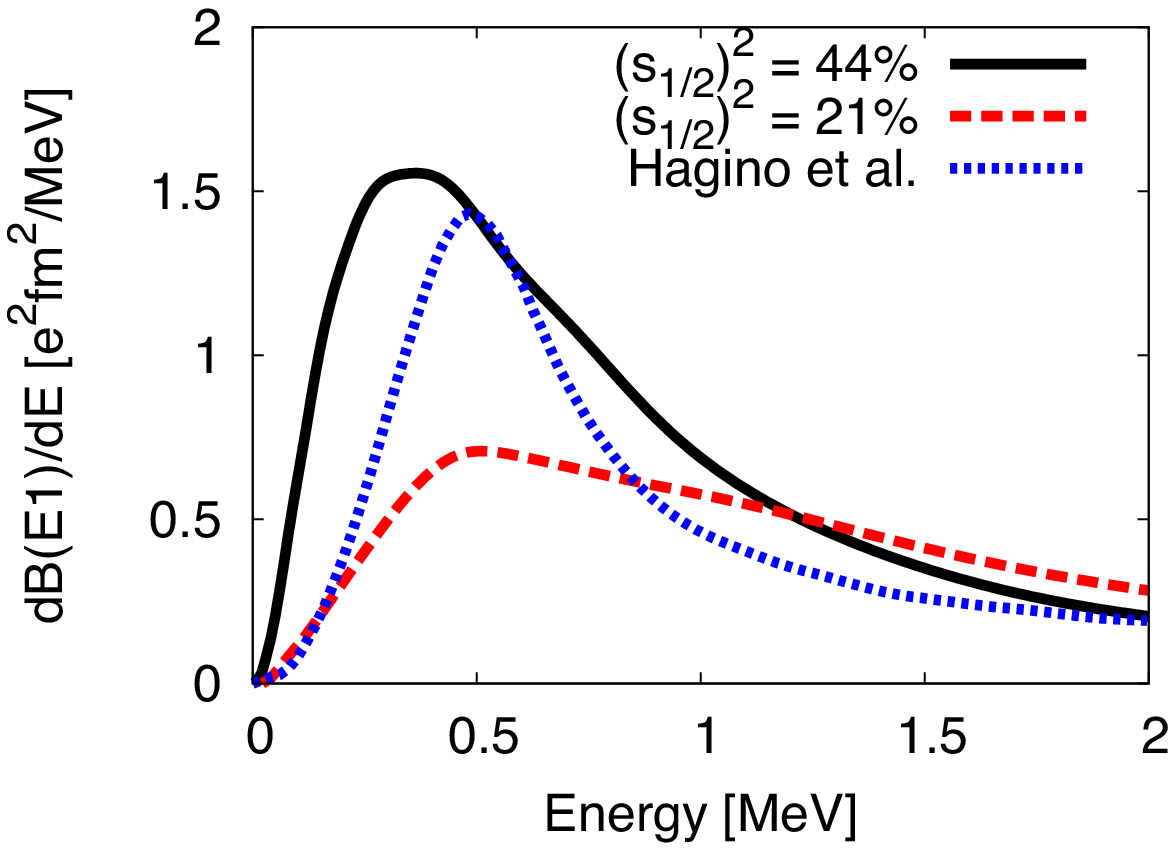}
\caption{\label{fig:dBE1} (Color online)
$E1$ strength distribution of $^{11}$Li with different ($s_{1/2}$)$^2$ components in the ground state.
The black (solid) line represents the present result shown in Fig.~\ref{fig:CS}.
The red (dashed) line denotes the restricted calculation with $(s_{1/2})^2=21$\%.
The blue (dotted) line is taken from Ref.~\cite{hagino09}.
}
\end{figure}

One of the characteristics of $^{11}$Li is quite a large $(s_{1/2})^2$ mixing of about 45\% in the ground state \cite{simon99}, which generates the halo structure as explained in \S~\ref{sec:11Li_cc}. 
To determine the effect of the large $s$-wave mixing on the Coulomb breakup strength, we calculate the $E1$ strength distributions using different ground-state wave functions of $^{11}$Li.
The calculated $E1$ distribution is shown in Fig.~\ref{fig:dBE1} as a red (dashed) line.
This result is obtained by the restricted coupled-channel calculation with a small $(s_{1/2})^2$ component of 21.0\%~\cite{myo03}.
The distribution shows small strength at the peak energy, the magnitude of which is about half of the original one with an $s$-wave mixing of 44\%.
The result indicates that $s$-wave mixing in the ground state plays a significant role in reproducing the low-lying enhancement in the breakup strength.
It is also suggested that a large $s$-wave mixing enhances the dineutron correlation in the $^{11}$Li ground state \cite{myo08}, as shown in Fig~\ref{fig:sec3_2n-density}.

It is important to clarify the effect of the correlations in $^9$Li on the $E1$ strength distribution of $^{11}$Li.
We compare the restricted coupled-channel calculation and that using the simple $^9$Li~+~$n$~+~$n$ model assuming an inert $^9$Li core~\cite{hagino09}.
Both wave functions contain almost the same amount of the $s$-wave component.
In two kinds of results, the red (dashed) and blue (dotted) lines shown in Fig.~\ref{fig:dBE1}, there exists a large difference in the strengths.
This is due to the fact that about 15\% of the strength in the coupled channel calculation escapes to the highly excited $^{11}$Li states processing the excited components of the $^9$Li core.

From these comparisons, it is summarized that the large $s$-wave mixing in the ground state of $^{11}$Li and the correlations in the $^9$Li core 
play the essential roles in reproducing the Coulomb breakup cross section, in particular, the simultaneous reproduction of the position and the magnitude of the low-lying enhancement.

\subsubsection{Invariant mass spectra for binary subsystems of $^\mathbf{6}$He and $^\mathbf{11}$Li}

To understand the mechanisms of Coulomb breakups of $^6$He and $^{11}$Li, we calculate the invariant mass spectra for binary subsystems such as core-$n$ and $n$-$n$
and discuss the correlations in the subsystems.
Using Eq.~(\ref{eq:cs_six}), we express the invariant mass spectra using the CSLS strength distribution as
\begin{equation}
\frac{d\sigma}{d\varepsilon} = \int \int d\vc{k} d\vc{K} \frac{d^6\sigma}{d\vc{k} d\vc{K}} \delta \left(\varepsilon - \frac{\hbar^2 k^2}{2\mu}\right),
\label{eq:inv}
\end{equation}
where $\varepsilon$ is the relative energy of the corresponding binary subsystem.

\begin{figure}[t]
\centering{\includegraphics[width=8.5cm,clip]{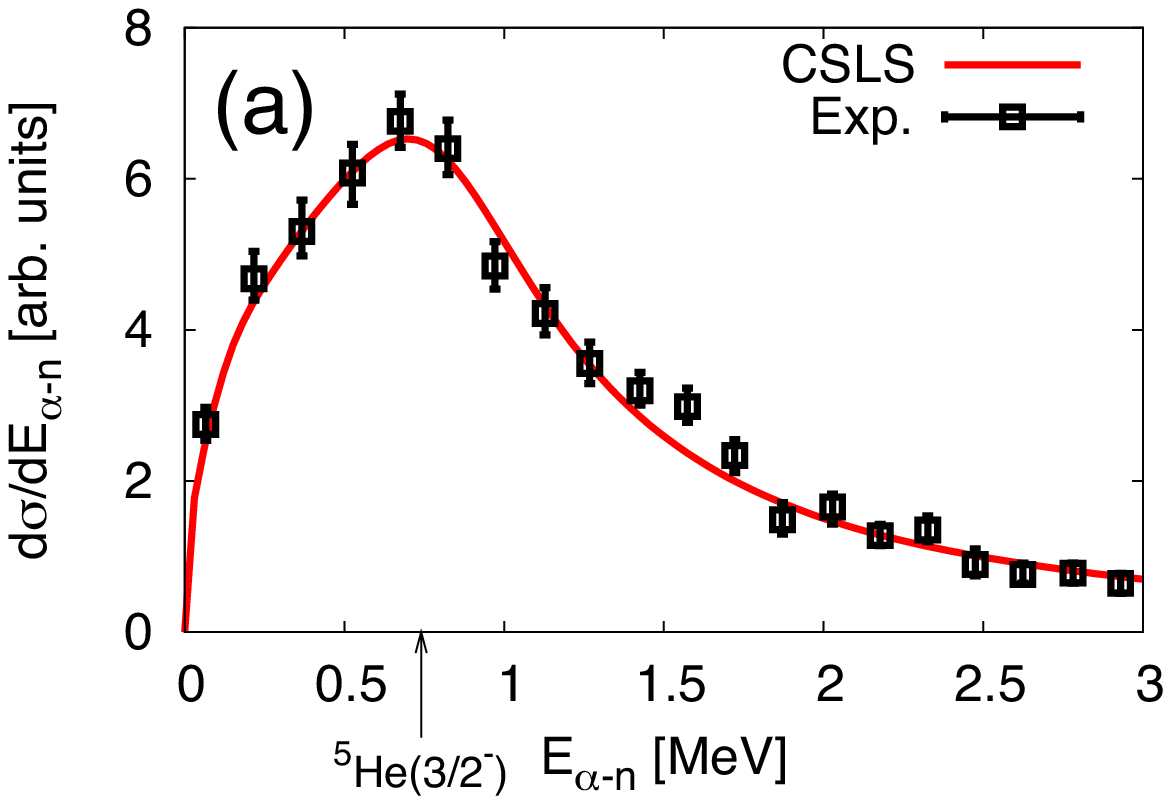}}
\hfill
\centering{\includegraphics[width=8.5cm,clip]{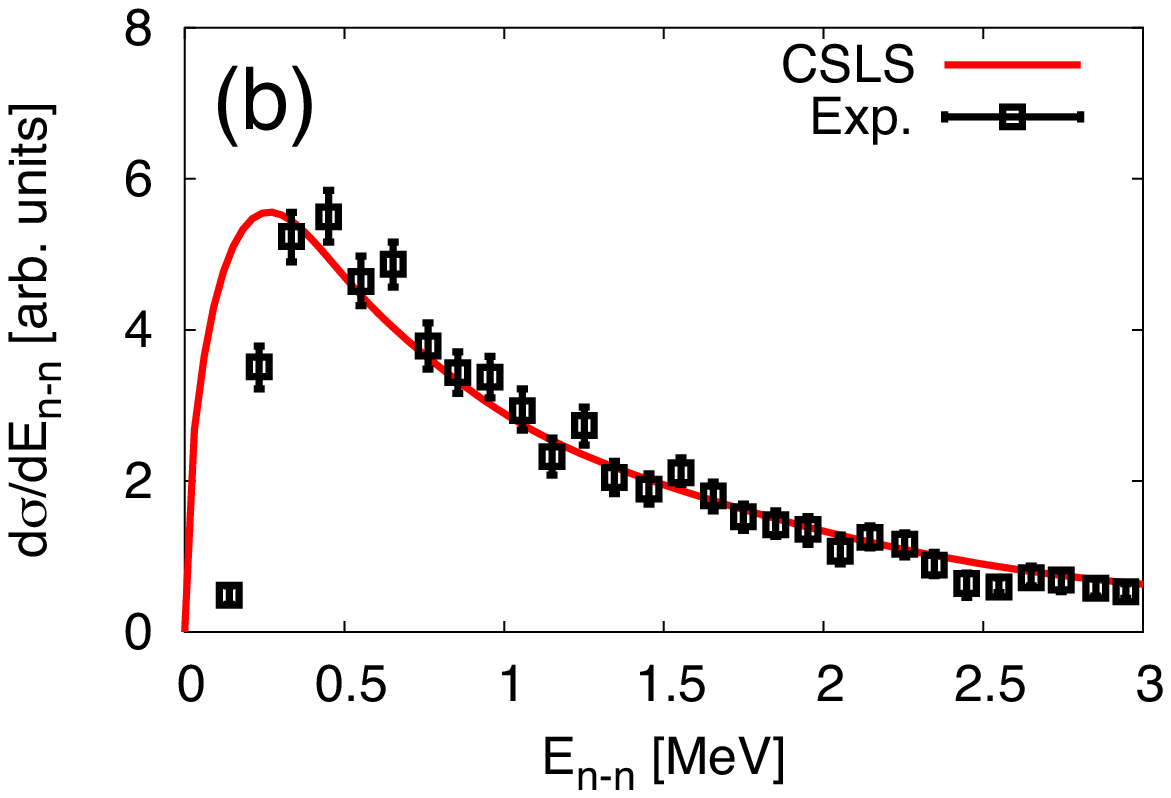}}
\caption{(Color online) 
Invariant mass spectra of the Coulomb breakup cross section of $^6$He with arbitrary units. 
The panels (a) and (b) represent the results with respect to
the $\alpha$-$n$ and $n$-$n$ binary subsystems, respectively. 
The open squares are the experimental data~\cite{aumann99}.
The arrow in the panel (a) indicates the $^5$He(3/2$^-$) resonance energy.}
\label{fig:CS_IMS}
\end{figure}

In Fig.~\ref{fig:CS_IMS}, we show the calculated invariant mass spectra for $^6$He in comparison with the experimental data \cite{aumann99}.
The panels (a) and (b) show the spectra for the $\alpha$-$n$ and $n$-$n$ subsystems, respectively.
These two spectra show good agreement with the experimental data.
This agreement indicates the reliability of the present CSLS method in investigating the subsystem correlation in the three-body Coulomb breakups.
For the $\alpha$-$n$ case (Fig.~\ref{fig:CS_IMS} (a)), it is found that the peak position of the strength coincides with the resonance energy of $^5$He(3/2$^-$); the position of coincidence is indicated by an arrow.
The $^5$He resonance is clearly confirmed in the invariant mass spectra.
This fact indicates the sequential breakup process of $^6$He via $^5$He+$n$ channel around the energy region via the $E1$ response.
For the $n$-$n$ case (Fig.~\ref{fig:CS_IMS} (b)), a low-lying enhancement is observed near the zero energy region, which comes from the $n$-$n$ virtual state.

\begin{figure}[t]
\centering\includegraphics[width=8.5cm,clip]{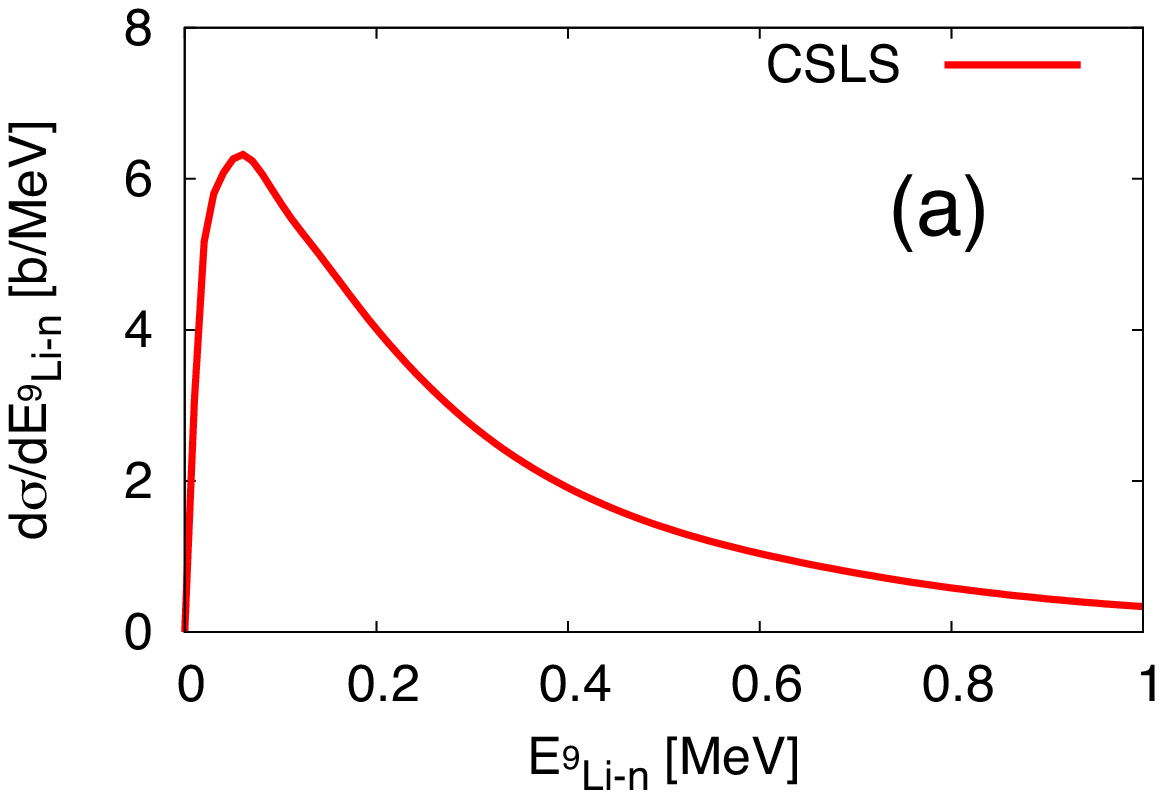}
\centering\includegraphics[width=8.5cm,clip]{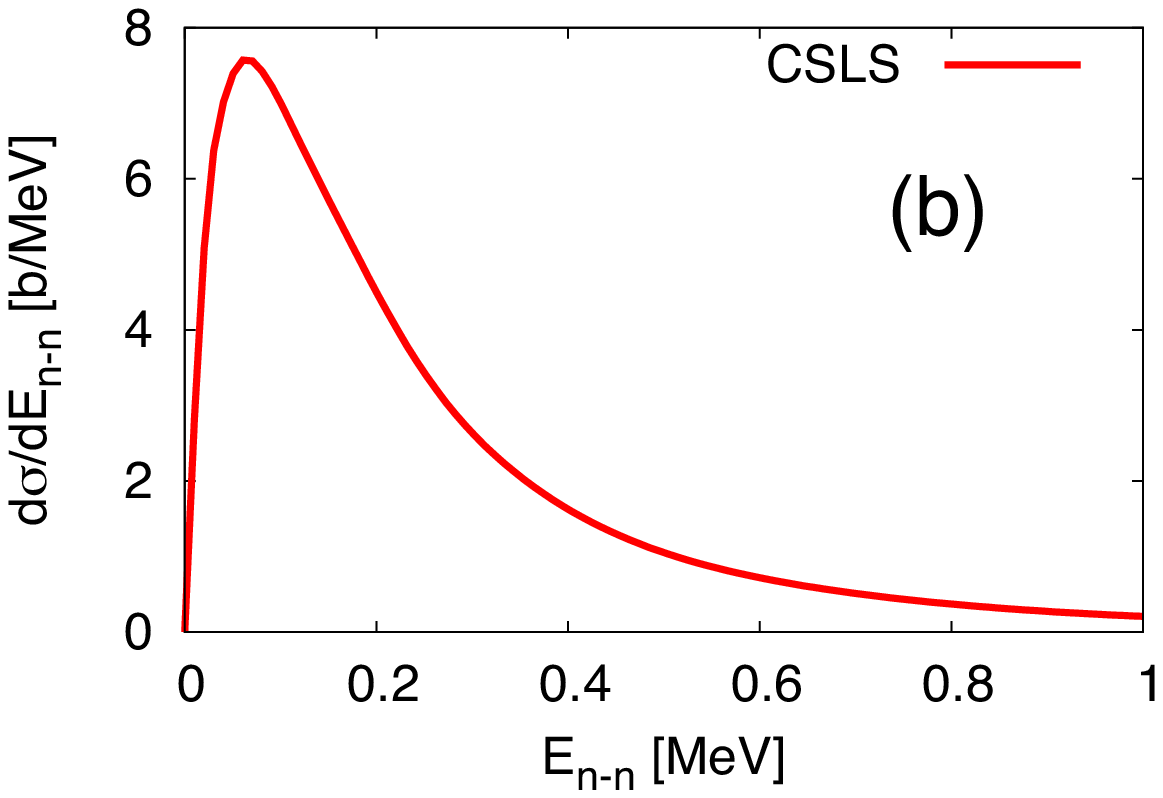}
\caption{\label{fig:invariant}(Color online)
Invariant mass spectra for $^9$Li-$n$ and $n$-$n$ binary subsystems. Panels (a) and (b) represent the results for the $^9$Li-$n$ and $n$-$n$ subsystems, respectively.}
\end{figure}

We also calculate the invariant mass spectra in the Coulomb breakup reaction of $^{11}$Li in Fig.~\ref{fig:invariant},
as functions of the relative energies of $^9$Li-$n$ and $n$-$n$ subsystems, respectively.
It is found that both spectra have sharp peak structures commonly below 0.1 MeV.
In Fig.~\ref{fig:invariant} (b), the peak seen in the $n$-$n$ case is caused obviously by the $n$-$n$ virtual state,similar to the $^6$He case
shown in Fig. \ref{fig:CS_IMS} (b).
\begin{figure}[t]
\centering\includegraphics[width=8.5cm,clip]{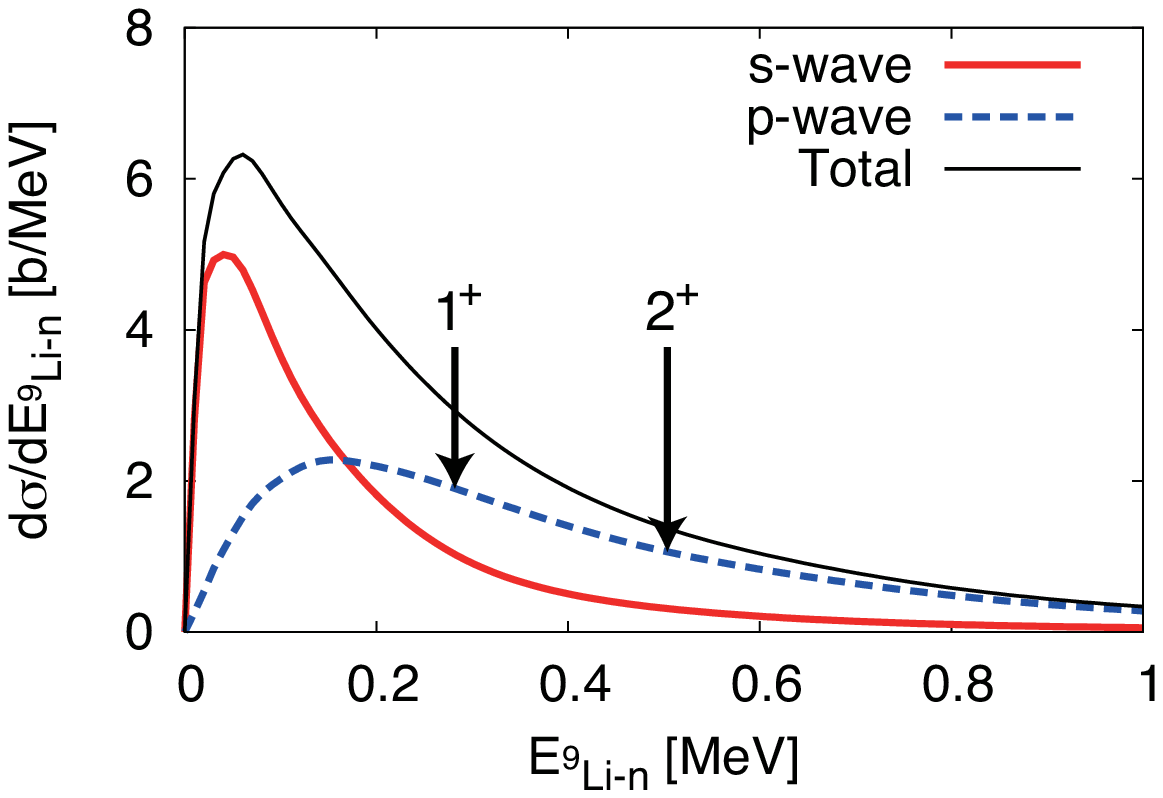}
\caption{\label{fig:inv_decomp}(Color online)
$s$-wave and $p$-wave components of the invariant mass spectra for $^9$Li-$n$. 
The black solid line indicates the sum of the components, which concedes with the result with FSI shown in Fig.~\ref{fig:invariant} (a).
The two arrows indicate the positions of the $p$-wave resonance energies of 1$^+$ and 2$^+$.}
\end{figure}

For the invariant mass spectra of the $^9$Li-$n$ subsystem, Hagino {\it et al.}~\cite{hagino09} use the simple $^9$Li~+~$n$~+~$n$ three-body model and reported that both the $s$-wave virtual state and the $p$-wave resonance of $^{10}$Li contribute to the spectra.
On the other hand, as shown in Fig.~\ref{fig:invariant} (a), our prediction considering the tensor and pairing correlation of $^9$Li is of a single prominent peak below 0.1 MeV.

We further decompose the spectra into the $s$-wave and $p$-wave components, as shown in Fig.~\ref{fig:inv_decomp}.
The results show that the $s$-wave component has a peak below 0.1 MeV, which comes from the virtual $s$-state of $^{10}$Li.
The $p$-wave component has a broad bump at around 0.15 MeV.
The bump energy does not correspond to the $p$-wave resonance energies in $^{10}$Li, which are indicated by two arrows in the figure.
This fact indicates that the $p$-wave contribution comes from the non-resonant continuum states of $^9$Li-$n$.
It is concluded that the shape of the $^9$Li-$n$ invariant mass spectra is mainly determined by the virtual $s$-state in $^{10}$Li, 
while the non-resonant $p$-wave contributes some amount in the spectra, which becomes dominant at energies higher than 0.2 MeV.
This conclusion contradicts the result in Ref.~\cite{hagino09}, in which the $p$-wave resonance has a sizable contribution to the strength.

The reason that the $p$-wave resonances are not observed in the present $^9$Li-$n$ invariant mass spectra can be understood as follows:
The $p$-wave resonances of $^{10}$Li are located at 0.275 MeV and 0.506 MeV for $1^+$ and $2^+$, respectively.
These energy positions are higher than the peak energy of around 0.25 MeV in the breakup cross section in Fig.~\ref{fig:CS}.
The relation between resonance and peak energies implies that the sequential breakup process via the $p$-wave resonances of $^{10}$Li is energetically not favored at around the peak energy of the cross section.
On the other hand, the breakup cross section calculated by Hagino {\it et al.}~\cite{hagino09} has a peak at around 0.5 MeV. 
In that case, the sequential breakup via the $p$-wave resonances of $^{10}$Li is favorably allowed and can exhibit a visible peak in the strength.
It should be noted that in Fig.~\ref{fig:inv_decomp}, the resonance energies of $1^+$ and $2^+$ are slightly changed from the values shown in Table~\ref{tab:sec3_10Li} since the $^9$Li-$n$ potential is modified to reproduce the recent observation of the two-neutron separation energy of $^{11}$Li~\cite{bachelet08}.

%%%%%%%%%%%%%%%%%%%%%%%%%%%%%%%%%%%%%%%%%%%%%%%%%%%%%%%%%
\subsection{Nuclear breakup reaction of $^\mathbf{6}$He}
From the results of the Coulomb breakup reactions of $^6$He and $^{11}$Li, it is shown that CSLS well describes the physical quantities associated with the three-body breakups.
We further apply CSLS to the nuclear breakup reaction of $^6$He.

In Coulomb breakup reactions, the $E1$ transition dominates the breakup process and can populate the specific excited states in the reactions.
Using the nuclear probes, one can access the other excited states. 
Considering both nuclear and Coulomb breakup reactions, one can obtain substantial information on the scattering properties of unstable nuclei.
In this review, we consider the case of $^6$He ($2^+$) for understanding the structure of the excited resonant states of two-neutron halo nuclei.
In fact, the $2^+_1$ resonance of $^6$He is strongly populated in the nuclear breakup, while this resonance is not favored in the Coulomb breakup reactions~\cite{aumann99}.

In the decay process of the $2^+_1$ resonance of $^6$He into $\alpha$ and two neutrons, the following decay modes are considered:
\begin{itemize}
\itemsep=0.0cm
\item[(1)] Emission of one neutron is followed by emission of the other, via the $^5$He(3/2$^-$) resonance.
\item[(2)] Two neutrons are emitted simultaneously (not via $^5$He(3/2$^-$) resonance), correlating with each other.
\item[(3)] Same as in case 2 but the two neutrons are emitted independently.
\end{itemize}

To investigate the decay modes of the $2^+_1$ resonance, we consider the $^6$He breakup reaction by $^{12}$C at 240 MeV/nucleon as a formation process of the $2^+_1$ resonance \cite{aumann99}.
We treat this process by means of the continuum-discretized coupled-channel method (CDCC)~\cite{kamimura86,austern87,yahiro12,kikuchi13b}.
After formation of the $2^+_1$ resonance, its decay into $\alpha$ and two neutrons is then investigated by CSLS.
The combination of CDCC and CSLS, CDCC$-$CSLS, is a powerful method to investigate the breakup reaction of unstable nuclei.

In CDCC with the pseudo state discretization method~\cite{matsumoto10}, the scattering is assumed to take place in the model space $\mathcal{P}$ defined by
\begin{equation}
\mathcal{P} = \sum_i \left| \chi_i \right\rangle \left\langle \chi_i \right|.
\end{equation}
Here $\chi_i$ is the wave function for the $i$-th eigenstate of $^6$He within the bound-state approximation of the three-body model, whose energy is given by $\epsilon_i$.
The CDCC $T$-matrix element to the $i$-th eigenstate, $\chi_i$, is given as
\begin{equation}
T^{\rm CDCC}_i = \bra \chi_i \psi_i^{(-)} (\vc{P}_i) | U - U^{\rm Coul}_{^6 {\rm He}} | \Psi^{(+)}_{\rm CDCC} \ket,
\end{equation}
where $\psi_i^{(-)}$ and $\vc{P}_i$ represent the final-state wave function and the asymptotic momentum, respectively, for the relative motion between $^6$He and $^{12}$C.
The initial scattering wave function $\Psi^{(+)}_{\rm CDCC}$ is obtained by solving the four-body Schr\"odinger equation in CDCC for $^6$He and $^{12}$C~\cite{matsumoto10}.
The potential $U$ is the sum of the optical potentials between the $^{12}$C target and the constituent particles in $^6$He.
In the calculation of the $T$-matrix, the Coulomb interaction between $^{12}$C and $^6$He, $U^{\rm Coul}_{^6{\rm He}}$, is subtracted from $U$.
Using the CDCC $T$-matrix element, the exact $T$-matrix element to a continuum state is well approximated by
\begin{equation}
\begin{split}
T (\vc{k},\vc{K},\vc{P}) &\approx \sum_i \bra \Psi^{(-)} (\vc{k},\vc{K}) | \chi_i \ket\ T_i^{\rm CDCC} \\
&= \sum_i f_i (\vc{k},\vc{K})\ T_i^{\rm CDCC},
\end{split}
\label{eq:Tmat}
\end{equation}
where $\vc{k}$ and $\vc{K}$ are the relative momenta in a Jacobi coordinate of the $\alpha$~+~$n$~+~$n$ three-body system.
The asymptotic momentum between the projectile and target is represented by $\vc{P}$.
To obtain the smoothing function $f_i(\vc{k},\vc{K})$, we use CSLS.
The exact scattering states of $^6$He, $\Psi^{(-)} (\vc{k},\vc{K})$, given in Eq.~(\ref{eq:CSLS_bra}).

To investigate the decay modes of the $2^+_1$ resonance of $^6$He, we calculate the double-differential cross section with respect to the relative energies, $\varepsilon_1$ and $\varepsilon_2$, of binary subsystems.
Using Eq.~(\ref{eq:Tmat}), the double-differential cross section is given as
\begin{equation}
\begin{split}
\frac{d^2\sigma}{d\varepsilon_1 d\varepsilon_2} =& \frac{(2\pi^4 \mu_P)}{\hbar^2 P_0} \int\int\int d\vc{k} d\vc{K} d\vc{P} \left| T(\vc{k},\vc{K},\vc{P}) \right|^2 \\
&\times \delta\left(E_{\rm tot }- \frac{\hbar^2P^2}{2\mu_P} - \varepsilon_1 - \varepsilon_2\right)\delta\left(\varepsilon_1 - \frac{\hbar^2 k^2}{2\mu}\right)\delta\left(\varepsilon_2 - \frac{\hbar^2 K^2}{2M}\right),
\end{split}
\end{equation}
where $P_0$ is the incident momentum of $^6$He in the center-of-mass system and $\mu_P$ is the reduced mass for the relative motion between $^6$He and $^{12}$C.

\begin{figure}[tb]
\includegraphics[width=8.5cm,clip]{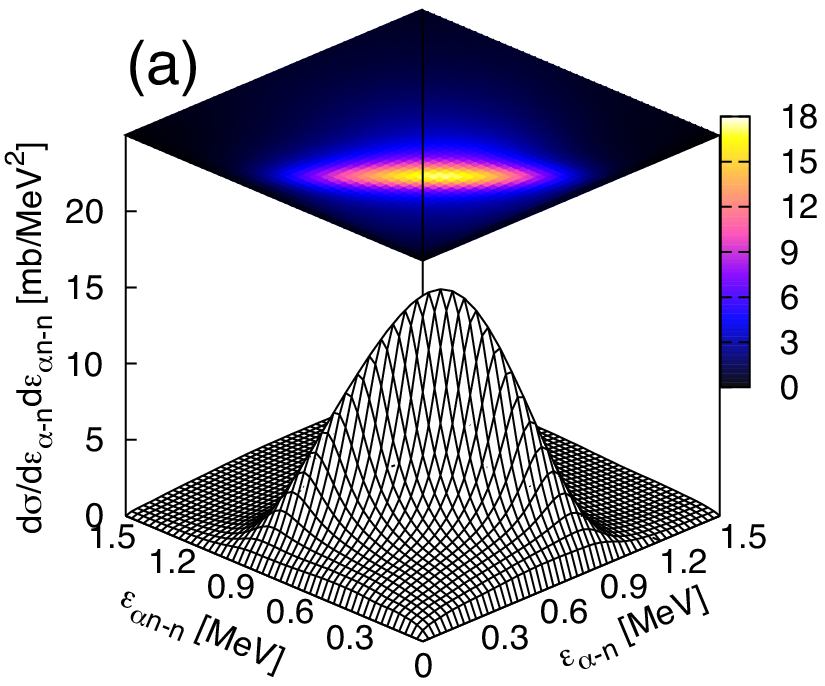}
\includegraphics[width=8.5cm,clip]{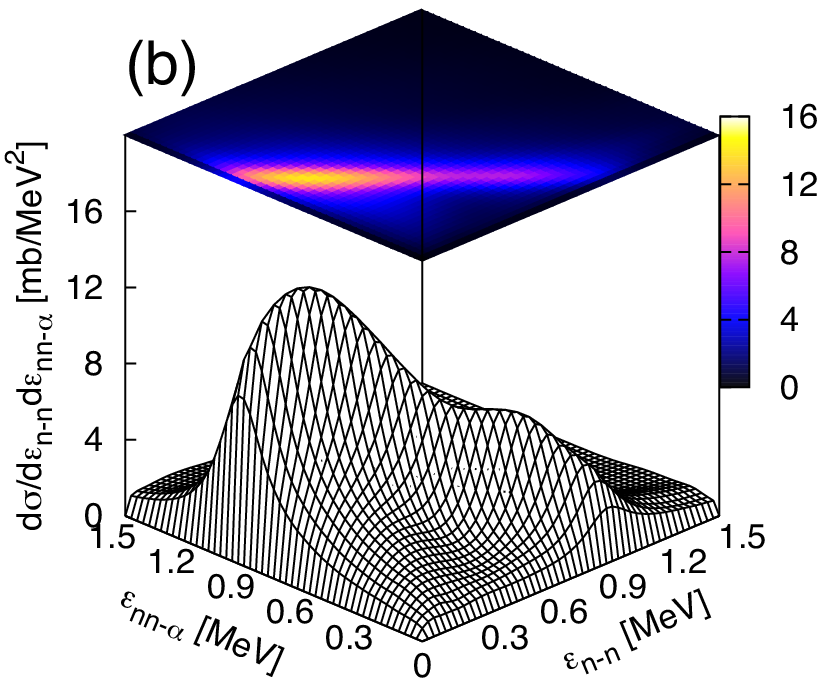}
\caption{\label{fig:2D_full}
(Color online) Double-differential cross sections of $^6$He by $^{12}$C at 240~MeV/nucleon.
Panels (a) and (b) represent the cross sections with respect to the subsystem energies of $\alpha$-$n$ and $n$-$n$, respectively (see the text for detail).}
\end{figure}

Figure~\ref{fig:2D_full} shows the double-differential cross section calculated by CDCC$-$CSLS.
In panel (a), the cross section is shown with respect to the energy between $\alpha$ and a neutron ($\varepsilon_{\alpha\mbox{-}n}$) and that between the other neutron and the c.m. of the $\alpha$-$n$ system ($\varepsilon_{\alpha n\mbox{-}n}$).
Similarly, the cross section with respect to the $n$-$n$ relative energy ($\varepsilon_{n\mbox{-}n}$) and the energy between the c.m. of the $2n$ system and $\alpha$ ($\varepsilon_{nn\mbox{-}\alpha}$) is shown in panel (b).
In both panels, the ridge structures can be clearly observed, corresponding to the total energy of the $\alpha+n+n$ three-body system, around 1.0~MeV.
This structure comes from the $2_1^+$ resonance of $^6$He obtained at 0.98~MeV above the three-body threshold with a decay width of 0.27~MeV.
The clear observation of the $2_1^+$ resonance is an important feature of the breakup process induced by $^{12}$C mainly due to nuclear interactions.

\begin{figure}[tb]
\begin{minipage}{0.48\textwidth}
\centering\includegraphics[width=8cm,clip]{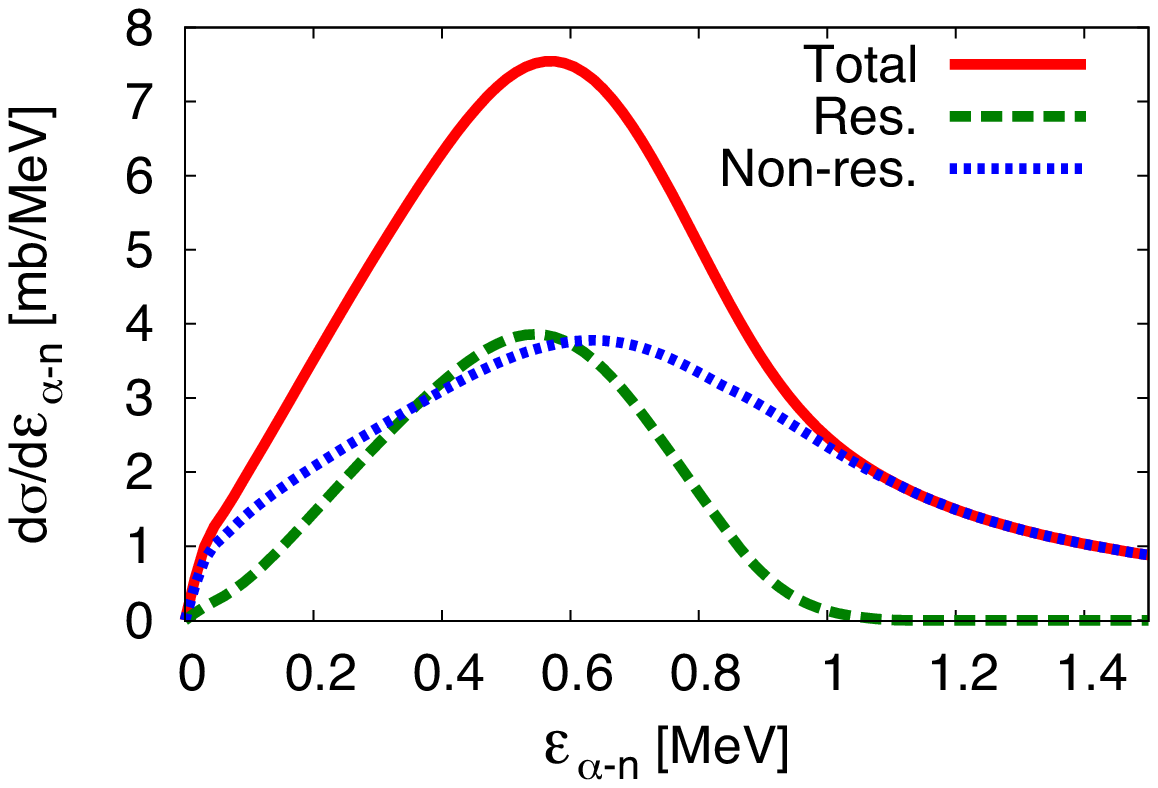}
\caption{\label{fig:inv1}
(Color online) Invariant mass spectra with respect to $\varepsilon_{\alpha\mbox{-}n}$ (solid line).
The dashed and dotted lines show the resonant and nonresonant contributions, respectively.}
\end{minipage}
\hfill
\begin{minipage}{0.48\textwidth}
\centering\includegraphics[width=8cm,clip]{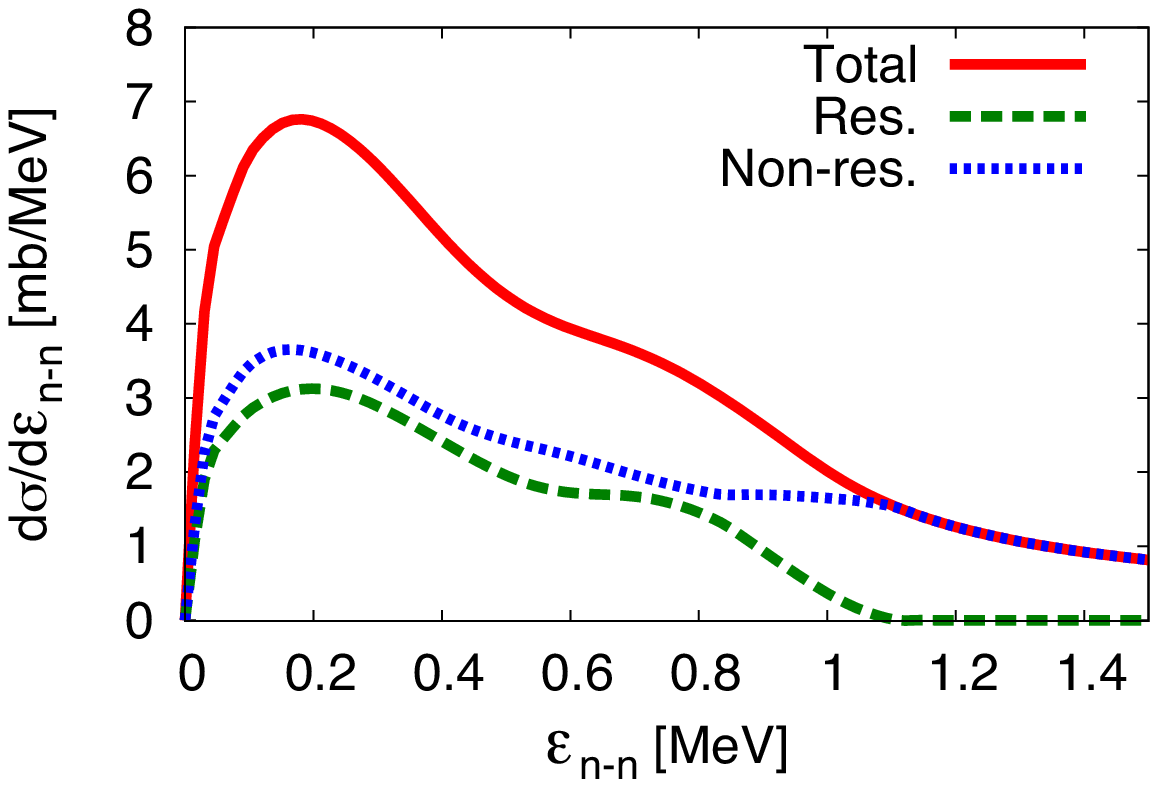}
\caption{\label{fig:inv2}
(Color online) Same as Fig.~\ref{fig:inv1} but for $d\sigma/d\varepsilon_{n\mbox{-}n}$.}
\end{minipage}
\end{figure}
We show the invariant mass spectra of the $2^+$ state and discuss the decay mode into the $\alpha+n+n$ system.
The invariant mass spectrum for a binary subsystem is calculated as
\begin{equation}
\left(\frac{d\sigma}{d\varepsilon_1}\right)_\text{Total} = \int \frac{d^2\sigma}{d\varepsilon_1 d\varepsilon_2} d\varepsilon_2.
\end{equation}
We also decompose the spectra into the contributions of resonant and non-resonant states;
\begin{equation}
\left(\frac{d\sigma}{d\varepsilon_1}\right)_\text{Total} = \left(\frac{d\sigma}{d\varepsilon_1}\right)_\text{Res.} + \left(\frac{d\sigma}{d\varepsilon_1}\right)_\text{Non-res.},
\end{equation}
where the resonant part is calculated by using a gate of width 0.27 MeV, centered on the $2^+_1$ resonance energy of 0.98 MeV.
The resonant part is given as
\begin{equation}
\left(\frac{d\sigma}{d\varepsilon_1}\right)_{\rm Res.} = \int_D \frac{d^2\sigma}{d\varepsilon_1 d\varepsilon_2}d\varepsilon_2, \qquad \{ D: 0.98-0.135\le \varepsilon_1 + \varepsilon_2 \le 0.98+0.135\}.
\end{equation}

The solid line in Fig.~\ref{fig:inv1} shows $\left(d\sigma/d\varepsilon_{\alpha\mbox{-}n}\right)_\text{Total}$ and the dashed line shows its resonant contribution.
The remaining dotted line is interpreted as the nonresonant contribution.
For the nonresonant part, the peak is confirmed at around 0.7~MeV corresponding to the energy of $^5$He(3/2$^-$). 
This indicates the sequential decay via the $^5$He(3/2$^-$), similar to the Coulomb breakup case \cite{kikuchi10} shown in Fig.~\ref{fig:CS_IMS}. 

For the resonant part in Fig.~\ref{fig:inv1}, a peak is found around 0.5~MeV, which is lower than the energy of the $^5$He(3/2$^-$) resonance.
Instead, the peak energy is about half the total energy $\varepsilon$ of the $2_1^+$ state, 
which indicates that two neutrons are emitted with an equal share of the total energy of the three-body system and the sequential decay via the $^5$He(3/2$^-$) resonance would be suppressed.
 
To pin down the decay mode of the two neutrons, we next discuss $\left(d\sigma/d\varepsilon_{n\mbox{-}n}\right)_\text{Total}$, shown in Fig.~\ref{fig:inv2}.
The resonant (dashed) and nonresonant (dotted) contributions are also shown in Fig.~\ref{fig:inv2}.
In the spectrum, we find two peaks indicating different types of correlations.
The first peak around 0.2~MeV suggests the so-called dineutron decay~\cite{bochkarev89}, in which the two neutrons are correlated by FSI.
This peak is also seen in the Coulomb breakup case \cite{kikuchi10} in Fig.~\ref{fig:CS_IMS}.
It should be noted that the dineutron decay does not directly imply the direct emission of a dineutron from the $2_1^+$ state.
The second peak (or shoulder) is around 0.8~MeV, which comes from the resonant part.
The $2n$ relative energy at the second peak of 0.8~MeV almost exhausts the total energy of the $2_1^+$ state ($\sim 1.0$~MeV).
Therefore, it suggests the back-to-back emission of two neutrons
It is also interesting that the experimental correlations~\cite{egorova12} in the decay of the mirror state $^6$Be$(2^+_1)$ are qualitatively similar to the present results.

The back-to-back emission of the $2_1^+$ state is an important finding of the present study with regard to the correlations among $\alpha$-$n$ and $n$-$n$ and FSI.
The $n$-$n$ interaction favors the virtual state of the $2n$ system, which corresponds to the first peak in $d\sigma/d\varepsilon_{n\mbox{-}n}$.
Furthermore, the $\alpha$-$n$ interaction that favors the sequential decay is also not important in the decay of the $2_1^+$ state, as shown in Fig.~\ref{fig:inv1}.
Therefore, one can conclude that the second peak in $d\sigma/d\varepsilon_{n\mbox{-}n}$ is free from
the FSI and directly reflects the structural property of the $2_1^+$ state of $^6$He.

\subsection{$\alpha$-$d$ scattering using $\alpha$~+~$p$~+~$n$ three-body model}
In $^6$Li, the $\alpha$~+~$d$ and $\alpha$~+~$p$~+~$n$ structures coexist in the low-excitation-energy region, since the thresholds of the $\alpha$~+~$d$ and $\alpha$~+~$p$~+~$n$ systems are closely located at the excitation energies of 1.47 and 3.70 MeV, respectively.
To investigate the scatterings associated with $^6$Li, it is necessary to theoretically obtain the three-body scattering solutions of $\alpha$~+~$p$~+~$n$.
We here describe the scattering states of $^6$Li by using CSLS and investigate the dynamical effects of $\alpha$~+~$p$~+~$n$ three-body structures on the $\alpha$~+~$d$ elastic scattering and the radiative capture reaction of $^2$H($\alpha$,$\gamma$)$^6$Li.
The effects of the deuteron breakup on the $d$-induced reactions have been extensively studied in the CDCC framework~\cite{kamimura86}.
In this work, we discuss the effects of not only the deuteron breakup but also the rearrangement to the $^5$He~+~$p$ and $^5$Li~+~$n$ channels on the above reactions \cite{kikuchi11}; the latter effect is not considered in CDCC.

We describe the scattering states of $^6$Li by using the $\alpha$~+~$p$~+~$n$ three-body model and CSLS.
For the $\alpha$~+~$d$ system, the asymptotic Hamiltonian is defined as
\begin{equation}
H_0 = h_d + T_{\rm rel} + V_{\rm Coul} (R),
\end{equation}
where $h_d$ is the internal Hamiltonian for deuteron.
The kinetic energy and the Coulomb interaction between $\alpha$ and deuteron are denoted as $T_{\rm rel}$ and $V_{\rm Coul}$, respectively.
The distance between the $\alpha$ particle and the center-of-mass of the deuteron is denoted as $R$.
The eigensolution of $H_0$ is expressed as
\begin{equation}
\Phi_0^{\ell J^\pi} (\vc{K},\vc{\xi}) = \left[\chi_d(\vc{r}) \otimes \phi^\ell_0 (\vc{K},\vc{R})\right]_{J^\pi},
\label{eq:asym_ad}
\end{equation}
where $\ell$ is the orbital angular momentum between $\alpha$ and deuteron and $J^\pi$ is the total spin and parity.
The relative momentum between the $\alpha$ particle and the deuteron is given as $\vc{K}$.
The set of relative coordinates is represented as $\vc{\xi} = (\vc{r},\vc{R})$.
The wave function $\chi_d$ is that of the deuteron.
The asymptotic relative wave function for the $\alpha$~+~$d$ system, $\phi_0^l$, is given as
\begin{equation}
\phi_0^\ell (\vc{K},\vc{R}) = (2\ell+1) i^\ell \frac{F_\ell(\eta,KR)}{KR} \sum_m Y_{\ell m}(\hat{\vc{K}}) Y^*_{\ell m}(\hat{\vc{R}}),
\end{equation}
where $F_\ell$ is a regular Coulomb wave function and $\eta$ is the Sommerfeld parameter.
Using Eqs.~(\ref{eq:CSLS_bra}) and (\ref{eq:asym_ad}), we obtain the scattering states of the $\alpha$~+~$d$ system $\Psi^{(-)}_{\ell J^\pi} (\vc{K})$.
To describe the $\alpha$~+~$p$~+~$n$ three-body components in the scattering process, we prepare the set of the eigenstates $\{\chi^\theta_\nu\}$ in Eq.~(\ref{eq:CSLS_bra}) 
as the solutions of the $\alpha$~+~$p$~+~$n$ three-body model with complex scaling.

%%%%%%%%%%%%%%%%%%%%%%%%%%%%%%%%%%%%%%%%%%%%%%%%%%%%%%%%%%%%%%%%%%%%%%%%%%%%%%%%
\subsubsection{Elastic phase shifts of $\alpha$~+~$d$ scatterings\label{sec:phs}}

\begin{figure}
\begin{minipage}{0.48\textwidth}
\centering\includegraphics[width=8.5cm]{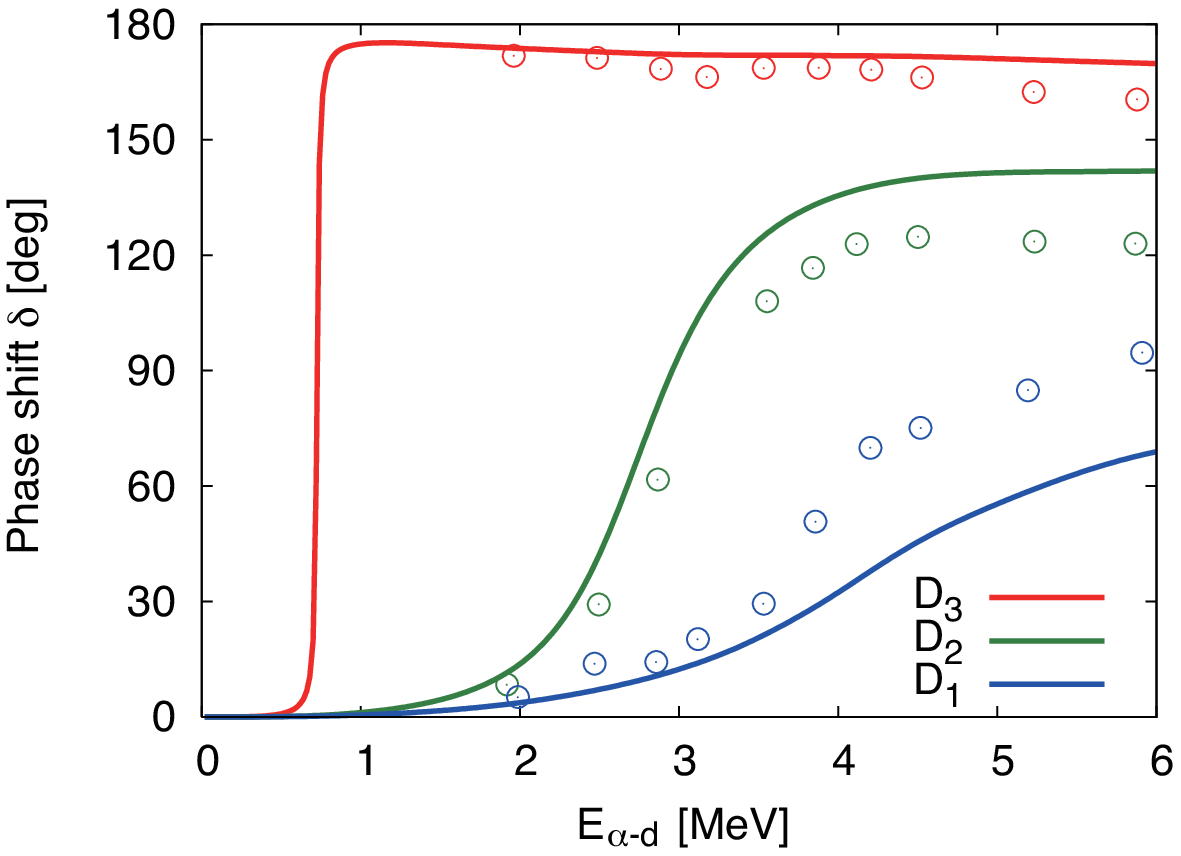}
\caption{\label{fig:phs_6Li} (Color online)
Elastic phase shifts for $\alpha$ + $d$ scatterings as functions of relative energies of the $\alpha$~+~$d$ system in comparison with the experimental data. The red, green, and blue lines show the calculated data for $D_3$, $D_2$, and $D_1$ scattering states, corresponding to the $3^+$, $2^+$, and $1^+$ states, respectively. The experimental data \cite{schmelzbach72,gruebler75} are shown as open circles; the colors of the circles correspond to the same states as the colors of lines (calculated data).}
\end{minipage}
\hfill
\begin{minipage}{0.48\textwidth}
\centering\includegraphics[width=8.5cm]{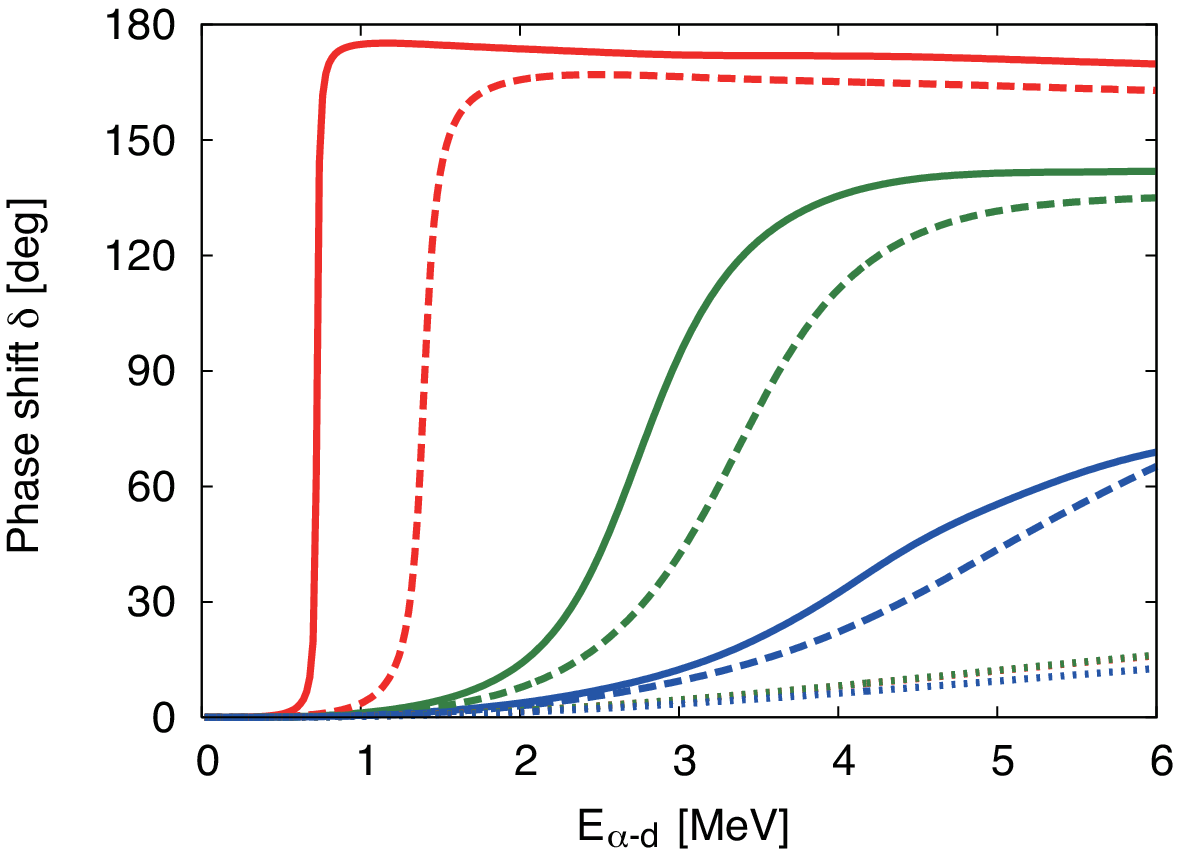}
\caption{\label{fig:phs_comp} (Color online)
Effects of the deuteron breakup and rearrangement on the phase shifts.
The solid lines indicate the full calculation results, same as in Fig.~\ref{fig:phs_6Li}. The dotted and dashed lines indicate the results of Elastic and Breakup calculations, respectively. Three dotted lines obtained for the Elastic calculation are almost identical to each other in the figure.
}
\end{minipage}
\end{figure}

In Fig.~\ref{fig:phs_6Li}, the $\alpha$~+~$d$ elastic phase shifts for the $D$-wave scatterings are shown in comparison with the observed data.
The calculated phase shifts well reproduce the trend of the observed data for the $1^+$, $2^+$, and $3^+$ states.
This agreement indicates the reliability of the three-body description of $^6$Li with CSLS.

We discuss the three-body effect on the $\alpha$~+~$d$ scatterings.
In particular, we concentrate on the effects of the deuteron breakup and of the rearrangement to the $^5$He~+~$p$ and $^5$Li~+~$n$ channels.
For this purpose, we show the following two types of results other than that shown in Fig.~\ref{fig:phs_6Li}.
One is the result in which only the elastic channel of $\alpha$~+~$d$ is taken into account in order to exclude the effects of the deuteron breakup and rearrangement, named as ``Elastic".
The other is the result in which the deuteron breakup channels are additionally included while the rearrangement channels of $^5$He~+~$p$ and $^5$Li~+~$n$ are excluded, named as ``Breakup".
This analysis enables us to estimate how the deuteron breakup and the rearrangement channels contribute to the $\alpha$~+~$d$ scattering observables.

We obtain the two types of results of the ``Elastic" and ``Breakup" cases by preparing different sets of eigenstates $\{\chi^\theta_\nu\}$ in Green's function in Eq.~(\ref{eq:GreenF}), since the $\alpha$~+~$p$~+~$n$ three-body structures are involved via $\{\chi^\theta_\nu\}$ in CSLS.
For the Elastic case, the wave function of the $p$~+~$n$ part in $^6$Li is fixed as a deuteron during the scattering.
We solve only the relative motion between the $\alpha$ particle and the deuteron and obtain the set of eigenstates $\{\chi^\theta_\nu\}$ of $^6$Li.

For the Breakup case, we take into account the excitations of the $p$-$n$ relative motion.
We first calculate the pseudo states of the $p$-$n$ system, which correspond to the ground and excited states.
Next, we solve the coupled-channel problem for the relative motion between the $\alpha$ particle and the $pn$ pseudo states, 
and obtain the set of $^6$Li eigenstates $\{\chi^\theta_\nu\}$.

Using the different sets of eigenstates $\{\chi^\theta_\nu\}$ in the Elastic and Breakup calculations, 
we obtain the two kinds of ECRs for the Elastic and Breakup ones, which are different from the original one used in the results in Fig.~\ref{fig:phs_6Li}.
In each ECR, the completeness relation in Eq.~(\ref{eq:sec2_ECR}) is satisfied.

The results are shown in Fig.~\ref{fig:phs_comp}.
In the results of the Elastic case (shown with dotted lines), the deuteron is retained in the ground state, and the phase shifts exhibit structureless behaviors.
On the other hand, the results of the Breakup case (shown with dashed lines) reflect the resonance behaviors for each state.
It is found that the deuteron breakup plays a significant role in the $\alpha$~+~$d$ scattering, as was suggested in Ref.~\cite{kamimura86}.
However, the resonance positions are still higher than the full calculations.
This difference suggests that the rearrangement effect shifts the resonance positions down by about 500 keV.
From the results in Fig.~\ref{fig:phs_comp}, it is found that the deuteron breakup has a significant role in producing the resonances of $^6$Li in the $\alpha$~+~$d$ scattering.
It is also found that the rearrangement channels of $^5$He~+~$p$ and $^5$Li~+~$n$ have a sizable role in determining the resonance positions.

%%%%%%%%%%%%%%%%%%%%%%%%%%%%%%%%%%%%%%%%%%%%%%%%%%%%%%%%%%%%
\subsubsection{Radiative capture cross section for $^6$Li}

%%%%%%%%%%%%%%%%%%%%%%%%%%%%%%%
\begin{figure}[t]
\begin{minipage}{0.48\textwidth}
\centering\includegraphics[width=8.5cm]{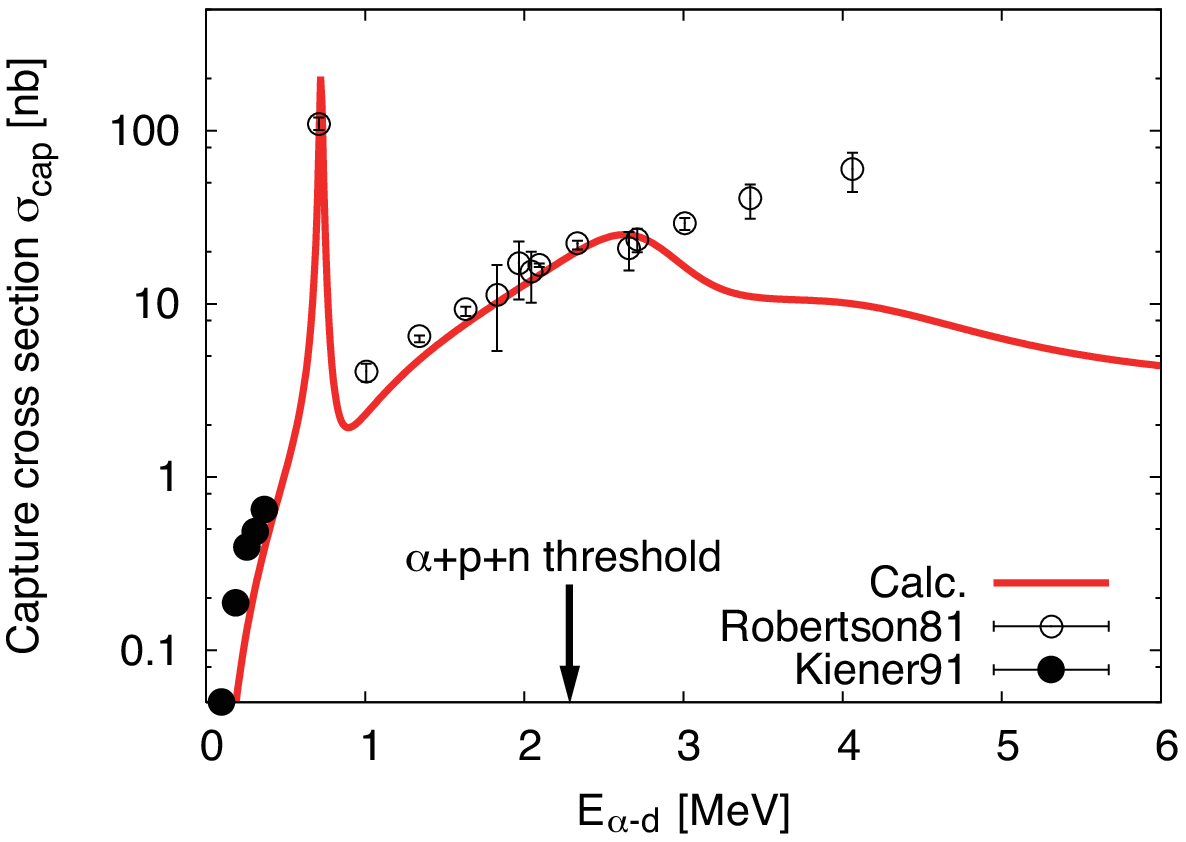}
\caption{\label{fig:pcs_6Li} (Color online)
Radiative capture cross section of $^2$H$(\alpha,\gamma)^6$Li in comparison with the observed data.
The red line shows the calculated cross section. The open and closed circles with error bars show the observed data taken from \cite{robertson81} and \cite{kiener91}, respectively. 
%The arrow indicates the $\alpha$~+~$p$~+~$n$ threshold energy.
}
\end{minipage}
\hfill
\begin{minipage}{0.48\textwidth}
\centering\includegraphics[width=8.5cm]{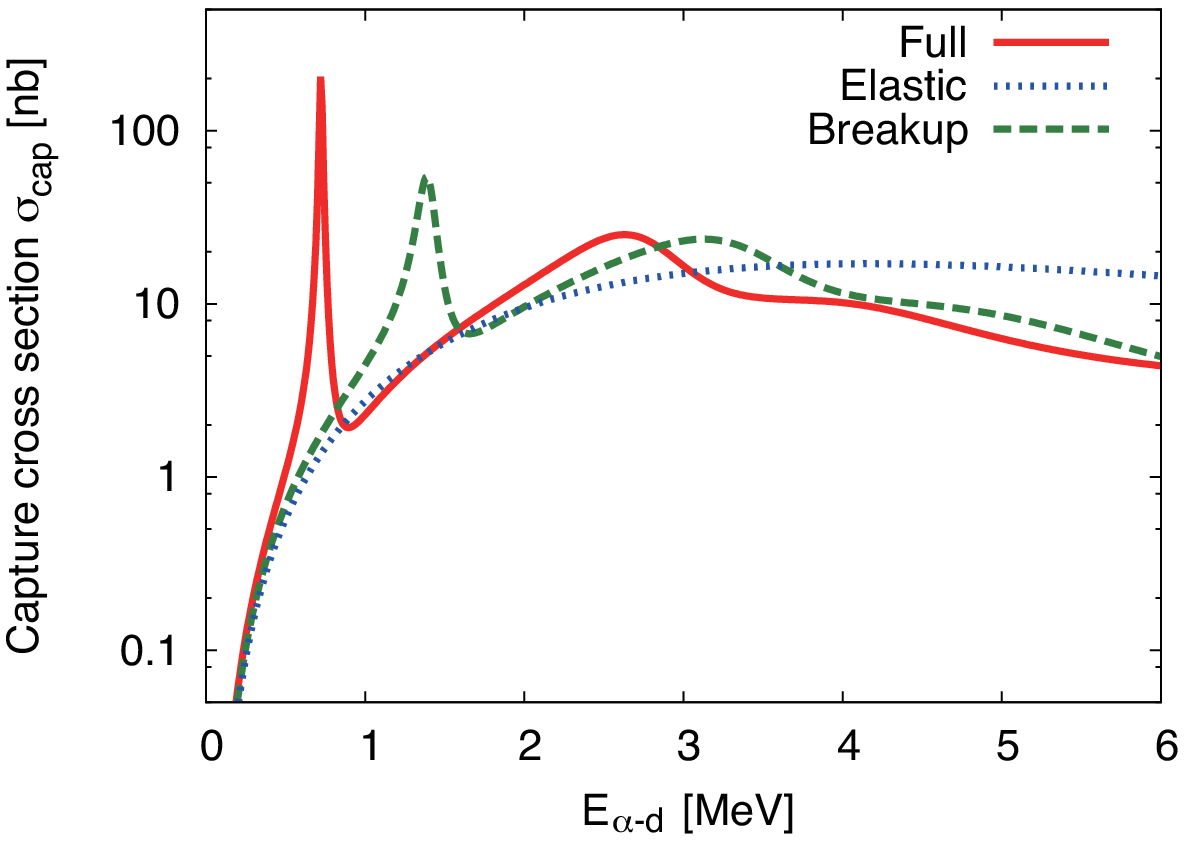}
\caption{\label{fig:pcs_comp} (Color online)
Effects of the deuteron breakup and the rearrangement on the radiative capture cross section. The red (solid) line is same as in Fig.~\ref{fig:pcs_6Li}. The blue (dotted) and green (dashed) lines present the cross sections for the cases of Elastic and Breakup, respectively.
}
\end{minipage}
\end{figure}
%%%%%%%%%%%%%%%%%%%%%%%%%%%%%%%

We investigate the effects of $\alpha$~+~$p$~+~$n$ three-body structures on the radiative capture reactions for $^6$Li.
The radiative capture cross section of $^2$H$(\alpha,\gamma)^6$Li, $\sigma_{\rm cap}$, is calculated by using the following relation:
\begin{equation}
\sigma_{\rm dis} (E) = \frac{2(2J_{\rm gs}+1)}{(2J_\alpha+1)(2J_d+1)}\frac{k_\gamma^2}{K^2}\sigma_{\rm cap} (E),
\label{eq:radcap}
\end{equation}
where $J_\alpha$ and $J_d$ represent the spin of the $\alpha$ particle and deuteron, respectively.
The term $k_\gamma$ is a wave number of the emitted photon, expressed as $k_\gamma = E_\gamma/\hbar c$.
The photodisintegration cross section is represented as $\sigma_{\rm dis}$ and calculated in the following equation.
\begin{equation}
\sigma_{\rm dis} (E) = \frac{4\pi^3}{75} \left(\frac{E_\gamma}{\hbar c}\right)^3 \int d\vc{K} \frac{d^3 B(E2)}{d\vc{K}} \delta \left( E - \frac{\hbar^2 K^2}{2M} + \varepsilon_d \right),
\label{eq:disint}
\end{equation}
where $M$ is the reduced mass corresponding to $\vc{K}$.
The photon energy is given as $E_\gamma = E+\varepsilon_d$, and $\varepsilon_d$ is the binding energy of the $^6$Li ground state measured from the $\alpha$~+~$p$~+~$n$ threshold.
In the present calculation of $\sigma_{\rm dis}$, we consider only the $E2$ transition, which is a dominant mode in the radiative capture of $^6$Li from the $\alpha$~+~$d$ system.
The $E2$ transition strength in Eq.~(\ref{eq:disint}) is calculated as
\begin{equation}
\frac{d^3B(E2)}{d\vc{K}} = \frac{1}{2J_0+1} \left|\bra \Psi^{(-)}_{\ell J^\pi} (\vc{K}) || O(E2) || \Psi_0 \ket\right|^2,
\label{eq:disE2}
\end{equation}
where $O(E2)$ is the $E2$ transition operator and $\Psi_0$ and $J_0$ are the wave function and total spin of the $^6$Li ground state, respectively.

We calculate the radiative capture cross section of $^2$H$(\alpha,\gamma)^6$Li, as shown in Fig.~\ref{fig:pcs_6Li}.
The results show totally reasonable agreement with the observed data in the energy region below the $\alpha$~+~$p$~+~$n$ threshold.
Above the energy of $E_{\alpha\mbox{-}d} = 3$ MeV, the cross section underestimates the observed three data pieces.
One of the possible approaches to overcome this underestimation is via transitions of a higher order, such as the $M1$ and multi-step transitions beyond the dominant $E2$ transition.
Another possibility is via transition with the $\alpha$~+~$p$~+~$n$ three-body components as intermediate states.

We discuss the effect of the $\alpha$~+~$p$~+~$n$ three-body structure on the radiative capture cross section.
We perform the same analysis for the cross sections as that performed for the phase shifts. The results are shown in Fig.~\ref{fig:pcs_comp}.
Similar to the phase shift case, the result without the two effects of the breakup and rearrangement (Elastic) show a structureless distribution.
With the inclusion of the deuteron breakup (Breakup), the radiative capture cross section shows the peaks and bumps corresponding to the resonance energies of $3^+_1$, $2^+_1$, and $1^+_2$ states; however, the resonance positions are slightly higher than the full result.
This difference is recovered by including the $^5$He~+~$p$ and $^5$Li~+~$n$ rearrangement channels.

\section{Summary and perspective}
 
The main aim of this review is to demonstrate the recent development of the complex scaling method (CSM) and its application to 
the descriptions of nuclear resonance and continuum states.
In unstable nuclei, most of the states are observed as unbound states due to the weak binding nature of the extra nucleons.
This situation demands a reliable theory for describing resonances satisfying the multi-particle decay condition.
The CSM is a powerful method for this purpose and can be easily applied to various nuclear models.
We have often applied the CSM to the nuclear cluster model,
because the relative motions between constituents are essential to generate resonances and can be solved accurately in the cluster model.

We have analyzed spectroscopy of the unstable nuclei on the basis of the cluster orbital shell model (COSM) consisting of core + nucleons with complex scaling, i.e. the CS-COSM.
Using the CS-COSM, we have presented the results of the many-body resonances appearing in light neutron-rich/proton-rich nuclei, including the predictions.
We have discussed the structure of resonances, such as single particle configurations and radius as well as energy eigenvalues.
This novel analysis of resonances is the advantage of the CSM because it enables to define the matrix elements of resonances like those of the bound states.
The further theoretical development to treat the many-body unbound states of nuclei on the basis of the shell model type approach would be desired.

We have also shown the reliability of the complex-scaled Green's function, which enables us to calculate the various strength functions
of electric responses, spectroscopic factors, and so on.
Green's function includes all kinds of information on not only resonances but also the non-resonant continuum states,
both of which contribute to the scattering observables.
Using Green's function, we can precisely extract the resonance contribution from the total strength.
This analysis clarifies the physical role of resonances in the observables.
We have also shown the application of the CSM to the calculation of the continuum level density, which is directly connected to the scattering solutions like the phase shifts.
We have discussed the role of resonance poles on the phase shifts in the two-body case.

The breakup reaction is an important phenomenon occurring in the unstable nuclei.
Two-neutron halo nuclei can be easily broken into three-body states owing to the small separation energy of extra neutrons in the ground states.
We have shown the Coulomb breakup strengths of two halo nuclei, $^6$He and $^{11}$Li, into core+$n$+$n$ final states.
For $^{11}$Li, we extended the theory to include the tensor and pairing correlations in the $^9$Li core.
These correlations are important to explain the halo formation in $^{11}$Li with a large amount of $s$-wave neutrons
and are treated in terms of the tensor-optimized shell model with the multi-configuration of the $^9$Li core.
As a result, the breakup strengths into the three-body scattering states are nicely described in the CSM.
This analysis enables us to examine the structure of halo nuclei in the many-body scattering states beyond the thresholds.

In the dynamics of three-body breakup reactions, it is useful to understand the correlations between constituents.
For this purpose, we have developed the method of complex-scaled solutions of the Lippmann-Schwinger equation.
By solving the three-body Lippmann-Schwinger equation with the CSM, we can obtain three-body scattering solutions.
We suitably extract the invariant mass spectra of any subsystems in the three-body breakup and examine the effect of binary resonances on the breakup process.
Recently, the continuum-discretized coupled channel method has been developed to use with the CSM, and the energy distribution of the $T$-matrix can now be obtained.
The CSM is expected to bring the unified description of the structures and reactions of nuclei.

Several theories have put forth the ab-initio description of resonances developed from the bare nucleon-nucleon interaction, such as few-body method \cite{carbonell14,horiuchi13} and the no-core shell model approach \cite{baroni2013}. 
Further extension to the many-body resonant and continuum states covering the wide range of mass numbers is the current task in this direction.

The CSM describes resonances with complex energy eigenvalues, the imaginary part of which represents the total decay width.
It is important to evaluate the partial decay widths of resonances for each decaying channel,
which provide useful information on decay properties of the resonances.
However, the theoretical foundation for describing the partial decay widths in the CSM has not yet been achieved.
It is desirable to develop a method for extracting the partial decay widths of the resonances in the CSM.
There is a theoretical development of obtaining the partial decay widths by using the continuity equation based on the time-dependent Schr\"odinger equation \cite{moiseyev11,goldzak10}.

In the theory of resonances, the matrix elements of the resonances are generally defined as complex values and easily obtained by using the CSM as well as eigenenergies.
The complex matrix elements of resonances exhibit useful information on the structure of resonance,  
but simultaneously result in the problem of the physical interpretation, which is not solved yet.

In addition to the resonances, the virtual states are a kind of unbound states, which often play an important role in nuclear structure.
The halo structure of $^{11}$Li is closely related to the presence of the virtual $s$-wave states in $^{10}$Li near the $^9$Li+$n$ threshold energy.
At present, it is not possible to obtain the virtual states directly in the CSM, different from resonance poles.
There are several methods of obtaining the virtual states such as the Jost function method \cite{masui00}, which, however, is limited to the two-body case.
It is desirable to develop a theoretical framework that can treat virtual states in the many-body case.

\section*{Acknowledgements}
The authors would like to acknowledge the collaborations with K. Ikeda, H. Toki, S. Aoyama, R. Suzuki, C. Kurokawa, B. G. Giraud, M. Odsuren
for the development of the complex scaling method in the application to the nuclear physics.

%%%%%%%%%%%%%%%%%%%%%%%%%%%%%%%%%%%%%%%%%%%%%%%%%%%%%%%%%%%%
\def\JL#1#2#3#4{ {{\rm #1}} \textbf{#2} (#3) #4}  % PPNP
\nc{\PR}[3]     {\JL{Phys. Rev.}{#1}{#2}{#3}}
\nc{\PRC}[3]    {\JL{Phys. Rev.~C}{#1}{#2}{#3}}
\nc{\PRA}[3]    {\JL{Phys. Rev.~A}{#1}{#2}{#3}}
\nc{\PRL}[3]    {\JL{Phys. Rev. Lett.}{#1}{#2}{#3}}
\nc{\NP}[3]     {\JL{Nucl. Phys.}{#1}{#2}{#3}}
\nc{\NPA}[3]    {\JL{Nucl. Phys.}{A#1}{#2}{#3}}
\nc{\PL}[3]     {\JL{Phys. Lett.}{#1}{#2}{#3}}
\nc{\PLB}[3]    {\JL{Phys. Lett.~B}{#1}{#2}{#3}}
\nc{\PTP}[3]    {\JL{Prog. Theor. Phys.}{#1}{#2}{#3}}
\nc{\PTPS}[3]   {\JL{Prog. Theor. Phys. Suppl.}{#1}{#2}{#3}}
\nc{\PRep}[3]   {\JL{Phys. Rep.}{#1}{#2}{#3}}
\nc{\JP}[3]     {\JL{J. of Phys.}{#1}{#2}{#3}}
\nc{\CMP}[3]    {\JL{Commun. Math. Phys.}{#1}{#2}{#3}}
\nc{\SPJ}[3]    {\JL{Sov.\ Phys.\ JETP}{#1}{#2}{#3}} 
\nc{\ZP}[3]     {\JL{Z.\ Phys.}{#1}{#2}{#3}}
\nc{\JMP}[3]    {\JL{J. Math. Phys.}{#1}{#2}{#3}}
\nc{\PPNP}[3]   {\JL{Prog. Part. Nucl. Phys.}{#1}{#2}{#3}}
\nc{\ANP}[3]    {\JL{Ann.\ of\ Phys.}{#1}{#2}{#3}}
\nc{\PTEP}[3]   {\JL{Prog. Theor. Exp. Phys.}{#1}{#2}{#3}}
\nc{\JPC}[3]     {\JL{J. Chem. Phys.}{#1}{#2}{#3}}
\nc{\andvol}[3] {{\it ibid.}\JL{}{#1}{#2}{#3}}
%%%%%%%%%%%%%%%%%%%%%%%%%%%%%%%%%%%%%%%%%%%%%%%%%%%%%%%%%%%%%


\begin{thebibliography}{99}
\itemsep -2pt 
% sec1 myo
\bibitem{tanihata85} I.~Tanihata~{\em et al.},~\PRL{55}{1985}{2676}.
\bibitem{tanihata13} I. Tanihata, H. Savajols, R. Kanungo, \PPNP{68}{2013}{215}.
\bibitem{tanihata96}  I. Tanihata,~\JP{G22}{1996}{157}.
\bibitem{lane1958} A.M. Lane, R. G. Thomas, \JL{Rev. Mod. Phys.}{30}{1958}{257}.
\bibitem{kapur1938} P. L. Kapur, R. K. Peierls, \JL{Proc. Roy. Soc. (London)}{A166}{1938}{277}.
\bibitem{wigner47} E.P. Wigner, L. Eisenbud, \PR{72}{1947}{29}.
\bibitem{ho83}       Y. K. Ho, \PRep{99}{1983}{1}.
\bibitem{moiseyev98} N. Moiseyev, \PRep{302}{1998}{211}.
\bibitem{moiseyev11} N. Moiseyev, {\it Non-Hermitian Quantum Mechanics} (Cambridge University Press, 2011) 
\bibitem{aguilar71}  J. Aguilar, J.~M. Combes, \CMP{22}{1971}{269}.
\bibitem{balslev71}  E. Balslev, J.~M. Combes, \CMP{22}{1971}{280}.
\bibitem{simon72}    B. Simon, \CMP{27}{1972}{1}. 
\bibitem{siegert39}  A.~J. Siegert, \PR{56}{1939}{750}. 
\bibitem{klaiman10} S. Klaiman, Moiseyev, \JL{J. Phys. B: At. Mo. Opt. Phys}{43}{2010}{185205}.
\bibitem{klaiman11} S. Klaiman, N. Hatano, \JPC{134}{2011}{154111}.
\bibitem{hatano14} N. Hatano, G. Ordonez, arXiv:1405.6683.
\bibitem{goldzak12} T. Goldzak, I. Gilary, N. Moiseyev, \JL{Mol. Phys.}{110}{2012}{537} and references therein.
\bibitem{pieper01} S.C. Pieper and R. B. Wiringa, \JL{Annu. Rev. Nucl. Part. Sci.}{51}{2001}{53}.

\bibitem{kikuchi10}  Y. Kikuchi, K. Kat\=o, T. Myo, M. Takashina, K. Ikeda, \PRC{81}{2010}{044308}. 
\bibitem{kikuchi11}  Y. Kikuchi, N. Kurihara, A. Wano, K. Kat\=o, T. Myo, M. Takashina, \PRC{84}{2011}{064610}.


\bibitem{gyarmati71} B. Gyarmati, T. Vertse, \NPA{160}{1971}{523}.
\bibitem{gyarmati72} B. Gyarmati, F. Krisztinkovics, T. Vertse, \PLB{41}{1972}{110}.
\bibitem{zeldovich61}Ya.~B. Zel'dovich, \SPJ{12}{542}{1961}.
\bibitem{romo68}     W. J. Romo, \NPA{116}{1968}{617}.
\bibitem{homma97}    M. Homma, T. Myo, K. Kat\=o, \PTP{97}{1997}{561}.
\bibitem{kruppa88} A. T. Kruppa, R. G. Lovas, B. Gyarmati, \PRC{37}{1988}{383}.
\bibitem{csoto94} A. Cs\'ot\'o, \PRC{49}{1994}{3035}. 
\bibitem{humblet61}  J. Humblet, L.~R. Rosenfeld, \NP{26}{1961}{529}.
\bibitem{berggren70} T. Berggren, \PLB{33}{1970}{547}.
\bibitem{berggren96} T. Berggren, \PLB{373}{1996}{1}.
\bibitem{berggren68} T. Berggren, \NPA{109}{1968}{265}.
\bibitem{berggren92} T. Berggren, P. Lind, \PRC{47}{1993}{768}.
\bibitem{betan02}  
R. Id Betan, R.~J. Liotta, N. Sandulescu, T. Vertse, \PRL{89}{2002}{042501};\\
R. Id Betan, R. J. Liotta, N. Sandulescu, T. Vertse, \PRC{67}{2003}{014322};\\
R. Id Betan, R. J. Liotta, N. Sandulescu, T. Vertse, \PLB{584}{2004}{48}.
\bibitem{michel02}
  N. Michel, W. Nazarewicz, M. P{\l}oszajczak, K. Bennaceur,
  Phys. Rev. Lett. {\bf 89} (2002) 042502;\\
  R. Id Betan, R. J. Liotta, N. Sandulescu and T. Vertse,
  Phys. Rev. Lett. {\bf 89} (2002) 042501;\\
  N. Michel, W. Nazarewicz, M. P{\l}oszajczak,  J. Oko{\l}owicz,
  Phys. Rev. C {\bf 67} (2003) 054311;\\
  N. Michel, W. Nazarewicz, M. P{\l}oszajczak,
  Phys. Rev. C {\bf 70} (2004) 064313.
\bibitem{okolowicz03} J. Oko{\l}owicz, M. P{\l}oszajczak, I. Rotter, Phys. Rep. {\bf 374} (2003) 271.
\bibitem{hagen05} G. Hagen, J. S. Vaagen, M. Hjorth-Jensen, J. of Phys. A {\bf 37} (2004) 8991;\\
G. Hagen, M. Hjorth-Jensen, J. S. Vaagen, \PRC{71}{2005}{044314}.
\bibitem{volya06} A.~Volya, V.~ Zelevinsky, Phys. Rev. C {\bf 74} (2006) 064314.
\bibitem{hagen07} G. Hagen, D. J. Dean, M. Hjorth-Jensen T. Papenbrock, Phys. Lett. {\bf B656} (2007) 169.
\bibitem{rotureau06}
  J. Rotureau, N. Michel, W. Nazarewicz, M. Ploszajczak, J. Dukelsky, Phys.Rev.Lett. {\bf 97} (2006) 110603.
\bibitem{jaganathen12}
  Y. Jaganathen, N. Michel, M. P{\l}oszajczak, Journal of Physics: Conference Series {\bf 403} (2012) 012022.
\bibitem{kurokawa05} C.~Kurokawa, K.~Kat\=o, Phys. Rev. {\bf C71} (2005) 021301; Nucl. Phys. {\bf A792} (2007) 87.
\bibitem{horiuchi12} H. Horiuchi, K. Ikeda, K. Kat\=o, \PTPS{192}{2012}{1}. % review
\bibitem{carbonell14} J. Carbonell, A. Deltuva, A.C. Fonseca, R. Lazauskas, \PPNP{74}{2014}{55}. %
\bibitem{charity10}  R. J. Charity {\em et al.}, \PRC{82}{2010}{041304}.
\bibitem{charity11}  R. J. Charity {\em et al.}, \PRC{84}{2011}{014320}.
\bibitem{myo11b}     T. Myo, Y. Kikuchi, K. Kat\=o, \PRC{84}{2011}{064306}. % 7B
\bibitem{myo12b}    T. Myo, Y. Kikuchi, K. Kat\=o, \PRC{85}{2012}{034338}. % C8
\bibitem{suzuki88}   Y. Suzuki, K. Ikeda, \PRC{38}{1988}{410}.
\bibitem{aoyama06}   S. Aoyama, T. Myo, K. Kat\=o, K. Ikeda, \PTP{116}{2006}{1}.
\bibitem{masui06} H. Masui, K. Kat\=o, K. Ikeda, \PRC{73}{2006}{034318}.
\bibitem{masui14}
  H. Masui, K. Kat\=o, N. Michel, M. P{\l}oszajczak, Phys. Rev. C {\bf 89} (2014) 044317.
\bibitem{masui12} H. Masui, K. Kat\=o, K. Ikeda \NPA{895}{2012}{1}.
\bibitem{myo98} T. Myo, A. Ohnishi, K. Kat\=o, \PTP{99}{1998}{801}.
\bibitem{matsumoto10} T. Matsumoto, K. Kat\=o, M. Yahiro, \PRC{82}{2010}{051602}.
\bibitem{ogata13}    K. Ogata, T. Myo, T. Furumoto, T. Matsumoto, M. Yahiro, \PRC{88}{2013}{024616}.
\bibitem{kruppa98} A. T. Kruppa, Phys. Lett. B {\bf 431} (1998), 384.
\bibitem{kruppa99} A. T. Kruppa, K. Arai, Phys. Rev. A {\bf 59} (1999), 3556.
\bibitem{suzuki05}   R.~Suzuki, T.~Myo, K. Kat\=o, \PTP{113}{2005}{1273}.
\bibitem{myo01}      T. Myo, K. Kat\=o, S. Aoyama, K. Ikeda, \PRC{63}{2001}{054313}.
\bibitem{kruppa07}   A.~T. Kruppa,~R. Suzuki, K. Kat\=o, \PRC{75}{2007}{044602}.
\bibitem{kikuchi09} Y Kikuchi, T. Myo, M. Takashina, K. Kat\=o, K. Ikeda, \PTP{122}{2009}{499}.
\bibitem{kikuchi13a} Y. Kikuchi, T. Myo, K. Kat\=o, K. Ikeda, \PRC{87}{2013}{034606}. 

% sec2 myo
\bibitem{hiyama03} E. Hiyama, Y. Kino, M. Kamimura, \PPNP{51}{2003}{223}.
\bibitem{ohtsubo13} S. Ohtsubo, Y. Fukushima, M. Kamimura, E. Hiyama, Prog. Theor. Exp. Phys. {\bf 2013} (2013) 073D02. 
\bibitem{ho12} Y. K. Ho, S. Kar, Few-Body Syst {\bf 53} (2012) 437.
\bibitem{garrido03} E. Garrido, D.V. Fedorovb, A.S. Jensen, \NPA{722}{2003}{221}.
\bibitem{myo09b}   T. Myo, R. Ando, K. Kat\=o, \PRC{80}{2009}{014315}. % 7He
\bibitem{newton60} R.~G. Newton, \JMP{1}{1960}{319}.
\bibitem{giraud03} B. G. Giraud, K. Kat\=o, \ANP{308}{2003}{115}.
\bibitem{giraud04} B. G. Giraud, K. Kat\=o, A. Ohnishi, \JP{A37}{2004}{11575}.
\bibitem{csoto90} A. Cs\'ot\'o, B. Gyarmati, A.~T. Kruppa, K.F. P\'al, N. Moiseyev, \PRA{41}{1990}{3469}. 
                   
% sec3 myo
\bibitem{myo07b}    T. Myo, K. Kat\=o, H. Toki, K. Ikeda, \PRC{76}{2007}{024305}. % 11Li
\bibitem{myo07a}     T. Myo, K. Kat\=o, K. Ikeda,~\PRC{76}{2007}{054309}. % 7He
\bibitem{myo10}      T. Myo, R. Ando, K. Kat\=o, \PLB{691}{2010}{150}.
\bibitem{myo09a}      T. Myo, H. Toki, K. Ikeda, \PTP{121}{2009}{511}. % TOSM
\bibitem{myo11a}      T. Myo, A. Umeya, H. Toki, K. Ikeda, \PRC{84}{2011}{034315}. % He-TOSM
\bibitem{kanada79}   H.~Kanada,~T.~Kaneko,~S.~Nagata, and~M.~Nomoto,~\PTP{61}{1979}{1327}.
\bibitem{tang78}     Y. C. Tang, M. LeMere, D. R. Thompson, \JL{Phys. Rep.}{47}{1978}{167}.
\bibitem{mougeot12}	X. Mougeot {\em et al.}, \PLB{718}{2012}{441}.
\bibitem{tanihata92} I.~Tanihata~{\em et al.},~\PL{B289}{1992}{261}.
\bibitem{alkazov97}  G. D. Alkhazov {\em et al.}, \PRL{78}{1997}{2313}.
\bibitem{kiselev05}  O. A. Kiselev {\em et al.}, Eur. Phys. J. A{\bf 25}, Suppl. 1 (2005) 215.
\bibitem{mueller07}  P. Mueller {\em et al.}, \PRL{99}{2007}{252501}.
\bibitem{skaza06}    F. Skaza {\em et al.},~\PRC{73}{2006}{044301}, and references therein.
\bibitem{beck07}  F. Beck {\em et al.},~\PLB{645}{2007}{128}.
\bibitem{myo03}      T. Myo, S. Aoyama, K. Kat\=o, K. Ikeda, \PLB{576}{2003}{281}.
\bibitem{korsheninnikov03} A.A. Korsheninnikov {\em et al.}, \PRL{90}{2003}{082501}.
\bibitem{adahchour06}A. Adahchour, P. Descouvemont, \PLB{639}{2006}{447}.
\bibitem{chulkov05}   C. V. Chulkov {\em et al.}, \NPA{759}{2005}{43}.
\bibitem{keeley07}   N. Keeley {\em et al.}, \PLB{646}{2007}{222}.
\bibitem{enyo07}     Y. Kanada-En'yo, \PRC{76}{2007}{044323}.
\bibitem{yamada08}   T. Yamada, Y. Funaki, H. Horiuchi, K. Ikeda, A. Tohsaki, \PTP{120}{2008}{1139}.
\bibitem{iwata00}    Y. Iwata {\em et al.}, \PRC{62}{2000}{064311}.
\bibitem{myo08} T. Myo, Y. Kikuchi, K. Kat\=o, H. Toki, K. Ikeda, \PTP{119}{2008}{561}.  % 11Li
\bibitem{saito77}    S. Saito, \PTPS{62}{1977}{11}.
\bibitem{myo05}      T. Myo, K. Kat\=o, K. Ikeda, \PTP{113}{2005}{763}.
\bibitem{myo12a}    T. Myo, A. Umeya, H. Toki, K. Ikeda, \PRC{86}{2012}{024318}. % Li-TOSM
\bibitem{sugimoto04}  S.~Sugimoto,~K. Ikeda and~H.~Toki, \NPA{740}{2004}{77}. 
\bibitem{ogawa06}  Y. Ogawa, H. Toki, S. Tamenaga, S. Sugimoto, K. Ikeda, \PRC{73}{2006}{034301}.
\bibitem{furutani80}  H. Furutani, H. Kanada, T. Kaneko, Si. Nagata, H. Nishioka, S. Okabe, S. Saito, T. Sakuda, M. Seya, \PTPS{68}{1980}{193}.
\bibitem{hasegawa71}  A. Hasegawa, S. Nagata, \PTP{45}{1971}{1786}.
\bibitem{akaishi04}  Y. Akaishi,~\NPA{738}{2004}{80}.
\bibitem{ikeda04}  K. Ikeda, S. Sugimoto, H. Toki, \NPA{738}{2004}{73}.
\bibitem{tanihata88}	I.~Tanihata~{\em et al.},~\PL{B206}{1988}{592}.
\bibitem{kato99}    K,~Kat\=o,~T.~Yamada, K.~Ikeda, \PTP{101}{1999}{119}.
\bibitem{teramond87} G.F. de T\'eramond, B. Gabioud, \PRC{36}{1987}{691}.
\bibitem{simon99}  H. Simon {\em et al.}, \PRL{83}{1999}{496}.
\bibitem{tostevin97}	J. A.~Tostevin, J. S. Al-Khalili, \NP{A616}{1997}{418c}.
\bibitem{dobrovolsy06}  A.~V.~Dobrovolsky {\em et al.},~\NP{A766}{2006}{1}.
\bibitem{sanchez06}  R. S\'anchez {\em et al.}, \PRL{96}{2006}{033002}.
\bibitem{puchalski06}  M. Puchalski, A. M. Moro, K. Pachucki, \PRL{97}{2006}{133001}.
\bibitem{esbensen92}	H.~Esbensen, G.~F.~Bertsch,~\NP{A542}{1992}{310}.
\bibitem{hagino05}  K.~Hagino, H. Sagawa, \PRC{72}{2005}{044321}.
\bibitem{matsuo05}  M.~Matsuo, K. Mizuyama, Y Serizawa, \PRC{71}{2005}{064326}.
\bibitem{zhukov93}  M. V. Zhukov, B. V. Danilin, D. V. Fedorov, J. M. Bang, I. J. Thompson, J. S. Vaagen, \PRep{231}{1993}{151}.
\bibitem{nielsen01}  E. Nielsen, D.V. Fedorov, A.S. Jensen, E. Garrido, \PRep{347}{2001}{373}.

% sec4 masui
\bibitem{volya09}
  A. Volya,
  Phys. Rev. C {\bf 79} (2009) 044308.
  
\bibitem{betan14}
  R.M. Id Betan,
  Phys. Lett. {\bf B730} (2014) 18.

\bibitem{masui07}
  H. Masui, K. Kat\=o, K. Ikeda, \PRC{75}{2007}{034316}.
  
\bibitem{yamamoto09} K. Yamamoto, H. Masui, K. Kat\=o, T. Wada, M. Ohta, \PTP{121}{2009}{375}.

\bibitem{gaudefroy12}
L. Gaudefroy {\em et al.}, Phys. Rev. Lett. {\bf 109} (2012) 202503.

\bibitem{tanaka10}  
K. Tanaka {\em et al.}, Phys. Rev. Lett.{\bf  104} (2010) 062701.

\bibitem{kobayashi12}
N. Kobayashi {\em et al.}, Phys. Rev. C {\bf 86}  (2012) 054604.

\bibitem{nakamura09}
T. Nakamura {\em et al.}, Phys. Rev. Lett. {\bf 103} (2009) 262501.
  
\bibitem{takechi12}  
M. Takechi {\em et al.}, Phys. Lett. B {\bf 707}  (2012) 357.

\bibitem{ozawa01b} A. Ozawa, T. Suzuki, I. Tanihata, Nucl. Phys. {\bf A693}  (2001) 32.
\bibitem{kanungo11} Kanungo {\em et al.},  Phys. Rev. C {\bf 84} (2011) 061304.  
\bibitem{nakada06}
  H. Nakada,
  Nucl. Phys. {\bf A764} (2006) 117.
  
\bibitem{abuibrahim09}
  B. Abu-Ibrahim, S. Iwasaki, W. Horiuchi, A. Kohama, Y. Suzuki,
  Jour. Phys. Soc. Japan {\bf 78}  (2009) 044201.

\bibitem{hagen09}
  G. Hagen, T. Papenbrock, D. J. Dean, M. Hjorth-Jensen, B. Velamur Asokan,
  Phys. Rev. C {\bf 80}, (2009) 021306.

\bibitem{kaneko91}  T. Kaneko, M. LeMere, Y. C. Tang, Phys. Rev. C {\bf 44} (1991) 1588.
\bibitem{ajzenberg86} F. Ajzenberg-Selove, Nucl. Phys. {\bf A460} (1986) 1.
\bibitem{ajzenberg87} F. Ajzenberg-Selove, Nucl. Phys. {\bf A475} (1987) 1.

\bibitem{fano61}
  U. Fano, \PR{124}{1961}{1866}.
  
\bibitem{mahaux69}
  C. Mahaux, H. Weidenm\"uller, {\it Shell Model Approaches to Nuclear Reactions}, North-Holand, Amsteldom, 1969.
  
\bibitem{bennaceur00}
  K. Bennaceur, F. Nowacki,  J. Oko{\l}owicz, M. P{\l}oszajczak, 
  Nucl. Phys. {\bf A671} (2000) 203.
  
\bibitem{gelfand61} I.M. Gel'fand and N.Ya. Vilenkin, 
{\em Generalized Functions}, Vol. 4, Academic Press, New York (1961);
K. Maurin, {\em Generalized Eigenfunction Expansions
  and Unitary Representations of Topological Groups}, 
Polish Scientific Publishers, Warsaw (1968);
A. Bohm, {\em The Rigged Hilbert Space and Quantum Mechanics},
 Lecture Notes in Physics 78, Springer, New York  (1978).

\bibitem{michel08}
N. Michel, W. Nazarewicz, M. P{\l}oszajczak, T. Vertse, J. Phys. G: Nucl. Part. Phys. {\bf 36}  (2008) 013101.

\bibitem{volkov65}
   A. B. Volkov,
  Nucl. Phys. {\bf 74} (1965) 33.

\bibitem{aoyama95} S.~Aoyama, S.~Mukai, K.~Kat\=o, K.~Ikeda, Prog.\ Theor.\  Phys.\ {\bf 93} (1995) 99.
\bibitem{kruppa14}
  A.T. Kruppa, G. Papadimitriou, W. Nazarewicz, N. Michel,
 Phys. Rev. C {\bf 89} (2014) 014330.

\bibitem{tilley93} D. R. Tilley, H. R. Weller, C. M. Cheves, Nucl. Phys. A {\bf 565} (1993) 1.
\bibitem{johnson67} C. H. Johnson, J. L. Fowler, Phys. Rev. {\bf 162} (1967) 890. 
\bibitem{blue65}  R. A. Blue, W. Haeberli,  Phys. Rev. {\bf 137}  (1965) B284.


\bibitem{dufour05} M. Dufour, P. Descouvemont, Phys. Rev. C {\bf 72} (2005) 015801.
\bibitem{igashira95} M. Igashira, Y. Nagai, K. Masuda, T. Ohsaki, H. Kitazawa, Astron. J. {\bf 441} (1995) L89. 
\bibitem{baye98} D. Baye, P. Descouvemont, M. Hesse,  Phys. Rev. C {\bf 58} (1998) 545.

\bibitem{morlock97}
R. Morlock, R. Kunz, A. Mayer, M. Jaeger, A. M\"uller, J. W. Hammer,
P. Mohr, H. Oberhummer, G. Staudt, V. K\"olle,
Phys. Rev. Lett. {\bf 79} (1997) 3837.  


 
% sec5 kato
\bibitem{levine69}  R. D. Levine, {\it Quantum Mechanics of Molecular Rate Processes} (Clarendon Press, Oxford, 1969), p. 101.
\bibitem{tsang75} T. Y. Tsang, T. A. Osborn, Nucl. Phys. {\bf A 247} (1975) 43.
\bibitem{osborn76} T. A. Osborn, T. Y. Tsang, Ann. of Phys. {\bf 101} (1976) 119.
\bibitem{arai99} K. Arai, A. T. Kruppa, \PRC{60}{1999}{064315}.
\bibitem{strutinsky67} V. M. Strutinsky, Nucl. Phys. {\bf A 95} (1967) 420.
\bibitem{shlomo92} S. Shlomo, Nucl. Phys. A {\bf 539} (1992) 17.
\bibitem{taylor72} J. R. Taylor, {\it Sacttering Theory: Tne Quantum Theory of Nonrelativistic Collisions} (Dover Pub., INC, New York, 1972), p.249. 
\bibitem{suzuki08} R. Suzuki, A. T. Kruppa, B. G. Giraud, K. Kat\=o, Prog. Theor. Phys. {\bf 119} (2008) 949.
\bibitem{odsuren14} M. Odsuren, K. Kat\=o, M. Aikawa, T. Myo \PRC{89}{2014}{034322}.
\bibitem{hoop66} B. Hoop Jr, H. H. Barschall, Nucl. Phys. {\bf 83}, 65 (1966); Th. Stammbach, R. L. Walter, Nucl. Phys. {\bf A180}, 225 (1972).
\bibitem{schmid61} E.~W.~Schmid, K.~Wildermuth, Nucl.\ Phys.\ {\bf 26} (1961), 463.
\bibitem{kukulin86} V. I. Kukulin, V. M. Krasnopol'sky, V. T. Voronchev, P. B. Sazonov, \NPA{417}{1984}{128}.
\bibitem{kato10} K. Kat\=o, C. Kurokawa, K. Arai, Tours Symposium on Nuclear Phyics and Astrophysics VII, Kobe, Japan, 16-20 November 2009, AIP Conference Proceedings {\bf 1238} (2010) 181.
\bibitem{itoh11} M. Itoh {\em et al.}, \PRC{84}{2011}{054308}.

% sec6 kikuchi
\bibitem{efros07} V D Efros, W Leidemann, G Orlandini, N Barnea, \JP{G34}{2007}{R459}. % LIT
\bibitem{aumann99} T. Aumann {\em et al.}, \PRC{59}{1999}{1252}.
\bibitem{wang02} J.~Wang {\em et al.}, \PRC{65}{2002}{034306}.
\bibitem{ieki93} K.~Ieki {\em et al.}, \PRL{70}{1993}{730}.
\bibitem{shimoura95} S.~Shimoura {\em et al.}, \PLB{348}{1995}{29}.
\bibitem{zinser97} M.~Zinser {\em et al.}, \NPA{619}{1997}{151}.
\bibitem{nakamura06} T.~Nakamura {\em et al.}, \PRL{96}{2006}{252502}.
\bibitem{bertulani88} C. A. Bertulani, G. Baur, Phys. Rep. {\bf 163} (1988) 299.
\bibitem{hagino09} K. Hagino, H. Sagawa, T. Nakamura, S. Shimoura, \PRC{80}{2009}{031301}.
\bibitem{bachelet08} C.~Bachelet {\em et al.}, \PRL{100}{2008}{182501}.
\bibitem{kamimura86} M. Kamimura, M. Yahiro, Y. Iseri, Y. Sakuragi, H. Kameyama, M. Kawai, \PTPS{89}{1986}{1}.
\bibitem{austern87} N. Austern, Y. Iseri, M. Kamimura, M. Kawai, G. Rawitscher, M. Yahiro \PRep{154}{1987}{125}.
\bibitem{yahiro12} M. Yahiro, K. Ogata, T. Matsumoto, K. Minomo, \PTEP{2012}{2012}{01A206}.
\bibitem{kikuchi13b} Y. Kikuchi, T. Matsumoto, K. Minomo, K. Ogata, \PRC{88}{2013}{021602}.
\bibitem{bochkarev89} O.~V.~Bochkarev {\em et al.}, \NPA{505}{1989}{215}.
\bibitem{egorova12} I.~A.~Egorova {\em et al.}, \PRL{109}{2012}{202502}.
\bibitem{schmelzbach72} P. A. Schmelzbach, W. Gr{\"u}ebler, V. K{\"o}nig, P. Marmier, \NPA{184}{1972}{193}.
\bibitem{gruebler75} W. Gr{\"u}ebler, P. A. Schmelzbach, V. K{\"o}nig, R. Risler, D. Boerma, \NPA{242}{1975}{265}.
\bibitem{robertson81} R. G. H. Robertson {\em et al.}, \PRL{47}{1981}{1867}.
\bibitem{kiener91} J. Kiener {\em et al.}, \PRC{44}{1991}{2195}. 
% sec7
\bibitem{horiuchi13} W. Horiuchi, Y. Suzuki \PRC{87}{2013}{034001}.
\bibitem{baroni2013} S. Baroni, P. Navr\'atil, S. Quaglioni, \PRL{110}{2013}{022505}.
\bibitem{masui00} H. Masui, S. Aoyama, T. Myo, K. Kat\=o, K. Ikeda, \NPA{673}{2000}{207}.
% moiseyev
\bibitem{goldzak10} T. Goldzak, I. Gilary, N. Moiseyev, \PRA{82}{2010}{052105}.
\end{thebibliography}
\end{document}